\documentstyle[12pt]{article}
\newcommand{\ret}{\nonumber \\}

\newcommand{\Section}[1]%
{\section{#1}\setcounter{equation}{0}%
\setcounter{theorem}{0}}
\newtheorem{theorem}{Theorem}
\newtheorem{lemma}[theorem]{Lemma}

\newtheorem{pro}[theorem]{Proposition}
\newtheorem{definition}[theorem]{Definition}

\newenvironment{proof}[1]%
{\par\noindent{\em #1:\ }}%
{~\rule{2mm}{2mm}\par\bigskip}
\begin{document}
%%%%%%%%%%%%%%%%%%%%%%%%%%%%%%%%%%%%%%%%%%%
\newpage\thispagestyle{empty}
{\topskip 2cm
\begin{center}
{\Large\bf Widths of the Hall Conductance Plateaus\\} 
\bigskip\bigskip\bigskip
{\Large Tohru Koma\\}
\bigskip\bigskip
{\small \it Department of Physics, Gakushuin University, 
Mejiro, Toshima-ku, Tokyo 171-8588, JAPAN}\\
\bigskip
{\small\tt e-mail: tohru.koma@gakushuin.ac.jp}
\end{center}
%This is also for double spacing
%\newpage
\vfil
\noindent
We study the charge transport of the noninteracting electron gas 
in a two-dimensional quantum Hall system with Anderson-type impurities 
at zero temperature. We prove that there exist localized states 
of the bulk order in the disordered-broadened Landau bands 
whose energies are smaller than a certain value determined 
by the strength of the uniform magnetic field. 
We also prove that, when the Fermi level lies in the localization regime, 
the Hall conductance is quantized to the desired integer 
and shows the plateau of the bulk order 
for varying the filling factor of the electrons rather than the Fermi level.  
\par
%%%%%%%%%%%%%%%%%%%%%%%%%%%%%%%%%%%%%%%%%%%%%%%%%%%%%%%%%%%%%%%%%%%%%%%%%%%
\noindent
\bigskip
\hrule
\bigskip
\noindent
{\bf KEY WORDS:} Quantum Hall effect; Landau Hamiltonian; strong magnetic field; 
Anderson localization, Hall conductance plateaus.\hfill
%%%%%%%%%%%%%%%%%%%%%%%%%%%%%%%%%%%%%%%%%%%%%%%%%%%%%%
\vfil}\newpage
%%%%%%%%%%%%%%%%%%%%%%%%%%%%%%%%%%%%%%%%%%%
%%%%%%%%%%%%%%%%%%%%%%%%%%%%%%%%%%%%%%%%%%%
\tableofcontents
\newpage
%%%%%%%%%%%%%%%%%%%%%%%%%%%%%%%%%%%%%%%%%%%
\Section{Introduction}

The two most remarkable facts of the integral quantum Hall effect \cite{PGBook} 
are the integrality of the Hall conductance and its robustness for varying 
the parameters such as the filling factor of the electrons and the strength of the disorder.  
The integrality is explained by the topological nature \cite{TKNN,Kohmoto} 
of the Hall conductance. The constancy of the Hall conductance is due to 
the Anderson localization of the wavefunctions of the electrons \cite{AA}.   

First of all we shall survey recent mathematical analysis of the quantum Hall effect. 
As for justification of the conductance formula leading to the topological 
invariant, satisfactory results have been obtained in the recent papers 
within the linear response approximation 
or an adiabatic limit of slowly applying an electric field \cite{ASY,Koma2,ES,Koma3,BGKS}. 
Avron, Seiler and Yaffe \cite{ASY} proved that a flux averaged charge 
transport\footnote{This is a non-trivial charge transport which is 
intrinsically different from the response to a static external field.} is 
quantized to an integer in the adiabatic limit under the assumption of 
a nonvanishing spectral gap above a non-degenerate ground state 
for a finite-volume interacting electron gas. 
In \cite{Koma2}, a static electric field with a regularized boundary condition 
was used as an external force  
to derive an electric current for a finite-volume interacting electron gas 
under the assumption of a nonvanishing spectral gap above 
the sector of the ground state(s). The resulting Hall conductance 
is equal to the universal conductance multiplied by the filling factor of 
the electrons in the infinite-volume limit. When the Fermi level lies in a spectral gap 
for a noninteracting electron gas on the whole plane ${\bf R}^2$, 
Elgart and Schlein \cite{ES} justified the Hall conductance formula 
which is written in terms of switch functions in the adiabatic limit. 
This formula was first introduced by Avron, Seiler and Simon \cite{ASS}.  
Without relying on the gap assumption, a general conductance formula was 
obtained for finite-volume interacting electron gases \cite{Koma3}. 
For the whole plane, Bouclet, Germinet, Klein and Schenker \cite{BGKS} obtained 
a Hall conductance formula for a random noninteracting electron gas 
with translation ergodicity under the assumption that the Fermi level falls 
into a localization regime. 

As to the localization and the related conductance plateaus, 
we refer only to a class of noninteracting electron gases because 
the localization of interacting electrons is still an unsolved problem. 
The existence of the localization at the edges of the disordered-broadened Landau 
bands was proved within a single-band approximation \cite{DMP},   
for a sufficiently strong magnetic field \cite{CH,Wang,GK}, 
or for a low density of the electrons at the band edges \cite{BCH}. 
The existence of the quantized Hall conductance plateaus was first 
proved by Kunz \cite{Kunz} under assumptions on a linear response formula of 
the conductance and on the band edge localization. The latter assumption on the localization 
can be removed for a tight-binding model. Namely, the constancy of the quantized Hall 
conductance was proved within the tight-binding approximation for varying the Fermi 
level \cite{BVS,AG}, or the strength of the potential \cite{RSB}. 
We should remark that, without relying on the translation ergodicity 
of the Hamiltonian, Elgart, Graf and Schenker \cite{EGS} proved 
the constancy of the quantized Hall conductance for a tight-binding case. 
For continuous models, Nakamura and Bellissard \cite{NB} proved that 
the states at the bottom of the spectrum do not contribute to the Hall conductance. 
Quite recently, Germinet, Klein and Schenker \cite{GKS} proved that the Hall conductance 
formula \cite{ASS} which is written in terms of switch functions 
shows a plateau for a random Landau Hamiltonian with translation ergodicity. 
In order to determine the integer of the quantized value of the Hall 
conductance, they further required the condition that 
the disordered-broadened Landau bands are disjoint, i.e., 
there exists a nonvanishing spectral gap between two neighboring Landau bands. 
However, the existence of the localized states at the band edges does not 
necessarily implies the appearance of the Hall conductance plateaus for 
varying the filling factor of the electrons because the density of the localized 
states may be vanishing in the infinite volume. In order to show the existence 
of such plateaus, we need to prove the existence of localized states 
of the bulk order. In passing, we remark that Wang \cite{Wang2} obtained  
the asymptotic expansion for the density of states in the large magnetic 
field limit.\footnote{In general, an asymptotic series does not give us any information 
for a fixed finite value of the parameter because the asymptotic series is not necessarily 
convergent. See, for example, Section~XII.3 of the book \cite{ReedSimonIV}. 
Thus the result of \cite{Wang2} dose not imply the existence 
of localized states of the bulk order for a fixed finite value of the magnetic field. 
See also the recent paper \cite{HM} for the difficulty of obtaining a lower bound for 
the density of states.} 

In this paper, we focus on the issue of proving the bulk order plateaus, 
and consider a noninteracting electron gas with Anderson-type impurities 
in a magnetic field in two dimensions at zero temperature. The centers of the bumps of 
the impurities form the triangular lattice. First we prove that there exist localized states 
of the bulk order in the disordered-broadened Landau bands 
whose energies are smaller than a certain value determined 
by the strength of the magnetic field. In order to obtain the Hall conductance 
as a linear response coefficient to an external electric field, 
we apply a time-dependent vector potential ${\bf A}_{\rm ex}(t)=(0,\alpha(t))$, 
where the function $\alpha(t)$ of time $t$ is given by (\ref{alphat}) in the 
next section. For $t\in [-T,0]$ with a large positive $T$, 
the corresponding electric field is adiabatically switched on, 
and for $t\ge 0$, the electric field becomes $(0,F)$ with 
the constant strength $F$.  
First we consider the finite, isolated system of an $L_x\times L_y$ rectangular box, 
and impose periodic boundary conditions for the wavefunctions 
with the help of the magnetic translation (\ref{magnetictranslation}), 
and then we take the infinite-volume limit.
The explicit expression of the conductance formula which we will 
use is given in ref.~\cite{Koma3}. 
We should remark that, when the system is translationally invariant,    
the constant Hall current flows on the torus without dissipation of energy 
as in ref.~\cite{Koma3}. 
We prove that, when the Fermi level lies in the localization regime, 
the Hall conductance is quantized to the desired integer 
and shows the plateau of the bulk order 
for varying the filling factor of the electrons. 
In our approach, we require neither the disjoint condition for 
the Landau bands nor the translation ergodicity\footnote{In a generic, realistic situation 
that there exist one- or two-dimensional objects such as dislocations in crystals and 
interfaces in semiconductors, we cannot expect that 
the system has translation ergodicity.} of the Hamiltonian 
which were assumed in \cite{GKS} as mentioned above.  Instead of these condition, 
we need the ``covering condition" that the whole plane ${\bf R}^2$ is covered by 
the supports of the bumps of the impurity potentials 
so that the sum of the bumps is strictly positive on the whole plane ${\bf R}^2$. 
This ``covering condition" is not required in \cite{GKS}. 

The present paper is organized as follows. In Section~\ref{model}, we describe the model, 
and state our main theorems. As preliminaries, 
we study the spectrum of the Hamiltonian without the random 
potential in Section~\ref{spectrumH0} and the site percolation on 
the triangular lattice of the impurities in Section~\ref{percolation}. 
In Section~\ref{IDER}, we obtain a decay bound for the resolvent (Green function) of 
a finite volume.  This bound becomes the initial data for the multi-scale analysis 
\cite{FS,vDK,CL} which is given in Section~\ref{MSA}. In order to prove constancy of the Hall 
conductance, we further need the fractional moment bound \cite{AENSS} for the resolvent. 
The bound is given in Section~\ref{FMB}. 
As preliminaries for proving the integrality and constancy of the Hall conductance, 
we study the finite volume Hall conductance in Section~\ref{Conplateaus}. 
The integrality of the Hall conductance is proved within the framework of 
``noncommutative geometry" \cite{ASS,BVS,AG} in Section~\ref{IndexApproach}, 
and the constancy is proved by using 
the homotopy argument \cite{BVS,RSB} in Section~\ref{Homotopy}. 
The widths of the Hall conductance plateaus and the corrections 
to the linear response formula are estimated in Sections~\ref{Widths} 
and \ref{CorrectionstoLRF}, respectively.  
Appendices~\ref{appendix:Wegner}--\ref{Proof9.7} are devoted to technical estimates. 
The standard Hall conductance formula which is given in Section~\ref{IndexApproach} 
is written in terms of the position operator of the electron. 
In Appendix~\ref{IndexSwitch}, we give a proof that this Hall conductance 
is equal to another Hall conductance \cite{ASS} 
which is written in terms of switch functions for a class of continuous models. 
%%%%%%%%%%%%%%%%%%%%%%%%%%%%%%%%%%%%%%%%%%%%%%
\Section{Model and the main results}
\label{model}

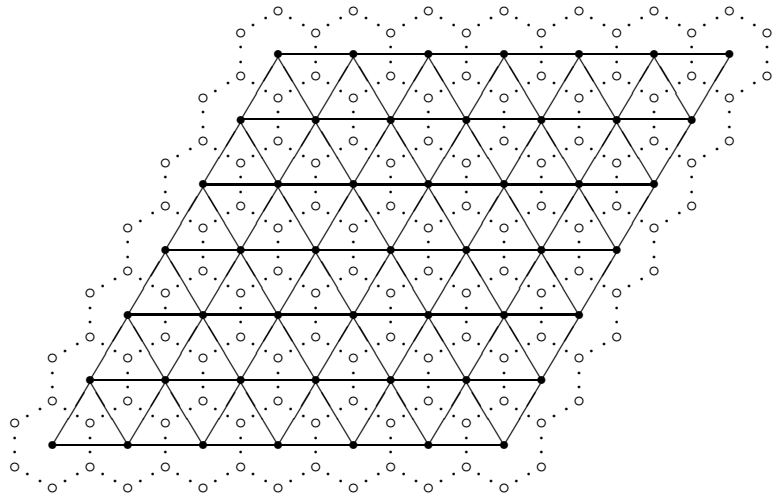
\begin{figure}[t]
\setlength{\unitlength}{1mm}
\begin{center}
\begin{picture}(100,57)(-10,-1.5)
\put(0,0){\line(1,0){60}}
\put(5,8.66025){\line(1,0){60}}
\put(10,17.3205){\line(1,0){60}}
\put(15,25.98076){\line(1,0){60}}
\put(20,34.6410){\line(1,0){60}}
\put(25,43.30127){\line(1,0){60}}
\put(30,51.96152){\line(1,0){60}}
%%%%%%%%%%%%%%%%%%%%%%%%%%%%%%%%%%
\put(0,0){\line(3,5){5.2}}
\put(5,8.66025){\line(3,5){5.2}}
\put(10,17.3205){\line(3,5){5.2}}
\put(15,25.98076){\line(3,5){5.2}}
\put(20,34.6410){\line(3,5){5.2}}
\put(25,43.30127){\line(3,5){5.2}}
%%%%%%%%%%%%%%%%%%%%%%%%%%%
\put(0,0){\circle*1}\put(5,8.66025){\circle*1}
\put(10,17.3205){\circle*1}
\put(15,25.98076){\circle*1}
\put(20,34.6410){\circle*1}
\put(25,43.30127){\circle*1}
\put(30,51.96152){\circle*1}
%%%%%%%%%%%%%%%%%%%%%%%%%%%%%%%%%
\put(10,0){\line(3,5){5.2}}
\put(15,8.66025){\line(3,5){5.2}}
\put(20,17.3205){\line(3,5){5.2}}
\put(25,25.98076){\line(3,5){5.2}}
\put(30,34.6410){\line(3,5){5.2}}
\put(35,43.30127){\line(3,5){5.2}}
%%%%%%%%%%%%%%%%%%%%%%%%%%%
\put(10,0){\circle*1}
\put(15,8.66025){\circle*1}
\put(20,17.3205){\circle*1}
\put(25,25.98076){\circle*1}
\put(30,34.6410){\circle*1}
\put(35,43.30127){\circle*1}
\put(40,51.96152){\circle*1}
%%%%%%%%%%%%%%%%%%%%%%%%%%%%%%%%%
\put(20,0){\line(3,5){5.2}}
\put(25,8.66025){\line(3,5){5.2}}
\put(30,17.3205){\line(3,5){5.2}}
\put(35,25.98076){\line(3,5){5.2}}
\put(40,34.6410){\line(3,5){5.2}}
\put(45,43.30127){\line(3,5){5.2}}
%%%%%%%%%%%%%%%%%%%%%%%%%%%
\put(20,0){\circle*1}
\put(25,8.66025){\circle*1}
\put(30,17.3205){\circle*1}
\put(35,25.98076){\circle*1}
\put(40,34.6410){\circle*1}
\put(45,43.30127){\circle*1}
\put(50,51.96152){\circle*1}
%%%%%%%%%%%%%%%%%%%%%%%%%%%%%%%%%
\put(30,0){\line(3,5){5.2}}
\put(35,8.66025){\line(3,5){5.2}}
\put(40,17.3205){\line(3,5){5.2}}
\put(45,25.98076){\line(3,5){5.2}}
\put(50,34.6410){\line(3,5){5.2}}
\put(55,43.30127){\line(3,5){5.2}}
%%%%%%%%%%%%%%%%%%%%%%%%%%%
\put(30,0){\circle*1}
\put(35,8.66025){\circle*1}
\put(40,17.3205){\circle*1}
\put(45,25.98076){\circle*1}
\put(50,34.6410){\circle*1}
\put(55,43.30127){\circle*1}
\put(60,51.96152){\circle*1}
%%%%%%%%%%%%%%%%%%%%%%%%%%%%%%%
\put(40,0){\line(3,5){5.2}}
\put(45,8.66025){\line(3,5){5.2}}
\put(50,17.3205){\line(3,5){5.2}}
\put(55,25.98076){\line(3,5){5.2}}
\put(60,34.6410){\line(3,5){5.2}}
\put(65,43.30127){\line(3,5){5.2}}
%%%%%%%%%%%%%%%%%%%%%%%%%%%
\put(40,0){\circle*1}
\put(45,8.66025){\circle*1}
\put(50,17.3205){\circle*1}
\put(55,25.98076){\circle*1}
\put(60,34.6410){\circle*1}
\put(65,43.30127){\circle*1}
\put(70,51.96152){\circle*1}
%%%%%%%%%%%%%%%%%%%%%%%%%%%%%%%
\put(50,0){\line(3,5){5.2}}
\put(55,8.66025){\line(3,5){5.2}}
\put(60,17.3205){\line(3,5){5.2}}
\put(65,25.98076){\line(3,5){5.2}}
\put(70,34.6410){\line(3,5){5.2}}
\put(75,43.30127){\line(3,5){5.2}}
%%%%%%%%%%%%%%%%%%%%%%%%%%%
\put(50,0){\circle*1}
\put(55,8.66025){\circle*1}
\put(60,17.3205){\circle*1}
\put(65,25.98076){\circle*1}
\put(70,34.6410){\circle*1}
\put(75,43.30127){\circle*1}
\put(80,51.96152){\circle*1}
%%%%%%%%%%%%%%%%%%%%%%%%%%%%%%%
\put(60,0){\line(3,5){5.2}}
\put(65,8.66025){\line(3,5){5.2}}
\put(70,17.3205){\line(3,5){5.2}}
\put(75,25.98076){\line(3,5){5.2}}
\put(80,34.6410){\line(3,5){5.2}}
\put(85,43.30127){\line(3,5){5.2}}
%%%%%%%%%%%%%%%%%%%%%%%%%%%
\put(60,0){\circle*1}
\put(65,8.66025){\circle*1}
\put(70,17.3205){\circle*1}
\put(75,25.98076){\circle*1}
\put(80,34.6410){\circle*1}
\put(85,43.30127){\circle*1}
\put(90,51.96152){\circle*1}
%%%%%%%%%%%%%%%%%%%%%%%%%%%%%%%
\put(10,0){\line(-3,5){5.2}}
\put(15,8.66025){\line(-3,5){5.2}}
\put(20,17.3205){\line(-3,5){5.2}}
\put(25,25.98076){\line(-3,5){5.2}}
\put(30,34.6410){\line(-3,5){5.2}}
\put(35,43.30127){\line(-3,5){5.2}}
%%%%%%%%%%%%%%%%%%%%%%%%%%%%%%
\put(20,0){\line(-3,5){5.2}}
\put(25,8.66025){\line(-3,5){5.2}}
\put(30,17.3205){\line(-3,5){5.2}}
\put(35,25.98076){\line(-3,5){5.2}}
\put(40,34.6410){\line(-3,5){5.2}}
\put(45,43.30127){\line(-3,5){5.2}}
%%%%%%%%%%%%%%%%%%%%%%%%%%%%%%
\put(30,0){\line(-3,5){5.2}}
\put(35,8.66025){\line(-3,5){5.2}}
\put(40,17.3205){\line(-3,5){5.2}}
\put(45,25.98076){\line(-3,5){5.2}}
\put(50,34.6410){\line(-3,5){5.2}}
\put(55,43.30127){\line(-3,5){5.2}}
%%%%%%%%%%%%%%%%%%%%%%%%%%%%%%
\put(40,0){\line(-3,5){5.2}}
\put(45,8.66025){\line(-3,5){5.2}}
\put(50,17.3205){\line(-3,5){5.2}}
\put(55,25.98076){\line(-3,5){5.2}}
\put(60,34.6410){\line(-3,5){5.2}}
\put(65,43.30127){\line(-3,5){5.2}}
%%%%%%%%%%%%%%%%%%%%%%%%%%%%%%
\put(50,0){\line(-3,5){5.2}}
\put(55,8.66025){\line(-3,5){5.2}}
\put(60,17.3205){\line(-3,5){5.2}}
\put(65,25.98076){\line(-3,5){5.2}}
\put(70,34.6410){\line(-3,5){5.2}}
\put(75,43.30127){\line(-3,5){5.2}}
%%%%%%%%%%%%%%%%%%%%%%%%%%%%%%
\put(60,0){\line(-3,5){5.2}}
\put(65,8.66025){\line(-3,5){5.2}}
\put(70,17.3205){\line(-3,5){5.2}}
\put(75,25.98076){\line(-3,5){5.2}}
\put(80,34.6410){\line(-3,5){5.2}}
\put(85,43.30127){\line(-3,5){5.2}}
%%%%%%%%%%%%%%%%%%%%%%%%%%%%%%%%%%%%%%%%%%%%%%%%%%%%%
\put(-5,2.8867){\circle1}
\put(-5,0.96225){\circle*0,3}
\put(-5,-0.96225){\circle*0,3}
\put(-5,-2.8867){\circle1}
%%%%%%%%%%%%%%%%%%%%%%%%%%%
\put(5,2.8867){\circle1}
\put(5,0.96225){\circle*0,3}
\put(5,-0.96225){\circle*0,3}
\put(5,-2.8867){\circle1}
%%%%%%%%%%%%%%%%%%%%%%%%%%%
\put(15,2.8867){\circle1}
\put(15,0.96225){\circle*0,3}
\put(15,-0.96225){\circle*0,3}
\put(15,-2.8867){\circle1}
%%%%%%%%%%%%%%%%%%%%%%%%%%%
\put(25,2.8867){\circle1}
\put(25,0.96225){\circle*0,3}
\put(25,-0.96225){\circle*0,3}
\put(25,-2.8867){\circle1}
%%%%%%%%%%%%%%%%%%%%%%%%%%%
\put(35,2.8867){\circle1}
\put(35,0.96225){\circle*0,3}
\put(35,-0.96225){\circle*0,3}
\put(35,-2.8867){\circle1}
%%%%%%%%%%%%%%%%%%%%%%%%%%%
\put(45,2.8867){\circle1}
\put(45,0.96225){\circle*0,3}
\put(45,-0.96225){\circle*0,3}
\put(45,-2.8867){\circle1}
%%%%%%%%%%%%%%%%%%%%%%%%%%%
\put(55,2.8867){\circle1}
\put(55,0.96225){\circle*0,3}
\put(55,-0.96225){\circle*0,3}
\put(55,-2.8867){\circle1}
%%%%%%%%%%%%%%%%%%%%%%%%%%%
\put(65,2.8867){\circle1}\put(65,0.96225){\circle*0,3}\put(65,-0.96225){\circle*0,3}
\put(65,-2.8867){\circle1}
%%%%%%%%%%%%%%%%%%%%%%%%%%%%%%%%%%%%%%%%%%%%%%%%%%%%%%%%%%%%%
\put(0,5.7735){\circle1}\put(-3.333,3.849){\circle*0.3}
\put(-1.6666,4.81125){\circle*0.3}
\put(3.333,3.849){\circle*0.3}
\put(1.6666,4.81125){\circle*0.3}
%%%%%%%%%%%%%%%%%%%%%%%%%%%%%%%%
\put(-3.333,-3.849){\circle*0.3}
\put(-1.6666,-4.81125){\circle*0.3}
\put(3.333,-3.849){\circle*0.3}
\put(1.6666,-4.81125){\circle*0.3}
\put(0,-5.7735){\circle1}
%%%%%%%%%%%%%%%%%%%%%%%%%%%%%%%%
\put(10,5.7735){\circle1}
\put(6.666,3.849){\circle*0.3}
\put(8.3333,4.81125){\circle*0.3}
\put(13.333,3.849){\circle*0.3}
\put(11.6666,4.81125){\circle*0.3}
%%%%%%%%%%%%%%%%%%%%%%%%%%%%%%%%
\put(6.666,-3.849){\circle*0.3}
\put(8.3333,-4.81125){\circle*0.3}
\put(13.333,-3.849){\circle*0.3}
\put(11.6666,-4.81125){\circle*0.3}
\put(10,-5.7735){\circle1}
%%%%%%%%%%%%%%%%%%%%%%%%%%%%%%%%
\put(20,5.7735){\circle1}
\put(16.666,3.849){\circle*0.3}
\put(18.3333,4.81125){\circle*0.3}
\put(23.333,3.849){\circle*0.3}
\put(21.6666,4.81125){\circle*0.3}
%%%%%%%%%%%%%%%%%%%%%%%%%%%%%%%%
\put(16.666,-3.849){\circle*0.3}
\put(18.3333,-4.81125){\circle*0.3}
\put(23.333,-3.849){\circle*0.3}
\put(21.6666,-4.81125){\circle*0.3}
\put(20,-5.7735){\circle1}
%%%%%%%%%%%%%%%%%%%%%%%%%%%%%%%%
\put(30,5.7735){\circle1}
\put(26.666,3.849){\circle*0.3}
\put(28.3333,4.81125){\circle*0.3}
\put(33.333,3.849){\circle*0.3}
\put(31.6666,4.81125){\circle*0.3}
%%%%%%%%%%%%%%%%%%%%%%%%%%%%%%%%
\put(26.666,-3.849){\circle*0.3}
\put(28.3333,-4.81125){\circle*0.3}
\put(33.333,-3.849){\circle*0.3}
\put(31.6666,-4.81125){\circle*0.3}
\put(30,-5.7735){\circle1}
%%%%%%%%%%%%%%%%%%%%%%%%%%%%%%%%
\put(40,5.7735){\circle1}
\put(36.666,3.849){\circle*0.3}
\put(38.3333,4.81125){\circle*0.3}
\put(43.333,3.849){\circle*0.3}
\put(41.6666,4.81125){\circle*0.3}
%%%%%%%%%%%%%%%%%%%%%%%%%%%%%%%%
\put(36.666,-3.849){\circle*0.3}
\put(38.3333,-4.81125){\circle*0.3}
\put(43.333,-3.849){\circle*0.3}
\put(41.6666,-4.81125){\circle*0.3}
\put(40,-5.7735){\circle1}
%%%%%%%%%%%%%%%%%%%%%%%%%%%%%%%%
\put(50,5.7735){\circle1}
\put(46.666,3.849){\circle*0.3}
\put(48.3333,4.81125){\circle*0.3}
\put(53.333,3.849){\circle*0.3}
\put(51.6666,4.81125){\circle*0.3}
%%%%%%%%%%%%%%%%%%%%%%%%%%%%%%%%
\put(46.666,-3.849){\circle*0.3}
\put(48.3333,-4.81125){\circle*0.3}
\put(53.333,-3.849){\circle*0.3}
\put(51.6666,-4.81125){\circle*0.3}
\put(50,-5.7735){\circle1}
%%%%%%%%%%%%%%%%%%%%%%%%%%%%%%%%
\put(60,5.7735){\circle1}
\put(56.666,3.849){\circle*0.3}
\put(58.3333,4.81125){\circle*0.3}
\put(63.333,3.849){\circle*0.3}
\put(61.6666,4.81125){\circle*0.3}
%%%%%%%%%%%%%%%%%%%%%%%%%%%%%%%%%%%%%%
\put(70,5.7735){\circle1}
\put(66.666,3.849){\circle*0.3}
\put(68.3333,4.81125){\circle*0.3}
%%%%%%%%%%%%%%%%%%%%%%%%%%%%%%%%%%%%%%%%%%%%%%%%%%%%%%%%%%%%%%%%
\put(56.666,-3.849){\circle*0.3}
\put(58.3333,-4.81125){\circle*0.3}
\put(63.333,-3.849){\circle*0.3}
\put(61.6666,-4.81125){\circle*0.3}\put(60,-5.7735){\circle1}
%%%%%%%%%%%%%%%%%%%%%%%%%%%%%%%%%%%%%%%%%%%%%%%%%%%%%%%%%%%%%%%%
\put(0,11.547){\circle1}\put(0,7.698){\circle*0.3}\put(0,9.6225){\circle*0.3}
\put(10,11.547){\circle1}\put(10,7.698){\circle*0.3}\put(10,9.6225){\circle*0.3}
\put(20,11.547){\circle1}\put(20,7.698){\circle*0.3}\put(20,9.6225){\circle*0.3}
\put(30,11.547){\circle1}\put(30,7.698){\circle*0.3}\put(30,9.6225){\circle*0.3}
\put(40,11.547){\circle1}\put(40,7.698){\circle*0.3}\put(40,9.6225){\circle*0.3}
\put(50,11.547){\circle1}\put(50,7.698){\circle*0.3}\put(50,9.6225){\circle*0.3}
\put(60,11.547){\circle1}\put(60,7.698){\circle*0.3}\put(60,9.6225){\circle*0.3}
\put(70,11.547){\circle1}\put(70,7.698){\circle*0.3}\put(70,9.6225){\circle*0.3}
%%%%%%%%%%%%%%%%%%%%%%%%%% 
\put(5,14.4337){\circle1}
\put(1.6666,12.50925){\circle*0.3}\put(3.3333,13.4715){\circle*0.3}
\put(8.3333,12.50925){\circle*0.3}\put(6.6666,13.4715){\circle*0.3}
\put(15,14.4337){\circle1}
\put(11.6666,12.50925){\circle*0.3}\put(13.3333,13.4715){\circle*0.3}
\put(18.3333,12.50925){\circle*0.3}\put(16.6666,13.4715){\circle*0.3}
\put(25,14.4337){\circle1}
\put(21.6666,12.50925){\circle*0.3}\put(23.3333,13.4715){\circle*0.3}
\put(28.3333,12.50925){\circle*0.3}\put(26.6666,13.4715){\circle*0.3}
\put(35,14.4337){\circle1}
\put(31.6666,12.50925){\circle*0.3}\put(33.3333,13.4715){\circle*0.3}
\put(38.3333,12.50925){\circle*0.3}\put(36.6666,13.4715){\circle*0.3}
\put(45,14.4337){\circle1}
\put(41.6666,12.50925){\circle*0.3}\put(43.3333,13.4715){\circle*0.3}
\put(48.3333,12.50925){\circle*0.3}\put(46.6666,13.4715){\circle*0.3}
\put(55,14.4337){\circle1}
\put(51.6666,12.50925){\circle*0.3}\put(53.3333,13.4715){\circle*0.3}
\put(58.3333,12.50925){\circle*0.3}\put(56.6666,13.4715){\circle*0.3}
\put(65,14.4337){\circle1}
\put(61.6666,12.50925){\circle*0.3}\put(63.3333,13.4715){\circle*0.3}
\put(68.3333,12.50925){\circle*0.3}\put(66.6666,13.4715){\circle*0.3}
\put(75,14.4337){\circle1}
\put(71.6666,12.50925){\circle*0.3}\put(73.3333,13.4715){\circle*0.3}
%%%%%%%%%%%%%%%%%%%%%%%%%%%%%%%%%%%%%%%%%%%%%%%%%%%%%%%%%%%%%%%%%%%%%%%
\put(5,20.20725){\circle1}
\put(5,18.28275){\circle*0.3}\put(5,16.3582){\circle*0.3}
\put(15,20.20725){\circle1}
\put(15,18.28275){\circle*0.3}\put(15,16.3582){\circle*0.3}
\put(25,20.20725){\circle1}
\put(25,18.28275){\circle*0.3}\put(25,16.3582){\circle*0.3}
\put(35,20.20725){\circle1}
\put(35,18.28275){\circle*0.3}\put(35,16.3582){\circle*0.3}
\put(45,20.20725){\circle1}
\put(45,18.28275){\circle*0.3}\put(45,16.3582){\circle*0.3}
\put(55,20.20725){\circle1}
\put(55,18.28275){\circle*0.3}\put(55,16.3582){\circle*0.3}
\put(65,20.20725){\circle1}
\put(65,18.28275){\circle*0.3}\put(65,16.3582){\circle*0.3}
\put(75,20.20725){\circle1}
\put(75,18.28275){\circle*0.3}\put(75,16.3582){\circle*0.3}
%%%%%%%%%%%%%%%%%%%%%%%%%%%%%%%%%%%%%%%%%%%%%%%%%%%%%%%%%%%%%%%%%
\put(10,23.094){\circle1}
\put(6.6666,21.1695){\circle*0.3}\put(8.3333,22.1317){\circle*0.3}
\put(13.3333,21.1695){\circle*0.3}\put(11.6666,22.1317){\circle*0.3}
\put(20,23.094){\circle1}
\put(16.6666,21.1695){\circle*0.3}\put(18.3333,22.1317){\circle*0.3}
\put(23.3333,21.1695){\circle*0.3}\put(21.6666,22.1317){\circle*0.3}
\put(30,23.094){\circle1}
\put(26.6666,21.1695){\circle*0.3}\put(28.3333,22.1317){\circle*0.3}
\put(33.3333,21.1695){\circle*0.3}\put(31.6666,22.1317){\circle*0.3}
\put(40,23.094){\circle1}
\put(36.6666,21.1695){\circle*0.3}\put(38.3333,22.1317){\circle*0.3}
\put(43.3333,21.1695){\circle*0.3}\put(41.6666,22.1317){\circle*0.3}
\put(50,23.094){\circle1}
\put(46.6666,21.1695){\circle*0.3}\put(48.3333,22.1317){\circle*0.3}
\put(53.3333,21.1695){\circle*0.3}\put(51.6666,22.1317){\circle*0.3}
\put(60,23.094){\circle1}
\put(56.6666,21.1695){\circle*0.3}\put(58.3333,22.1317){\circle*0.3}
\put(63.3333,21.1695){\circle*0.3}\put(61.6666,22.1317){\circle*0.3}
\put(70,23.094){\circle1}
\put(66.6666,21.1695){\circle*0.3}\put(68.3333,22.1317){\circle*0.3}
\put(73.3333,21.1695){\circle*0.3}\put(71.6666,22.1317){\circle*0.3}
\put(80,23.094){\circle1}
\put(76.6666,21.1695){\circle*0.3}\put(78.3333,22.1317){\circle*0.3}
%%%%%%%%%%%%%%%%%%%%%%%%%%%%%%%%%%%%%%%%%%%%%%%%%%%%%%%%%%%%%%%%%%%%%
\put(10,28.8675){\circle1}
\put(10,25.0185){\circle*0.3}\put(10,26.94301){\circle*0.3}
\put(20,28.8675){\circle1}
\put(20,25.0185){\circle*0.3}\put(20,26.94301){\circle*0.3}
\put(30,28.8675){\circle1}
\put(30,25.0185){\circle*0.3}\put(30,26.94301){\circle*0.3}
\put(40,28.8675){\circle1}
\put(40,25.0185){\circle*0.3}\put(40,26.94301){\circle*0.3}
\put(50,28.8675){\circle1}
\put(50,25.0185){\circle*0.3}\put(50,26.94301){\circle*0.3}
\put(60,28.8675){\circle1}
\put(60,25.0185){\circle*0.3}\put(60,26.94301){\circle*0.3}
\put(70,28.8675){\circle1}
\put(70,25.0185){\circle*0.3}\put(70,26.94301){\circle*0.3}
\put(80,28.8675){\circle1}
\put(80,25.0185){\circle*0.3}\put(80,26.94301){\circle*0.3}
%%%%%%%%%%%%%%%%%%%%%%%%%%%%%%%%%%%%%%%%%%%%%%%%%%%%%%%%%%%%%%%%%
\put(15,31.7542){\circle1}
\put(11.6666,29.8297){\circle*0.3}\put(13.3333,30.79201){\circle*0.3}
\put(18.3333,29.8297){\circle*0.3}\put(16.6666,30.79201){\circle*0.3}
\put(25,31.7542){\circle1}
\put(21.6666,29.8297){\circle*0.3}\put(23.3333,30.79201){\circle*0.3}
\put(28.3333,29.8297){\circle*0.3}\put(26.6666,30.79201){\circle*0.3}
\put(35,31.7542){\circle1}
\put(31.6666,29.8297){\circle*0.3}\put(33.3333,30.79201){\circle*0.3}
\put(38.3333,29.8297){\circle*0.3}\put(36.6666,30.79201){\circle*0.3}
\put(45,31.7542){\circle1}
\put(41.6666,29.8297){\circle*0.3}\put(43.3333,30.79201){\circle*0.3}
\put(48.3333,29.8297){\circle*0.3}\put(46.6666,30.79201){\circle*0.3}
\put(55,31.7542){\circle1}
\put(51.6666,29.8297){\circle*0.3}\put(53.3333,30.79201){\circle*0.3}
\put(58.3333,29.8297){\circle*0.3}\put(56.6666,30.79201){\circle*0.3}
\put(65,31.7542){\circle1}
\put(61.6666,29.8297){\circle*0.3}\put(63.3333,30.79201){\circle*0.3}
\put(68.3333,29.8297){\circle*0.3}\put(66.6666,30.79201){\circle*0.3}
\put(75,31.7542){\circle1}
\put(71.6666,29.8297){\circle*0.3}\put(73.3333,30.79201){\circle*0.3}
\put(78.3333,29.8297){\circle*0.3}\put(76.6666,30.79201){\circle*0.3}
\put(85,31.7542){\circle1}
\put(81.6666,29.8297){\circle*0.3}\put(83.3333,30.79201){\circle*0.3}
%%%%%%%%%%%%%%%%%%%%%%%%%%%%%%%%%%%%%%%%%%%%%%%%%%%%%%%%%%%%%%%%%%%%%
\put(15,37.52776){\circle1}
\put(15,33.6787){\circle*0.3}\put(15,35.6032){\circle*0.3}
\put(25,37.52776){\circle1}
\put(25,33.6787){\circle*0.3}\put(25,35.6032){\circle*0.3}
\put(35,37.52776){\circle1}
\put(35,33.6787){\circle*0.3}\put(35,35.6032){\circle*0.3}
\put(45,37.52776){\circle1}
\put(45,33.6787){\circle*0.3}\put(45,35.6032){\circle*0.3}
\put(55,37.52776){\circle1}
\put(55,33.6787){\circle*0.3}\put(55,35.6032){\circle*0.3}
\put(65,37.52776){\circle1}
\put(65,33.6787){\circle*0.3}\put(65,35.6032){\circle*0.3}
\put(75,37.52776){\circle1}
\put(75,33.6787){\circle*0.3}\put(75,35.6032){\circle*0.3}
\put(85,37.52776){\circle1}
\put(85,33.6787){\circle*0.3}\put(85,35.6032){\circle*0.3}
%%%%%%%%%%%%%%%%%%%%%%%%%%%%%%%%%%%%%%%%%%%%%%%%%%%%%%%%%%%%%%%%%
\put(20,40.4145){\circle1}
\put(16.6666,38.49){\circle*0.3}\put(18.3333,39.4522){\circle*0.3}
\put(23.3333,38.49){\circle*0.3}\put(21.6666,39.4522){\circle*0.3}
\put(30,40.4145){\circle1}
\put(26.6666,38.49){\circle*0.3}\put(28.3333,39.4522){\circle*0.3}
\put(33.3333,38.49){\circle*0.3}\put(31.6666,39.4522){\circle*0.3}
\put(40,40.4145){\circle1}
\put(36.6666,38.49){\circle*0.3}\put(38.3333,39.4522){\circle*0.3}
\put(43.3333,38.49){\circle*0.3}\put(41.6666,39.4522){\circle*0.3}
\put(50,40.4145){\circle1}
\put(46.6666,38.49){\circle*0.3}\put(48.3333,39.4522){\circle*0.3}
\put(53.3333,38.49){\circle*0.3}\put(51.6666,39.4522){\circle*0.3}
\put(60,40.4145){\circle1}
\put(56.6666,38.49){\circle*0.3}\put(58.3333,39.4522){\circle*0.3}
\put(63.3333,38.49){\circle*0.3}\put(61.6666,39.4522){\circle*0.3}
\put(70,40.4145){\circle1}
\put(66.6666,38.49){\circle*0.3}\put(68.3333,39.4522){\circle*0.3}
\put(73.3333,38.49){\circle*0.3}\put(71.6666,39.4522){\circle*0.3}
\put(80,40.4145){\circle1}
\put(76.6666,38.49){\circle*0.3}\put(78.3333,39.4522){\circle*0.3}
\put(83.3333,38.49){\circle*0.3}\put(81.6666,39.4522){\circle*0.3}
\put(90,40.4145){\circle1}
\put(86.6666,38.49){\circle*0.3}\put(88.3333,39.4522){\circle*0.3}
%%%%%%%%%%%%%%%%%%%%%%%%%%%%%%%%%%%%%%%%%%%%%%%%%%%%%%%%%%%%%%%%%%
\put(20,46.188){\circle1}
\put(20,42.339){\circle*0.3}\put(20,44.2635){\circle*0.3}
\put(30,46.188){\circle1}
\put(30,42.339){\circle*0.3}\put(30,44.2635){\circle*0.3}
\put(40,46.188){\circle1}
\put(40,42.339){\circle*0.3}\put(40,44.2635){\circle*0.3}
\put(50,46.188){\circle1}
\put(50,42.339){\circle*0.3}\put(50,44.2635){\circle*0.3}
\put(60,46.188){\circle1}
\put(60,42.339){\circle*0.3}\put(60,44.2635){\circle*0.3}
\put(70,46.188){\circle1}
\put(70,42.339){\circle*0.3}\put(70,44.2635){\circle*0.3}
\put(80,46.188){\circle1}
\put(80,42.339){\circle*0.3}\put(80,44.2635){\circle*0.3}
\put(90,46.188){\circle1}
\put(90,42.339){\circle*0.3}\put(90,44.2635){\circle*0.3}
%%%%%%%%%%%%%%%%%%%%%%%%%%%%%%%%%%%%%%%%%%%%%%%%%%%%%%%%%%%%%
\put(25,49.0747){\circle1}
\put(21.6666,47.15){\circle*0.3}\put(23.3333,48.1125){\circle*0.3}
\put(28.3333,47.15){\circle*0.3}\put(26.6666,48.1125){\circle*0.3}
\put(35,49.0747){\circle1}
\put(31.6666,47.15){\circle*0.3}\put(33.3333,48.1125){\circle*0.3}
\put(38.3333,47.15){\circle*0.3}\put(36.6666,48.1125){\circle*0.3}
\put(45,49.0747){\circle1}
\put(41.6666,47.15){\circle*0.3}\put(43.3333,48.1125){\circle*0.3}
\put(48.3333,47.15){\circle*0.3}\put(46.6666,48.1125){\circle*0.3}
\put(55,49.0747){\circle1}
\put(51.6666,47.15){\circle*0.3}\put(53.3333,48.1125){\circle*0.3}
\put(58.3333,47.15){\circle*0.3}\put(56.6666,48.1125){\circle*0.3}
\put(65,49.0747){\circle1}
\put(61.6666,47.15){\circle*0.3}\put(63.3333,48.1125){\circle*0.3}
\put(68.3333,47.15){\circle*0.3}\put(66.6666,48.1125){\circle*0.3}
\put(75,49.0747){\circle1}
\put(71.6666,47.15){\circle*0.3}\put(73.3333,48.1125){\circle*0.3}
\put(78.3333,47.15){\circle*0.3}\put(76.6666,48.1125){\circle*0.3}
\put(85,49.0747){\circle1}
\put(81.6666,47.15){\circle*0.3}\put(83.3333,48.1125){\circle*0.3}
\put(88.3333,47.15){\circle*0.3}\put(86.6666,48.1125){\circle*0.3}
\put(95,49.0747){\circle1}
\put(91.6666,47.15){\circle*0.3}\put(93.3333,48.1125){\circle*0.3}
%%%%%%%%%%%%%%%%%%%%%%%%%%%%%%%%%%%%%%%%%%%%%%%%%%%%%%%%%%%%%%%%%%%%
\put(25,54.8482){\circle1}
\put(25,50.999){\circle*0.3}\put(25,52.9237){\circle*0.3}
\put(35,54.8482){\circle1}
\put(35,50.999){\circle*0.3}\put(35,52.9237){\circle*0.3}
\put(45,54.8482){\circle1}
\put(45,50.999){\circle*0.3}\put(45,52.9237){\circle*0.3}
\put(55,54.8482){\circle1}
\put(55,50.999){\circle*0.3}\put(55,52.9237){\circle*0.3}
\put(65,54.8482){\circle1}
\put(65,50.999){\circle*0.3}\put(65,52.9237){\circle*0.3}
\put(75,54.8482){\circle1}
\put(75,50.999){\circle*0.3}\put(75,52.9237){\circle*0.3}
\put(85,54.8482){\circle1}
\put(85,50.999){\circle*0.3}\put(85,52.9237){\circle*0.3}
\put(95,54.8482){\circle1}
\put(95,50.999){\circle*0.3}\put(95,52.9237){\circle*0.3}
%%%%%%%%%%%%%%%%%%%%%%%%%%%%%%%%%%%%%%%%%%%%%%%%%%%%%%%%%
\put(30,57.735){\circle1}
\put(26.6666,55.8105){\circle*0.3}\put(28.3333,56.7727){\circle*0.3}
\put(33.3333,55.8105){\circle*0.3}\put(31.6666,56.7727){\circle*0.3}
\put(40,57.735){\circle1}
\put(36.6666,55.8105){\circle*0.3}\put(38.3333,56.7727){\circle*0.3}
\put(43.3333,55.8105){\circle*0.3}\put(41.6666,56.7727){\circle*0.3}
\put(50,57.735){\circle1}
\put(46.6666,55.8105){\circle*0.3}\put(48.3333,56.7727){\circle*0.3}
\put(53.3333,55.8105){\circle*0.3}\put(51.6666,56.7727){\circle*0.3}
\put(60,57.735){\circle1}
\put(56.6666,55.8105){\circle*0.3}\put(58.3333,56.7727){\circle*0.3}
\put(63.3333,55.8105){\circle*0.3}\put(61.6666,56.7727){\circle*0.3}
\put(70,57.735){\circle1}
\put(66.6666,55.8105){\circle*0.3}\put(68.3333,56.7727){\circle*0.3}
\put(73.3333,55.8105){\circle*0.3}\put(71.6666,56.7727){\circle*0.3}
\put(80,57.735){\circle1}
\put(76.6666,55.8105){\circle*0.3}\put(78.3333,56.7727){\circle*0.3}
\put(83.3333,55.8105){\circle*0.3}\put(81.6666,56.7727){\circle*0.3}
\put(90,57.735){\circle1}
\put(86.6666,55.8105){\circle*0.3}\put(88.3333,56.7727){\circle*0.3}
\put(93.3333,55.8105){\circle*0.3}\put(91.6666,56.7727){\circle*0.3}
%%%%%%%%%%%%%%%%%%%%%%%%%%%%%%%%%%%%%%%%%%%%%%%%%%%%%%%
\end{picture}
\end{center}
\caption{The parallelogram in the triangular lattice with its dual, 
hexagonal lattice.}
\label{trianglelattice}
\end{figure}

Consider a two-dimensional electron system with Anderson-type impurities 
in a uniform magnetic field $(0,0,B)$ perpendicular to the $x$-$y$ plane 
in which the electron is confined. For simplicity we assume that 
the electron does not have the spin degrees of freedom. 
The Hamiltonian is given by 
\begin{equation}
H_\omega=H_0+V_\omega
\label{Hamomega} 
\end{equation}
with the unperturbed Hamiltonian,  
\begin{equation}
H_0=\frac{1}{2m_e}({\bf p}+e{\bf A})^2+V_0,
\label{Ham0} 
\end{equation}
and with a random potential $V_\omega$, 
where ${\bf p}:=-i\hbar\nabla$ with the Planck constant $\hbar$, and 
$-e$ and $m_e$ are, respectively, the charge of electron and 
the mass of electron; ${\bf A}$ and $V_0$ are, respectively, a vector potential 
and an electrostatic potential. 
The system is defined on a rectangular box 
\begin{equation}
\Lambda^{\rm sys}:=[-L_x/2,L_x/2]\times[-L_y/2,L_y/2]\subset{\bf R}^2
\label{Lamsys}
\end{equation}
with the periodic boundary conditions. The vector potential 
${\bf A}=(A_x,A_y)$ consists of two parts as ${\bf A}={\bf A}_{\rm P}+{\bf A}_0$,
where ${\bf A}_0({\bf r})=(-By,0)$ which gives the uniform magnetic 
field and the vector potential ${\bf A}_{\rm P}$ satisfies 
the periodic boundary condition, 
\begin{equation}
{\bf A}_{\rm P}(x+L_x,y)={\bf A}_{\rm P}(x,y+L_y)={\bf A}_{\rm P}(x,y). 
\end{equation}
This condition for ${\bf A}_{\rm P}$ implies that the corresponding magnetic flux piercing 
the rectangular box $\Lambda^{\rm sys}$ is vanishing. Therefore the total magnetic flux 
is given by $BL_xL_y$ from the vector potential ${\bf A}_0$ only. 
We assume that the components of the vector potential ${\bf A}_{\rm P}$ are 
continuously differentiable on ${\bf R}^2$.
Further we assume that the electrostatic potential $V_0$ satisfies 
the periodic boundary condition, 
\begin{equation}
V_0(x+L_x,y)=V_0(x,y+L_y)=V_0(x,y), 
\end{equation}
and $\Vert V_0^+\Vert_\infty+\Vert V_0^-\Vert_\infty\le v_0<\infty$ 
with some positive constant $v_0$ 
which is independent of the system sizes $L_x,L_y$. 
Here $V_0^\pm=\max\{\pm V_0,0\}$. 
As a random potential $V_\omega$, we consider an Anderson-type impurity 
potential, 
\begin{equation}
V_\omega({\bf r})=\sum_{{\bf z}\in{\bf L}^2}\lambda_{\bf z}(\omega)
u({\bf r}-{\bf z}),
\label{Vrandom}
\end{equation}
for ${\bf r}:=(x,y)\in {\bf R}^2$. The constants 
$\{\lambda_{\bf z}(\omega)|\ {\bf z}\in{\bf L}^2\}$ form a family of independent, 
identically distributed random variables on the two-dimensional triangular 
lattice ${\bf L}^2\subset{\bf R}^2$ with the lattice constant $a>0$.
The common distribution of the random variables has a density $g\ge0$ which 
has compact support, i.e., ${\rm supp}\ g\subset[\lambda_{\rm min},\lambda_{\rm max}]$ 
with $\lambda_{\rm min}<0<\lambda_{\rm max}$. 
Further the density $g$ satisfies the following conditions: 
\begin{equation}
g\in L^\infty({\bf R})\cap C({\bf R})\ \ \ \mbox{and}\quad  
\int_{-\lambda_-}^{\lambda_+} g(\lambda)d\lambda>1/2
\label{assumptiong}
\end{equation}
with two positive numbers $\lambda_+$ and $\lambda_-$. 
We consider two cases: (i)  a small $\lambda_-$ 
and (ii) a small $\lambda_+$. 
We assume that the condition (\ref{assumptiong}) holds 
for both of the two cases. 
If the density $g$ is an even function of $\lambda$ and 
is concentrated near $\lambda=0$, this requirement holds. We take 
\begin{equation}
{\bf L}^2=\left\{\left.{\bf z}=m{\bf a}_1+n{\bf a}_2\right|
(m,n)\in{\bf Z}^2\right\}
\end{equation}
with the two primitive translation vectors, ${\bf a}_1=(a,0)$ and  
${\bf a}_2=\left(a/2,\sqrt{3}a/2\right)$. 
The triangular lattice is imbedded in ${\bf R}^2$ such that 
each face is an equilateral triangle as described in Fig.~\ref{trianglelattice}.
We also consider its dual, hexagonal lattice which is defined as follows. 
Choose a vertex of the dual lattice at the center of gravity of 
each triangle, i.e., the intersection of the bisectors of the sides of 
the triangle. For the edges of the dual lattice, take the line segments 
along these same bisectors, and connecting the centers of gravity 
of adjacent triangles.  
%%%%%%%%%%%%%%%%%%%%%%%%%%%%%%%%%%%%%%%%%%%%%%%%%%%%%%%

We assume that the bump $u$ of the single-site potential in (\ref{Vrandom})   
satisfies the following conditions: $0\le u\in L^\infty({\bf R}^2)$, 
\begin{equation}
u({\bf r})=0\quad\mbox{for}\ \ |{\bf r}|\ge r_u\quad\mbox{with a constant}\ \ 
r_u\in\left(\sqrt{3}a/3,\sqrt{3}a/2\right),
\label{uout}
\end{equation}
and 
\begin{equation}
u({\bf r})\ge u_0>0\quad \mbox{for}\ {\bf r}\ \mbox{in the face of the hexagon  
with the center ${\bf r}=0$ of gravity}.
\label{unoflat}
\end{equation}
Here $u_0$ is a positive constant. 
The first condition (\ref{uout}) implies that the single-site potentials $u$ 
has compact support, and overlap with only nearest neighbor $u$. 
The next condition (\ref{unoflat}) 
implies that the whole space ${\bf R}^2$ is covered by the supports of 
the bumps $\{u(\cdot-{\bf z})\}_{{\bf z}\in{\bf L}^2}$ of the impurities so that 
\begin{equation}
\sum_{{\bf z}\in{\bf L}^2}u({\bf r}-{\bf z})\ge u_0
\quad\mbox{for any ${\bf r}\in{\bf R}^2$}.
\end{equation}
We should remark that this ``covering condition" is needed for 
estimating the number of the localized states and for applying 
the fractional moment method \cite{AENSS}, which yields 
a decaying bound for the resolvent.     
 
Clearly the random potential $V_\omega$ of (\ref{Vrandom}) 
does not necessarily satisfy the periodic boundary condition, 
\begin{equation}
V_\omega(x+L_x,y)=V_\omega(x,y+L_y)=V_\omega(x,y), 
\label{PBCVomega}
\end{equation}
without a special relation between the lattice constant $a$ and 
the system sizes $L_x,L_y$. 
Therefore we will replace the random potential $V_\omega$ with ${\tilde V}_\omega$ 
which is slightly different from $V_\omega$ in a neighborhood of the boundaries 
so that ${\tilde V}_\omega$ satisfies the periodic boundary condition (\ref{PBCVomega}).
Before proceeding further, we check that the boundary effect due to 
this procedure is almost negligible and does not affect the following argument. 
Write   
\begin{equation}
L_x^{\rm P}/2=N_xa\quad \mbox{and}\quad 
L_y^{\rm P}/2=N_y\cdot\sqrt{3}a/2
\quad\mbox{with positive integers}\ N_x, N_y. 
\end{equation}
When we take the sizes to be $L_x=L_x^{\rm P}$ and $L_y=L_y^{\rm P}$,  
the periodic boundary condition is automatically satisfied 
without replacing the random potential. However, for a given lattice constant $a$, 
the sizes do not necessarily satisfy the flux quantization condition, 
$L_x^{\rm P}L_y^{\rm P}=2\pi M\ell_B^2$, which we need in the following 
argument. Here $M$ is a positive integer, and 
$\ell_B$ is the so-called magnetic length defined as 
$\ell_B:=\sqrt{\hbar/(eB)}$. In a generic situation, we have 
\begin{equation}
2\pi M\ell_B^2<L_x^{\rm P}L_y^{\rm P}<2\pi(M+1)\ell_B^2\quad
\mbox{with some positive integer }\ M.
\end{equation}
In order to recover the flux quantization condition, we change the size a little bit 
in the $y$ direction. Namely we choose the sizes as 
$L_x=L_x^{\rm P}$ and $L_y=L_y^{\rm P}-\delta L_y$ with a small $\delta L_y$. 
Substituting this into $L_xL_y=2\pi M\ell_B^2$, we have 
\begin{equation}
0<\delta L_y<{2\pi\ell_B^2}/{L_x}.
\label{deltaLybound} 
\end{equation}
Notice $L_y<L_y^{\rm P}$, and consider the triangles 
which overlap with the upper boundary of $\Lambda^{\rm sys}$. 
We replace these equilateral triangles with the isosceles triangles 
of the height $(\sqrt{3}a/2-\delta L_y)$. 
{From} the above bound (\ref{deltaLybound}) for $\delta L_y$, 
the height of the isosceles triangles is slightly shorter than  
the height $\sqrt{3}a/2$ of the equilateral triangles 
for a sufficiently large $L_x$. In the same way as in (\ref{Vrandom}), 
we put the impurity potentials on the center of the gravity of the isosceles triangles. 
Because of the bound (\ref{deltaLybound}) for $\delta L_y$, 
the effect of this procedure is negligibly small for a sufficiently large size $L_x$. 
When the boundary effect plays an essential role as 
in Section~\ref{IndexApproach} below, 
we denote by ${\tilde V}_{\omega,\Lambda}$ 
this resulting random potential for a region $\Lambda$.  
Otherwise, we will often use the same notation $V_\omega$ for short.  

When ${\bf A}_{\rm P}=0$, we require the differentiability, 
$V=V_0+V_\omega\in C^2$, in addition to the above conditions,   
in order to obtain the exponential decay bound for 
the resolvent $(H_\omega-z)^{-1}$ in Appendix~\ref{appendix:decayresolventLandau}.   

As mentioned above, we require the flux quantization condition, 
$L_xL_y=2\pi M\ell_B^2$, with a sufficiently large positive integer 
$M$. The number $M$ is exactly equal to 
the number of the states in a single Landau level of the single-electron 
Hamiltonian in the simple uniform magnetic field with no electrostatic potential. 
This condition $L_xL_y=2\pi M\ell_B^2$ for the sizes $L_x,L_y$ is 
convenient for imposing the following periodic boundary conditions:
For an electron wavefunction $\varphi$, we impose 
periodic boundary conditions, 
\begin{equation}
t^{(x)}(L_x)\varphi({\bf r})=\varphi({\bf r}) \quad\mbox{and}\quad
t^{(y)}(L_y)\varphi({\bf r})=\varphi({\bf r}),
\label{singlePBC}
\end{equation}
where $t^{(x)}(\cdots)$ and $t^{(y)}(\cdots)$ are magnetic translation 
operators \cite{Zak} defined as 
\begin{equation}
t^{(x)}(x')f(x,y)=f(x-x',y), \qquad t^{(y)}(y')f(x,y)
=\exp[iy'x/\ell_B^2]f(x,y-y')
\label{magnetictranslation}
\end{equation}
for a function $f$ on ${\bf R}^2$. 

In order to measure the conductance, we introduce 
the time-dependent vector field ${\bf A}_{\rm ex}(t)=(0,\alpha(t))$ 
with 
\begin{equation}
\alpha(t)=-Ft\times\cases{e^{\eta t}, & $t\le 0$;\cr
                           1, & $t>0$.\cr}
\label{alphat}
\end{equation}
Here $F$ is the strength of the electric field, and $\eta>0$ 
is a small adiabatic parameter. The $y$-component of 
the corresponding external electric field 
is given by 
\begin{equation}
E_{{\rm ex},y}(t)=-\frac{\partial}{\partial t}\alpha(t)
=\cases{F(1+\eta t)e^{\eta t}, & $t\le 0$;\cr
                           F, & $t>0$.\cr}
\end{equation}
The time-dependent Hamiltonian is given by 
\begin{equation}
H_\omega(t)=\frac{1}{2m_e}[{\bf p}+e{\bf A}+e{\bf A}_{\rm ex}(t)]^2
+V_0+V_\omega. 
\end{equation}
The velocity operator is 
\begin{equation}
{\bf v}(t)=(v_x(t),v_y(t))=\frac{1}{m_e}[{\bf p}+e{\bf A}+e{\bf A}_{\rm ex}(t)].
\label{v(t)}
\end{equation}
Let $U(t,t_0)$ be the time evolution operator. We choose 
the initial time $t=t_0=-T$ with a large $T>0$. 
Then the total current density is given by 
\begin{equation}
{\bf j}_{\rm tot}(t)=(j_{{\rm tot},x}(t),j_{{\rm tot},y}(t))=-\frac{e}{L_xL_y}
{\rm Tr}\ U^\dagger(t,t_0){\bf v}(t)U(t,t_0)P_{\rm F}
\quad\mbox{for }\ t\ge 0, 
\end{equation}
where $P_{\rm F}$ is the projection on energies smaller than the 
Fermi energy $E_{\rm F}$. This total current density is decomposed into \cite{Koma3} 
the initial current density ${\bf j}_0$ and the induced current density 
${\bf j}_{\rm ind}(t)$ as ${\bf j}_{\rm tot}(t)={\bf j}_0+{\bf j}_{\rm ind}(t)$. 
Here the initial current density ${\bf j}_0$ is given by 
\begin{equation}
{\bf j}_0=-\frac{e}{L_xL_y}
{\rm Tr}\ \frac{1}{m_e}[{\bf p}+e{\bf A}]P_{\rm F}.
\end{equation}
Further the induced current density ${\bf j}_{\rm ind}(t)$ is decomposed into 
the linear part and the nonlinear part in the strength $F$ as  
\begin{equation}
{\bf j}_{\rm ind}(t)=(\sigma_{{\rm tot},xy}(t),\sigma_{{\rm tot},yy}(t))F
+{\bf j}_{\rm ind}^{\ \prime}(t)\quad \mbox{with}\quad 
{\bf j}_{\rm ind}^{\ \prime}(t)=o(F),
\label{currentinddecom}
\end{equation}
where the coefficients, $\sigma_{{\rm tot},sy}(t)$ for $s=x,y$, of 
the linear term are the total conductance which are written 
\begin{equation}
\sigma_{{\rm tot},sy}(t)=\sigma_{sy}+\delta\sigma_{sy}(t)  
\end{equation}
with the small corrections, $\delta\sigma_{sy}(t)$, due to the initial 
adiabatic process, and $o(F)$ denotes a quantity $q$ satisfying 
$q/F\rightarrow 0$ as $F\rightarrow 0$.  
Since the present system has no electron-electron interaction, 
the order estimate for the nonlinear part ${\bf j}_{\rm ind}^{\ \prime}(t)$ 
in (\ref{currentinddecom}) holds also in the infinite volume limit \cite{Koma3,BGKS,Kato}
with the same form of the linear part of the induced current. 
But the nonlinear part ${\bf j}_{\rm ind}^{\ \prime}(t)$ depends on 
the adiabatic parameters $\eta$ and $T$. Therefore we cannot take 
the adiabatic limit $T\uparrow \infty$ and $\eta\downarrow 0$ for 
the nonlinear part ${\bf j}_{\rm ind}^{\ \prime}(t)$ of the induced current. 

Now we describe our main theorems. Let $\nu=N/M$ be the filling factor of 
the electrons for a finite volume, where $N$ is the number of the electrons, 
and write $\omega_c=eB/m_e$ for the cyclotron frequency.
Consider first the case with ${\bf A}_{\rm P}=0$ in the infinite volume limit.  

\begin{theorem}
\label{maintheorem1}
Assume that the filling factor $\nu$ satisfies $n-1<\nu\le n$ 
with a positive integer $n$. 
Then there exist positive constants, $B_0(n)$ and $v_0$, such that 
there appear localized states of the bulk order around 
the energy ${\cal E}_{n-1}=(n-1/2)\hbar\omega_c$, i.e., the $n$-th Landau band 
center, for any magnetic field $B>B_0(n)$ and for any potential $V_0\in C^2$ 
satisfying $\Vert V_0^+\Vert_\infty+\Vert V_0^-\Vert_\infty\le v_0$. 
Further, when the Fermi level lies in the localization regime, 
the conductances are quantized as 
\begin{equation}
\sigma_{xy}=-\frac{e^2}{h}\times\cases{n & for the upper localization regime \cr 
                   (n-1) & for the lower localization regime \cr}
\quad \mbox{and}\quad \sigma_{yy}=0,
\label{conductancequantization}
\end{equation} 
and exhibit the plateaus for varying the filling factor. With probability one, 
there exist positive constants, ${\cal C}_j(\omega)<\infty$, $j=1,2,3$, such that  
the corrections $\delta\sigma_{sy}(t)$ due to the initial adiabatic 
process satisfy  
\begin{equation}
\left|\delta\sigma_{sy}(t)\right|\le 
[{\cal C}_1(\omega)+{\cal C}_2(\omega)T]e^{-\eta T}
+{\cal C}_3(\omega)\eta^{1/13},
\label{sigmacorrections} 
\end{equation}
and that the expectation ${\bf E}[{\cal C}_j]$ of 
the positive constants ${\cal C}_j(\omega)$ 
is finite for $j=1,2,3$, i.e., ${\bf E}[{\cal C}_j]<\infty$. 
Here the constant ${\cal C}_j(\omega)$ itself without the expectation may 
depend on the random event $\omega$ of the random potential $V_\omega$. 
\end{theorem}  

In the case with ${\bf A}_{\rm P}\ne 0$ in the infinite volume limit, 
we require a strong disorder because of a technical reason. 
(See Appendix~\ref{section:decayR} for details.) 
We take $u=\hbar\omega_c{\hat u}$ with a fixed, dimensionless function 
${\hat u}$ for the random potential $V_\omega$ of (\ref{Vrandom}). 
This potential behaves as $\Vert u\Vert_\infty\sim{\rm Const.}\times B$ 
for a large $B$. For this random potential, we have:    

\begin{theorem}
\label{maintheorem2}
Assume that the filling factor $\nu$ satisfies $n-1<\nu\le n$. 
Then there exist positive constants, $B_0(n),\alpha_0(n)$ and $w_0$, such that 
there appear localized states of the bulk order 
around the energy ${\cal E}_{n-1}=(n-1/2)\hbar\omega_c$
for any magnetic field $B>B_0(n)$ and for any vector potential 
${\bf A}_{\rm P}$ satisfying $\Vert|{\bf A}_{\rm P}|\Vert_\infty\le 
\alpha_0(n)B^{1/2}$ 
and for the function ${\hat u}$ satisfying $\Vert{\hat u}\Vert_\infty\le w_0$. 
Further, when the Fermi level lies in the localization regime, 
the conductances, $\sigma_{xy}$ and $\sigma_{yy}$, 
are quantized as in (\ref{conductancequantization}) 
and exhibit the plateaus for varying the filling factor. 
With probability one, the corrections $\delta\sigma_{sy}(t)$ 
due to the initial adiabatic process satisfy 
a bound having the same form as that of the bound (\ref{sigmacorrections}). 
\end{theorem}

\noindent
{\bf Remark:} 
1. In the conditions in the first theorem, we can also take $u=\hbar\omega_c{\hat u}$ 
for the random potential with a small norm $\Vert{\hat u}\Vert_\infty$. 
\smallskip

\noindent
2. In the conditions in the second theorem, we can take $V_0$ which behaves as  
$\Vert V_0\Vert_\infty\sim{\rm Const.}\times B$ for a large $B$ with a small 
positive constant, instead of a fixed potential $V_0$.  

\noindent
3. These two theorems do not necessarily state that both of 
the upper and lower band edges exhibit the localized states of the 
bulk order. Namely a localization region of the bulk order may appears 
only at one side for a single Landau band.   
\smallskip

\noindent
4. We do not require the assumption that the $n$-th Landau band is separated 
from the rest of the spectrum by two spectral gaps. 
\smallskip

\noindent
5. In the previous analyses \cite{CH,Wang,GK,GKS}, they considered the Hamiltonian 
having the form with ${\bf A}_{\rm P}=0$ and $V_0=0$. The analyses rely on 
the special properties of the unperturbed Hamiltonian. For example, they use 
the explicit forms of the integral kernel of the projections onto 
the Landau levels. The extension to the case 
with ${\bf A}_{\rm P}\ne 0$ and $V_0\ne 0$ needs additional, non-trivial analyses 
for localization.   
In addition, we do not require a periodicity of the potentials ${\bf A}_{\rm P},V_0$ 
with a finite period. Therefore the Hamiltonian $H_\omega$ does not need to be 
translation ergodic. 
\smallskip

\noindent
6. The widths of the plateaus can be estimated as we will show 
in Section~\ref{Widths} below. In particular, when ${\bf A}_{\rm P}=0$ and $V_0=0$, 
the ratio of the localized states to the total number $M$ of the states 
in the single Landau level tends to one as the strength $B$ of the magnetic field 
goes to infinity. This implies that our estimate 
for the widths of the plateaus shows the optimal, expected 
value in this limit.  

%%%%%%%%%%%%%%%%%%%%%%%%%%%%%%%%%%%%%%%%%%%%%
\Section{Spectral gaps of the Hamiltonian $H_0$}
\label{spectrumH0}

In order to show the localization for the disordered-broadened Landau bands, 
we first need to check the condition for the appearance of the spectral gaps 
in the spectrum of the Hamiltonian $H_0$ 
of (\ref{Ham0}) with a generic, bounded potential $V_0$. 

First we recall the simplest Landau Hamiltonian for a single electron 
only in the uniform magnetic field. The Hamiltonian is given by   
\begin{equation}
H_{\rm L}=\frac{1}{2m_e}({\bf p}+e{\bf A}_0)^2.
\label{hamHL} 
\end{equation}
We assume that the electron is confined to 
the same finite rectangular box $\Lambda^{\rm sys}$ 
of (\ref{Lamsys}) as the box for the Hamiltonian $H_\omega$ of 
(\ref{Hamomega}), and impose the periodic boundary conditions (\ref{singlePBC}) 
for the wavefunctions with the flux quantization condition $L_xL_y=2\pi M\ell_B^2$.  
Then the energy eigenvalues ${\cal E}_n$ of $H_{\rm L}$ 
are given by\footnote{See, for example, Refs.~\cite{Koma2,Koma1}.}
\begin{equation}
{\cal E}_n:=\left(n+\frac{1}{2}\right)\hbar\omega_c
\quad\mbox{for }\ n=0,1,2,\ldots.
\label{pureLandauenergy}
\end{equation}

The Hamiltonian $H_0$ of (\ref{Ham0}) on the finite box $\Lambda^{\rm sys}$ is written  
\begin{eqnarray}
H_0&=&\frac{1}{2m_e}({\bf p}+e{\bf A}_0+e{\bf A}_{\rm P})^2+V_0\ret
   &=&H_{\rm L}+\frac{e}{2m_e}{\bf A}_{\rm P}\cdot({\bf p}+e{\bf A}_0)+
\frac{e}{2m_e}({\bf p}+e{\bf A}_0)\cdot{\bf A}_{\rm P}
+\frac{e^2}{2m_e}\left|{\bf A}_{\rm P}\right|^2+V_0.
\end{eqnarray}
Using the Schwarz inequality, one has   
\begin{equation}
\left|\left(\psi,{\bf A}_{\rm P}\cdot({\bf p}+e{\bf A}_0)\psi\right)\right|
\le\left\Vert|{\bf A}_{\rm P}|\right\Vert_\infty
\sqrt{\left(\psi,({\bf p}+e{\bf A}_0)^2\psi\right)}
\end{equation}
for the normalized vector $\psi$ in the domain of the Hamiltonian. From this inequality, 
the energy expectation can be evaluated as  
\begin{eqnarray}
\left(\psi,H_0\psi\right)&\le&\left(\psi,H_{\rm L}\psi\right)
+\frac{\sqrt{2}e}{\sqrt{m_e}}\left\Vert|{\bf A}_{\rm P}|\right\Vert_\infty
\sqrt{\left(\psi,H_{\rm L}\psi\right)}+\frac{e^2}{2m_e}
\left\Vert|{\bf A}_{\rm P}|\right\Vert_\infty^2+\left\Vert V_0^+\right\Vert_\infty
\end{eqnarray}
and 
\begin{eqnarray}
\left(\psi,H_0\psi\right)&\ge&\left(\psi,H_{\rm L}\psi\right)
-\frac{\sqrt{2}e}{\sqrt{m_e}}\left\Vert|{\bf A}_{\rm P}|\right\Vert_\infty
\sqrt{\left(\psi,H_{\rm L}\psi\right)}
-\left\Vert V_0^-\right\Vert_\infty,
\label{lowerboundexpectation}
\end{eqnarray}
where $V_0^{\pm}=\max\left\{\pm V_0,0\right\}$. 
Let us denote by ${\cal E}_{n,+}^{\rm edge}$ and ${\cal E}_{n,-}^{\rm edge}$, 
respectively,  
the upper and lower edges of the $n+1$-th Landau band which is broadened 
by the potentials $V_0$ and ${\bf A}_{\rm P}$. 
{From} the standard argument about the min-max principle,\footnote{See, for 
example, Section~XIII.1 of the book \cite{ReedSimonIV} by M.~Reed and B.~Simon.}
one has 
\begin{equation}
{\cal E}_{n,+}^{\rm edge}\le {\cal E}_n
+\frac{\sqrt{2}e}{\sqrt{m_e}}\left\Vert|{\bf A}_{\rm P}|\right\Vert_\infty
\sqrt{{\cal E}_n}+\frac{e^2}{2m_e}
\left\Vert|{\bf A}_{\rm P}|\right\Vert_\infty^2+\left\Vert V_0^+\right\Vert_\infty
\label{Eedge+}
\end{equation}
for $n=0,1,2,\ldots$. For the lower edge, we assume 
\begin{equation}
\frac{e}{\sqrt{2m_e}}\left\Vert|{\bf A}_{\rm P}|\right\Vert_\infty
\le \sqrt{\frac{1}{2}\hbar\omega_c}.
\label{monotone}
\end{equation}
Then the right-hand side of the bound (\ref{lowerboundexpectation}) 
is a strictly monotone increasing function of the expectation
$(\psi,H_{\rm L}\psi)$. Therefore, the same argument yields 
\begin{equation}
{\cal E}_{n,-}^{\rm edge}\ge {\cal E}_n
-\frac{\sqrt{2}e}{\sqrt{m_e}}\left\Vert|{\bf A}_{\rm P}|\right\Vert_\infty
\sqrt{{\cal E}_n}-\left\Vert V_0^-\right\Vert_\infty
\label{Eloweredgebound} 
\end{equation}
for $n=0,1,2,\ldots$. If this right-hand side with the index $n+1$ is strictly 
larger than the right-hand side of (\ref{Eedge+}) with the index $n$, 
then there exists a spectral gap above the Landau band with the index $n$, i.e., 
${\cal E}_{n+1,-}^{\rm edge}>{\cal E}_{n,+}^{\rm edge}$. This gap condition 
can be written as   
\begin{equation}
\hbar\omega_c>
\frac{\sqrt{2}e}{\sqrt{m_e}}\left\Vert|{\bf A}_{\rm P}|\right\Vert_\infty
\left(\sqrt{{\cal E}_{n+1}}+\sqrt{{\cal E}_n}\right)
+\frac{e^2}{2m_e}
\left\Vert|{\bf A}_{\rm P}|\right\Vert_\infty^2+\left\Vert V_0^+\right\Vert_\infty
+\left\Vert V_0^-\right\Vert_\infty.
\label{gapcondition} 
\end{equation}
Clearly this is stronger than the condition (\ref{monotone}) for 
the vector potential ${\bf A}_{\rm P}$. Therefore we have no need to take 
into account the condition (\ref{monotone}). 

%%%%%%%%%%%%%%%%%%%%%%%%%%%%%%%%%%%%%%%%%%%%%
\Section{Site percolation on the triangular lattice}
\label{percolation}

The classical motion of the electron is forbidden 
in the regions that the strength of the potential is smaller than 
the deviation of the energy of the electron from the Landau 
energies ${\cal E}_n$ of (\ref{pureLandauenergy}). 
In those regions, the Green function of the electron decays exponentially. 
In order to get the decay bound for the Green function, 
we study the distribution of those classically forbidden regions. 
We reformulate this problem 
as a site percolation problem on the triangular lattice. 
The idea of using percolation is due to Combes and Hislop \cite{CH} 
or Wang \cite{Wang}. But both of their random potentials are different from 
the present potential which we require 
for estimating the number of the localized states.   

We begin with setting up site percolation on the triangular lattice ${\bf L}^2$ 
for the present random potential. 
We say that the site ${\bf z}\in{\bf L}^2$ is occupied if 
$\lambda_{\bf z}(\omega)\in(-\lambda_-,\lambda_+)$. 
The probability $p$ that a site ${\bf z}$ is occupied is given by 
\begin{equation}
p=\int_{-\lambda_-}^{\lambda_+}g(\lambda)d\lambda.
\label{p} 
\end{equation}
The assumption (\ref{assumptiong}) implies $p>p_c=1/2$. Here $p_c$ is the critical 
probability which equals $1/2$ for the present site percolation 
on the triangular lattice \cite{Kesten,Grimmett}. 
A path of ${\bf L}^2$ is a sequence ${\bf z}_0,{\bf z}_1,\ldots,{\bf z}_n$ of 
sites ${\bf z}_j$ such that all of the adjacent two site 
${\bf z}_j,{\bf z}_{j+1}$ are corresponding to a side of a unit triangle. 
If ${\bf z}_0={\bf z}_n$, then we say that the path is closed, and we call a closed 
path a circuit.   
If all of the site ${\bf z}_j$ of the path are occupied, then we say  
that the path is occupied. Similarly we define an unoccupied path, 
an occupied circuit, etc. We denote by $P_p(A)$ the probability that 
an event $A$ occurs. 

Let $\Pi_{\ell,\ell'}$ be a parallelogram with the lengths $\ell a,\ell' a$ of 
the sides in the triangular lattice. See Fig.~\ref{trianglelattice}. 
More precisely, it is given by 
\begin{equation}
\Pi_{\ell,\ell'}:=\left\{m{\bf a}_1+n{\bf a}_2\ \left|\ |m|\le\frac{\ell}{2},
|n|\le \frac{\ell'}{2},
(m,n)\in{\bf Z}^2\right.\right\}.
\end{equation}
Here we take $\ell,\ell'$ even integers for simplicity. 
Consider the event ${\bar A}_{\ell,\ell'}$ 
that there exists a unoccupied path in the parallelogram $\Pi_{\ell,\ell'}$  
joining a site on the lower side with the length $\ell$ to a site on 
the upper side. Since the connectivity between 
two sites with an unoccupied path decays exponentially for 
$p>p_c$, the probability $P_p({\bar A}_{\ell,\ell'})$ that 
the event ${\bar A}_{\ell,\ell'}$ occurs is bounded as 
\begin{equation}
P_p({\bar A}_{\ell,\ell'})\le {\rm Const.}\times \ell\exp[-m_p\ell']
\quad\mbox{for}\ p>p_c=1/2,
\end{equation}
where $m_p$ is a positive function of $p$. 
Let $A_{\ell',\ell}$ be the event that there exists an occupied path 
in the parallelogram $\Pi_{\ell,\ell'}$  
joining a site on the left side with the length $\ell'$ to a site on 
the right side. Then this event $A_{\ell',\ell}$ is the complementary 
event of ${\bar A}_{\ell,\ell'}$ because of the structure of the triangular lattice.  
Immediately,   
\begin{equation}
P_p(A_{\ell',\ell})+P_p({\bar A}_{\ell,\ell'})=1.  
\end{equation}
Combining this with the above inequality, one has 
\begin{equation}
P_p(A_{\ell',\ell})\ge 1-{\rm Const.}\times \ell\exp[-m_p\ell']
\quad\mbox{for}\ p>p_c=1/2.
\label{PpAellellbound}
\end{equation}

Consider a parallelogram-shaped region consisting of hexagons such that   
\begin{equation}
\Lambda_{\ell+1,\ell'+1}^{\rm para}({\bf z}_0)
:=\bigcup_{{\bf z}\in \Pi_{\ell,\ell'}}h_{{\bf z}+{\bf z}_0},
\label{paralleloLambda}
\end{equation}
where $h_{\bf z}$ is the region of the face of the hexagon with  
the center ${\bf z}$, including the six sides of the hexagon, and 
${\bf z}_0$ is the center of the region 
$\Lambda_{\ell+1,\ell'+1}^{\rm para}({\bf z}_0)$. 
The region has the jagged boundary as in Fig.~\ref{trianglelattice}. 
Moreover we define an annular region as 
\begin{equation}
\Lambda_{3\ell,3\ell'}^{\rm annu}({\bf z}_0)
:=\Lambda_{3\ell,3\ell'}^{\rm para}({\bf z}_0)\backslash
\Lambda_{\ell,\ell'}^{\rm para}({\bf z}_0),
\end{equation}
where both $\ell$ and $\ell'$ take an odd integer. 

Let us consider the event $D_{\ell,\ell'}$ in 
$\Lambda_{3\ell,3\ell'}^{\rm annu}({\bf z}_0)\cap{\bf L}^2$ that there exists 
an occupied circuit ${\cal C}$ encircling the inside region 
$\Lambda_{\ell,\ell'}^{\rm para}({\bf z}_0)$.  
Then, from the inequality (\ref{PpAellellbound}) and 
FKG inequality,\footnote{See, for example, the book \cite{Grimmett}.}  
the probability $P_p(D_{\ell,\ell'})$ for this event satisfies 
\begin{eqnarray}
P_p(D_{\ell,\ell'})&\ge& \left[P_p(A_{\ell',3\ell})\right]^2
\left[P_p(A_{\ell,3\ell'})\right]^2\ret
&\ge& 1-{\rm Const.}\times \left\{\ell\exp[-m_p\ell']
+\ell'\exp[-m_p\ell]\right\}\ret
& &\quad \mbox{for}\ p>p_c=1/2.
\label{PpDbound}
\end{eqnarray}
Therefore the event of an occupied circuit occurs with 
the probability nearly equal to one for large $\ell,\ell'$. 

We denote by $b_{j,j+1}$ the side $\overline{{\bf z}_j{\bf z}_{j+1}}$ of 
a unit triangle, i.e.,  
\begin{equation}
b_{j,j+1}:=\{{\bf r}=\lambda{\bf z}_j+(1-\lambda){\bf z}_{j+1}|\> \lambda\in
[0,1]\}.
\end{equation}
We define the region ${\cal R}_{j,j+1}$ including the side $b_{j,j+1}$ as 
\begin{equation}
{\cal R}_{j,j+1}:=\left\{{\bf r}|{\rm dist}({\bf r},b_{j,j+1})\le r_1\right\}
\quad\mbox{with}\quad r_1:=\frac{\sqrt{3}}{2}a-r_u,
\label{widthcond}
\end{equation}
where $r_u$ is given in the condition (\ref{uout}) for the bump $u$. 
Further we define the ribbon region ${\cal R}_{\cal C}$ 
associated a circuit ${\cal C}$ by 
\begin{equation}
{\cal R}_{\cal C}:=\bigcup_{b_{j,j+1}\in{\cal C}}{\cal R}_{j,j+1}. 
\end{equation} 
Clearly $r_1$ is strictly positive from the condition 
$r_u\in(\sqrt{3}a/3,\sqrt{3}a/2)$ of (\ref{uout}), 
and the ribbon region ${\cal R}_{\cal C}$ has a nonzero width $2r_1$. 

\begin{pro}
\label{pro:occucircuit}
There appears an occupied circuit ${\cal C}$ in the annular 
region $\Lambda_{3\ell,3\ell'}^{\rm annu}({\bf z}_0)$
with a probability larger than 
\begin{equation}
P^{\rm perc}:=1-C^{\rm perc}\left\{\ell\exp[-m_p\ell']
+\ell'\exp[-m_p\ell]\right\},
\label{Pperc}
\end{equation}
where $C^{\rm perc}$ is the positive constant in the right-hand side of 
the above bound (\ref{PpDbound}), and $p>p_c=1/2$ is given by (\ref{p}). 
Further the following hold for 
the ribbon region ${\cal R}_{\cal C}$ associated 
with the occupied circuit ${\cal C}$:   
\begin{equation}
{\rm dist}\left({\cal R}_{\cal C},
\partial\Lambda_{3\ell,3\ell'}^{\rm annu}({\bf z}_0)\right)
=r_u-\frac{\sqrt{3}}{3}a=:r_2>0
\label{defr2}
\end{equation}
and 
\begin{equation}
-\lambda_-u_1\le V_\omega({\bf r})\le \lambda_+u_1\quad 
\mbox{for}\ {\bf r}\in{\cal R}_{\cal C}.
\label{restrictVomega} 
\end{equation}
Here $\partial\Lambda_{3\ell,3\ell'}^{\rm annu}({\bf z}_0)$ is the boundary 
of the annular region $\Lambda_{3\ell,3\ell'}^{\rm annu}({\bf z}_0)$, 
and $u_1:=2\Vert u\Vert_\infty$.
\end{pro}

\begin{proof}{Proof}
The lower bound (\ref{Pperc}) of the probability is nothing but 
the right-hand side of (\ref{PpDbound}). 
The positivity (\ref{defr2}) of $r_2$ follows from the condition (\ref{uout}) of 
the bump $u$ of the random potential $V_\omega$, 
and the bound (\ref{restrictVomega}) follows from the condition (\ref{uout}) and 
the definition (\ref{Vrandom}) of the random potential $V_\omega$. 
\end{proof}

%%%%%%%%%%%%%%%%%%%%%%%%%%%%%%%%%%%%%%%%%%%%%
\Section{Initial decay estimate for the resolvent}
\label{IDER}

Now let us estimate the decay of the resolvent (Green function) 
for a finite parallelogram-shaped region. The resulting decay bound in 
Proposition~\ref{proinidecay} below will become 
the initial data for the multi-scale analysis to obtain 
the decay bounds for the resolvent in larger scales in the next section. 
However, by using the multi-scale analysis, 
we cannot get a similar decay bound for 
the resolvent for two arbitrary points in the infinite-volume limit. 
On the other hand, the fractional moment method leads us to 
a decay bound for a fractional moment of the resolvent in 
the infinite-volume limit. Actually, as we will see in Section~\ref{FMB}, 
the initial decay estimate of this section yields such a decay bound. 
But the resolvent itself without taking a fractional moment cannot be 
evaluated by the method \cite{AENSS}. 
Due to technical reason related to these observations, 
we need both multi-scale analysis and fractional moment analysis, 
in order to prove the existence of the conductance plateaus with 
a bulk order width. 

Although the method in this section is basically the same as in the previous papers 
\cite{CH,Wang} as mentioned in the preceding section, 
we need more detailed analysis about the magnetic field dependence of the 
decay bounds, in order to estimate the number of the localized states which 
yield the Hall conductance plateau with a bulk order width. 

Fix the random variables $\lambda_{{\bf z}+{\bf z}_0}$ with 
${\bf z}\in{\bf L}^2\backslash\Pi_{3\ell-1,3\ell'-1}$. 
Here $\ell$ and $\ell'$ are odd integers larger than 1. 
Consider the parallelogram-shaped region $\Lambda_{3\ell,3\ell'}^{\rm para}({\bf z}_0)$ 
centered at ${\bf z}_0$, and 
assume $\Lambda_{3\ell,3\ell'}^{\rm para}({\bf z}_0)\subset\Lambda^{\rm sys}$ 
with a sufficiently large box $\Lambda^{\rm sys}$ of (\ref{Lamsys}). 
We write $\Lambda_{3\ell,3\ell'}=\Lambda_{3\ell,3\ell'}^{\rm para}({\bf z}_0)$ for short. 
Further we consider the Hamiltonian $H_\omega$ restricting to the region 
$\Lambda_{3\ell,3\ell'}$ with the Dirichlet boundary conditions. 
The Hamiltonian is written as 
\begin{equation}
H_{\Lambda_{3\ell,3\ell'}}
=\frac{1}{2m_e}({\bf p}+e{\bf A})^2+\left.V_0\right|_{\Lambda_{3\ell,3\ell'}}+
{\hat V}_{\omega,3\ell,3\ell'}+\delta V_{\omega,3\ell,3\ell'},
\label{ham3ell3ell}
\end{equation}
where we have decomposed the random potential $V_\omega$ into two parts,  
\begin{equation}
{\hat V}_{\omega,3\ell,3\ell'}({\bf r})=\sum_{{\bf z}\in\Pi_{3\ell-1,3\ell'-1}}
\lambda_{{\bf z}+{\bf z}_0}(\omega)u({\bf r}-{\bf z}_0-{\bf z}). 
\end{equation}
and $\delta V_{\omega,3\ell,3\ell'}=
\left.V_\omega\right|_{\Lambda_{3\ell,3\ell'}}
-{\hat V}_{\omega,3\ell,3\ell'}$. 
Clearly, the first part ${\hat V}_{\omega,3\ell,3\ell'}$ of the random potential 
is determined by only the random variables $\lambda_{{\bf z}+{\bf z}_0}(\omega)$ 
at the sites ${\bf z}+{\bf z}_0$ lying in the parallelogram-shaped region 
$\Lambda_{3\ell,3\ell'}^{\rm para}({\bf z}_0)$, and so it is independent of 
the outside random variables. Further we have 
\begin{equation}
\sum_{{\bf z}\in\Pi_{3\ell-1,3\ell'-1}}
u({\bf r}-{\bf z}_0-{\bf z})\ge u_0\quad\mbox{for any }\ {\bf r}\in\Lambda_{3\ell,3\ell'} 
\end{equation}
{from} the assumption (\ref{unoflat}). This condition will be useful to 
obtain the Wegner estimate \cite{Wegner} for the density of the states. 
(See Appendix~\ref{appendix:Wegner} for the Wegner estimate.) 

On the other hand, the random potential $\delta V_{\omega,3\ell,3\ell'}$ which is 
supported by only the region near the boundary of $\Lambda_{3\ell,3\ell'}$, 
depends on the the outside random variables. 
Following \cite{CH}, we absorb this term into the operator $W(\chi)$ of 
(\ref{defW0}) below which will appear in the geometric resolvent equation. 
(See Appendix~\ref{MSAProofs} for details.) Thus we consider the Hamiltonian, 
\begin{equation}
{\hat H}_{\Lambda_{3\ell,3\ell'}}
=\frac{1}{2m_e}({\bf p}+e{\bf A})^2+\left.V_0\right|_{\Lambda_{3\ell,3\ell'}}+
{\hat V}_{\omega,3\ell,3\ell'},
\label{hamMSA}
\end{equation}
without the potential $\delta V_{\omega,3\ell,3\ell'}$, 
instead of the Hamiltonian $H_{\Lambda_{3\ell,3\ell'}}$ of (\ref{ham3ell3ell}).  

Assume that the energy $E\in{\bf R}$ satisfies the condition,
\begin{equation}
{\cal E}_{n,+}^{\rm edge}+\lambda_+u_1<E<
{\cal E}_{n+1,-}^{\rm edge}-\lambda_-u_1\quad\mbox{with }\ u_1=2\Vert u\Vert_\infty.
\label{energyEingap}  
\end{equation}
We write the resolvent as $R_{3\ell,3\ell'}=R_{3\ell,3\ell'}(E+i\varepsilon)
=\left({\hat H}_{\Lambda_{3\ell,3\ell'}}-E-i\varepsilon\right)^{-1}$ 
with $\varepsilon\in{\bf R}$. For $\delta\in(0,r_2)$, consider the region, 
$\Lambda_{3\ell,3\ell'}^\delta:=
\left\{\left.{\bf r}\in\Lambda_{3\ell,3\ell'}\right|
{\rm dist}({\bf r},\partial\Lambda_{3\ell,3\ell'})>\delta\right\}$, 
where $r_2$ is given by (\ref{defr2}), and $\partial\Lambda_{3\ell,3\ell'}$ is 
the boundary of the region $\Lambda_{3\ell,3\ell'}$. 
Let $\chi_{3\ell,3\ell'}^\delta$ be a $C^2$, positive cut-off function which satisfies 
\begin{equation}
\left.\chi_{3\ell,3\ell'}^\delta\right|_{\Lambda_{3\ell,3\ell'}^\delta}=1
\quad\mbox{and}\quad 
{\rm supp}\, \left|\nabla\chi_{3\ell,3\ell'}^\delta\right|\subset\Lambda_{3\ell,3\ell'}
\backslash\Lambda_{3\ell,3\ell'}^\delta.  
\end{equation}
We denote by $\chi_{\ell,\ell'}$ the characteristic function of 
the region $\Lambda_{\ell,\ell'}^{\rm para}({\bf z}_0)$.
 
The purpose of this section is to estimate the decay of 
$W(\chi_{3\ell,3\ell'}^\delta)R_{3\ell,3\ell'}\chi_{\ell,\ell'}$, where 
\begin{equation}
W(\chi)=\left[({\bf p}+e{\bf A})^2/(2m_e),\chi\right]
\label{defW0} 
\end{equation}
for a $C^2$ function $\chi$. Note that 
\begin{eqnarray}
\left\Vert W(\chi_{3\ell,3\ell'}^\delta)R_{3\ell,3\ell'}\chi_{\ell,\ell'}\right\Vert
&\le&\frac{\hbar^2}{2m_e}\left\Vert(\Delta\chi_{3\ell,3\ell'}^\delta)
R_{3\ell,3\ell'}\chi_{\ell,\ell'}\right\Vert\ret
&+&\frac{\hbar}{m_e}\sum_{i=x,y}\left\Vert(\partial_i\chi_{3\ell,3\ell'}^\delta)
(p_i+eA_i)R_{3\ell,3\ell'}\chi_{\ell,\ell'}\right\Vert,
\label{normWRchi}
\end{eqnarray}
where we have written $\nabla=(\partial_x,\partial_y)$, and used  
\begin{equation}
W(\chi)=-\frac{i\hbar}{m_e}({\bf p}+e{\bf A})\cdot\nabla\chi
+\frac{\hbar^2}{2m_e}\Delta\chi
=-\frac{i\hbar}{m_e}\nabla\chi\cdot({\bf p}+e{\bf A})
-\frac{\hbar^2}{2m_e}\Delta\chi.
\label{defW}
\end{equation}

We write ${\cal R}$ for the ribbon region ${\cal R}_{\cal C}$ 
in Proposition~\ref{pro:occucircuit} for short. 
Let $\epsilon$ be a small positive number, and let 
${\cal C}_\epsilon:=\{{\bf r}\in{\cal R}|{\rm dist}({\bf r},{\cal C})<\epsilon/2\}$, 
so that the region ${\cal C}_\epsilon$ has the width $\epsilon$, 
where ${\cal C}$ is the occupied, closed path in Proposition~\ref{pro:occucircuit}.
Let $\chi_1$ be a $C^2$, positive cut-off function satisfying 
$\chi_1\left|_{\Lambda_{\ell,\ell'}}\right.=1$ and 
${\rm supp}|\nabla\chi_1|\subset{\cal C}_\epsilon$, 
where we have written $\Lambda_{\ell,\ell'}$ for 
$\Lambda_{\ell,\ell'}^{\rm para}({\bf z}_0)$ for short. 
Since we can take $\chi_1$ to satisfy 
$(\partial_i\chi_{3\ell,3\ell'}^\delta)\chi_1=0$ from the definition 
of $\chi_{3\ell,3\ell'}^\delta$, one has  
\begin{equation}
(\partial_i\chi_{3\ell,3\ell'}^\delta)(p_i+eA_i)R_{3\ell,3\ell'}\chi_{\ell,\ell'}
=D_iR_{3\ell,3\ell'}\chi_1\chi_{\ell,\ell'}
=-D_iR_{3\ell,3\ell'}W(\chi_1)R_{3\ell,3\ell'}\chi_{\ell,\ell'},
\label{chi2Rchiell}
\end{equation}
where we have written $D_i=(\partial_i\chi_{3\ell,3\ell'}^\delta)(p_i+eA_i)$.  
In order to estimate this right-hand side, we define the $\epsilon$ border of 
the ribbon region ${\cal R}$ by 
${\cal R}_\epsilon:=\{{\bf r}\in{\cal R}|{\rm dist}({\bf r},\partial{\cal R})<\epsilon\}$, 
and define $r_3:={\rm dist}({\cal R}_\epsilon,{\cal C}_\epsilon)>0$. 
We choose a small parameter $\epsilon$ so that 
the distance $r_3$ becomes strictly positive. 
Further we introduce two $C^2$, positive cut-off functions, 
${\tilde \chi}_{\cal R}^{\epsilon/2}$ and $\chi_{\cal R}^\epsilon$, 
which satisfy the following conditions: 
\begin{equation}
\left.{\tilde \chi}_{\cal R}^{\epsilon/2}
\right|_{{\cal R}\backslash{\cal R}_{\epsilon/2}}=1,\quad
{\rm supp}\left|\nabla{\tilde \chi}_{\cal R}^{\epsilon/2}\right|
\subset{\cal R}_{\epsilon/2},
\end{equation}
and 
\begin{equation}
\left.\chi_{\cal R}^\epsilon\right|_{{\cal R}\backslash{\cal R}_{\epsilon}}=1,
\quad
\left.\chi_{\cal R}^\epsilon\right|_{{\cal R}_{\epsilon/2}}=0,
\quad
\mbox{and}\quad
{\rm supp}|\nabla\chi_{\cal R}^\epsilon|\subset{\cal R}_\epsilon\backslash
{\cal R}_{\epsilon/2}. 
\end{equation}
Consider the Hamiltonian, 
\begin{equation}
H_{\cal R}:=\frac{1}{2m_e}({\bf p}+e{\bf A})^2+V_{\cal R},
\label{hamR}
\end{equation}
on the finite rectangular box $\Lambda^{\rm sys}$ of (\ref{Lamsys}),  
where we impose the periodic boundary conditions (\ref{singlePBC}) 
with the flux quantization condition $L_xL_y=2\pi M\ell_B^2$,  
and the potential is given by 
$V_{\cal R}={\tilde \chi}_{\cal R}^{\epsilon/2}(V_0+V_\omega)$. 
Then one has the geometric resolvent equation, 
$R_{3\ell,3\ell'}\chi_{\cal R}^\epsilon=
\chi_{\cal R}^\epsilon R_{\cal R}-R_{3\ell,3\ell'}W(\chi_{\cal R}^\epsilon)R_{\cal R}$ 
with $R_{\cal R}:=(H_{\cal R}-E-i\epsilon)^{-1}$, where we have used 
${\tilde \chi}_{\cal R}^{\epsilon/2}\chi_{\cal R}^\epsilon=\chi_{\cal R}^\epsilon$ 
which is easily obtained from the definitions. 
Using this equation and $D_i\chi_{\cal R}^\epsilon=0$, 
the right-hand side of (\ref{chi2Rchiell}) is written as 
\begin{eqnarray}
-D_iR_{3\ell,3\ell'}W(\chi_1)R_{3\ell,3\ell'}
\chi_{\ell,\ell'}&=&-D_iR_{3\ell,3\ell'}\chi_{\cal R}^\epsilon 
W(\chi_1)R_{3\ell,3\ell'}\chi_{\ell,\ell'}\ret
&=&D_iR_{3\ell,3\ell'}W(\chi_{\cal R}^\epsilon)R_{\cal R}W(\chi_1)
R_{3\ell,3\ell'}\chi_{\ell,\ell'}.
\label{chi2RWRchi}
\end{eqnarray}
Consequently, one obtains 
\begin{equation}
D_iR_{3\ell,3\ell'}\chi_{\ell,\ell'}
=D_iR_{3\ell,3\ell'}W(\chi_{\cal R}^\epsilon)R_{\cal R}W(\chi_1)
R_{3\ell,3\ell'}\chi_{\ell,\ell'}.
\label{DRchi}
\end{equation}
In the same way,  
\begin{equation}
(\Delta\chi_{3\ell,3\ell'}^\delta)R_{3\ell,3\ell'}\chi_{\ell,\ell'}
=(\Delta\chi_{3\ell,3\ell'}^\delta)R_{3\ell,3\ell'}
W(\chi_{\cal R}^\epsilon)R_{\cal R}W(\chi_1)
R_{3\ell,3\ell'}\chi_{\ell,\ell'}.
\label{trchiRchi}
\end{equation}

If the random potential $V_\omega$ satisfies the condition (\ref{restrictVomega}), 
then the energy $E$ satisfying the condition (\ref{energyEingap}) is in the spectral 
gap of the Hamiltonian $H_{\cal R}$. 
Therefore we can apply the Combes-Thomas method \cite{CT} to evaluate decay of 
the resolvent $R_{\cal R}$ in the right-hand sides of (\ref{DRchi}) and 
(\ref{trchiRchi}). We write ${\cal A}={\rm supp}|\nabla\chi_{\cal R}^\epsilon|$ 
and ${\cal B}={\cal C}_\epsilon$, and denote 
by $\chi_{\cal A}$ and $\chi_{\cal B}$ the characteristic function 
of ${\cal A}$ and ${\cal B}$, respectively. 
The resolvent $R_{\cal R}$  decays as 
\begin{equation}
\left\Vert\chi_{\cal A}R_{\cal R}\chi_{\cal B}\right\Vert
\le C_1^{(n)}e^{-\beta r_3}
\label{Rbjdecay}
\end{equation}
with a probability larger than $P^{\rm perc}$ of (\ref{Pperc}) 
in Proposition~\ref{pro:occucircuit}, 
where $C_1^{(n)}$ and $\beta$ are positive constants. 
The derivation of this decay bound and the explicit parameter dependence 
of $C_1^{(n)}$ and of $\beta$ for the present model are given 
in Appendix~\ref{section:decayR}. In the application of the decay bound, 
we choose the cut-off function $\chi_{\cal R}^\epsilon$ so that 
the region ${\rm supp}|\nabla\chi_{\cal R}^\epsilon|$ has a smooth boundary. 

\begin{lemma}
\label{lemma:pRbounds}
Let $R=(H_\omega-E-i\varepsilon)^{-1}$ with generic, bounded potentials 
$V_0, V_\omega$ and $E,\varepsilon\in{\bf R}$ satisfying 
$E\notin\sigma(H_\omega)$ or $\varepsilon\ne 0$, and 
let $\mbox{\boldmath $\alpha$}=(\alpha_x,\alpha_y)$ be a vector-valued $C^1$ function. 
Then
\begin{equation}
\Vert\mbox{\boldmath $\alpha$}\cdot({\bf p}+e{\bf A})R\Vert
\le 2\sqrt{2m_e}\Vert R\Vert^{1/2}(1+f_{E,R})^{1/2}
\max_{i=x,y}\{\Vert\alpha_i\Vert_\infty\},
\label{chipRbound}
\end{equation}
\begin{equation}
\left\Vert(p_i+eA_i)R({\bf p}+e{\bf A})\cdot\mbox{\boldmath $\alpha$}\right\Vert\le
2m_e\Vert|\mbox{\boldmath $\alpha$}|\Vert_\infty(1+f_{E,R}) 
\label{pRpchibound}
\end{equation}
and 
\begin{equation}
\left\Vert R
({\bf p}+e{\bf A})\cdot\mbox{\boldmath $\alpha$}\right\Vert
\le\sqrt{2m_e}\Vert|\mbox{\boldmath $\alpha$}|\Vert_\infty
\Vert R\Vert^{1/2}(1+f_{E,R})^{1/2},
\label{pRRpbound}
\end{equation}
where we have written 
\begin{equation}
f_{E,R}=[|E|+\Vert(V_0^-+V_\omega^-)\Vert_\infty]\Vert R\Vert.
\label{deffER}
\end{equation}
\end{lemma}
The proof is given in Appendix~\ref{Proof5.1}. 
{From} (\ref{defW}) and these bounds of Lemma~\ref{lemma:pRbounds}, one has 
\begin{equation}
\left\Vert W(\chi_1)R_{3\ell,3\ell'}\right\Vert
\le f_1(|E|,\Vert R_{3\ell,3\ell'}\Vert),
\label{Wchi1Rbound}
\end{equation}
\begin{equation}
\left\Vert R_{3\ell,3\ell'}W(\chi_{\cal R}^\epsilon)\right\Vert
\le f_2(|E|,\Vert R_{3\ell,3\ell'}\Vert)
\label{RWchibphibound}
\end{equation}
and
\begin{equation}
\left\Vert(p_i+eA_i)R_{3\ell,3\ell'}W(\chi_{\cal R}^\epsilon)\right\Vert
\le f_3(|E|,\Vert R_{3\ell,3\ell'}\Vert),
\label{pRWchibphibound}
\end{equation}
for the operators in the right-hand sides of (\ref{DRchi}) and (\ref{trchiRchi}),  
where the functions, $f_1,f_2$ and $f_3$, are given by  
\begin{eqnarray}
f_1(|E|,\Vert R\Vert)&=&\frac{\hbar^2}{2m_e}\Vert\Delta\chi_1\Vert_\infty
\Vert R\Vert\ret
&+&2\hbar\sqrt{\frac{2}{m_e}}
\left\{\Vert R\Vert+\left[|E|+\Vert(V_0^-+V_\omega^-)\Vert_\infty\right]
\Vert R\Vert^2\right\}^{1/2}
\max_{i=x,y}\{\Vert\partial_i\chi_1\Vert_\infty\},\ret
\end{eqnarray}
\begin{eqnarray}
f_2(|E|,\Vert R\Vert)&=&\frac{\hbar^2}{2m_e}
\Vert\Delta\chi_{\cal R}^\epsilon\Vert_\infty
\Vert R\Vert\ret
&+&\hbar\sqrt{\frac{2}{m_e}}
\left\{\Vert R\Vert +\left[|E|+\Vert(V_0^-+V_\omega^-)\Vert_\infty\right]
\Vert R\Vert^2\right\}^{1/2}
\Vert|\nabla\chi_{\cal R}^\epsilon|\Vert_\infty
\end{eqnarray}
and
\begin{eqnarray}
f_3(|E|,\Vert R\Vert)&=&\frac{\hbar^2}{\sqrt{2m_e}}
\Vert\Delta\chi_{\cal R}^\epsilon\Vert_\infty
\left\{\Vert R\Vert +\left[|E|+\Vert(V_0^-+V_\omega^-)\Vert_\infty\right]
\Vert R\Vert^2\right\}^{1/2}\ret
&+&2\hbar\Vert|\nabla\chi_{\cal R}^\epsilon|\Vert_\infty
\left\{1+\left[|E|+\Vert(V_0^-+V_\omega^-)\Vert_\infty\right]\Vert R\Vert\right\}.
\end{eqnarray}

The norm $\Vert R_{3\ell,3\ell'}\Vert$ of the resolvent in these upper 
bounds can be evaluated by using the Wegner estimate. 
See Appendix~\ref{appendix:Wegner} for details. 
{From} the resulting Theorem~\ref{theorem:Wegner}, we have that, 
for any $\delta E>0$, 
\begin{equation}
\Vert R_{3\ell,3\ell'}\Vert\le (\delta E)^{-1}
\label{WegnerRbound}
\end{equation}
with a probability larger than 
\begin{equation}
1-C_{\rm W}K_3\Vert g\Vert_\infty\delta E
\left|\Lambda_{3\ell,3\ell'}^{\rm para}({\bf z}_0)\right|,
\label{WegnerProbability} 
\end{equation}
where $C_{\rm W}$ is a positive constant, and the positive constant $K_3$ 
is given by (\ref{defK3}) in the theorem. 

\begin{pro}
\label{proinidecay}
For any $E$ satisfying the gap condition (\ref{energyEingap}), 
and for any $\delta E>0$, the following bound is valid: 
\begin{eqnarray}
& &\sup_{\varepsilon\ne 0}\left\Vert
W(\chi_{3\ell,3\ell'}^\delta)R_{3\ell,3\ell'}(E+i\varepsilon)\chi_{\ell,\ell'}\right\Vert
\le C_1^{(n)}e^{-\beta r_3}f_1(|E|,(\delta E)^{-1})\ret
&\times&\left[\frac{\hbar^2}{2m_e}\Vert\Delta\chi_{3\ell,3\ell'}^\delta\Vert_\infty
f_2(|E|,(\delta E)^{-1})+
\frac{2\hbar}{m_e}\max_{i=x,y}\Vert\partial_i\chi_{3\ell,3\ell'}^\delta\Vert_\infty
f_3(|E|,(\delta E)^{-1})\right]
\label{chi2Rchibound}
\end{eqnarray}
with probability at least 
\begin{equation}
P^{\rm ini}:=1-\left\{C^{\rm perc}\left\{\ell\exp[-m_p\ell']
+\ell'\exp[-m_p\ell]\right\}+C_{\rm W}K_3\Vert g\Vert_\infty\delta E
\left|\Lambda_{3\ell,3\ell'}^{\rm para}({\bf z}_0)\right|\right\}. 
\end{equation}
\end{pro}
\begin{proof}{Proof}
Combining (\ref{DRchi}), (\ref{trchiRchi}), (\ref{Rbjdecay}), (\ref{Wchi1Rbound}), 
(\ref{RWchibphibound}), (\ref{pRWchibphibound}) and (\ref{WegnerRbound}), 
the right-hand side of (\ref{normWRchi}) is estimated. 
Using (\ref{Pperc}), (\ref{WegnerProbability}) and 
the inequality ${\rm Prob}(A\cap B)\ge{\rm Prob}(A)+{\rm Prob}(B)-1$, 
the probability is estimated. 
\end{proof}

Since we can take the ribbon region ${\cal R}$ satisfying 
${\rm supp}\ \delta V_{\omega,3\ell,3\ell'}\cap{\cal R}\subset\partial{\cal R}$, 
we can obtain a similar bound for $\left\Vert\delta V_{\omega,3\ell,3\ell'}
R_{3\ell,3\ell'}(E+i\varepsilon)\chi_{\ell,\ell'}\right\Vert$ to (\ref{chi2Rchibound}). 
Fix the ratio $\ell'/\ell$. For simplicity we take $\ell'=\ell$. 
Fix $\xi>4$. We choose $\ell=\ell_0$ to satisfy    
\begin{equation}
2C^{\rm perc}\ell_0\exp[-m_p\ell_0]\le\ell_0^{-\xi}/2,
\label{initialell0condition}
\end{equation}
and choose $\delta E$ in (\ref{WegnerRbound}) so that 
\begin{equation}
C_{\rm W}K_3\Vert g\Vert_\infty\delta E
\left|\Lambda_{3\ell_0,3\ell_0}^{\rm para}({\bf z}_0)\right|=\ell_0^{-\xi}/2.
\label{condition:deltaE} 
\end{equation}
Clearly these implies $P^{\rm ini}\ge 1-\ell_0^{-\xi}$. 
Therefore, if we can find a large $\beta$ in (\ref{chi2Rchibound}) 
so that the right-hand side of (\ref{chi2Rchibound}) with $\ell=\ell'=\ell_0$ 
becomes small, then we have 
\begin{equation}
{\rm Prob}\left[\sup_{\varepsilon\ne 0}\left\Vert
{\tilde W}_{3\ell_0}^\delta R_{3\ell_0,3\ell_0}(E+i\varepsilon)
\chi_{\ell_0,\ell_0}\right\Vert\le e^{-\gamma_0\ell_0}\right]\ge 1-\ell_0^{-\xi}
\label{initialdecay}
\end{equation}
with some $\gamma_0>0$, where ${\tilde W}_{3\ell}^\delta:=W(\chi_{3\ell,3\ell}^\delta)
-\delta V_{\omega,3\ell,3\ell}\chi_{3\ell,3\ell}^\delta$. 
 
For the norm $\left\Vert\chi_{\ell,\ell'}R_{3\ell,3\ell'}(E+i\varepsilon)
W(\chi_{3\ell,3\ell'}^\delta)\right\Vert$, 
one can obtain a similar bound to (\ref{chi2Rchibound}) in the same way. 
As a result, we have 
\begin{equation}
{\rm Prob}\left[\sup_{\varepsilon\ne 0}\left\Vert\chi_{\ell_0,\ell_0}
R_{3\ell_0,3\ell_0}(E+i\varepsilon)\left({\tilde W}_{3\ell_0}^\delta\right)^\ast
\right\Vert\le e^{-\gamma_0\ell_0}\right]\ge 1-\ell_0^{-\xi}
\label{initialdecay2}
\end{equation}
under the same assumption. 

Let us study the condition which realizes such a large $\beta$. 
Consider first the case with ${\bf A}_{\rm P}=0$, and 
fix the size $\ell=\ell_0$ of the box to satisfy 
the condition (\ref{initialell0condition}). 
Further, fix $\Delta{\cal E}>0$ and ${\hat \delta}_\pm>0$.  
Assume that the energy $E\in{\bf R}$ satisfies 
\begin{equation}
{\cal E}_{n-1}+\Vert V_0^+\Vert_\infty+\lambda_+u_1+{\hat\delta}_-\hbar\omega_c\le E
\le{\cal E}_n-\Vert V_0^-\Vert_\infty-\lambda_-u_1-\Delta{\cal E}
\label{Eatlowerbandedge} 
\end{equation}
or
\begin{equation}
{\cal E}_n+\Vert V_0^+\Vert_\infty+\lambda_+u_1+\Delta{\cal E}\le E
\le{\cal E}_{n+1}-\Vert V_0^-\Vert_\infty-\lambda_-u_1-2{\hat\delta}_+\hbar\omega_c,
\label{Eatupperbandedge} 
\end{equation}
where ${\cal E}_{-1}=-\infty$. We call the energy interval 
satisfying the condition (\ref{Eatlowerbandedge}) the lower localization 
regime around the band center ${\cal E}_n$, and call 
the interval for the condition (\ref{Eatupperbandedge}) 
the upper localization regime. 
These conditions imply that we cannot treat the energy $E$ near 
the Landau level ${\cal E}_n$. 
{From} the assumption ${\bf A}_{\rm P}=0$, we have 
\begin{equation}
{\cal E}_{n-1,+}^{\rm edge}\le {\cal E}_{n-1}+\Vert V_0^+\Vert_\infty,
\quad\mbox{and}\quad 
{\cal E}_{n,-}^{\rm edge}\ge {\cal E}_n-\Vert V_0^-\Vert_\infty.
\end{equation} 
{From} these bounds, one has 
\begin{equation}
{\cal E}_{n-1,+}^{\rm edge}+\lambda_+u_1+{\hat\delta}_-\hbar\omega_c\le E
\le{\cal E}_{n,-}^{\rm edge}-\lambda_-u_1-\Delta{\cal E} 
\end{equation}
or
\begin{equation}
{\cal E}_{n,+}^{\rm edge}+\lambda_+u_1+\Delta{\cal E}\le E
\le{\cal E}_{n+1,-}^{\rm edge}-\lambda_-u_1-2{\hat\delta}_+\hbar\omega_c, 
\end{equation}
and so the energy $E$ satisfies the condition (\ref{energyEingap}). 
Since the random potential $V_\omega$ satisfies the condition (\ref{restrictVomega}) on 
the ribbon region ${\cal R}$, the energy $E$ satisfies the 
condition in Theorem~\ref{theorem:decayLandauresolvent} as  
\begin{equation}
{\cal E}_{n-1}+\Vert(V_0+V_\omega)^+\Vert_\infty+{\hat\delta}_-\hbar\omega_c\le E
\le{\cal E}_{n}-\Vert(V_0+V_\omega)^-\Vert_\infty-\Delta{\cal E} 
\end{equation}
or
\begin{equation}
{\cal E}_n+\Vert(V_0+V_\omega)^+\Vert_\infty+\Delta{\cal E}\le E
\le{\cal E}_{n+1}-\Vert(V_0+V_\omega)^-\Vert_\infty-2{\hat\delta}_+\hbar\omega_c 
\end{equation}
As a result, we can take $\beta={\tilde \kappa}_n\ell_B^{-1}\propto\sqrt{B}$ 
for $B\ge B_{0,1}^{(n)}$, 
and the constant $C_1^{(n)}$ in (\ref{chi2Rchibound}) is independent of $B$. 
Here ${\tilde \kappa}_n$ and $B_{0,1}^{(n)}$ are positive constants 
which depend only on the index $n$ of the Landau level. 

On the other hand, one can choose $\delta E$ to satisfy 
$(\delta E)^{-1}\sim {\rm Const.}\times B$ for a large $B$ 
from the condition (\ref{condition:deltaE})
and $K_3\sim {\rm Const.}\times B$ for a large $B$. 
The asymptotic behavior of $K_3$ is easily derived from the expression (\ref{defK3}). 
See the remark below Theorem~\ref{theorem:Wegner}. 
Combining these observations with the bound (\ref{chi2Rchibound}), 
we reach the result that there exists a large, positive $B_2^{(n)}$ such that the statement 
(\ref{initialdecay}) holds for all $B\ge B_2^{(n)}$. 
The positive constant $B_2^{(n)}$ depends only on the index $n$ of 
the Landau level. 

We have fixed the size $\ell$ of the box to $\ell=\ell_0$ at the first, 
and chosen a large strength $B$ of the magnetic field. 
However, fixing the initial size of the box is not convenient for applying 
the multi-scale analysis to the present system. 
In fact, we must choose a sufficiently large $\ell_0$ for the initial size to satisfy 
a certain condition which depends on the strength $B$ of the magnetic 
field in the analysis. Clearly, changing the initial size $\ell_0$ to 
a larger value is not allowed for a fixed $B$ because  
the right-hand side of the bound (\ref{chi2Rchibound}) depends 
on the size $\ell_0$ of the box through $(\delta E)^{-1}\propto\ell_0^{\xi+2}$. 
In order to avoid 
this technical difficulty, we take  $\ell_0$ to be a function of 
$B$ as 
\begin{equation}
\ell_0=\ell_0(B)={\hat\ell}_0\left[\sqrt{B/B_2^{(n)}}\right]_{\rm odd}^\le,
\label{defell_0(B)}
\label{ell0B}
\end{equation}
where $[x]_{\rm odd}^\le$ denotes the largest odd integer 
which is smaller than or equal to $x$, and $B_2^{(n)}$  
is the lower bound for the strength $B$ of the magnetic field
which is determined in the above; the odd integer ${\hat\ell}_0$ 
is chosen so that $\ell_0(B_2^{(n)})$ satisfies 
the condition (\ref{initialell0condition}).

{From} the definition (\ref{defell_0(B)}) and $\beta={\tilde \kappa}_n\ell_B^{-1}$, 
one has  
\begin{equation}
e^{-\beta r_3}\le e^{-{\tilde \gamma}_0^{(n)}\ell_0}
\end{equation}
with ${\tilde\gamma}_0^{(n)}={\tilde\kappa}_nr_3/(\ell_{B_2^{(n)}}{\hat\ell}_0)$. 
Using this, (\ref{defell_0(B)}) and $(\delta E)^{-1}\sim{\rm Const.}\times 
B\ell_0^{\xi+2}$ for a large $B$,  
the right-hand side of the bound (\ref{chi2Rchibound}) is bounded from above by 
\begin{equation}
C\ell_0^{2\xi+11}e^{-{\tilde \gamma}_0^{(n)}\ell_0}
=\exp\left[-\left\{{\tilde\gamma}_0^{(n)}-
(\log C)/\ell_0-(2\xi+11)(\log\ell_0)/\ell_0\right\}\ell_0\right]
\end{equation}
for a large $B$. Here $C$ is the positive constant. Thus 
there exists a large, positive $B_0^{(n)}$ such that the statement 
(\ref{initialdecay}) holds for any $B\ge B_0^{(n)}$. 
The positive constant $B_0^{(n)}$ depends only on the index $n$ of 
the Landau level. In addition to this, we can choose the constant 
$\gamma_0=\gamma_0^{(n)}$ 
so that $\gamma_0^{(n)}$ is independent of $B$ and 
of the initial size $\ell_0=\ell_0(B)$. Actually $\gamma_0^{(n)}$ depends only 
on the index $n$ of the Landau level. 

Next consider the case with ${\bf A}_{\rm P}\ne 0$. 
In this case, we also take $\ell_0=\ell_0(B)$ of (\ref{defell_0(B)}). 
The decay bound (\ref{iniresolventbound0}) for the resolvent 
in Theorem~\ref{theorem:decayLandauresolvent} was 
the key to the above argument. However, for ${\bf A}_{\rm P}\ne 0$, 
we must rely on the different, weaker bound (\ref{Rdecaybound}) 
in Theorem~\ref{theorem:Rdecaybound}. 
In fact, we cannot obtain a similar bound to (\ref{iniresolventbound0}) 
because of a technical reason. 
To begin with, let us see the difference between the two bounds. 
Let $E$ be the energy in the spectrum $\sigma(H_\omega)$ 
of the Hamiltonian $H_\omega$ of the whole system, 
and let $\sigma(H_{\cal R})$ be the spectrum of the Hamiltonian $H_{\cal R}$ 
of (\ref{hamR}) having the local potential $V_{\cal R}$ supported by 
the ribbon region ${\cal R}$. 
Then the distance between $E$ and $\sigma(H_{\cal R})$ is at most 
of order of $\Vert V_\omega\Vert_\infty$. Namely,
\begin{equation}
{\rm dist}(\sigma(H_{\cal R}),E)=\min\{|E_+-E|,|E-E_-|\}\le\Vert V_\omega\Vert_\infty, 
\end{equation}
where $(E_-,E_+)$ is the spectral gap of $H_{\cal R}$. Substituting this into 
the expression (\ref{generalbeta}) of $\beta$ in the bound (\ref{Rdecaybound}) for 
the resolvent, we have that the parameter $\beta$ is at most 
of ${\cal O}(1)$ for a large strength $B$ of the magnetic field. 
Thus we cannot realize a large $\beta$ by taking only a large $B$. 

In order to realize a large $\beta$, we require a strong disorder, 
together with the strong magnetic field. To this end, 
we take $u=\hbar\omega_c{\hat u}$ with a fixed, dimensionless function ${\hat u}$
for the random potential $V_\omega$ of (\ref{Vrandom}), and choose 
$C_0=\hbar\omega_c{\hat C}_0$ with a fixed, dimensionless constant ${\hat C}_0$ 
as the constant in (\ref{def:C0}).  
Fix ${\hat\delta}_\pm>0$. Assume that the energy $E$ satisfies 
\begin{equation}
{\tilde{\cal E}}_{n,+}^{\rm edge}+\hbar\omega_c{\hat\delta}_-
\le E\le{\tilde{\cal E}}_{n+1,-}^{\rm edge}-\hbar\omega_c{\hat\delta}_+,
\label{energyconditiongeneral}  
\end{equation}
where 
\begin{equation}
{\tilde{\cal E}}_{n,+}^{\rm edge}={\cal E}_n+\frac{\sqrt{2}e}{\sqrt{m_e}}
\Vert|{\bf A}_{\rm P}|\Vert_\infty\sqrt{{\cal E}_n}
+\frac{e^2}{2m_e}\Vert|{\bf A}_{\rm P}|\Vert_\infty^2
+\Vert V_0^+\Vert_\infty+\lambda_+u_1
\end{equation}
and 
\begin{equation}
{\tilde{\cal E}}_{n+1,-}^{\rm edge}={\cal E}_{n+1}-\frac{\sqrt{2}e}{\sqrt{m_e}}
\Vert|{\bf A}_{\rm P}|\Vert_\infty\sqrt{{\cal E}_{n+1}}
-\Vert V_0^-\Vert_\infty-\lambda_-u_1. 
\end{equation}
The condition (\ref{energyconditiongeneral}) implies that we cannot treat 
the energy $E$ in the interval $[{\tilde{\cal E}}_{n,-}^{\rm edge}
-\hbar\omega_c{\hat\delta}_+,{\tilde{\cal E}}_{n,+}^{\rm edge}
+\hbar\omega_c{\hat\delta}_-]$ near the Landau level ${\cal E}_n$. 
{From} (\ref{Eedge+}), (\ref{Eloweredgebound}) and (\ref{restrictVomega}), we have 
\begin{equation}
E_-\le {\cal E}_{n,+}^{\rm edge}+\lambda_+u_1\le {\tilde{\cal E}}_{n,+}^{\rm edge}
\quad
\mbox{and}
\quad
E_+\ge{\cal E}_{n+1,-}^{\rm edge}-\lambda_-u_1\ge{\tilde{\cal E}}_{n+1,-}^{\rm edge}.
\end{equation} 
Therefore we obtain $E_+-E\ge \hbar\omega_c{\hat\delta}_+$ and 
$E-E_-\ge\hbar\omega_c{\hat\delta}_-$. 
Substituting these into the expression (\ref{generalbeta}) of $\beta$, 
we have 
\begin{equation}
\beta\ge\frac{\sqrt{2m_e}}{\hbar}
\sqrt{\frac{(\hbar\omega_c)^3{\hat C}_0
{\hat\delta}_+{\hat\delta}_-}{C_0(E-E_-)+16(E_++{\tilde C}_0)(E_-+{\tilde C}_0)}}
={\cal O}(B^{1/2})\quad\mbox{for a large }B, 
\end{equation}
where we have used $C_0=\hbar\omega_c{\hat C}_0$ with a fixed constant 
${\hat C}_0$. In this case, one can easily have $K_3={\cal O}(1)$ 
for a large $B$, from (\ref{defcalU}), (\ref{noflatregion}), (\ref{K0n0}), 
(\ref{def:EmaxDelta}) and (\ref{defK3}). Therefore we can 
choose $(\delta E)^{-1}\sim{\rm Const.}\times\ell_0^{\xi+2}$ 
to satisfy the condition (\ref{condition:deltaE}). 
Consequently the same statement (\ref{initialdecay}) holds for a strong magnetic field 
and for a strong potential $u={\cal O}(B)$. 
 
%%%%%%%%%%%%%%%%%%%%%%%%%%%%%%%%%%%%%%%%%%%%%%%%%%%%
\Section{Multi-scale analysis}
\label{MSA}

Starting from the initial decay estimates (\ref{initialdecay}) and 
(\ref{initialdecay2}) for the resolvent, 
we derive similar estimates for larger scales 
without losing too much. The main results of this section 
are given in Lemmas~\ref{lemma:relationell'ell} and \ref{lemma:lksequence} below. 
The proofs are given in Appendix~\ref{MSAProofs}. 
We stress again that these results for large but finite volumes never 
yield a similar decay bound for two arbitrary points in the whole plane ${\bf R}^2$.  
As to the resolvent in the infinite volume, 
we will rely on the fractional moment method in the next section.  
The multi-scale analysis given here is a simplified 
version \cite{CL,CH1,KSS} of \cite{FS}. 
Although the method itself is well known, we must carefully handle  
the magnetic field dependence of the decay bound for the resolvent again. 

Let $\ell$ be an odd integer larger than 1, and denote by 
$\Lambda_\ell({\bf z})=\Lambda_{\ell,\ell}^{\rm para}({\bf z})$ 
the parallelogram box with sidelength $\ell$ and with center 
${\bf z}\in\Gamma_\ell:=\ell{\bf L}^2=\{m\ell{\bf a}_1+n\ell{\bf a}_2|\ 
m,n\in{\bf Z}\}$.
The distance between two lattice sites in $\Gamma_\ell$ is defined by 
$|{\bf z}|=\max\{|m|\ell,|n|\ell\}$. 
Fix a small $\delta>0$. Let $\chi_\ell({\bf z})$ be the characteristic function 
of the region 
$\Lambda_\ell({\bf z})$, and let $\chi_{3\ell}^\delta({\bf z})$ be 
a $C^3$, positive cut-off function satisfying 
\begin{equation}
\left.\chi_{3\ell}^\delta({\bf z})\right|_{\Lambda_{3\ell}^\delta({\bf z})}=1, \quad 
\mbox{and}\quad {\rm supp}\left|\nabla\chi_{3\ell}^\delta({\bf z})\right|
\subset\Lambda_{3\ell}({\bf z})\backslash\Lambda_{3\ell}^\delta({\bf z}),
\end{equation}
where $\Lambda_{3\ell}^\delta({\bf z}):=\{{\bf r}\in\Lambda_{3\ell}({\bf z})|\ 
{\rm dist}({\bf r},\partial\Lambda_{3\ell}({\bf z})>\delta\}$. We write  
\begin{equation}
R_{3\ell,{\bf z}}(E+i\varepsilon)=\left({\hat H}_{\Lambda_{3\ell}({\bf z})}
-E-i\varepsilon\right)^{-1},
\quad
\mbox{and} 
\quad
{\tilde W}_{3\ell}^\delta({\bf z})=W(\chi_{3\ell}^\delta({\bf z}))
-\delta V_{\omega,3\ell,3\ell}\chi_{3\ell}^\delta({\bf z}).
\end{equation}
Here the Hamiltonian ${\hat H}_{\Lambda_{3\ell}({\bf z})}$ is given by (\ref{hamMSA}). 

\begin{definition}
A parallelogram box $\Lambda_{3\ell}({\bf z})$ is called $\gamma$-good for some 
$\gamma>0$ if the following two bounds hold:  
\begin{equation}
\sup_{\varepsilon\ne 0}\left\Vert {\tilde W}_{3\ell}^\delta({\bf z})
R_{3\ell,{\bf z}}(E+i\varepsilon)\chi_\ell({\bf z})\right\Vert\le e^{-\gamma\ell}
\end{equation}
and 
\begin{equation}
\sup_{\varepsilon\ne 0}\left\Vert\chi_\ell({\bf z}) 
R_{3\ell,{\bf z}}(E+i\varepsilon)\left({\tilde W}_{3\ell}^\delta({\bf z})\right)^\ast
\right\Vert\le e^{-\gamma\ell}.
\end{equation}
\end{definition}

\noindent
{\bf Remark:} The probability ${\rm Prob}[\Lambda_{3\ell}({\bf z})
\mbox{ is $\gamma$-good}]$ is independent of the center ${\bf z}$. 

\begin{lemma}
\label{lemma:relationell'ell}
Let $\ell,\ell'$ be odd integers larger than $1$ such that $\ell'$ is a multiple 
of $\ell$ and satisfies $\ell'>4\ell$. 
Assume ${\rm Prob}[\Lambda_{3\ell}(\cdots)\mbox{ is $\gamma$-good}]\ge 
1-\eta$ with a small $\eta>0$. Then  
\begin{equation}
{\rm Prob}[\Lambda_{3\ell'}(\cdots)\mbox{ is $\gamma'$-good}]\ge 1-\eta'
\end{equation}
with
\begin{equation}
\eta'=\left({5\ell'}/{\ell}\right)^4\eta^2+\left(\ell'\right)^{-\xi}/2
\label{eta'}
\end{equation}
and with
\begin{equation}
\gamma'=\gamma\left(1-{4\ell}/{\ell'}\right)-\ell^{-1}\log(c_0K_3^2|E|)
-{(\ell')}^{-1}(2s+7)\log\ell'.
\label{gamma'}
\end{equation}
Here $c_0$ is a positive constant. 
\end{lemma}

We define a sequence of monotone increasing length scales $\ell_k$ as 
\begin{equation}
\ell_{k+1}=\ell_k\left[\ell_k^{1/2}\right]_{\rm odd}^\ge
\quad\mbox{for }k=0,1,2,\ldots,
\label{defellk+1}
\end{equation}
where $[x]_{\rm odd}^\ge$ denotes the smallest odd integer which 
is larger than or equal to $x$. Clearly we have $\ell_{k+1}\ge\ell_k^{3/2}$ 
and $\ell_{k+1}>4\ell_k$ for all $k$ if the initial scale $\ell_0$ is large enough. 

\begin{lemma}
\label{lemma:lksequence}
Take $\ell_0=\ell_0(B)$ which is given by (\ref{ell0B}), i.e., the function of 
the strength $B$ of the magnetic field. 
Then there exists a minimum strength $B_0>0$ of the magnetic field such that 
\begin{equation}
{\rm Prob}[\Lambda_{3\ell_k}(\cdots)\mbox{ is $\gamma_\infty$-good}]\ge 
1-(\ell_k)^{-\xi}
\end{equation}
with some $\gamma_\infty>0$ for any $B>B_0$ and for any $k$. 
\end{lemma}

%%%%%%%%%%%%%%%%%%%%%%%%%%%%%%%%%%%%%%%%%%%%%%%%%%%%%%%
\Section{Fractional moment bound for the resolvent}
\label{FMB}

As mentioned at the beginning of the preceding section, 
the multi-scale analysis has the disadvantage 
for the decay estimate of the resolvent in the infinite volume. 
In order to compensate for the disadvantage, we rely on the fractional 
moment method. The key points of the method are that the fractional moment 
of the resolvent is finite due to the resonance-diffusing effect of the 
disorder, and satisfies a ``correlation inequality" \cite{AENSS}.  
But the resolevnt itself without taking the fractional 
moment cannot be evaluated by the method, as we already mentioned above. 
The aim of this section is to obtain the decay bound (\ref{DRI}) below for 
the fractional moment of the resolvent for the present system, 
following ref.~\cite{AENSS}. 
Further, the decay bound (\ref{DRI}) so obtained yields 
the decay bound (\ref{PFD}) below for the Fermi sea projection $P_{\rm F}$.  
In the article \cite{AENSS}, the authors showed that a decay estimate of 
a resolvent in the multiscale analysis yields a fractional moment bound for 
the resolvent. 
In this section, we obtain the fractional moment bound more directly 
from the initial decay estimate which was studied in Section~\ref{IDER}. 
Actually, the initial decay estimate was the initial data for 
the multi-scale analysis in Section~\ref{MSA}. 

Consider the present system described by the Hamiltonian $H_\omega$ 
with the random potential $V_\omega$ on a finite region $\Lambda$ or 
the infinite plane ${\bf R}^2$. 
Let $\chi_{\cal A},\chi_{\cal B}$ be the characteristic functions 
of the sets ${\cal A},{\cal B}$ with a compact support, 
respectively. Then the fractional moment bound for the resolvent is 
\begin{equation}
\sup_{\varepsilon\ne 0}{\bf E}
\left\Vert\chi_{\cal A}(E_{\rm F}+i\varepsilon-H_\omega)^{-1}
\chi_{\cal B}\right\Vert^s\le{\rm Const.}e^{-\mu r},
\label{DRI}
\end{equation}
where ${\bf E}$ is the expectation with respect to 
the random variables of the potential $V_\omega$, 
$s\in(0,1/3)$, $\mu$ is a positive constant, and we have written 
$r={\rm dist}({\cal A},{\cal B})$. 

We denote by $s_\ell({\bf u})$ the square box centered at 
${\bf u}=(u_1,u_2)\in{\bf R}^2$ with the sidelength $\ell$, i.e., 
$s_\ell({\bf u})=\{{\bf r}=(x,y)\in{\bf R}^2|\max\{|x-u_1|,|y-u_2|\}\le\ell/2\}$. 
Consider the Hamiltonian 
\begin{equation}
H_{s_L({\bf z}_0)}=\frac{1}{2m_e}({\bf p}+e{\bf A})^2
+\left.(V_0+V_\omega)\right|_{s_L({\bf z}_0)}
\end{equation}
on the square region $s_L({\bf z}_0)$ centered at ${\bf z}_0\in{\bf L}^2$ 
with the sidelength $L$, 
where we impose the Dirichlet boundary conditions, and write the resolvent as  
\begin{equation}
R_L=R_L(E+i\varepsilon)
=\left(H_{s_L({\bf z}_0)}-E-i\varepsilon\right)^{-1}
\end{equation}
for $E,\varepsilon\in{\bf R}$. 
We write $\chi_{\tilde r}({\bf z})$ for the characteristic function of 
the square box $s_{\tilde r}({\bf z})$ with 
the sidelength ${\tilde r}:=\sqrt{3}a/2$ for ${\bf z}\in{\bf L}^2$. 
Let $\delta{\cal A}_L({\bf z}_0)=s_{L-3{\tilde r}}({\bf z}_0)
\backslash s_{L-23{\tilde r}}({\bf z}_0)$, 
and let ${\bf z}'\in\delta{\cal A}_L({\bf z}_0)\cap{\bf L}^2$. 
In order to obtain the fractional moment bound for the resolvent, 
we want to evaluate 
\begin{equation}
\sup_{\varepsilon\ne 0}{\bf E}
\left[\left\Vert\chi_{\tilde r}({\bf z}')R_L(E+i\varepsilon)
\chi_{\tilde r}({\bf z}_0)\right\Vert^s\right]\quad\mbox{for}\ \ s\in(0,1/3). 
\end{equation}

In the same way as in Section~\ref{percolation}, 
one can find a ribbon region ${\cal R}$ such that the conditions 
(\ref{widthcond}) and (\ref{restrictVomega}) are satisfied with probability larger than 
\begin{equation}
P^{\rm perc}=1-C^{\rm perc}Le^{-m_pL}\quad
\mbox{with two positive constants,}\ C^{\rm perc}\ \mbox{and}\ m_p,
\end{equation}
and that the ribbon ${\cal R}$ encircles the square box $s_{\tilde r}({\bf z}_0)$, 
and that the following two conditions are satisfied: 
${\rm dist}({\cal R},s_{\tilde r}({\bf z}_0))>0$ and 
${\rm dist}({\cal R},s_{\tilde r}({\bf z}'))>0$ 
for all ${\bf z}'\in\delta{\cal A}_L({\bf z}_0)\cap{\bf L}^2$. 
Further, we can find a $C^2$, positive cut-off function $\chi_1$ such that 
$\chi_1|_{s_{\tilde r}({\bf z}_0)}=1$ 
and ${\rm supp}|\nabla\chi_1|\subset{\cal C}_\epsilon$, where 
the region ${\cal C}_\epsilon$ near the center ${\cal C}$ of the 
ribbon ${\cal R}$ is the same as in Section~\ref{percolation}.  
Since we can choose $\chi_1$ so that $\chi_{\tilde r}({\bf z}')\chi_1=0$, 
we have 
\begin{equation}
\chi_{\tilde r}({\bf z}')R_L\chi_{\tilde r}({\bf z}_0)=
\chi_{\tilde r}({\bf z}')R_L\chi_1\chi_{\tilde r}({\bf z}_0)
=\chi_{\tilde r}({\bf z}')R_LW(\chi_1)R_L
\chi_{\tilde r}({\bf z}_0).
\end{equation}
In the same way as in Section~\ref{percolation}, 
we can take the $C^2$, positive cut-off function $\chi_{\cal R}^\epsilon$ 
and the resolvent $R_{\cal R}$ for the Hamiltonian $H_{\cal R}$. 
Therefore the right-hand side can be further written as  
\begin{eqnarray}
\chi_{\tilde r}({\bf z}')R_L\chi_{\tilde r}({\bf z}_0)
&=&\chi_{\tilde r}({\bf z}')R_LW(\chi_1)R_L
\chi_{\tilde r}({\bf z}_0)\ret
&=&\chi_{\tilde r}({\bf z}')R_L\chi_{\cal R}^\epsilon
W(\chi_1)R_L\chi_{\tilde r}({\bf z}_0)\ret
&=&-\chi_{\tilde r}({\bf z}')R_LW(\chi_{\cal R}^\epsilon)
R_{\cal R}W(\chi_1)R_L\chi_{\tilde r}({\bf z}_0), 
\end{eqnarray}
where we have used $\chi_{\cal R}^\epsilon|_{{\cal R}\backslash{\cal R}_\epsilon}=1$, 
$\chi_{\tilde r}({\bf z}_0)\chi_{\cal R}^\epsilon=0$ and 
the geometric resolvent equation, 
$R_L\chi_{\cal R}^\epsilon=\chi_{\cal R}^\epsilon R_{\cal R}
-R_LW(\chi_{\cal R}^\epsilon)R_{\cal R}$. 
Using the bounds (\ref{Wchi1Rbound}) and (\ref{RWchibphibound}) 
for the resolvent $R_L$ instead of $R_{3\ell,3\ell'}$, we obtain 
\begin{eqnarray}
\left\Vert\chi_{\tilde r}({\bf z}')R_L\chi_{\tilde r}({\bf z}_0)\right\Vert
&\le&\left\Vert
\chi_{\tilde r}({\bf z}')R_LW(\chi_{\cal R}^\epsilon)
\chi_{\cal A}R_{\cal R}\chi_{\cal B}W(\chi_1)R_L\chi_{\tilde r}({\bf z}_0)
\right\Vert\ret
&\le&\left\Vert R_LW(\chi_{\cal R}^\epsilon)\right\Vert
\left\Vert\chi_{\cal A}R_{\cal R}\chi_{\cal B}\right\Vert
\left\Vert W(\chi_1)R_L\right\Vert\ret
&\le&f_1(|E|,\Vert R_L\Vert)f_2(|E|,\Vert R_L\Vert)
\left\Vert\chi_{\cal A}R_{\cal R}\chi_{\cal B}\right\Vert,
\end{eqnarray} 
where ${\cal A}={\rm supp}|\nabla\chi_{\cal R}^\epsilon|$ and 
${\cal B}={\cal C}_\epsilon\supset{\rm supp}|\nabla\chi_1|$. 
In the same way as in Section~\ref{percolation}, we have 
\begin{equation}
\left\Vert\chi_{\cal A}R_{\cal R}\chi_{\cal B}\right\Vert
\le C_1^{(n)}e^{-\beta r_3}
\end{equation}
with probability larger than $(1-C^{\rm perc}Le^{-m_pL})$, 
where $C_1^{(n)}$ and $\beta$ are the corresponding positive constants. 
By using the Wegner estimate, the norm of the resolvent can be also evaluated as 
$\Vert R_L\Vert\le(\delta E)^{-1}$ with probability larger than 
$(1-C_W\Vert g\Vert_\infty\delta E L^2)$. We choose $L$ to satisfy 
\begin{equation}
C^{\rm perc}Le^{-m_pL}\le(L/a)^{-{\tilde \xi}}/2
\end{equation}
with a positive number ${\tilde \xi}$ which we will determine below, 
and choose $\delta E$ so that 
\begin{equation}
C_W\Vert g\Vert_\infty\delta E L^2
=(L/a)^{-{\tilde \xi}}/2.
\end{equation}
Then we have 
\begin{equation}
\left\Vert\chi_{\tilde r}({\bf z}')R_L\chi_{\tilde r}({\bf z}_0)\right\Vert
\le C_1^{(n)}f_1(|E|,(\delta E)^{-1})f_2(|E|,(\delta E)^{-1})
e^{-\beta r_3}
\label{goodeventineq}
\end{equation}
with probability larger than $(1-(L/a)^{-{\tilde \xi}})$. 
We denote by $D_L$ the set of the events $\omega$ 
satisfying the above inequality (\ref{goodeventineq}). 
Note that 
\begin{eqnarray}
{\bf E}\left[
\left\Vert\chi_{\tilde r}({\bf z}')R_L\chi_{\tilde r}({\bf z}_0)\right\Vert^s
\right]&\le&
{\bf E}\left[
\left\Vert\chi_{\tilde r}({\bf z}')R_L\chi_{\tilde r}({\bf z}_0)\right\Vert^s
{\bf I}(D_L)\right]\ret
&+&
{\bf E}\left[
\left\Vert\chi_{\tilde r}({\bf z}')R_L\chi_{\tilde r}({\bf z}_0)\right\Vert^s
{\bf I}(D_L^c)\right],
\label{EchiRLchidecom}
\end{eqnarray}
where ${\bf I}(A)$ is the indicator function of an event $A$. The first term 
in the right-hand side is estimated as 
\begin{equation}
{\bf E}\left[
\left\Vert\chi_{\tilde r}({\bf z}')R_L\chi_{\tilde r}({\bf z}_0)\right\Vert^s
{\bf I}(D_L)\right]
\le\left[C_1^{(n)}f_1(|E|,(\delta E)^{-1})f_2(|E|,(\delta E)^{-1})\right]^s
e^{-s\beta r_3}.
\label{EchiRLchidecom1}
\end{equation}
Using the H\"older inequality with $s<t<1$, the second term is estimated as 
\begin{eqnarray}
{\bf E}\left[
\left\Vert\chi_{\tilde r}({\bf z}')R_L\chi_{\tilde r}({\bf z}_0)\right\Vert^s
{\bf I}(D_L^c)\right]
&\le&{\bf E}\left[
\left\Vert\chi_{\tilde r}({\bf z}')R_L\chi_{\tilde r}({\bf z}_0)\right\Vert^t
\right]^{s/t}{\bf E}\left[{\bf I}(D_L^c)\right]^{1-s/t}\ret
&\le&{\rm Const.}B^s\times(L/a)^{-(1-s/t){\tilde \xi}},
\label{EchiRLchidecom2}
\end{eqnarray}
where $B$ is the strength of the magnetic field, and 
we have used the fractional moment bound eq.~(3.19) in \cite{AENSS}, and 
${\bf E}\left[{\bf I}(D_L^c)\right]\le(L/a)^{-{\tilde \xi}}$. 
The factor $B^s$ comes from a careful but easy calculation in  
the fractional moment bound. The positive constant depends only on 
the index of the Landau level in the condition (\ref{energyEingap}) for 
the energy $E$.  

We take $L$ to be a function of $B$ as $L=L(B)={\rm Const.}\times B^{1/2}$. 
Then the argument of Section~\ref{IDER} yields 
$\beta r_3={\rm Const.}\times B^{1/2}$. 
We choose $t$ and ${\tilde\xi}$ to satisfy $(1-s/t){\tilde\xi}>3+12s$. 
Combining these, (\ref{EchiRLchidecom}), (\ref{EchiRLchidecom1}) and 
(\ref{EchiRLchidecom2}), we obtain that the quantity, 
\begin{equation}
B^{5s}L^3\sup_{\varepsilon\ne 0}{\bf E}
\left[\left\Vert\chi_{\tilde r}({\bf z}')R_L(E+i\varepsilon)
\chi_{\tilde r}({\bf z}_0)\right\Vert^s\right], 
\end{equation}
becomes small for a sufficiently large $B$ of the strength of the magnetic 
field. This implies that the finite-volume criteria\footnote{We should remark 
the following: The condition ${\bf z}'\in\delta{\cal A}_L({\bf z}_0)
\cap{\bf L}^2$ is slightly different from that in Theorem~1.2 of \cite{AENSS}. 
In fact, our argument relies on Lemma~4.1 of \cite{AENSS}.} of Theorem~1.2 
in \cite{AENSS} is satisfied for magnetic fields 
whose strength $B$ is larger than a positive $B_0$. The factor $B^{5s}$ comes from 
the $B$-dependence of the constant in eq.~(1.18) of 
the criteria.
Thus the fractional moment bound (\ref{DRI}) for the resolvent holds 
for such a large magnetic field. 

Let $P_{\rm F}$ be the projection on energies smaller than 
the Fermi energy $E_{\rm F}$, and 
let $\chi_{\cal A},\chi_{\cal B}$ be the characteristic functions 
of the sets ${\cal A},{\cal B}$ with a compact support, 
respectively. The following Lemma~\ref{lemma:PFD} is due to \cite{AENSS}. 
In order to make the paper self-contained, we give a proof which is 
slightly different from that in \cite{AENSS}. 

\begin{lemma}
\label{lemma:PFD}
The following bound holds: 
\begin{equation}
{\bf E}\left\Vert\chi_{\cal A}P_{\rm F}\chi_{\cal B}\right\Vert
\le{\rm Const.}\exp[-\mu\>{\rm dist}({\cal A},{\cal B})].
\label{PFD}
\end{equation}
\end{lemma}

\begin{proof}{Proof}
Write $R(z)=(z-H_\omega)^{-1}$. Using the contour integral, one has  
\begin{eqnarray}
\chi_{\cal A}P_{\rm F}\chi_{\cal B}&=&\frac{1}{2\pi i}
\int_{E_0}^{E_{\rm F}}dE\>\chi_{\cal A}R(E+iy_-)\chi_{\cal B}
+\frac{1}{2\pi i}\int_{y_-}^{y_+}idy\>
\chi_{\cal A}R(E_{\rm F}+iy)\chi_{\cal B}\ret
&+&\frac{1}{2\pi i}\int_{E_{\rm F}}^{E_0}dE\>
\chi_{\cal A}R(E+iy_+)\chi_{\cal B}
+\frac{1}{2\pi i}\int_{y_+}^{y_-}idy\>
\chi_{\cal A}R(E_0+iy)\chi_{\cal B},
\label{CPFC} 
\end{eqnarray}
where $E_0$ is a real constant satisfying $H_\omega>E_0$. The integral near 
the Fermi energy is justified because the operator norm limit, 
$\lim_{\varepsilon\downarrow 0}
\chi_{\cal A}R(E\pm i\varepsilon)\chi_{\cal B}$, 
exists \cite{AENSS,KSS,Branges} almost surely for almost every energy $E\in{\bf R}$. 
We can choose finite $E_0$ and $y_\pm$ so that 
\begin{equation}
\left\Vert\chi_{\cal A}R(E_0+iy)
\chi_{\cal B}\right\Vert\le{\rm Const.}e^{-\mu r}\quad \mbox{for any real}\ y
\end{equation}
and 
\begin{equation}
\left\Vert\chi_{\cal A}R(E+iy_\pm)
\chi_{\cal B}\right\Vert\le{\rm Const.}e^{-\mu r}\quad\mbox{for }\ E\in[E_0,E_{\rm F}]
\end{equation}
with the same decay constant $\mu$. See Appendix~\ref{Decaygeneral} for details. 
Therefore it is enough to evaluate the second integral in the right-hand side 
of (\ref{CPFC}). It is written 
\begin{equation}
\mathop{\lim\inf}_{\varepsilon_n\rightarrow 0}
\int_{I(\varepsilon_n)}dy\>\chi_{\cal A}R(E_{\rm F}+iy)\chi_{\cal B},
\end{equation}
where $\{\varepsilon_n\}_n$ is a decreasing sequence, and we have written 
$I(\varepsilon_n)=[y_-,y_+]\backslash(-\varepsilon_n,\varepsilon_n)$. Using Fatou's lemma, 
we have 
\begin{equation}
{\bf E}\left\Vert\int_{y_-}^{y_+}dy\>
\chi_{\cal A}R(E_{\rm F}+iy)\chi_{\cal B}\right\Vert
\le\mathop{\lim\inf}_{\varepsilon_n\rightarrow 0}{\bf E}\left\Vert 
\int_{I(\varepsilon_n)}dy\>
\chi_{\cal A}R(E_{\rm F}+iy)\chi_{\cal B}\right\Vert.
\end{equation}
This right-hand side is evaluated as 
\begin{eqnarray}
{\bf E}\left\Vert\int_{I(\varepsilon_n)}dy\>
\chi_{\cal A}R(E_{\rm F}+iy)\chi_{\cal B}\right\Vert
&\le&{\bf E}\int_{I(\varepsilon_n)}dy\>
\left\Vert
\chi_{\cal A}R(E_{\rm F}+iy)\chi_{\cal B}\right\Vert\ret
&\le&{\bf E}\int_{I(\varepsilon_n)}dy\>
\left\Vert
\chi_{\cal A}R(E_{\rm F}+iy)\chi_{\cal B}\right\Vert^s
\left\Vert
\chi_{\cal A}R(E_{\rm F}+iy)\chi_{\cal B}\right\Vert^{1-s}\ret
&\le&{\bf E}\int_{I(\varepsilon_n)}dy\>\left\Vert
\chi_{\cal A}R(E_{\rm F}+iy)\chi_{\cal B}\right\Vert^s
|y|^{s-1}\ret&\le&{\rm Const.}s^{-1}(|y_+|^s+|y_-|^s)e^{-\mu r},
\end{eqnarray}
where we have used Fubini-Tonelli theorem and the decay bound (\ref{DRI}) 
for the resolvent. This yields the desired result. 
\end{proof}

%%%%%%%%%%%%%%%%%%%%%%%%%%%%%%%%%%%%%%%%%%%%%%%%%%%%%%
\Section{Finite volume Hall conductance}
\label{Conplateaus}

We recall the previous results of the linear response coefficients \cite{Koma3}. 
The total conductance for finite volume and for $t\ge 0$ is written 
\begin{equation}
\sigma_{{\rm tot},sy}(t)=\cases{\sigma_{xy}+\gamma_{xy}\cdot t 
+\delta\sigma_{xy}(t), & for $s=x$;\cr
\gamma_{yy}\cdot t+\delta\sigma_{yy}(t), & for $s=y$. \cr}
\label{sigmatot}
\end{equation}
Our goal is to give the proof of all the statements of 
Theorems~\ref{maintheorem1} and \ref{maintheorem2}. 
Namely, when the Fermi level lies in the localization regime, 
the Hall conductance $\sigma_{xy}$ is quantized to the integer 
as in (\ref{conductancequantization}), and 
both of the acceleration coefficients $\gamma_{sy}$ vanish, and 
the corrections $\delta\sigma_{sy}(t)$ due to the initial adiabatic process 
are small as in the bound (\ref{sigmacorrections}). 
For this purpose, we first treat the Hall conductance $\sigma_{xy}$, and 
prepare some technical lemmas for the Hall conductance $\sigma_{xy}$ 
for finite volume in this section.  

In the following, we write $\Lambda=\Lambda^{\rm sys}$ for short. 
The explicit form of the Hall conductance $\sigma_{xy}$ for 
the finite region $\Lambda$ is given by \cite{Koma3} 
\begin{equation}
\sigma_{xy}=-\frac{i\hbar e^2}{L_xL_y}{\rm Tr}\>
P_{{\rm F},\Lambda}[P_{x,\Lambda},P_{y,\Lambda}],
\label{sigmaxyLambda}
\end{equation}
where $P_{{\rm F},\Lambda}$ is the corresponding Fermi sea projection and 
\begin{equation}
P_{s,\Lambda}=\frac{1}{2\pi i}\int_\gamma dz R_\Lambda(z)v_s
R_\Lambda(z)\quad\mbox{for }\ s=x,y.
\label{CrepPs}
\end{equation}
Here $v_s$ are the velocity operators, i.e., $(v_x,v_y)={\bf v}(t=0)$  
for ${\bf v}(t)$ of (\ref{v(t)}), and 
$R_\Lambda=(z-H_{\omega,\Lambda})^{-1}$ with the finite-volume Hamiltonian 
$H_{\omega,\Lambda}$ with the periodic boundary conditions; 
the closed path $\gamma$ encircles 
the energy eigenvalues below the Fermi level $E_{\rm F}$.  

Take two rectangular regions $\Omega$ and $\Lambda'$ so that 
the following conditions are satisfied: 
\begin{equation}
\Omega\subset\Lambda'\subset\Lambda, \quad 
{\rm dist}(\Omega,\partial\Lambda')=\delta L/2\quad\mbox{and }\quad
{\rm dist}(\Lambda',\partial\Lambda)=\delta L/2,
\label{distOmegaLambdaLambdap}
\end{equation}
where we have taken the width $\delta L$ of the boundary regions as 
$\delta L=a\left({L}/{a}\right)^\kappa$ with $\kappa\in(1/2,1)$ and 
with $L=\max\{L_x,L_y\}$. Here $a$ is the lattice 
constant of the triangular lattice ${\bf L}^2$. 
Clearly we can take $\Omega$ satisfying $|\Omega|={\cal O}(L^2)$ and 
$|\Lambda\backslash\Omega|={\cal O}(L\delta L)$. 
We decompose $\sigma_{xy}$ into two parts as 
$\sigma_{xy}=\sigma_{xy}^{\rm in}+\sigma_{xy}^{\rm out}$ with 
\begin{equation}
\sigma_{xy}^{\rm in}=-\frac{i\hbar e^2}{L_xL_y}{\rm Tr}\>\chi_\Omega
P_{{\rm F},\Lambda}[P_{x,\Lambda},P_{y,\Lambda}]\chi_\Omega
\end{equation}
and 
\begin{equation}
\sigma_{xy}^{\rm out}=-\frac{i\hbar e^2}{L_xL_y}{\rm Tr}\>\chi_\Omega^c
P_{{\rm F},\Lambda}[P_{x,\Lambda},P_{y,\Lambda}]\chi_\Omega^c,
\end{equation}
where $\chi_\Omega$ is the characteristic function of $\Omega$, 
and $\chi_\Omega^c=1-\chi_\Omega$.  
We choose $k$ so that 
\begin{equation}
8\ell_ka\le\delta L<8\ell_{k+1}a,
\label{deltaLbound}
\end{equation}
where $\ell_k$ is a length scale in the sequence $\{\ell_k\}_k$ 
which is determined by 
the recursive equation (\ref{defellk+1}). 

\begin{lemma}
\label{lemma:ARBdecay}
Let ${\cal A},{\cal B}$ be subsets of $\Lambda$. 
If ${\rm dist}({\cal A},{\cal B})\ge 7\ell_k a/2$, then 
the following bound is valid:
\begin{equation}
\Vert\chi_{\cal A}R_\Lambda(z)\chi_{\cal B}\Vert
\le{\rm Const.}L^{\kappa(\xi-2)+4}\exp\left[-\mu_\infty L^{2\kappa/3}\right]
\end{equation}
with probability larger than $(1-{\rm Const.}L^{-2[\kappa(\xi+2)-3]/3})$, 
where $\chi_{\cal A},\chi_{\cal B}$ are, respectively, the characteristic functions 
of ${\cal A},{\cal B}$, and $\mu_\infty$ is a positive constant. 
\end{lemma}

\begin{proof}{Proof}
In order to prove the statement of Lemma~\ref{lemma:ARBdecay}, 
we rely on the argument of the multiscale analysis in Section~\ref{MSA}. 
Therefore we use the same notations, $\Lambda_\ell({\bf z}), \chi_\ell({\bf z}), 
\chi_{3\ell}^\delta({\bf z})$, etc. 
{From} the assumption ${\rm dist}({\cal A},{\cal B})\ge 7\ell_k a/2$, 
there is a sublattice ${\bf L}_{\cal B}$ of $\ell_k{\bf L}^2$ such that 
${\cal B}\subset\bigcup_{{\bf u}\in{\bf L}_{\cal B}}\Lambda_{\ell_k}({\bf u})$ 
and that ${\rm dist}({\cal A},\Lambda_{3\ell_k}({\bf u}))>0$ for 
all ${\bf u}\in{\bf L}_{\cal B}$. 
Using the adjoint of the geometric resolvent equation, 
\begin{equation}
R_\Lambda(z)\chi_{3\ell_k}^\delta({\bf u})
=\chi_{3\ell_k}^\delta({\bf u})R_{3\ell_k,{\bf u}}(z)
+R_\Lambda(z)\left({\tilde W}_{3\ell_k}^\delta({\bf u})\right)^\ast
R_{3\ell_k,{\bf u}}(z),
\end{equation}
we have 
\begin{eqnarray}
\chi_{\cal A}R_\Lambda(z)\chi_{\ell_k}({\bf u}) 
&=&\chi_{\cal A}\chi_{3\ell_k}^\delta({\bf u})R_{3\ell_k,{\bf u}}(z)
\chi_{\ell_k}({\bf u})
+\chi_{\cal A}R_\Lambda(z)\left({\tilde W}_{3\ell_k}^\delta(0)\right)^\ast
R_{3\ell_k,{\bf u}}(z)\chi_{\ell_k}({\bf u})\ret
&=&\chi_{\cal A}R_\Lambda(z)\left({\tilde W}_{3\ell_k}^\delta({\bf u})\right)^\ast
R_{3\ell_k,{\bf u}}(z)\chi_{\ell_k}({\bf u})
\end{eqnarray}
for ${\bf u}\in{\bf L}_{\cal B}$, where we have used 
$\chi_{\cal A}\chi_{3\ell_k}^\delta({\bf u})=0$ which 
follows from ${\rm dist}({\cal A},\Lambda_{3\ell_k}({\bf u}))>0$. This yields 
\begin{equation}
\left\Vert\chi_{\cal A}R_\Lambda(z)\chi_{\ell_k}({\bf u})\right\Vert
\le\Vert R_\Lambda(z)\Vert \Vert\left({\tilde W}_{3\ell_k}^\delta({\bf u})\right)^\ast
R_{3\ell_k,{\bf u}}\chi_{\ell_k}({\bf u})\Vert. 
\label{chitRchibound}
\end{equation}

On the other hand, we can prove the bound which is given by replacing 
the operator ${\tilde W}_{3\ell'}^\delta({\bf u})$ with its adjoint 
in the bound (\ref{WRchi3ellpbound}) in the same way as in the proof 
of Lemma~\ref{lemma:WRchibound}. 
Therefore the argument of Lemma~\ref{lemma:lksequence} yields  
\begin{equation}
\left\Vert\left({\tilde W}_{3\ell_k}^\delta({\bf u})\right)^\ast
R_{3\ell_k,{\bf u}}\chi_{\ell_k}({\bf u})\right\Vert
\le e^{-\gamma_\infty\ell_k}
\end{equation}
with the probability larger than $(1-\ell_k^{-\xi})$. 
{From} the Wegner estimate (\ref{Wegnerestimate}), 
one has $\Vert R_\Lambda\Vert\le 
C_WK_3\Vert g\Vert_\infty \ell_k^{\xi}\left|\Lambda\right|$ 
with probability larger than $(1-\ell_k^{-\xi})$. 
{From} (\ref{fracellk}), we have $\delta L<8\ell_{k+1}a<16\ell_k^{3/2}a$. 
Immediately, $\left({\delta L}/{16a}\right)^{2/3}\le \ell_k$. 
Substituting these inequalities into (\ref{chitRchibound}), we have 
\begin{equation}
\left\Vert\chi_{\cal A}R_\Lambda(z)\chi_{\ell_k}({\bf u})\right\Vert
\le{\rm Const.}\times \ell_k^\xi L^2\exp\left[-\mu_\infty L^{2\kappa/3}\right]
\label{decaywavefunction} 
\end{equation}
with probability larger than $(1-2\ell_k^{-\xi})$, where $\mu_\infty$ is 
the corresponding positive constant. The set ${\cal B}$ is covered by 
the sets $\Lambda_{\ell_k}({\bf u})$. 
The number $|{\bf L}_{\cal B}|$ is at most ${\cal O}(L^2/\ell_k^2)$. 
Let $M_\Lambda$ denote the event that  
the bound (\ref{decaywavefunction}) holds for all of the site 
${\bf u}\in\ell_k{\bf L}$ satisfying $\Lambda_{\ell_k}\cap\Lambda\ne\emptyset$.  
Clearly the probability ${\rm Prob}(M_\Lambda)$ 
that the event $M_\Lambda$ occurs, is larger than 
$(1-{\rm Const.}\ell_k^{-\xi}L^2\ell_k^{-2})$. 
{From} these observations, one can easily show the statement of the lemma.   
\end{proof}
 
Let $\delta$ be a small positive number, 
and define  
$\Lambda^\delta:=\{{\bf r}\in\Lambda|\ {\rm dist}({\bf r},\partial\Lambda)>\delta\}$. 
Then one has $\Lambda\backslash\Lambda^{\delta/2}=
\{{\bf r}\in\Lambda|\ {\rm dist}({\bf r},\partial\Lambda)\le\delta/2\}$. 
Let $\chi_\Lambda^\delta\in C^2(\Lambda)$ be a positive cutoff function 
satisfying the following two conditions:   
\begin{equation}
\left.\chi_\Lambda^\delta\right|_{\Lambda^\delta}=1
\quad\mbox{and}\quad
\left.\chi_\Lambda^\delta\right|_{\Lambda\backslash\Lambda^{\delta/2}}=0.
\end{equation}

\begin{lemma}
\label{lemma:sigmaxyinrep}
The Hall conductance for the bulk region is written  
\begin{equation}
\sigma_{xy}^{\rm in}=\frac{e^2}{h}\frac{2\pi i}{L_xL_y}
{\rm Tr}\>\chi_\Omega P_{{\rm F},\Lambda}[[P_{{\rm F},\Lambda},x],
[P_{{\rm F},\Lambda},y]]\chi_\Omega
+{\cal O}\left(\exp\left[-\mu_\infty'L^{2\kappa/3}\right]\right) 
\label{sigmaxyinfinite}
\end{equation}
with probability larger than $(1-{\rm Const.}L^{-2[\kappa(\xi+2)-3]/3})$, 
where $\mu_\infty'$ is a positive constant. 
\end{lemma}

\noindent
{\bf Remark:} Since $\kappa(\xi+2)-3>0$ from their definitions, 
the Hall conductance $\sigma_{xy}^{\rm in}$ for 
the bulk region in the infinite volume limit is given by 
\begin{equation}
\sigma_{xy}^{\rm in}=\frac{e^2}{h}\lim_{L\uparrow\infty}
{\cal I}(P_{{\rm F},\Lambda};\Omega),
\label{sigmaxyininfty} 
\end{equation}
with probability one if the limit in the right-hand side exists. 
Here we have written 
\begin{equation}
{\cal I}(P_{{\rm F},\Lambda};\Omega)=\frac{2\pi i}{|\Omega|}
{\rm Tr}\>\chi_\Omega P_{{\rm F},\Lambda}[[P_{{\rm F},\Lambda},x],
[P_{{\rm F},\Lambda},y]]\chi_\Omega.
\label{def:IPFL}
\end{equation}

\begin{proof}{Proof}
Using the contour integral representation as in (\ref{CrepPs}), one has 
\begin{equation}
{\rm Tr}\>\chi_\Omega P_{{\rm F},\Lambda}P_{x,\Lambda}P_{y,\Lambda}
=\frac{1}{(2\pi i)^2}\int_\gamma dz_1\int_\gamma dz_2\>
{\rm Tr}\>\chi_\Omega R_\Lambda(z_1)R_\Lambda(z_2)v_xR_\Lambda(z_2)
P_{y,\Lambda}\chi_\Omega.
\end{equation}
The integrand is decomposed into two parts as     
\begin{eqnarray}
& &{\rm Tr}\>\chi_\Omega R_\Lambda(z_1)R_\Lambda(z_2)v_xR_\Lambda(z_2)
P_{y,\Lambda}\chi_\Omega\ret
&=&{\rm Tr}\>\chi_\Omega R_\Lambda(z_1)R_\Lambda(z_2)v_x
\chi_\Lambda^\delta R_\Lambda(z_2)P_{y,\Lambda}\chi_\Omega\ret
&+&{\rm Tr}\>\chi_\Omega R_\Lambda(z_1)R_\Lambda(z_2)v_x
(1-\chi_\Lambda^\delta)R_\Lambda(z_2)P_{y,\Lambda}\chi_\Omega.
\end{eqnarray}
For the second term in the right-hand side, we have  
\begin{eqnarray}
& &\left\Vert\chi_\Omega R_\Lambda(z_1)R_\Lambda(z_2)
v_x(1-\chi_\Lambda^\delta)R_\Lambda(z_2)\right\Vert\ret
&\le&\left\Vert\chi_\Omega R_\Lambda(z_1)\right\Vert
\left\Vert\chi_{\Lambda'}R_\Lambda(z_2)v_x(1-\chi_\Lambda^\delta)
R_\Lambda(z_2)\right\Vert\ret
&+&\left\Vert\chi_\Omega R_\Lambda(z_1)(1-\chi_{\Lambda'})\right\Vert
\left\Vert R_\Lambda(z_2)v_x(1-\chi_\Lambda^\delta)
R_\Lambda(z_2)\right\Vert. 
\end{eqnarray}
{From} (\ref{chipRbound}), (\ref{distOmegaLambdaLambdap}), 
(\ref{deltaLbound}) and Lemma~\ref{lemma:ARBdecay}, this gives  
the small contribution.  
{From} this and the contour integral representation, it is sufficient to 
consider  
\begin{equation}
{\rm Tr}\>\chi_\Omega R_\Lambda(z_1)R_\Lambda(z_2)v_x\chi_\Lambda^\delta
R_\Lambda(z_2)R_\Lambda(z_3)v_y\chi_\Lambda^\delta R_\Lambda(z_3)\chi_\Omega. 
\end{equation}
Using the identity, 
$v_x\chi_\Lambda^\delta=({i}/{\hbar})[H_\omega,x]\chi_\Lambda^\delta$, 
one has 
\begin{equation}
R_\Lambda(z_2)v_x\chi_\Lambda^\delta R_\Lambda(z_2)=\frac{i}{\hbar}
[R_\Lambda(z_2),x\chi_\Lambda^\delta]
-\frac{i}{\hbar}R_\Lambda(z_2)xW(\chi_\Lambda^\delta)R_\Lambda(z_2).
\end{equation}
In the same way, the second term in the right-hand side leads to 
the small correction. The statement of the lemma follows from 
these observations.  
\end{proof}

We denote by ${\bf Z}_b^2$ the rectangular lattice $\{(b_1n_1,b_2n_2)|(n_1,n_2)\in{\bf Z}^2\}$ 
with a pair $b=(b_1,b_2)$ of lattice constants, and denote by 
$({\bf Z}_b^2)^\ast$ the dual lattice, i.e., 
$({\bf Z}_b^2)^\ast={\bf Z}_b^2-(b_1,b_2)/2$. 
Let $s_b({\bf u})$ be the $b_1\times b_2$ rectangular box centered at 
${\bf u}=(u_1,u_2)\in{\bf Z}_b^2$, and $\chi_b({\bf u})$ the characteristic function 
of $s_b({\bf u})$. When we consider the characteristic function $\chi_b({\bf u})$ 
on the region $\Lambda$, we restrict $\chi_b({\bf u})$ to $\Lambda$.  

\begin{lemma}
\label{lemma:ETrbounded}
Let ${\bf u}\in{\bf Z}_b^2$ satisfying $s_b({\bf u})\cap\Lambda\ne\emptyset$. 
Then there exists a positive 
constant $C$ which is independent of the location ${\bf u}$ 
and of the size $|\Lambda|$ such that 
\begin{equation}
{\bf E}\left[\left|{\rm Tr}\>\chi_b({\bf u})P_{{\rm F},\Lambda}
[P_{{\rm F},\Lambda},\sharp][P_{{\rm F},\Lambda},\sharp]\chi_b({\bf u})\right|\right]<C,
\end{equation}
where $\sharp$ is either $x$ or $y$. 
\end{lemma}

\begin{proof}{Proof}
Note that 
\begin{eqnarray}
& &{\bf E}\left[\left|{\rm Tr}\>\chi_b({\bf u})P_{{\rm F},\Lambda}
[P_{{\rm F},\Lambda},x][P_{{\rm F},\Lambda},y]\chi_b({\bf u})\right|\right]\ret
&\le&\mathop{\sum_{{\bf v},{\bf w}\in{\bf Z}_b^2:}}_{s_b({\bf v})
\cap\Lambda\ne\emptyset,\>s_b({\bf w})\cap\Lambda\ne\emptyset}
{\bf E}\left[\left|{\rm Tr}\>\chi_b({\bf u})P_{{\rm F},\Lambda}\chi_b({\bf v})
[P_{{\rm F},\Lambda},x]\chi_b({\bf w})[P_{{\rm F},\Lambda},y]\chi_b({\bf u})\right|\right].
\end{eqnarray}
In order to estimate the summand in this right-hand side, 
we introduce the two component function $(x^b,y^b)$ of ${\bf r}$ which is defined by  
$(x^b,y^b)=(u_1,u_2)$ for ${\bf r}\in s_b({\bf u})$ 
with ${\bf u}\in{\bf Z}_b^2$. Using this function, one has  
\begin{equation}
[P_{{\rm F},\Lambda},x]=[P_{{\rm F},\Lambda},(x-x^b)]
+[P_{{\rm F},\Lambda},x^b]
\quad\mbox{and}\quad
[P_{{\rm F},\Lambda},y]=[P_{{\rm F},\Lambda},(y-y^b)]
+[P_{{\rm F},\Lambda},y^b].
\end{equation}
Further we note that, for any bounded operator $A$,  
\begin{eqnarray}
\left|{\rm Tr}\>\chi_b({\bf u})P_{{\rm F},\Lambda}
\chi_b({\bf v})A\right|&\le&
\sqrt{{\rm Tr}\>\chi_b({\bf u})P_{{\rm F},\Lambda}\chi_b({\bf u})}
\cdot\sqrt{{\rm Tr}\>AA^\ast\chi_b({\bf v})P_{{\rm F},\Lambda}\chi_b({\bf v})}\ret
&\le&\Vert A\Vert 
\sqrt{{\rm Tr}\>\chi_b({\bf u})P_{{\rm F},\Lambda}\chi_b({\bf u})}
\cdot\sqrt{{\rm Tr}\>\chi_b({\bf v})P_{{\rm F},\Lambda}\chi_b({\bf v})}\ret
&\le&{\rm Const.}\Vert A\Vert,  
\end{eqnarray}
where we have used 
\begin{eqnarray}
{\rm Tr}\>\chi_\varepsilon({\bf u})P_{\rm F}\chi_\varepsilon({\bf u})
&=&{\rm Tr}\>\chi_\varepsilon({\bf u})\frac{1}{H_0+{\cal C}_0}(H_0+{\cal C}_0) 
P_{\rm F}(H_0+{\cal C}_0)\frac{1}{H_0+{\cal C}_0}\chi_\varepsilon({\bf u})\ret
&=&{\rm Tr}\>\chi_\varepsilon({\bf u})\frac{1}{H_0+{\cal C}_0}(H_\omega-V_\omega+{\cal C}_0) 
P_{\rm F}(H_\omega-V_\omega+{\cal C}_0)\frac{1}{H_0+{\cal C}_0}\chi_\varepsilon({\bf u})\ret
&\le&{\rm Const.}{\rm Tr}\>\chi_\varepsilon({\bf u})\left(\frac{1}{H_0+{\cal C}_0}\right)^2
\chi_\varepsilon({\bf u})<\infty.
\label{localTr} 
\end{eqnarray}
Here we have taken a real number ${\cal C}_0$ satisfying $H_0+{\cal C}_0>0$, 
and used $\Vert H_\omega P_{\rm F}\Vert<\infty$.
Therefore the statement of the lemma follows from the norm bounds such as 
\begin{equation}
\left\Vert\chi_b({\bf v})[P_{{\rm F},\Lambda},x^b]\chi_b({\bf w})\right\Vert
\le |w_1-v_1|\Vert\chi_b({\bf v})P_{{\rm F},\Lambda}\chi_b({\bf w})\Vert,
\end{equation}
\begin{equation}
\left\Vert\chi_b({\bf v})[P_{{\rm F},\Lambda},(x-x^b)]\chi_b({\bf w})\right\Vert
\le{\rm Const.}\Vert\chi_b({\bf v})P_{{\rm F},\Lambda}\chi_b({\bf w})\Vert
\end{equation}
and 
\begin{eqnarray}
& &{\bf E}\left[\Vert\chi_b({\bf v})P_{{\rm F},\Lambda}\chi_b({\bf w})\Vert
\Vert\chi_b({\bf w})P_{{\rm F},\Lambda}\chi_b({\bf u})\Vert\right]\ret
&\le&
\left\{{\bf E}\left[\Vert\chi_b({\bf v})P_{{\rm F},\Lambda}
\chi_b({\bf w})\Vert^2\right]\right\}^{1/2}
\left\{{\bf E}\left[\Vert\chi_b({\bf w})P_{{\rm F},\Lambda}
\chi_b({\bf u})\Vert^2\right]\right\}^{1/2}\ret
&\le&{\rm Const.}e^{-\mu|{\bf v}-{\bf w}|/2}e^{-\mu|{\bf w}-{\bf u}|/2},
\end{eqnarray}
where we have used Schwarz's inequality, $\Vert\chi_b({\bf v})P_{{\rm F},\Lambda}
\chi_b({\bf w})\Vert\le 1$, and the bound (\ref{PFD}) for the Fermi sea 
projection. 
\end{proof}

In the same way, we obtain

\begin{lemma}
\label{lemma:ETrbound2}
Let ${\bf u},{\bf v}\in{\bf Z}_b^2$ satisfying $s_b({\bf u})\cap\Lambda\ne\emptyset$ 
and $s_b({\bf v})\cap\Lambda\ne\emptyset$. 
Then there exists a positive 
constant $C$ which is independent of the locations ${\bf u},{\bf v}$ 
and of the size $|\Lambda|$ such that 
\begin{equation}
{\bf E}\left[\left|{\rm Tr}\>\chi_b({\bf u})P_{{\rm F},\Lambda}
[P_{{\rm F},\Lambda},\sharp][P_{{\rm F},\Lambda},\sharp]\chi_b({\bf u})\cdot
{\rm Tr}\>\chi_b({\bf v})P_{{\rm F},\Lambda}
[P_{{\rm F},\Lambda},\sharp][P_{{\rm F},\Lambda},\sharp]
\chi_b({\bf v})\right|\right]<C,
\label{TrDCPF2}
\end{equation}
where $\sharp$ is either $x$ or $y$. 
\end{lemma}

Using the magnetic translations and the argument in the proof of 
Lemma~\ref{lemma:sigmaxyinrep}, 
the Hall conductance for the boundary region is written as 
\begin{equation}
\sigma_{xy}^{\rm out}=\frac{e^2}{h}\frac{2\pi i}{L_xL_y}
\sum_{\Omega'}{\rm Tr}\>\chi_{\Omega'}P_{{\rm F},\Lambda}'[[P_{{\rm F},\Lambda}',x],
[P_{{\rm F},\Lambda}',y]]\chi_{\Omega'}
+{\cal O}\left(\exp\left[-\mu_\infty'L^{2\kappa/3}\right]\right) 
\label{sigmaxyoutfinite}
\end{equation}
with probability larger than $(1-{\rm Const.}L^{-2[\kappa(\xi+2)-3]/3})$, 
where $\Omega'$ is the translate of a portion of the boundary region 
$\Lambda\backslash\Omega$, and $P_{{\rm F},\Lambda}'$ is the corresponding 
translate of the Fermi sea projection. From this result, the Hall conductance 
$\sigma_{xy}^{\rm out}$ for the boundary region is vanishing 
in a limit $L\uparrow\infty$ with probability one: 

\begin{theorem}
\label{theorem:sigmaxyoutzero}
There exists a sequence $\{L_n\}_n$ of the system sizes $L=L_n$ such that 
$|\sigma_{xy}^{\rm out}|\rightarrow 0$ as $L_n\rightarrow\infty$ 
for almost every $\omega$. 
\end{theorem}

\begin{proof}{Proof}
Write
\begin{equation} 
I'(L)=\frac{1}{L_xL_y}
\sum_{\Omega'}\left|{\rm Tr}\>\chi_{\Omega'}P_{{\rm F},\Lambda}'[[P_{{\rm F},\Lambda}',x],
[P_{{\rm F},\Lambda}',y]]\chi_{\Omega'}\right|.
\end{equation}
By Lemma~\ref{lemma:ETrbounded}, we have 
${\bf E}[I'(L)]\rightarrow 0$ as $L\rightarrow\infty$. Combining this and 
the inequality ${\bf E}[I'(L)]\ge \varepsilon{\rm Prob}(I'(L)>\varepsilon)$, 
we can find a sequence $\{L_n\}_n$ of the system sizes $L=L_n$ such that 
$\{L_n\}_n$ satisfies the following two conditions: 
\begin{equation}
\sum_n{\rm Prob}(I'(L_n)>\varepsilon_n)<\infty
\quad 
\mbox{and} 
\quad
\sum_n L_n^{-2[\kappa(\xi+2)-3]/3}<\infty,
\label{Probcond}
\end{equation}
where $\{\varepsilon_n\}_n$ is a sequence satisfying $\varepsilon_n\rightarrow 0$ 
as $n\rightarrow \infty$. The application of Borel-Cantelli theorem yields 
that for almost every $\omega$, there exists a number $n_0(\omega)$ which may depend on 
$\omega$ such that 
$I'(L_n)\le \varepsilon_n$ for all $n\ge n_0(\omega)$ and that 
the finite size correction for $\sigma_{xy}^{\rm out}$ is evaluated by  
${\cal O}\left(\exp\left[-\mu_\infty'L_n^{2\kappa/3}\right]\right)$ 
for all $n\ge n_0(\omega)$. By combining this with the expression (\ref{sigmaxyoutfinite}) 
of $\sigma_{xy}^{\rm out}$, the statement of the theorem is proved. 
\end{proof}

%%%%%%%%%%%%%%%%%%%%%%%%%%%%%%%%%%%%%%%%%%%%%%%%%%%%%%%%%%%%%%%%%%%%%%%
\Section{Integrality of the Hall conductance\hfill\break\hfill
---Index theoretical approach---}
\label{IndexApproach}

In this section, integrality of the Hall conductance is proved 
by using the index theoretical method \cite{ASS,BVS,AG}. 

When we apply the method of \cite{ASS} using a pair index of two 
projections to a concrete example  
of a continuous random model such as the present system, 
there arises a problem that 
we need a decay bound for the integral kernel 
of the Fermi sea projection whose Fermi energy lies in the localization regime. 
But getting such a decay bound is very difficult, and so this problem is still unsolved. 
Recently, Germinet, Klein and Schenker \cite{GKS} proved the constancy of the Hall 
conductance for a random Landau Hamiltonian which is translation ergodic, 
without relying on a decay bound for the integral kernel 
of the Fermi sea projection. In their proof, 
they used a consequence of the multiscale analysis which is related to 
multiplicity of the eigenvalues of the Hamiltonian, for  
the Hall conductance formula\footnote{The explicit form of the Hall conductance 
formula using switch functions is given in Appendix~\ref{IndexSwitch}. 
We also discuss the relation between 
this and the standard Hall conductance formula using the position operator 
instead of the switch functions.} which is expressed 
in terms of switch functions instead of the position operator of the electron. 
This Hall conductance formula 
was justified \cite{ES} within the linear response approximation 
under the assumption on a spectral gap above the Fermi level.  
The integer of the quantized value of 
the Hall conductance can be determined under the assumption that   
the disordered-broadened 
Landau bands are disjoint, i.e., there exists a nonvanishing spectral gap 
between two neighboring Landau bands. 

In our approach, we assume neither the above disjoint condition 
for the Landau bands nor the periodicity of the potentials 
$V_0$ and ${\bf A}_{\rm P}$ which implies a translation ergodic Hamiltonian. 
But we must require the ``covering condition" (\ref{unoflat}) 
which is not required in \cite{GKS}. 
This condition is needed to estimate the number of the localized states and 
to obtain a decay bound  \cite{AENSS} for a fractional moment of the resolvent. 
In order to circumvent the above problem about a decay estimate for 
the integral kernel of the Fermi sea projection, we introduce 
a partition of unity which is a collection of the characteristic functions 
of a small rectangular boxes. 

Let $s_\varepsilon({\bf u})$ be the 
$\varepsilon_1\times\varepsilon_2$ rectangular box ceneterd at 
${\bf u}\in({\bf Z}_\varepsilon^2)^\ast$ with the pair  
$\varepsilon=(\varepsilon_1,\varepsilon_2)$ of the sidelengths,  
and $\chi_\varepsilon({\bf u})$ the characteristic function.  
We take $N_1\varepsilon_1=a$ and $N_2\varepsilon_2=\sqrt{3}a/2$ with 
large positive integers $N_1,N_2$ and with the lattice constant $a$ of 
the triangular lattice ${\bf L}^2$ so that the set of all the boxes 
$s_\varepsilon({\bf u})$ is invariant under the lattice translations of 
the triangular lattice ${\bf L}^2$. 
When we take the limit $\varepsilon_1,\varepsilon_2\downarrow 0$, 
we keep the ratio $\varepsilon_1/\varepsilon_2$ finite. 
We introduce a unitary operator, 
\begin{equation}
U_{\bf a}^\varepsilon
=\sum_{{\bf u}\in({\bf Z}_\varepsilon^2)^\ast}
\chi_\varepsilon({\bf u})
\exp[i\theta_{\bf a}({\bf u})] \quad\mbox{for}\ \ {\bf a}\in{\bf Z}_\varepsilon^2, 
\end{equation}
where $\theta_{\bf a}({\bf u})$ is the angle 
of sight from ${\bf a}$ to ${\bf u}$, 
i.e., ${\rm arg}({\bf u}-{\bf a})$ in the terminology of the complex plane. 
Consider the operator, $T:=P_{\rm F}-U_{\bf a}^\varepsilon 
P_{\rm F}\left(U_{\bf a}^\varepsilon\right)^\ast$. 

\begin{lemma}
For fixed parameters $\varepsilon_1,\varepsilon_2$, 
${\bf E}\left[{\rm Tr}\>|T|^3\right]<\infty$. 
\end{lemma}
\begin{proof}{Proof}
Note that 
\begin{eqnarray}
T=P_{\rm F}-U_{\bf a}^\varepsilon P_{\rm F}
\left(U_{\bf a}^\varepsilon\right)^\ast
&=&\sum_{{\bf u},{\bf v}\in({\bf Z}_\varepsilon^2)^\ast}
\chi_\varepsilon({\bf u})\left[
P_{\rm F}-U_{\bf a}^\varepsilon P_{\rm F}\left(U_{\bf a}^\varepsilon\right)^\ast
\right]\chi_\varepsilon({\bf v})\ret
&=&\sum_{{\bf u},{\bf v}\in({\bf Z}_\varepsilon^2)^\ast}
\left[1-e^{i\theta_{\bf a}({\bf u})-i\theta_{\bf a}({\bf v})}\right]
\chi_\varepsilon({\bf u})P_{\rm F}\chi_\varepsilon({\bf v}). 
\end{eqnarray}
Define $t_{{\bf u},{\bf v}}:=1-e^{i\theta_{\bf a}({\bf u})
-i\theta_{\bf a}({\bf v})}$ and $T_{{\bf u},{\bf v}}:=t_{{\bf u},{\bf v}}
\chi_\varepsilon({\bf u})P_{\rm F}\chi_\varepsilon({\bf v})$. 
Following the idea of \cite{AG}, we introduce 
$T_{{\bf u},{\bf v}}^{({\bf b})}=T_{{\bf u},{\bf v}}\delta_{{\bf u}-{\bf b},{\bf v}}$. 
Clearly, one has $\sum_{{\bf b}}T_{{\bf u},{\bf v}}^{({\bf b})}=T_{{\bf u},{\bf v}}$ 
and  
\begin{eqnarray}
\left({T^{({\bf b})}}^\ast T^{({\bf b})}\right)_{{\bf u},{\bf v}}
&=&\sum_{{\bf w}}T_{{\bf w},{\bf u}}^\ast\delta_{{\bf w}-{\bf b},{\bf u}}
T_{{\bf w},{\bf v}}\delta_{{\bf w}-{\bf b},{\bf v}}\ret
&=&T_{{\bf u+b},{\bf u}}^\ast T_{{\bf u+b},{\bf u}}\delta_{{\bf u},{\bf v}}\ret
&=&|t_{{\bf u+b},{\bf u}}|^2\chi_\varepsilon({\bf u})P_{\rm F}\chi_\varepsilon({\bf u+b})
P_{\rm F}\chi_\varepsilon({\bf u})\delta_{{\bf u},{\bf v}}.
\end{eqnarray}
Using these identities and Minkowski's inequality, one obtains 
\begin{eqnarray}
\left({\bf E}\>{\rm Tr}|T|^3\right)^{1/3}
&\le&\sum_{\bf b}\left({\bf E}\>{\rm Tr}\>|T^{({\bf b})}|^3\right)^{1/3}\ret
&\le&\sum_{\bf b}\left\{\sum_{\bf u}|t_{{\bf u+b},{\bf u}}|^3
{\bf E}\left[{\rm Tr}\>\left|\chi_\varepsilon({\bf u})P_{\rm F}\chi_\varepsilon({\bf u+b})
P_{\rm F}\chi_\varepsilon({\bf u})\right|^{3/2}\right]\right\}^{1/3}.
\label{TrT3bound} 
\end{eqnarray}
Since the inequality,
\begin{equation}
\left|1-e^{i\theta_{\bf a}({\bf u})-i\theta_{\bf a}({\bf v})}\right|
\le\frac{2|{\bf u-v}|}{|{\bf u-a}|},
\label{2sinthetadiffbound}
\end{equation}
holds as in \cite{ASS}, one obtains
\begin{equation}
\sum_{\bf u}|t_{{\bf u+b},{\bf u}}|^3\le 2^3
\sum_{\bf u}\frac{|{\bf b}|^3}{|{\bf u-a}|^3}
\le{\rm Const.}|{\bf b}|^3.
\label{ineqt3} 
\end{equation}
Note that 
\begin{eqnarray}
& &{\rm Tr}\>\left|\chi_\varepsilon({\bf u})P_{\rm F}\chi_\varepsilon({\bf u+b})
P_{\rm F}\chi_\varepsilon({\bf u})\right|^{3/2}\ret
&\le&\sqrt{{\rm Tr}\>\chi_\varepsilon({\bf u})P_{\rm F}\chi_\varepsilon({\bf u+b})
P_{\rm F}\chi_\varepsilon({\bf u})}\ret
&\times&\sqrt{{\rm Tr}\>\chi_\varepsilon({\bf u})P_{\rm F}\chi_\varepsilon({\bf u+b})
P_{\rm F}\chi_\varepsilon({\bf u})P_{\rm F}\chi_\varepsilon({\bf u+b})
P_{\rm F}\chi_\varepsilon({\bf u})}\ret
&\le&\sqrt{{\rm Tr}\>\chi_\varepsilon({\bf u})P_{\rm F}\chi_\varepsilon({\bf u})}\cdot
\left\Vert\chi_\varepsilon({\bf u})P_{\rm F}\chi_\varepsilon({\bf u+b})\right\Vert
\sqrt{{\rm Tr}\>\chi_\varepsilon({\bf u})P_{\rm F}\chi_\varepsilon({\bf u})}\ret
&\le&\left\Vert\chi_\varepsilon({\bf u})P_{\rm F}\chi_\varepsilon({\bf u+b})\right\Vert
{\rm Tr}\>\chi_\varepsilon({\bf u})P_{\rm F}\chi_\varepsilon({\bf u}).
\label{TrCPCPC3/2}
\end{eqnarray}
{From} this, the decay bound (\ref{PFD}) for the Fermi sea projection 
and (\ref{localTr}), we have 
\begin{equation}
{\bf E}\left[{\rm Tr}\>\left|\chi_\varepsilon({\bf u})P_{\rm F}\chi_\varepsilon({\bf u+b})
P_{\rm F}\chi_\varepsilon({\bf u})\right|^{3/2}\right]
\le{\rm Const.}e^{-\mu|{\bf b}|}
\end{equation}
Combining this, (\ref{TrT3bound}) and (\ref{ineqt3}) yields 
\begin{equation}
\left({\bf E}\>{\rm Tr}\>|T|^3\right)^{1/3}\le{\rm Const.}\sum_{\bf b}|{\bf b}|
e^{-\mu|{\bf b}|/3}<\infty.
\end{equation}
\end{proof}

The result implies that the operator $T^3$ is trace class for almost 
every $\omega$. Thus we can define the relative index \cite{ASS}, 
\begin{equation}
{\rm Index}(P_{\rm F},U_{\bf a}^\varepsilon P_{\rm F}(U_{\bf a}^\varepsilon)^\ast)
:={\rm Tr}\left(P_{\rm F}-U_{\bf a}^\varepsilon P_{\rm F}(U_{\bf a}^\varepsilon)^\ast
\right)^3,
\end{equation} 
for the pair of the projections. This right-hand side takes an integer value 
as proved in \cite{ASS}. 

Let $\ell_{\rm P}$ be a large positive integer. 
Let ${\bf A}^{\rm LP}\in C^1({\bf R}^2)$ be a periodic 
vector potential satisfying the periodicity, 
\begin{equation}
{\bf A}^{\rm LP}({\bf r}+\ell_{\rm P}{\bf a}_1)=
{\bf A}^{\rm LP}({\bf r}+\ell_{\rm P}{\bf a}_2)=
{\bf A}^{\rm LP}({\bf r}),
\end{equation}
and $V_0^{\rm LP}$ a periodic electrostatic potential satisfying the same periodicity, 
\begin{equation}
V_0^{\rm LP}({\bf r}+\ell_{\rm P}{\bf a}_1)=
V_0^{\rm LP}({\bf r}+\ell_{\rm P}{\bf a}_2)=
V_0^{\rm LP}({\bf r}),
\end{equation}
where ${\bf a}_j$ are the primitive translation vectors of the triangular lattice 
${\bf L}^2$. 
In order to prove integrality of the Hall conductance, 
we consider the Hamiltonian,   
\begin{equation}
H_\omega^{\rm LP}=\frac{1}{2m_e}[{\bf p}+e({\bf A}^{\rm LP}+{\bf A}_0)]^2
+V_0^{\rm LP}+V_\omega,
\label{HamLP}
\end{equation}
on the whole plane ${\bf R}^2$. Namely this Hamiltonian is obtained 
by replacing ${\bf A}_{\rm P}, V_0$ with ${\bf A}^{\rm LP}, V_0^{\rm LP}$   
in the Hamiltonian $H_\omega$ of (\ref{Hamomega}). 
We choose the integer $\ell_{\rm P}$ so that 
the unit cell of the large triangular lattice 
$\ell_{\rm P}{\bf L}^2$ contains the rectangular region 
$\Lambda^{\rm sys}$ of (\ref{Lamsys}) 
on which the present finite Hall system is defined. 

We note that the magnetic translations act on the random potential $V_\omega$ as 
the corresponding translation, and that the pair index does not depend on 
the location ${\bf a}$ of the flux \cite{ASS}. 
Therefore the pair index for the Hamiltonian $H_\omega^{\rm LP}$ 
is an invariant function of the randomness under  
the lattice translations of the triangular lattice $\ell_P{\bf L}^2$. Further, 
since the index is measurable \cite{ASS} and integrable 
with respect to the random variables, 
Birkhoff's ergodic theorem implies that \cite{AG} 
the value of the pair index does not fluctuate in the sense that 
it takes an integer given by its mean for almost every random potentials.  
But the value of the integer may depend on the period $\ell_{\rm P}$. 

As in \cite{ASS}, the relation between the pair index, 
${\rm Index}(P_{\rm F},U_{\bf a}^\varepsilon P_{\rm F}(U_{\bf a}^\varepsilon)^\ast)$, 
and the Fredholm index, ${\rm Index}(P_{\rm F}U_{\bf a}^\varepsilon P_{\rm F})$, of 
the operator $P_{\rm F}U_{\bf a}^\varepsilon P_{\rm F}$ in range of $P_{\rm F}$ 
is given by 
\begin{equation}
{\rm Index}(P_{\rm F}U_{\bf a}^\varepsilon P_{\rm F})=
-{\rm Index}(P_{\rm F},U_{\bf a}^\varepsilon P_{\rm F}(U_{\bf a}^\varepsilon)^\ast).
\end{equation} 
Consider another unitary operator, 
\begin{equation}
U_{\bf a}:=\frac{x+iy-(a_1+ia_2)}{|x+iy-(a_1+ia_2)|}.
\end{equation}
Clearly, one has $P_{\rm F}U_{\bf a}P_{\rm F}=P_{\rm F}U_{\bf a}^\varepsilon P_{\rm F}+
P_{\rm F}(U_{\bf a}-U_{\bf a}^\varepsilon)P_{\rm F}$. 
Since the operator of the second term is compact, stability theory\footnote{See, 
for example, the book \cite{STFI}.}  
of the Fredholm indices implies that $P_{\rm F}U_{\bf a}P_{\rm F}$ becomes  
a Fredholm operator, too, and that the index is invariant under the compact 
perturbation. Thus we have   
\begin{equation} 
{\rm Index}(P_{\rm F}U_{\bf a}P_{\rm F})
={\rm Index}(P_{\rm F}U_{\bf a}^\varepsilon P_{\rm F})
=-{\rm Index}(P_{\rm F},U_{\bf a}^\varepsilon P_{\rm F}(U_{\bf a}^\varepsilon)^\ast).
\end{equation}
In consequence, the pair index does not depend on 
the parameters $\varepsilon_1,\varepsilon_2$. 

Following \cite{EGS}, we obtain the expression (\ref{Index4}) below 
with (\ref{def:calIvarepsilon}) for the pair index. 
The expression leads to the well-known Hall conductance formula \cite{BVS} 
which is written in terms of the position operator of the electron. 

To begin with, we note that 
\begin{eqnarray}
& &{\rm Tr}\left(P_{\rm F}-U_{\bf a}^\varepsilon 
P_{\rm F}(U_{\bf a}^\varepsilon)^\ast
\right)^3\ret
&=&\sum_{{\bf u},{\bf v},{\bf w}}{\rm Tr}\>
\chi_\varepsilon({\bf u})(P_{\rm F}-U_{\bf a}^\varepsilon P_{\rm F}
(U_{\bf a}^\varepsilon)^\ast)\chi_\varepsilon({\bf v})
(P_{\rm F}-U_{\bf a}^\varepsilon P_{\rm F}
(U_{\bf a}^\varepsilon)^\ast)\chi_\varepsilon({\bf w})\ret
& &\qquad\qquad\qquad\times(P_{\rm F}-U_{\bf a}^\varepsilon P_{\rm F}
(U_{\bf a}^\varepsilon)^\ast)\chi_\varepsilon({\bf u})\ret
&=&\sum_{{\bf u},{\bf v},{\bf w}}t_{{\bf u},{\bf v}}t_{{\bf v},{\bf w}}
t_{{\bf w},{\bf u}}{\rm Tr}\>\chi_\varepsilon({\bf u})P_{\rm F}\chi_\varepsilon({\bf v})
P_{\rm F}\chi_\varepsilon({\bf w})P_{\rm F}\chi_\varepsilon({\bf u}). 
\end{eqnarray}
Since the index is independent of the location ${\bf a}$ of the flux \cite{ASS}, 
one has 
\begin{equation}
{\rm Index}(P_{\rm F}U_{\bf a}P_{\rm F})
=-\frac{1}{{\cal V}_\ell}\sum_{{\bf a}\in\Lambda_\ell}
\sum_{{\bf u},{\bf v},{\bf w}}t_{{\bf u},{\bf v}}t_{{\bf v},{\bf w}}
t_{{\bf w},{\bf u}}{\rm Tr}\>\chi_\varepsilon({\bf u})P_{\rm F}\chi_\varepsilon({\bf v})
P_{\rm F}\chi_\varepsilon({\bf w})P_{\rm F}\chi_\varepsilon({\bf u}),
\label{Index1}
\end{equation}
where $\Lambda_\ell=\varepsilon_1\{-\ell,-\ell+1,\ldots,\ell\}
\times\varepsilon_2\{-\ell',-\ell'+1,\ldots,\ell'\}\subset
{\bf Z}_\varepsilon^2$, and 
${\cal V}_\ell=(2\ell+1)(2\ell'+1)$. We choose $\ell'$ so that the ratio $\ell'/\ell$ 
is finite. 

\begin{lemma}
\label{lemma:Index2}
There exists a sequence $\{\ell_n\}_n$ such that for almost every $\omega$, 
the index is written 
\begin{equation}
{\rm Index}(P_{\rm F}U_{\bf a}P_{\rm F})
=-\lim_{\ell_n\uparrow\infty}
\frac{1}{{\cal V}_{\ell_n}}\sum_{{\bf u}\in\Lambda_{\ell_n}^\ast}
\sum_{{\bf v},{\bf w}}\sum_{{\bf a}\in{\bf Z}_\varepsilon^2}
t_{{\bf u},{\bf v}}t_{{\bf v},{\bf w}}t_{{\bf w},{\bf u}}
S_{\bf u,v,w,u}
\label{Index2}
\end{equation}
with
\begin{equation}
S_{\bf u,v,w,u}:={\rm Tr}\>\chi_\varepsilon({\bf u})P_{\rm F}\chi_\varepsilon({\bf v})
P_{\rm F}\chi_\varepsilon({\bf w})P_{\rm F}\chi_\varepsilon({\bf u}),
\label{defS}
\end{equation}
where the lattice $\Lambda_\ell^\ast$ is given by 
\begin{equation}
\Lambda_\ell^\ast=\varepsilon_1\{-\ell+1/2,-\ell+3/2,\ldots,\ell-1/2\}
\times\varepsilon_2\{-\ell'+1/2,-\ell'+3/2,\ldots,\ell'-1/2\}.
\label{Lambdaellast}
\end{equation} 
\end{lemma}
The proof is given in Appendix~\ref{IndexProof}. Using Connes' area formula \cite{Connes}, 
\begin{equation} 
\sum_{{\bf a}\in{\bf Z}_\varepsilon^2}
t_{{\bf u},{\bf v}}t_{{\bf v},{\bf w}}t_{{\bf w},{\bf u}}
=\frac{2\pi i}{\varepsilon_1\varepsilon_2}({\bf v}-{\bf u})\times({\bf w}-{\bf u}), 
\end{equation}
the index of (\ref{Index2}) is written 
\begin{eqnarray}
& &{\rm Index}(P_{\rm F}U_{\bf a}P_{\rm F})\ret
&=&-\lim_{|\Omega|\uparrow\infty}
\frac{2\pi i}{\left|\Omega\right|}\sum_{{\bf u}\in\Lambda_{\ell_n}^\ast}
\sum_{{\bf v},{\bf w}}
({\bf v}-{\bf u})\times({\bf w}-{\bf u})
{\rm Tr}\>\chi_\varepsilon({\bf u})P_{\rm F}\chi_\varepsilon({\bf v})
P_{\rm F}\chi_\varepsilon({\bf w})P_{\rm F}\chi_\varepsilon({\bf u}),\ret
\label{Index3}
\end{eqnarray}
where $\Omega$ is the $L_1\times L_2$ rectangular box centered at ${\bf r}=0$ with 
the sidelengths $L_1=2\ell_n\varepsilon_1$ and $L_2=2\ell_n'\varepsilon_2$. 
Note that  
\begin{equation}
({\bf v}-{\bf u})\times({\bf w}-{\bf u})=({\bf v}-{\bf w})\times({\bf w}-{\bf u})
=(v_2-w_2)(w_1-u_1)-(v_1-w_1)(w_2-u_2).
\label{idexpro}
\end{equation}
Further we have 
\begin{eqnarray}
& &(v_2-w_2)(w_1-u_1)
{\rm Tr}\>\chi_\varepsilon({\bf u})P_{\rm F}\chi_\varepsilon({\bf v})
P_{\rm F}\chi_\varepsilon({\bf w})P_{\rm F}\chi_\varepsilon({\bf u})\ret
&=&{\rm Tr}\>\chi_\varepsilon({\bf u})P_{\rm F}
\chi_\varepsilon({\bf v})[X_2^\varepsilon,P_{\rm F}]\chi_\varepsilon({\bf w})
[X_1^\varepsilon,P_{\rm F}]\chi_\varepsilon({\bf u}), 
\end{eqnarray}
where the two-component function $(X_1^\varepsilon,X_2^\varepsilon)$ of 
${\bf r}=(x,y)$ is given by $(X_1^\varepsilon,X_2^\varepsilon)=(u_1,u_2)$ 
for ${\bf r}$ in 
the $\varepsilon_1\times\varepsilon_2$ rectangular box $s_\varepsilon({\bf u})$ 
centered at ${\bf u}=(u_1,u_2)$. From these observations, the index is written 
\begin{equation}
{\rm Index}(P_{\rm F}U_{\bf a}P_{\rm F})
=\lim_{|\Omega|\uparrow\infty}{\cal I}^\varepsilon(P_{\rm F};\Omega,\ell_P)
\label{Index4}
\end{equation}
for almost every $\omega$, where we have written 
\begin{equation}
{\cal I}^\varepsilon(P_{\rm F};\Omega,\ell_P)=
\frac{2\pi i}{\left|\Omega\right|}
{\rm Tr}\>\chi_\Omega P_{\rm F}
[[P_{\rm F},X_1^\varepsilon],[P_{\rm F},X_2^\varepsilon]]\chi_\Omega.
\label{def:calIvarepsilon}
\end{equation}
{From} the proof of Lemma~\ref{lemma:Index2}, we obtain 
\begin{equation}
{\bf E}\left[\left|{\rm Index}(P_{\rm F}U_{\bf a}P_{\rm F})
-{\cal I}^\varepsilon(P_{\rm F};\Omega,\ell_P)\right|\right]
\rightarrow 0\quad\mbox{as }\ |\Omega|\uparrow\infty.
\label{EIndexIvarepsilonapprox}
\end{equation}
We also write 
\begin{equation}
{\cal I}(P_{\rm F};\Omega,\ell_P):=\frac{2\pi i}{\left|\Omega\right|}
{\rm Tr}\>\chi_\Omega P_{\rm F}
[[P_{\rm F},X_1],[P_{\rm F},X_2]]\chi_\Omega,
\end{equation}
where $(X_1,X_2)=(x,y)$. This right-hand side is nothing but 
the well-known form of the Hall conductance formula \cite{BVS}. 

\begin{lemma}
\label{lemma:diffIIvarepsilon}
The following bound is valid: 
\begin{equation}
{\bf E}\left[\left|
{\cal I}(P_{\rm F};\Omega,\ell_P)-{\cal I}^\varepsilon(P_{\rm F};\Omega,\ell_P)
\right|\right]\le{\rm Const.}|\varepsilon|,
\end{equation}
where $|\varepsilon|=\sqrt{\varepsilon_1^2+\varepsilon_2^2}$, 
and the positive constant in the right-hand side is independent 
of $\Omega$ and $\ell_P$. 
\end{lemma}

\begin{proof}{Proof}
Note that  
\begin{eqnarray}
& &\left|{\rm Tr}\>\chi_\Omega P_{\rm F}
[P_{\rm F},X_1][P_{\rm F},X_2]\chi_\Omega
-{\rm Tr}\>\chi_\Omega P_{\rm F}
[P_{\rm F},X_1^\varepsilon][P_{\rm F},X_2^\varepsilon]\chi_\Omega\right|\ret
&\le&\left|{\rm Tr}\>\chi_\Omega P_{\rm F}
[P_{\rm F},(X_1-X_1^\varepsilon)][P_{\rm F},X_2]\chi_\Omega\right|
+\left|{\rm Tr}\>\chi_\Omega P_{\rm F}
[P_{\rm F},X_1^\varepsilon][P_{\rm F},(X_2-X_2^\varepsilon)]\chi_\Omega\right|,\ret
\end{eqnarray}
\begin{equation}
[P_{\rm F},X_2]=[P_{\rm F},(y-y^b)]+[P_{\rm F},y^b]
\quad\mbox{and}\quad 
[P_{\rm F},X_1^\varepsilon]
=[P_{\rm F},(X_1^\varepsilon-x^b)]+[P_{\rm F},x^b].
\end{equation} 
Therefore we can prove the statement of the theorem in the same way as 
in the proof of Lemma~\ref{lemma:ETrbounded}. 
\end{proof}

Since we can apply Borel-Cantelli 
theorem as in the proof of Theorem~\ref{theorem:sigmaxyoutzero}
to this result, Lemma~\ref{lemma:diffIIvarepsilon} 
yields that there exists a sequence 
$\{\varepsilon_n=(\varepsilon_{1,n},\varepsilon_{2,n})\}_n$ of 
$\varepsilon=\varepsilon_n$ 
satisfying $\varepsilon_{j,n}\rightarrow 0$ as $n\rightarrow\infty$, for 
$j=1,2$, such that  
\begin{equation}
\left|
{\cal I}(P_{\rm F};\Omega,\ell_P)-{\cal I}^{\varepsilon_n}(P_{\rm F};\Omega,\ell_P)
\right|\rightarrow 0 \quad\mbox{as }\ n\rightarrow \infty
\label{Index5}
\end{equation}
almost surely for any fixed large $\Omega$. 

Let $\{\delta_n\}_{n=1}^\infty$ be a sequence of positive numbers $\delta_n$ 
satisfying $\delta_n\rightarrow 0$ as $n\rightarrow \infty$, and 
let $\{p_n\}_{n=1}^\infty$ be a sequence of positive numbers $p_n<1$ satisfying 
$\sum_n p_n<\infty$. Relying on Lemma~\ref{lemma:diffIIvarepsilon}, 
we choose $\varepsilon=\varepsilon_n=
(\varepsilon_{1,n},\varepsilon_{2,n})$ for each $n$ so that 
\begin{equation}
{\rm Prob}\left[\left|{\cal I}(P_{\rm F};\Omega,\ell_P)
-{\cal I}^{\varepsilon_n}(P_{\rm F};\Omega,\ell_P)
\right|>\delta_n/2\right]\le p_n.
\end{equation}
Further, for this $\varepsilon=\varepsilon_n$, 
we can choose a sufficiently large $\Omega=\Omega_n$ so that 
\begin{equation}
{\rm Prob}\left[\left|{\rm Index}(P_{\rm F}U_{\bf a}P_{\rm F})
-{\cal I}^{\varepsilon_n}(P_{\rm F};\Omega_n,\ell_P)\right|
>\delta_n/2\right]\le p_n,
\end{equation}
from the proof of Lemma~\ref{lemma:Index2}. The application of Borel-Cantelli 
theorem yields that for almost every $\omega$, 
there exists a number $n_0(\omega)$ such that 
for all $n\ge n_0(\omega)$, the following two inequalities are valid: 
\begin{equation}
\left|{\cal I}(P_{\rm F};\Omega_n,\ell_P)
-{\cal I}^{\varepsilon_n}(P_{\rm F};\Omega_n,\ell_P)
\right|\le\delta_n/2.
\end{equation}
and 
\begin{equation}
\left|{\rm Index}(P_{\rm F}U_{\bf a}P_{\rm F})
-{\cal I}^{\varepsilon_n}(P_{\rm F};\Omega_n,\ell_P)\right|
\le \delta_n/2.
\end{equation}
These two inequalities imply   
\begin{eqnarray}
& &\left|{\rm Index}(P_{\rm F}U_{\bf a}P_{\rm F})
-{\cal I}(P_{\rm F};\Omega_n,\ell_P)\right|\ret
&\le&\left|{\rm Index}(P_{\rm F}U_{\bf a}P_{\rm F})
-{\cal I}^{\varepsilon_n}(P_{\rm F};\Omega_n,\ell_P)\right|
+\left|{\cal I}^{\varepsilon_n}(P_{\rm F};\Omega_n,\ell_P)
-{\cal I}(P_{\rm F};\Omega_n,\ell_P)\right|\ret
&\le&\delta_n/2+\delta_n/2=\delta_n.
\label{indexbound} 
\end{eqnarray}
This result is summarized as the following index theorem: 

\begin{theorem}
\label{theorem:standindcond}
For a fixed period $\ell_P$ and for almost every $\omega$, 
there exists a sequence $\{\Omega_n\}_n$ of the regions $\Omega=\Omega_n$ 
going to ${\bf R}^2$ as $n\rightarrow\infty$ such that 
\begin{equation}
{\rm Index}(P_{\rm F}U_{\bf a}P_{\rm F})=\lim_{\Omega_n\uparrow{\bf R}^2}
\frac{2\pi i}{\left|\Omega_n\right|}
{\rm Tr}\>\chi_{\Omega_n}P_{\rm F}
[[P_{\rm F},X_1],[P_{\rm F},X_2]]\chi_{\Omega_n}.
\end{equation}
\end{theorem}

For taking the infinite-volume limit $\Lambda\uparrow{\bf R}^2$, 
we want to take a sequence $\{\Lambda_n\}_n$ of 
the finite region $\Lambda=\Lambda_n$ of the present system so that 
the condition (\ref{distOmegaLambdaLambdap}) is satisfied for  
$\Lambda=\Lambda_n\supset\Omega=\Omega_n$.
Then we must take the limit $\ell_P\uparrow\infty$ 
together with the limit $\Omega_n\uparrow{\bf R}^2$ 
so that the unit cell of the lattice $\ell_P{\bf L}^2$ includes 
the region $\Lambda=\Lambda_n$ of the present system.
In the following, we consider a sequence  
$\{\Lambda_n,\Omega_n,\ell_{{\rm P},n}\}_n$ which satisfies 
this requirement. 

Since both of the key bounds 
in the proofs of Lemmas~\ref{lemma:Index2} and \ref{lemma:diffIIvarepsilon}
do not depend on the period $\ell_{\rm P}$, we can take the limit $\ell_P\uparrow\infty$ 
together with the limit $\Omega_n\uparrow{\bf R}^2$ in the above argument 
for Theorem~\ref{theorem:standindcond} 
so that the above requirement is satisfied. 
But both of ${\rm Index}(P_{\rm F}U_{\bf a}P_{\rm F})$ and 
${\cal I}(P_{\rm F};\Omega_n,\ell_P)$ may go to infinity 
as $\ell_P\uparrow\infty$. First let us prove that this case does not occur. 
{From} the argument of the proof of Lemma~\ref{lemma:ETrbounded}, 
one can easily show that the expectation value 
${\bf E}[|{\cal I}(P_{\rm F};\Omega,\ell_P)|]$ is bounded uniformly in 
$\Omega$ and $\ell_{\rm P}$. 
Combining this with Fatou's lemma, we have 
\begin{equation}
{\bf E}\left[\mathop{\lim\inf}_{\Omega\uparrow{\bf R}^2,\ell_{\rm P}\uparrow\infty}
|{\cal I}(P_{\rm F};\Omega,\ell_P)|\right]\le
\mathop{\lim\inf}_{\Omega\uparrow{\bf R}^2,\ell_{\rm P}\uparrow\infty}
{\bf E}\left[|{\cal I}
(P_{\rm F};\Omega,\ell_P)|\right]<\infty. 
\end{equation}
This implies that for almost every $\omega$, there exists a sequence 
$\{\Omega_n(\omega),\ell_{{\rm P},n}(\omega)\}_n$ of 
the pair $\{\Omega,\ell_{\rm P}\}$ 
such that $\{\Omega_n(\omega)\}$ is a subsequence of the sequence $\{\Omega_n\}$ 
of Theorem~\ref{theorem:standindcond}, and that $\lim_{n\uparrow\infty}{\cal I}
(P_{\rm F};\Omega_n(\omega),\ell_{{\rm P},n}(\omega))$ exists. 
Here we should stress that the sequence 
$\{\Omega_n(\omega),\ell_{{\rm P},n}(\omega)\}_n$ may depend on 
the random event $\omega$. 
On the other hand, the inequality (\ref{indexbound}) holds for 
a large pair $\{\Omega,\ell_P\}=\{\Omega_n(\omega),
\ell_{{\rm P},n}(\omega)\}$. These observations imply 
that for a fixed $\omega$, the index ${\rm Index}(P_{\rm F}U_{\bf a}P_{\rm F})$ 
converges to an integer as $n\uparrow\infty$, too. 
But, since the index does not depend on $\omega$ 
as mentioned above, we can write $\{\ell_{{\rm P},n}\}_n$ for 
the sequence $\{\ell_{{\rm P},n}(\omega)\}_n$ 
by dropping the $\omega$ dependence, and obtain the result that 
the following limit exists and is constant for almost every $\omega$:   
\begin{equation}
{\rm Index}_\infty(P_{\rm F}U_{\bf a}P_{\rm F}):=
\lim_{\ell_{{\rm P},n}\uparrow\infty}
{\rm Index}(P_{\rm F}U_{\bf a}P_{\rm F}).
\end{equation}

Newly we choose $\{\Omega,\ell_P\}=\{\Omega_n,\ell_{{\rm P},n}\}$  
in the inequality (\ref{indexbound}). Then, since the index converges to 
the integer as $n\uparrow\infty$, we obtain    

\begin{theorem}
\label{theorem:indexinftyellP}
For almost every $\omega$, 
there exists a sequence $\{\Omega_n,\ell_{{\rm P},n}\}_n$ 
of the pair $\{\Omega=\Omega_n,\ell_{\rm P}=\ell_{{\rm P},n}\}$ 
such that the following relation holds:  
\begin{equation}
{\rm Index}_\infty(P_{\rm F}U_{\bf a}P_{\rm F})
=\lim_{\Omega_n\uparrow{\bf R}^2,\ell_{{\rm P},n}\uparrow\infty}
\frac{2\pi i}{\left|\Omega_n\right|}
{\rm Tr}\>\chi_{\Omega_n}P_{\rm F}
[[P_{\rm F},X_1],[P_{\rm F},X_2]]\chi_{\Omega_n}.
\end{equation}
\end{theorem}

\begin{theorem}
\label{theorem:indexRIF}
There exists a subsequence $\{\Omega_n,\ell_{{\rm P},n}\}_n$ of 
the sequence of the preceding Theorem~\ref{theorem:indexinftyellP} such that  
\begin{eqnarray}
{\rm Index}_\infty(P_{\rm F}U_{\bf a}P_{\rm F})
&=&\lim_{\Omega_n\uparrow{\bf R}^2,\ell_{{\rm P},n}\uparrow\infty}
\frac{2\pi i}{\left|\Omega_n\right|}
{\rm Tr}\>\chi_{\Omega_n} P_{\rm F}
[[P_{\rm F},X_1],[P_{\rm F},X_2]]\chi_{\Omega_n}\ret
&=&\lim_{\Omega_n\uparrow{\bf R}^2}
\frac{2\pi i}{|\Omega_n|}{\rm Tr}\>
\chi_{\Omega_n} P_{{\rm F},\Lambda_n}
[[P_{{\rm F},\Lambda_n},x],[P_{{\rm F},\Lambda_n},y]]\chi_{\Omega_n}
\end{eqnarray}
with probability one. Here we take $\Lambda=\Lambda_n$ and 
$\ell_{\rm P}=\ell_{{\rm P},n}$ 
so that the region $\Lambda=\Lambda_n$ of the present system 
satisfies the condition (\ref{distOmegaLambdaLambdap}) for each $\Omega=\Omega_n$, 
and the unit cell of the lattice $\ell_{{\rm P},n}{\bf L}^2$ includes 
the region $\Lambda_n$.
\end{theorem}

Theorem~\ref{theorem:indexRIF} follows from the following lemma: 

\begin{lemma}
\label{lemma:indexRIF}
Under the same assumption as in Theorem~\ref{theorem:indexRIF}, we have 
\begin{equation}
{\bf E}\left[\left|{\cal I}(P_{\rm F};\Omega_n,\ell_{{\rm P},n})
-{\cal I}(P_{{\rm F},\Lambda_n};\Omega_n)
\right|\right]\rightarrow 0\quad\mbox{as }\ 
n\rightarrow\infty. 
\end{equation}
\end{lemma}
The proof is given in Appendix~\ref{Proof9.7}. 

\begin{theorem}
\label{theorem:sigmaxyindexinfty}
There exists a sequence $\{L_n\}_n$ of the system 
sizes $L=L_n$ such that the Hall conductance $\sigma_{xy}$ 
in the infinite volume limit exists and is quantized to an integer as  
\begin{equation}
\sigma_{xy}=\frac{e^2}{h}
{\rm Index}_\infty(P_{\rm F}U_{\bf a}P_{\rm F})
\end{equation}
for almost every $\omega$. 
\end{theorem}

\begin{proof}{Proof}
From Theorems~\ref{theorem:indexinftyellP} 
and \ref{theorem:indexRIF}, there exists a sequence of 
$\{\Lambda_n,\Omega_n,\ell_{{\rm P},n}\}_n$ of the triplet 
$\{\Lambda,\Omega,\ell_{\rm P}\}$ 
such that the following three conditions are satisfied: 
(i) the condition (\ref{distOmegaLambdaLambdap}) is satisfied for  
$\Lambda=\Lambda_n\supset\Omega=\Omega_n$, 
(ii) the unit cell with the period $\ell_{{\rm P},n}$ includes 
the region $\Lambda=\Lambda_n$ of the system with the linear size $L=L_n'$,  
and (iii) for almost every $\omega$, the following formula holds: 
\begin{equation}
{\rm Index}_\infty(P_{\rm F}U_{\bf a}P_{\rm F})
=\lim_{n\uparrow\infty}\frac{2\pi i}{|\Omega_n|}{\rm Tr}\>
\chi_{\Omega_n}P_{{\rm F},\Lambda_n}
[[P_{{\rm F},\Lambda_n},x],[P_{{\rm F},\Lambda_n},y]]\chi_{\Omega_n}.
\end{equation}
Take a subsequence $\{L_n\}_n$ of $\{L_n'\}_n$ so that 
the sequence $\{L_n\}_n$ of the system sizes satisfies 
the two conditions of (\ref{Probcond}) in the proof of Lemma~\ref{theorem:sigmaxyoutzero}. 
Then, for almost every $\omega$, the contribution $\sigma_{xy}^{\rm out}$ 
of the Hall conductance from the boundary region is vanishing as $n\uparrow\infty$, 
and the correction of the Hall conductance $\sigma_{xy}^{\rm in}$ 
of (\ref{sigmaxyinfinite}) for the bulk region is also vanishing 
in this limit. Combining this with (\ref{sigmaxyinfinite}) yields the desired result. 
\end{proof}

%%%%%%%%%%%%%%%%%%%%%%%%%%%%%%%%%%%%%%%%%%%%%%%%%%%%%%%%%%%
\Section{Constancy of the Hall conductance\hfill\break\hfill
---Homotopy argument---}
\label{Homotopy}

In this section, we prove that the Hall conductance $\sigma_{xy}$ 
is constant as long as both the strengths of the potentials and the Fermi energy 
vary in the localization regime. 

%%%%%%%%%%%%%%%%%%%%%%%%%%%%%%%%%%%%%%%%%%%%%%%%%%%%%%%%%%%%
\subsection{Changing the strengths of the potentials}

Consider first changing the strengths of the potentials 
$V_0^{\rm LP}, {\bf A}^{\rm LP}, V_\omega$ in the Hamiltonian $H_\omega^{\rm LP}$ 
of (\ref{HamLP}) on the whole plane ${\bf R}^2$. 
In order to prove constancy of the Hall conductance, 
we extend the homotopy argument of \cite{RSB} for 
lattice models to continuous models by relying on the fractional 
moment bound \cite{AENSS} for the resolvent.  
As a byproduct, we prove that the quantized value of the Hall conductance 
is independent of the period $\ell_{\rm P}$ of the potentials 
$V_0^{\rm LP},{\bf A}^{\rm LP}$. See Theorem~\ref{theorem:indpellP} below. 

Since all the cases can be handled in the same way, 
we consider only the case where the strength of 
the vector potential ${\bf A}_{\rm P}$ varies. 
We denote by $P_{\rm F}'$ the Fermi sea projection 
for the Hamiltonian $H_\omega'$ with the vector potential 
${\bf A}_{\rm P}'={\bf A}_{\rm P}+\delta{\bf A}_{\rm P}$ with 
a small change $\Vert|\delta{\bf A}_{\rm P}|\Vert_\infty$. 
Since the index does not depend on $\omega$, we have 
\begin{eqnarray}
\left|{\rm Index}(P_{\rm F}'U_{\bf a}P_{\rm F}')-
{\rm Index}(P_{\rm F}U_{\bf a}P_{\rm F})\right|&=&
{\bf E}\left[\left|{\rm Index}(P_{\rm F}'U_{\bf a}P_{\rm F}')-
{\rm Index}(P_{\rm F}U_{\bf a}P_{\rm F})\right|\right]\ret
&\le&{\bf E}\left[|{\rm Index}(P_{\rm F}'U_{\bf a}P_{\rm F}')-
{\cal I}^\varepsilon(P_{\rm F}';\Omega,\ell_P)|\right]\ret
&+&{\bf E}\left[\left|{\cal I}^\varepsilon(P_{\rm F}';\Omega,\ell_P)
-{\cal I}^\varepsilon(P_{\rm F};\Omega,\ell_P)\right|\right]\ret
&+&{\bf E}\left[\left|{\cal I}^\varepsilon(P_{\rm F};\Omega,\ell_P)
-{\rm Index}(P_{\rm F}U_{\bf a}P_{\rm F})\right|\right].
\end{eqnarray}
{From} (\ref{EIndexIvarepsilonapprox}), 
the first and the third terms in the right-hand side become small 
for a large $\Omega$. Therefore it is sufficient to show 
that the second term become small for a small change 
$\Vert|\delta{\bf A}_{\rm P}|\Vert_\infty$ of the vector potential. 

Relying on the expression given by the right-hand side of (\ref{Index3}), 
let us estimate the difference,
\begin{eqnarray}
& &{\rm Tr}\>\chi_\varepsilon({\bf u})P_{\rm F}'\chi_\varepsilon({\bf v})
P_{\rm F}'\chi_\varepsilon({\bf w})P_{\rm F}'\chi_\varepsilon({\bf u})
-{\rm Tr}\>\chi_\varepsilon({\bf u})P_{\rm F}\chi_\varepsilon({\bf v})
P_{\rm F}\chi_\varepsilon({\bf w})P_{\rm F}\chi_\varepsilon({\bf u})\ret 
&=&{\rm Tr}\>\chi_\varepsilon({\bf u})\Delta P_{\rm F}\chi_\varepsilon({\bf v})
P_{\rm F}'\chi_\varepsilon({\bf w})P_{\rm F}'\chi_\varepsilon({\bf u})
+{\rm Tr}\>\chi_\varepsilon({\bf u})P_{\rm F}\chi_\varepsilon({\bf v})
\Delta P_{\rm F}\chi_\varepsilon({\bf w})P_{\rm F}'\chi_\varepsilon({\bf u})\ret
&+&
{\rm Tr}\>\chi_\varepsilon({\bf u})P_{\rm F}\chi_\varepsilon({\bf v})
P_{\rm F}\chi_\varepsilon({\bf w})\Delta P_{\rm F}\chi_\varepsilon({\bf u}),
\end{eqnarray}
where $\Delta P_{\rm F}=P_{\rm F}'-P_{\rm F}$. 
The first term in the right-hand side is estimated as  
\begin{eqnarray}
\left|{\rm Tr}\>\chi_\varepsilon({\bf u})\Delta P_{\rm F}\chi_\varepsilon({\bf v})
P_{\rm F}'\chi_\varepsilon({\bf w})P_{\rm F}'\chi_\varepsilon({\bf u})\right|
&\le&\sqrt{{\rm Tr}\>\chi_\varepsilon({\bf u})\Delta P_{\rm F}\chi_\varepsilon({\bf v})
P_{\rm F}'\chi_\varepsilon({\bf v})\Delta P_{\rm F}\chi_\varepsilon({\bf u})}\ret
&\times&\sqrt{{\rm Tr}\>\chi_\varepsilon({\bf u})P_{\rm F}'\chi_\varepsilon({\bf w})
P_{\rm F}'\chi_\varepsilon({\bf w})P_{\rm F}'\chi_\varepsilon({\bf u})}\ret
&\le&{\rm Const.}\Vert\chi_\varepsilon({\bf u})\Delta P_{\rm F}\chi_\varepsilon({\bf v})
\Vert
\Vert\chi_\varepsilon({\bf u})P_{\rm F}'\chi_\varepsilon({\bf w})\Vert\ret
\label{1stTRrDPPpPp}
\end{eqnarray}
by using Schwarz's inequality and the bound (\ref{localTr}). 
Similarly, the second term is estimated as   
\begin{eqnarray}
& &\left|{\rm Tr}\>\chi_\varepsilon({\bf u})P_{\rm F}\chi_\varepsilon({\bf v})
\Delta P_{\rm F}\chi_\varepsilon({\bf w})P_{\rm F}'\chi_\varepsilon({\bf u})\right|\ret
&\le&\sqrt{{\rm Tr}\>\chi_\varepsilon({\bf u})P_{\rm F}\chi_\varepsilon({\bf u})}\ret
&\times&\sqrt{{\rm Tr}\>\chi_\varepsilon({\bf u})P_{\rm F}'\chi_\varepsilon({\bf w})
\Delta P_{\rm F}\chi_\varepsilon({\bf v})P_{\rm F}\chi_\varepsilon({\bf v})
\Delta P_{\rm F}\chi_\varepsilon({\bf w})P_{\rm F}'\chi_\varepsilon({\bf u})}\ret
&\le&{\rm Const.}\Vert\chi_\varepsilon({\bf u})P_{\rm F}'\chi_\varepsilon({\bf w})
\Delta P_{\rm F}\chi_\varepsilon({\bf v})\Vert\ret
&\le&{\rm Const.}\Vert\chi_\varepsilon({\bf u})P_{\rm F}'\chi_\varepsilon({\bf w})
\Vert
\Vert\chi_\varepsilon({\bf w})\Delta P_{\rm F}\chi_\varepsilon({\bf v})\Vert.
\end{eqnarray}
The third term can be handled in the same way as for the first term. 

The contour representation of the Fermi sea projection yields 
\begin{equation}
\left\Vert\chi_\varepsilon({\bf u})[P_{\rm F}'-P_{\rm F}]
\chi_\varepsilon({\bf v})\right\Vert
\le I_{{\bf u},{\bf v}}^{(1)}+I_{{\bf u},{\bf v}}^{(2)}+
I_{{\bf u},{\bf v}}^{(+)}+I_{{\bf u},{\bf v}}^{(-)}
\end{equation}
with
\begin{equation}
I_{{\bf u},{\bf v}}^{(1)}=\frac{1}{2\pi}\int_{y_-}^{y_+}dy
\left\Vert\chi_\varepsilon({\bf u})[R'(E_{\rm F}+iy)
-R(E_{\rm F}+iy)]\chi_\varepsilon({\bf v})\right\Vert,
\end{equation}
\begin{equation}
I_{{\bf u},{\bf v}}^{(2)}=\frac{1}{2\pi}\int_{y_-}^{y_+}dy
\left\Vert\chi_\varepsilon({\bf u})[R'(E_0+iy)
-R(E_0+iy)]\chi_\varepsilon({\bf v})\right\Vert
\end{equation}
and
\begin{equation}
I_{{\bf u},{\bf v}}^{(\pm)}=\frac{1}{2\pi}\int_{E_0}^{E_{\rm F}}dE
\left\Vert\chi_\varepsilon({\bf u})[R'(E+iy_\pm)-
R(E+iy_\pm)]\chi_\varepsilon({\bf v})\right\Vert,
\end{equation}
where $R'(z)=(z-H_\omega')^{-1}$ and $R(z)=(z-H_\omega)^{-1}$. 

First let us estimate the last three integrals except for $I_{{\bf u},{\bf v}}^{(1)}$. 
Note that $R'(z)-R(z)=-R'(z)\delta H_\omega R(z)$, where 
\begin{equation}
\delta H_\omega=\frac{e}{2m_e}[\delta{\bf A}\cdot({\bf p}+e{\bf A})
+({\bf p}+e{\bf A})\cdot\delta{\bf A}]+\frac{e^2}{2m_e}|\delta{\bf A}|^2.
\end{equation}
Since all the contributions in the perturbation $\delta H_\omega$ 
can be handled in the same way, we consider only 
\begin{eqnarray}
& &\chi_\varepsilon({\bf u})R'(z)
\delta A_s(p_s+eA_s)R(z)\chi_\varepsilon({\bf v})\ret
&=&\sum_{{\bf u}'}\chi_\varepsilon({\bf u})R'(z){\tilde\chi}_b({\bf u}')
\delta A_s(p_s+eA_s)\chi_b^\delta({\bf u}')
R(z)\chi_\varepsilon({\bf v})
\label{deltaAps}
\end{eqnarray}
as a typical one. Here $\{\chi_b^\delta({\bf u})\}_{\bf u}$ is the partition 
of unity which is given in the proof of Lemma~\ref{lemma:pchiRchibound} in 
Appendix~\ref{Proof9.7}, and ${\tilde \chi}_b({\bf u})$ is 
the characteristic function of the support of $\chi_b^\delta({\bf u})$. 
The norm is estimated as  
\begin{eqnarray}
& &\left\Vert\chi_\varepsilon({\bf u})R'(z)
\delta A_s(p_s+eA_s)R(z)\chi_\varepsilon({\bf v})\right\Vert\ret
&\le&\Vert\delta A_s\Vert_\infty\sum_{{\bf u}'}
\left\Vert\chi_\varepsilon({\bf u})R'(z){\tilde\chi}_b({\bf u}')
\right\Vert\left\Vert
(p_s+eA_s)\chi_b^\delta({\bf u}')
R(z)\chi_\varepsilon({\bf v})\right\Vert\ret
&\le&{\rm Const.}\Vert\delta A_s\Vert_\infty\sum_{{\bf u}'}
\left\Vert\chi_\varepsilon({\bf u})R'(z){\tilde\chi}_b({\bf u}')\right\Vert\ret
& &\quad\quad\times\left[\left\Vert{\tilde\chi}_b({\bf u}')
R(z)\chi_\varepsilon({\bf v})\right\Vert+
{\rm Const.}\left\Vert{\tilde\chi}_b({\bf u}')
R(z)\chi_\varepsilon({\bf v})\right\Vert^{1/2}\right],
\end{eqnarray}
where we have used Lemma~\ref{lemma:pchiRchibound} 
for getting the second inequality. Since ${\rm dist}(\sigma(H_\omega),z)>0$ 
in the present situation, 
all the norms about the resolvent $R(z)$ decay exponentially at the large distance.  
Therefore we obtain 
\begin{equation}
I_{{\bf u},{\bf v}}^{(\sharp)}\le {\rm Const.}\Vert|\delta{\bf A}_{\rm P}|\Vert_\infty
\exp[-\mu'|{\bf u}-{\bf v}|] 
\end{equation}
with a positive constant $\mu'$, where $\sharp=2,\pm$. 
Consider the contribution from (\ref{1stTRrDPPpPp}) because the rest can be treated 
in the same way. The corresponding contribution is estimated by 
\begin{eqnarray}
& &\frac{\Vert|\delta{\bf A}_{\rm P}|\Vert_\infty}{|\Omega|}
\sum_{{\bf u}\in\Lambda_\ell^\ast}\sum_{{\bf v},{\bf w}}
|{\bf u}-{\bf v}||{\bf u}-{\bf w}|e^{-\mu'|{\bf u}-{\bf v}|}
{\bf E}[\Vert\chi_\varepsilon({\bf u})P_{\rm F}'\chi_\varepsilon({\bf w})\Vert]\ret
&\le&{\rm Const.}\Vert|\delta{\bf A}_{\rm P}|\Vert_\infty, 
\end{eqnarray}
where we have used the decay bound (\ref{PFD}) for the Fermi sea projection.  

Let $s\in(0,1/3)$. The rest of the integrals is written 
\begin{eqnarray}
I_{{\bf u},{\bf v}}^{(1)}&=&\frac{1}{2\pi}\int_{y_-}^{y_+}dy
\left\Vert\chi_\varepsilon({\bf u})[R'(E_{\rm F}+iy)-R(E_{\rm F}+iy)]
\chi_\varepsilon({\bf v})\right\Vert^{s/3}\ret
& &\qquad\quad\times
\left\Vert\chi_\varepsilon({\bf u})[R'(E_{\rm F}+iy)-R(E_{\rm F}+iy)]
\chi_\varepsilon({\bf v})\right\Vert^{1-s/3}\ret
&\le&\frac{1}{\pi}\int_{y_-}^{y_+}dy
\left\Vert\chi_\varepsilon({\bf u})[R'(E_{\rm F}+iy)-R(E_{\rm F}+iy)]
\chi_\varepsilon({\bf v})\right\Vert^{s/3}|y|^{s/3-1},
\end{eqnarray}
where we have used the inequality $(\sum_j a_j)^s\le\sum_j a_j^s$ for 
$s\in(0,1)$ and $a_j\ge 0$, and the inequality 
$\Vert R^\sharp(E_{\rm F}+iy)\Vert\le|y|^{-1}$ for $R^\sharp=R',R$. 
For the norm of the operator in the integrand, 
the contribution from the term (\ref{deltaAps}) can be estimated as  
\begin{eqnarray}
& &\left\Vert\chi_\varepsilon({\bf u})
R'(E_{\rm F}+iy)\delta A_s(p_s+eA_s)R(E_{\rm F}+iy)
\chi_\varepsilon({\bf v})\right\Vert^{s/3}\ret
&\le&{\rm Const.}\Vert \delta A_s\Vert_\infty\sum_{{\bf u}'}\left\{
\left\Vert\chi_\varepsilon({\bf u})
R'(E_{\rm F}+iy){\tilde\chi}_b({\bf u}')\right\Vert^{s/3}
\left\Vert{\tilde\chi}_b({\bf u}')R(E_{\rm F}+iy)
\chi_\varepsilon({\bf v})\right\Vert^{s/3}\right.\ret
&+&\left.{\rm Const.}
\left\Vert\chi_\varepsilon({\bf u})
R'(E_{\rm F}+iy){\tilde\chi}_b({\bf u}')\right\Vert^{s/3}
\left\Vert{\tilde\chi}_b({\bf u}')R(E_{\rm F}+iy)
\chi_\varepsilon({\bf v})\right\Vert^{s/6}\right\},
\label{I1integrandbound}
\end{eqnarray}
where we have used Lemma~\ref{lemma:pchiRchibound} for getting the inequality. 

Consider the contribution from the first term in the summand in the right-hand 
side of (\ref{I1integrandbound}) because the second term can be handled 
in the same way. The corresponding contribution 
from (\ref{1stTRrDPPpPp}) is estimated by 
\begin{eqnarray}
& &\frac{\Vert|\delta{\bf A}_{\rm P}|\Vert_\infty}{|\Omega|}
\sum_{{\bf u}\in\Lambda_\ell^\ast}\sum_{{\bf v},{\bf w}}\sum_{{\bf u}'}
|{\bf u}-{\bf v}||{\bf u}-{\bf w}|{\bf E}\int_{y_-}^{y_+}dy|y|^{s/3-1}\ret
&\times&\left\Vert\chi_\varepsilon({\bf u})
R'(E_{\rm F}+iy){\tilde\chi}_b({\bf u}')\right\Vert^{s/3}
\left\Vert{\tilde\chi}_b({\bf u}')R(E_{\rm F}+iy)
\chi_\varepsilon({\bf v})\right\Vert^{s/3}
\Vert\chi_\varepsilon({\bf u})P_{\rm F}'\chi_\varepsilon({\bf w})\Vert.\ret
\label{I1estimate}
\end{eqnarray}
Using H\"older's inequality, we have 
\begin{eqnarray}
& &{\bf E}\left[\left\Vert\chi_\varepsilon({\bf u})
R'(E_{\rm F}+iy){\tilde\chi}_b({\bf u}')\right\Vert^{s/3}
\left\Vert{\tilde\chi}_b({\bf u}')R(E_{\rm F}+iy)
\chi_\varepsilon({\bf v})\right\Vert^{s/3}
\Vert\chi_\varepsilon({\bf u})P_{\rm F}'\chi_\varepsilon({\bf w})\Vert\right]\ret
&\le&\left\{{\bf E}\left[\left\Vert\chi_\varepsilon({\bf u})
R'(E_{\rm F}+iy){\tilde\chi}_b({\bf u}')\right\Vert^s\right]\right\}^{1/3}
\left\{{\bf E}\left[\left\Vert{\tilde\chi}_b({\bf u}')R(E_{\rm F}+iy)
\chi_\varepsilon({\bf v})\right\Vert^s\right]\right\}^{1/3}\ret
& &\qquad\qquad\times\left\{{\bf E}\left[\Vert\chi_\varepsilon({\bf u})P_{\rm F}'
\chi_\varepsilon({\bf w})\Vert^3\right]\right\}^{1/3}\ret
&\le&{\rm Const.}e^{-\mu|{\bf u}-{\bf u}'|/3}e^{-\mu|{\bf u}'-{\bf v}|/3}
e^{-\mu|{\bf u}-{\bf w}|/3},
\label{Holder3E} 
\end{eqnarray}
where we have used the decay bounds (\ref{DRI}), (\ref{PFD}) and 
$\Vert\chi_\varepsilon({\bf u})P_{\rm F}'\chi_\varepsilon({\bf w})\Vert\le 1$.
Relying on Fatou's lemma and Fubini-Tonelli theorem, and 
substituting the bound (\ref{Holder3E}) into (\ref{I1estimate}), 
we can obtain the desired result. 

Thus the index is constant as long as the strengths of the potentials 
vary in the localization regime. 
In order to describe our statement more precisely, we recall 
the definitions of the lower and upper localization regimes 
which are given by the intervals 
(\ref{Eatlowerbandedge}) and (\ref{Eatupperbandedge}), respectively, 
in the case of ${\bf A}_{\rm P}=0$. 
In the case of ${\bf A}_{\rm P}\ne 0$, the corresponding condition 
is given by (\ref{energyconditiongeneral}). See also Section~\ref{Widths}.  
We continuously change the strength of 
all the potentials, $V_0,V_\omega$ and ${\bf A}_{\rm P}$, 
starting from the special point, $\Vert V_0\Vert_\infty=\Vert V_\omega\Vert_\infty=0$ and 
$\Vert|{\bf A}_{\rm P}|\Vert_\infty=0$, for a fixed Fermi energy $E_{\rm F}$. 
Then, if the Fermi energy $E_{\rm F}$ is staying the lower or upper localization regime, 
the index is equal to the special case that all the potentials are vanishing. 
This result also implies that the index does not depend on 
the period $\ell_P$ of the potentials. 
In consequence, Theorem~\ref{theorem:sigmaxyindexinfty} is refined as  

\begin{theorem}
\label{theorem:indpellP}
Suppose that the Fermi energy $E_{\rm F}$ lies in the localization regime 
around the $n$-th Landau energy ${\cal E}_{n-1}=(n-1/2)\hbar\omega_c$. 
There exists a sequence $\{L_{x,n},L_{y,n}\}_n$ of the system 
sizes such that the Hall conductance $\sigma_{xy}$ 
in the infinite volume limit exists and is quantized to an integer as 
\begin{equation}
\sigma_{xy}=-\frac{e^2}{h}\times
\cases{n & for the upper localization regime\cr
       n-1 & for the lower localization regime\cr} 
\end{equation}
with probability one.  
\end{theorem}

\noindent
{\bf Remark:} 1. When the strength of one of the potentials becomes 
sufficiently large for a fixed strength of the magnetic field, 
the localization regimes become empty in our definition. 
Thus we need the condition that the strengths of the potentials are weak, 
compared to the strength of the magnetic field.  
\smallskip

\noindent
2. The number of the states in a localization regime will be proved to 
be of bulk order for the weak potentials, 
compared to the strength of the magnetic field in Section~\ref{Widths}. 
\smallskip

\noindent
3. We do not require any assumption 
on the tails $\lambda\in[\lambda_{\rm min},-\lambda_-]
\cup[\lambda_+,\lambda_{\max}]$ of the coupling constant 
of the random potential $V_\omega$.  
Therefore we allow the possibility that the spectral gap between 
two neiboring disordered-broadened Landau bands vanishes owing to 
the tails of the random potential. 

%%%%%%%%%%%%%%%%%%%%%%%%%%%%%%%%%%%%%%%%%%%%%%%%%%%
\subsection{Changing the Fermi level}

Next let us prove the constancy of the Hall conductance 
for changing the Fermi level $E_{\rm F}$. 

The Hall conductance $\sigma_{xy}$ of (\ref{sigmaxyLambda}) is written 
\begin{equation}
\sigma_{xy}=-\frac{i\hbar e^2}{L_xL_y}\ 
\sum_{m,n:E_m<E_F<E_n}\ 
\left[\frac{\left(\varphi_m,v_x\varphi_n\right)
\left(\varphi_n,v_y\varphi_m\right)}{(E_m-E_n)^2}-(x\leftrightarrow y)\right]
\label{sigmaxysum}
\end{equation}
in terms of the eigenvector $\varphi_n$ of 
the single electron Hamiltonian $H_\omega$ with 
the eigenvalue $E_n$, $n=1,2,\ldots$, on the box $\Lambda^{\rm sys}$. 
We take the energy eigenvalues $E_n$ in increasing order, 
repeated according to multiplicity.  

Consider changing the number of the electrons below the Fermi level 
from $N$ to $N'$ in the localization regime. Without loss of generality, 
we can assume $N'>N$. We denote by $E_{\rm F}$ and $E_{\rm F}'$ 
the corresponding two Fermi energies for $N$ and $N'$ electrons, respectively.  
The sum in the right-hand side of (\ref{sigmaxysum}) for $N'$ electrons 
is written as 
\begin{eqnarray}
& &\sum_{m,n:E_m<E_F'<E_n}\ 
\left[\frac{\left(\varphi_m,v_x\varphi_n\right)
\left(\varphi_n,v_y\varphi_m
\right)}{(E_m-E_n)^2}-(x\leftrightarrow y)\right]\ret
&=&\sum_{m\le N}\sum_{n=N'+1}^\infty
\left[\frac{\left(\varphi_m,v_x\varphi_n\right)
\left(\varphi_n,v_y\varphi_m
\right)}{(E_m-E_n)^2}-(x\leftrightarrow y)\right]\ret
&+&\sum_{m=N+1}^{N'}\sum_{n=N'+1}^\infty
\left[\frac{\left(\varphi_m,v_x\varphi_n\right)
\left(\varphi_n,v_y\varphi_m
\right)}{(E_m-E_n)^2}-(x\leftrightarrow y)\right]\ret
&=&\sum_{m\le N}\sum_{n=N+1}^\infty
\left[\frac{\left(\varphi_m,v_x\varphi_n\right)
\left(\varphi_n,v_y\varphi_m
\right)}{(E_m-E_n)^2}-(x\leftrightarrow y)\right]\ret
&-&\sum_{m\le N}\sum_{n=N+1}^{N'}
\left[\frac{\left(\varphi_m,v_x\varphi_n\right)
\left(\varphi_n,v_y\varphi_m
\right)}{(E_m-E_n)^2}-(x\leftrightarrow y)\right]\ret
&+&\sum_{m=N+1}^{N'}\sum_{n=N'+1}^\infty
\left[\frac{\left(\varphi_m,v_x\varphi_n\right)
\left(\varphi_n,v_y\varphi_m
\right)}{(E_m-E_n)^2}-(x\leftrightarrow y)\right].
\end{eqnarray}
The first double sum in the right-hand side of the second equality 
leads the Hall conductance $\sigma_{xy}$ for $N$ electrons. 
Therefore it is sufficient to estimate the other two double sums.  
These two sums are compactly written as 
\begin{equation}
\sum_{m=N+1}^{N'}\ \sum_{n\le N\ {\rm and}\ n>N'}
\left[\frac{\left(\varphi_m,v_x\varphi_n\right)
\left(\varphi_n,v_y\varphi_m
\right)}{(E_m-E_n)^2}-(x\leftrightarrow y)\right].
\label{localizedsum}
\end{equation}
In consequence, the difference between the two Hall conductances for 
$N$ and $N'$ electrons is written 
\begin{equation}
\Delta\sigma_{xy}^{\rm loc}=-\frac{i\hbar e^2}{L_xL_y}
{\rm Tr}\>\Delta P_\Lambda^{\rm loc}[P_{x,\Lambda},P_{y,\Lambda}]
\Delta P_\Lambda^{\rm loc},
\end{equation}
where $\Delta P_\Lambda^{\rm loc}$ is the spectral projection 
onto the localization regime, and 
\begin{equation}
P_{s,\Lambda}=\frac{1}{2\pi i}\int_\gamma dz R_\Lambda(z)v_s R_\Lambda(z),
\end{equation}
with the resolvent $R_\Lambda(z)=(z-H_{\omega,\Lambda})^{-1}$ for 
the present Hamiltonian $H_{\omega,\Lambda}$ 
on the box $\Lambda=\Lambda^{\rm sys}$. Here the closed path $\gamma$ encircles 
the energy eigenvalues of the ``localized" states.  
In the same way as in Lemma~\ref{lemma:sigmaxyinrep} and Eq.~(\ref{sigmaxyoutfinite}), 
we have 
\begin{equation}
\Delta\sigma_{xy}^{\rm loc}
=\frac{e^2}{h}{\cal I}^{\rm loc}(\Delta P_\Lambda^{\rm loc})+\delta(L)
+{\cal O}\left(\exp\left[-\mu'L^{2\kappa/3}\right]\right)
\label{DeltasimgaxylocInd}
\end{equation}
with probability larger than $(1-{\rm Const.}L^{-2[\kappa(\xi+2)-3]/3})$, 
where 
\begin{equation}
{\cal I}^{\rm loc}(\Delta P_\Lambda^{\rm loc})
=\frac{2\pi i}{|\Omega|}{\rm Tr}\>
\chi_\Omega \Delta P_\Lambda^{\rm loc}[[\Delta P_\Lambda^{\rm loc},x],
[\Delta P_\Lambda^{\rm loc},y]]\chi_\Omega,
\end{equation}
and $\delta(L)$ is the correction which comes from the boundary 
region $\Lambda\backslash\Omega$. 
In the same way as in the proof of Lemma~\ref{theorem:sigmaxyoutzero}, 
we can show that ${\bf E}[\delta(L)]\rightarrow 0$ as 
$L\rightarrow \infty$. 

\begin{lemma}
\label{lemma:Ilocbound}
The following bound is valid: 
\begin{equation}
{\bf E}\left|{\cal I}^{\rm loc}(\Delta P_\Lambda^{\rm loc})\right|
\le{\rm Const.}{\Delta E}^{1/2},
\end{equation}
where $\Delta E=E_{\rm F}'-E_{\rm F}$.  
\end{lemma}

\begin{proof}{Proof}
Using Schwarz's inequality, one has 
\begin{equation}
{\bf E}\left|{\cal I}^{\rm loc}(\Delta P_\Lambda^{\rm loc})\right|
\le\frac{2\pi}{|\Omega|}\sqrt{{\bf E}
[{\rm Tr}\>\chi_\Omega \Delta P_\Lambda^{\rm loc}\chi_\Omega]
\cdot{\bf E}
[{\rm Tr}\>\chi_\Omega A^\ast\Delta P_\Lambda^{\rm loc}A
\chi_\Omega]},
\end{equation}
where we have written $A=[[\Delta P_\Lambda^{\rm loc},x],
[\Delta P_\Lambda^{\rm loc},y]]$. From the Wegner estimate (\ref{Wegnerestimate}), 
one has 
\begin{equation}
\frac{1}{|\Omega|}{\bf E}\left[{\rm Tr}\>\chi_\Omega\Delta P_\Lambda^{\rm loc}
\chi_\Omega\right]
\le\frac{1}{|\Omega|}{\bf E}\left[
{\rm Tr}\>\chi_\Lambda\Delta P_\Lambda^{\rm loc}\chi_\Lambda\right]
\le{\rm Const.}\Delta E.
\end{equation}
Therefore it is sufficient to show that the quantity 
$|\Omega|^{-1}{\bf E}[{\rm Tr}\>\chi_\Omega A^\ast\Delta P_\Lambda^{\rm loc}A
\chi_\Omega]$ is bounded. Schwarz's inequality yields  
\begin{eqnarray}
{\rm Tr}\>\chi_b({\bf u})A^\ast
\Delta P_\Lambda^{\rm loc}A\chi_b({\bf u})
&=&\sum_{{\bf v},{\bf w}}
{\rm Tr}\>\chi_b({\bf u})A^\ast\chi_b({\bf v})
\Delta P_\Lambda^{\rm loc}\chi_b({\bf w})A\chi_b({\bf u})\ret
&\le&\sum_{{\bf v},{\bf w}}\sqrt{{\rm Tr}\>\chi_b({\bf u})A^\ast\chi_b({\bf v})
\Delta P_\Lambda^{\rm loc}\chi_b({\bf v})A\chi_b({\bf u})}\ret
& &\qquad\times\sqrt{{\rm Tr}\>\chi_b({\bf u})A^\ast\chi_b({\bf w})
\Delta P_\Lambda^{\rm loc}\chi_b({\bf w})A\chi_b({\bf u})}\ret
&\le&{\rm Const.}\sum_{{\bf v},{\bf w}}
\left\Vert\chi_b({\bf v})A\chi_b({\bf u})\right\Vert\cdot
\left\Vert\chi_b({\bf w})A\chi_b({\bf u})\right\Vert, 
\end{eqnarray}
where we have used the bound (\ref{localTr}) for getting the second inequality. 
One can show that the expectation value of the right-hand side 
is finite in the same way as in the proof of Lemma~\ref{lemma:ETrbounded}. 
\end{proof}

We denote by $M_\Lambda'$ the event $M_\Lambda$ for 
the Fermi energy $E_{\rm F}'$ in the proof of Lemma~\ref{lemma:ARBdecay}. 
Note that 
\begin{eqnarray}
& &{\bf E}\left[\left|{\rm Index}(P_{\rm F}U_{\bf a}P_{\rm F})-
(h/e^2)\sigma_{xy}^{\rm in}\right|{\bf I}(M_\Lambda\cap M_\Lambda')\right]\ret
&\le&{\bf E}\left[\left|{\rm Index}(P_{\rm F}U_{\bf a}P_{\rm F})-
{\cal I}^\varepsilon(P_{\rm F};\Omega,\ell_P)\right|\right]\ret
&+&{\bf E}\left[\left|{\cal I}^\varepsilon(P_{\rm F};\Omega,\ell_P)-
{\cal I}(P_{\rm F};\Omega,\ell_P)\right|\right]
+{\bf E}\left[\left|{\cal I}(P_{\rm F};\Omega,\ell_P)-
{\cal I}(P_{{\rm F},\Lambda};\Omega)\right|\right]\ret
&+&{\bf E}\left[\left|{\cal I}(P_{{\rm F},\Lambda};\Omega)
-(h/e^2)\sigma_{xy}^{\rm in}\right|{\bf I}(M_\Lambda\cap M_\Lambda')\right].
\end{eqnarray}
{From} (\ref{sigmaxyinfinite}), (\ref{EIndexIvarepsilonapprox}) 
and Lemmas~\ref{lemma:diffIIvarepsilon} 
and \ref{lemma:indexRIF}, all the terms in the right-hand side become small 
for large $|\Omega|, L$ and for a small $\varepsilon$.  
Further, from (\ref{sigmaxyoutfinite}), the proof 
of Lemma~\ref{theorem:sigmaxyoutzero} and (\ref{DeltasimgaxylocInd}), 
we have 
\begin{eqnarray}
{\bf E}\left[\left|{\sigma_{xy}^{\rm in}}'-\sigma_{xy}^{\rm in}\right|
{\bf I}(M_\Lambda\cap M_\Lambda')\right]
&\le&{\bf E}\left[|\Delta\sigma_{xy}^{\rm loc}|{\bf I}(M_\Lambda\cap M_\Lambda')\right]
+{\bf E}\left[\left|{\sigma_{xy}^{\rm out}}'\right|
{\bf I}(M_\Lambda\cap M_\Lambda')\right]\ret
&+&{\bf E}\left[\left|\sigma_{xy}^{\rm out}\right|
{\bf I}(M_\Lambda\cap M_\Lambda')\right]\ret
&\le&\frac{e^2}{h}{\bf E}\left|{\cal I}^{\rm loc}(\Delta P_\Lambda^{\rm loc})\right|
+(\mbox{small correction}).
\end{eqnarray}
{From} these observations and the fact that the indices are constant for 
almost every $\omega$, we obtain  
\begin{eqnarray}
& &\left|{\rm Index}(P_{\rm F}'U_{\bf a}P_{\rm F}')
-{\rm Index}(P_{\rm F}U_{\bf a}P_{\rm F})\right|
{\bf E}\left[{\bf I}(M_\Lambda\cap M_\Lambda')\right]\ret
&\le&{\bf E}\left[\left|{\rm Index}(P_{\rm F}'U_{\bf a}P_{\rm F}')
-{\rm Index}(P_{\rm F}U_{\bf a}P_{\rm F})\right|
{\bf I}(M_\Lambda\cap M_\Lambda')\right]\ret
&\le&{\bf E}\left|{\cal I}^{\rm loc}(\Delta P_\Lambda^{\rm loc})\right|
+(\mbox{small correction})\ret
&\le&{\rm Const.}\Delta E^{1/2}+(\mbox{small correction}), 
\end{eqnarray}
where we have used Lemma~\ref{lemma:Ilocbound} for getting the last inequality. 
This implies that the index must be constant for a small change $\Delta E$ 
of the Fermi energy in the localization regime because 
the index is equal to an integer for almost every $\omega$. 

%%%%%%%%%%%%%%%%%%%%%%%%%%%%%%%%%%%%%%%%%%%%%%%%%%%%%%%%%%%%%%%%
\Section{Widths of the Hall conductance plateaus}
\label{Widths}

In this section, 
we prove that the widths of the Hall conductance plateaus are of the bulk order 
under certain conditions for the potentials, 
by estimating the number of the localized states. 
The conditions are realized for weak potentials as we will see in this section. 

Consider first the case with ${\bf A}_{\rm P}=0$. 
To begin with, we note that, 
when the strength of the random potential $V_\omega$ continuously increases 
from $\lambda=0$ to $\lambda\in[-\lambda_-,\lambda_+]
\subset[\lambda_{\rm min},\lambda_{\rm max}]$, 
the energies $E$ of the $n+1$-th Landau band are broadened into the interval, 
\begin{equation}
-\Vert V_0^-\Vert_\infty-\lambda_-u_1\le E-{\cal E}_n
\le \Vert V_0^+\Vert_\infty+\lambda_+u_1.
\end{equation}

Let ${\hat\delta}$ be a small positive parameter.  
For the lower region of the band, we choose 
$\lambda_+=\lambda_+^{\rm low}\le\lambda_{\rm max}$, 
$\lambda_-=\lambda_-^{\rm low}$ and 
${\hat \delta}_-={\hat\delta}$ 
in the condition (\ref{Eatlowerbandedge}) so that 
the pair $(\lambda_+,\lambda_-)=(\lambda_+^{\rm low},\lambda_-^{\rm low})$ 
satisfies the condition (\ref{assumptiong}) with a small 
$\lambda_-=\lambda_-^{\rm low}$. Then the condition (\ref{Eatlowerbandedge}) for 
the energy $E$ leading to a localized state becomes 
\begin{equation}
{\cal E}_{n-1}+\Vert V_0^+\Vert_\infty+\lambda_+^{\rm low}u_1
+{\hat\delta}\hbar\omega_c\le E\le 
{\cal E}_n-\Vert V_0^-\Vert_\infty-\lambda_-^{\rm low}u_1-\Delta{\cal E}.
\label{lowlocalized}
\end{equation}
We call this interval the lower localization regime. 
For the upper region of the band, we choose 
$\lambda_+=\lambda_+^{\rm up}$, $\lambda_-=\lambda_-^{\rm up}\ge-\lambda_{\rm min}$ and  
${\hat\delta}_+={\hat\delta}/2$ 
in the condition (\ref{Eatupperbandedge}) 
so that 
the pair $(\lambda_+,\lambda_-)=(\lambda_+^{\rm up},\lambda_-^{\rm up})$ 
satisfies the condition (\ref{assumptiong}) 
with a small $\lambda_+=\lambda_+^{\rm up}$. 
The condition (\ref{Eatupperbandedge}) for localization is 
\begin{equation}
{\cal E}_n+\Vert V_0^+\Vert+\lambda_+^{\rm up}u_1+\Delta{\cal E}
\le E\le {\cal E}_{n+1}-\Vert V_0^-\Vert_\infty-\lambda_-^{\rm up}u_1
-{\hat\delta}\hbar\omega_c.
\label{uplocalized}
\end{equation}
We call this interval the upper localization regime. 
We require that the positive constants, 
${\hat\delta}$, $\lambda_+^{\rm low}$ and $\lambda_-^{\rm up}$, satisfy  
\begin{equation}
\Vert V_0^+\Vert_\infty+\Vert V_0^-\Vert_\infty
+(\lambda_+^{\rm low}+\lambda_-^{\rm up})u_1+2{\hat\delta}\hbar\omega_c
<\hbar\omega_c
\label{bandcond}
\end{equation}
so that the lower and upper localization regimes overlap with each other.  
This condition is satisfied for a large 
strength $B$ of the magnetic field for fixed strengths of the potentials. 
We stress that we allow the possibility that the spectral gap between 
two neighboring disordered-broadened Landau bands vanishes owing to the tails 
$\lambda\in[\lambda_{\rm min},-\lambda_-^{\rm up}]
\cup[\lambda_+^{\rm low},\lambda_{\rm max}]$ of 
the coupling constants. 
In this situation, all the states in the $n+1$-th broadened Landau band 
are localized except for the energies $E$ satisfying 
$-\delta E_-\le E-{\cal E}_n\le\delta E_+$,
where $\delta E_-=\Vert V_0^-\Vert_\infty+\lambda_-^{\rm low}u_1+\Delta{\cal E}$ 
and $\delta E_+=\Vert V_0^+\Vert_\infty+\lambda_+^{\rm up}u_1+\Delta{\cal E}$. 
In other words, the number of the extended states can be bounded by 
the number of the energy eigenvalues $E$ 
satisfying $-\delta E_-\le E-{\cal E}_n\le\delta E_+$. 

In order to obtain the upper bound for the number of the extended states, 
consider first the special case with $V_0=0$ in the Hamiltonian 
$H_0$ of (\ref{Ham0}). Namely, the Hamiltonian $H_0$ is equal to 
the simplest Landau Hamiltonian $H_{\rm L}$ of (\ref{hamHL}), 
combining with the present assumption ${\bf A}_{\rm P}=0$.  
In this case, the number of the extended states can be estimated 
with probability nearly equal to one for the sufficiently large volume\footnote{
See, for example, Chap.~VI of the book \cite{CL}.} 
by using the Wegner estimate (\ref{Wegnerestimate}). 
When $V_0\ne 0$, the deviation of the energy eigenvalues is bounded from above by 
$\Vert V_0^+\Vert_\infty$ and from below by $-\Vert V_0^-\Vert_\infty$.  
{From} these observations and the min-max principle, 
we can estimate the number $N_{\rm ext}$ of the extended 
states which appear only near the center of the band as 
\begin{equation}
N_{\rm ext}\le C_{\rm W}^{(0)}K_3^{(0)}\Vert g\Vert_\infty
(\delta E_++\delta E_-+\Vert V_0^+\Vert_\infty+\Vert V_0^-\Vert_\infty)
|\Lambda|
\end{equation}
in the case with $V_0\ne 0$, where $C_{\rm W}^{(0)}$ 
and $K_3^{(0)}$ are the positive constants for the case of $V_0=0$.  
Since the total number of the states in the $n+1$-th Landau band 
is given by $M=|\Lambda|eB/(2\pi\hbar)$, the number $N_{\rm loc}$ of 
the localized states in the band is evaluated as 
\begin{equation}
N_{\rm loc}=M-N_{\rm ext}\ge B|\Lambda|
\left(\frac{e}{2\pi\hbar}-C_{\rm W}^{(0)}\frac{K_3^{(0)}}{B}\Vert g\Vert_\infty
\delta E\right)
\label{Nloc}
\end{equation}
with $\delta E=2\left(\Vert V_0^+\Vert_\infty+\Vert V_0^-\Vert_\infty\right)
+(\lambda_-^{\rm low}+\lambda_+^{\rm up})u_1+2\Delta{\cal E}$. 
We note that $K_3^{(0)}/B\sim {\rm Const.}$ for a large $B$ from the remark 
below Theorem~\ref{theorem:Wegner}.  
Thus, if the strength of the potential $V_0$ is sufficiently weak, we can choose 
the parameters $\lambda_-^{\rm low},\lambda_+^{\rm up},\Delta{\cal E}$ 
so that the right-hand side (\ref{Nloc}) is strictly positive for any 
large magnetic field. This implies that  
the number $N_{\rm loc}$ is of order of the bulk.  
In order to discuss the case for a strong random potential which behaves like 
$\Vert u\Vert_\infty\sim B$ for a strong magnetic field, 
we recall $u_1=2\Vert u\Vert_\infty$. We also have $K_3^{(0)}={\cal O}(1)$ 
which was already obtained at the end of Section~\ref{IDER}.
From these and the same argument, 
we can also get the lower bound for the number of the localized states, 
i.e., the width of the Hall conductance. 

Let us see that the above estimate for the widths of the plateaus gives 
the optimal value in the limit $B\uparrow\infty$ for $V_0=0$. 
{From} the above bounds, we have 
\begin{equation}
\frac{N_{\rm ext}}{M}\le{\rm Const.}
\left[(\lambda_-^{\rm low}+\lambda_+^{\rm up})u_1+2\Delta{\cal E}\right].
\end{equation}
{From} the argument about the initial decay estimate for 
the resolvent in Section~\ref{IDER}, we can take the three parameters, 
$\lambda_-^{\rm low},\lambda_+^{\rm up},\Delta{\cal E}$, 
so as to go to zero in the strong magnetic field limit $B\uparrow\infty$. 
Thus the density of the extended states in the Landau level 
is vanishing in the limit. 

Next consider the case with ${\bf A}_{\rm P}\ne 0$. 
The method to show the existence of the Hall conductance plateau with 
the width of bulk order is basically the same as 
in the above case with ${\bf A}_{\rm P}=0$, except for considering 
strong potentials.
We assume that the bump $u$ of the random potential $V_\omega$ 
is written $u=\hbar\omega_c{\hat u}$ with a fixed, dimensionless 
function ${\hat u}$, and that the vector potential ${\bf A}_{\rm P}$ 
satisfies $\Vert|{\bf A}_{\rm P}|\Vert_\infty\le\alpha_0B^{1/2}$ 
with a small, positive constant $\alpha_0$. 
Instead of the condition (\ref{bandcond}), 
we require that the corresponding positive constants, 
${\hat\delta}$, $\lambda_+^{\rm low}$ 
and $\lambda_-^{\rm up}$, satisfy  
\begin{eqnarray}
& &\Vert V_0^+\Vert_\infty+\Vert V_0^-\Vert_\infty
+(\lambda_+^{\rm low}+\lambda_-^{\rm up})u_1+{\hat\delta}\hbar\omega_c\ret
&+&\frac{\sqrt{2e}}{\sqrt{m_e}}
\Vert|{\bf A}_{\rm P}|\Vert_\infty
\left(\sqrt{{\cal E}_n}+\sqrt{{\cal E}_{n+1}}\right)
+\frac{e^2}{2m_e}\Vert|{\bf A}_{\rm P}|\Vert_\infty^2
<\hbar\omega_c.
\end{eqnarray}
Similarly, for the lower localization regime, we choose 
$\lambda_+=\lambda_+^{\rm low}\le\lambda_{\rm max}$, $\lambda_-=\lambda_-^{\rm low}$ and 
${\hat \delta}_+={\hat \delta}_-={\hat\delta}/2$ 
in the condition (\ref{energyconditiongeneral}) so that 
the pair $(\lambda_+,\lambda_-)=(\lambda_+^{\rm low},\lambda_-^{\rm low})$ 
satisfies the condition (\ref{assumptiong}) with a small 
$\lambda_-=\lambda_-^{\rm low}$. 
Then all the states in the $n+1$-th broadened Landau band with the energies 
$E\le{\cal E}_n-\delta E_-$ are localized, 
where 
\begin{equation}
\delta E_-=\Vert V_0^-\Vert_\infty+\lambda_-^{\rm low}u_1
+\frac{1}{2}{\hat\delta}\hbar\omega_c
+\frac{\sqrt{2}e}{\sqrt{m_e}}\Vert|{\bf A}_{\rm P}|\Vert_\infty
\sqrt{{\cal E}_n}.
\end{equation} 
For the upper localization regime, we choose 
$\lambda_+=\lambda_+^{\rm up}$, $\lambda_-=\lambda_-^{\rm up}\ge-\lambda_{\rm min}$ and  
${\hat\delta}_+={\hat\delta}_-={\hat\delta}/2$ 
in the same condition (\ref{energyconditiongeneral}) so that 
the pair $(\lambda_+,\lambda_-)=(\lambda_+^{\rm up},\lambda_-^{\rm up})$ 
satisfies the condition (\ref{assumptiong}) 
with a small $\lambda_+=\lambda_+^{\rm up}$. 
Then all the states in the $n+1$-th broadened Landau band with the energies 
$E\ge{\cal E}_n+\delta E_+$ are localized, where 
\begin{equation}
\delta E_+=\Vert V_0^+\Vert_\infty+\lambda_+^{\rm up}u_1
+\frac{1}{2}{\hat\delta}\hbar\omega_c
+\frac{\sqrt{2}e}{\sqrt{m_e}}\Vert|{\bf A}_{\rm P}|\Vert_\infty
\sqrt{{\cal E}_n}+\frac{e^2}{2m_e}\Vert|{\bf A}_{\rm P}|\Vert_\infty^2. 
\end{equation}
The corresponding $\delta E$ in (\ref{Nloc}) is given by 
\begin{equation}
\delta E=2\left(\Vert V_0^+\Vert_\infty+\Vert V_0^-\Vert_\infty\right)
+(\lambda_+^{\rm up}+\lambda_-^{\rm low})u_1
+{\hat\delta}\hbar\omega_c
+\frac{4\sqrt{2}e}{\sqrt{m_e}}\Vert|{\bf A}_{\rm P}|\Vert_\infty
\sqrt{{\cal E}_n}+\frac{e^2}{m_e}\Vert|{\bf A}_{\rm P}|\Vert_\infty^2.
\end{equation}
Consequently there appears the Hall conductance plateau with the width of 
the bulk order for a fixed potential $V_0$, for a strong magnetic field, and 
for small parameters, $\lambda_+^{\rm up},\lambda_-^{\rm low},{\hat\delta},
\alpha_0$. 

%%%%%%%%%%%%%%%%%%%%%%%%%%%%%%%%%%%%%%%%%%%%%%%%%%%%%%%%%%%%%%%%%%%%%%%%%
\Section{Corrections to the Linear response formula} 
\label{CorrectionstoLRF}

The aim of this section is to prove that 
both of the acceleration coefficients $\gamma_{uy}$, $u=x,y$, 
in the linear response formula (\ref{sigmatot}) are vanishing 
in the infinite volume limit, and that  
the corrections $\delta\sigma_{uy}(t)$, $u=x,y$, due to the initial 
adiabatic process in (\ref{sigmatot}) satisfy the bound (\ref{sigmacorrections}).  

We recall the expression of the acceleration coefficients \cite{Koma3}, 
\begin{equation}
\gamma_{uy}=\frac{e^2}{L_xL_y}
\left[\frac{N}{m_e}\delta_{u,y}+{\rm Tr}\>v_u(P_{y,\Lambda}P_{{\rm F},\Lambda}
+P_{{\rm F},\Lambda}P_{y,\Lambda})\right]\quad\mbox{for }\ u=x,y.
\label{gammasy}
\end{equation}
Using the partition of unity, $\{\chi_b({\bf u})\}_{\bf u}$, which was 
introduced in Section~\ref{Conplateaus}, we have  
\begin{equation}
{\rm Tr}\>v_u(P_{y,\Lambda}P_{{\rm F},\Lambda}
+P_{{\rm F},\Lambda}P_{y,\Lambda})=
\sum_{\bf u}\left[{\rm Tr}\>v_uP_{y,\Lambda}\chi_b({\bf u})P_{{\rm F},\Lambda}
+{\rm Tr}\>v_uP_{{\rm F},\Lambda}\chi_b({\bf u})P_{y,\Lambda}\right].
\label{vsPyP}
\end{equation}

First let us consider the first term 
${\rm Tr}\>v_uP_{y,\Lambda}\chi_b({\bf u})P_{{\rm F},\Lambda}$ 
in the summand in the right-hand side. 
Since we can shift the location of the box $s_b({\bf u})$ by 
using the magnetic translations, we can assume 
${\rm dist}({\bf u},\partial\Lambda)={\cal O}(L)$. 
Write $R_\Lambda(z)=(z-H_{\omega,\Lambda})^{-1}$. Note that 
\begin{eqnarray}
{\rm Tr}\>v_uP_{y,\Lambda}\chi_b({\bf u})P_{{\rm F},\Lambda}
&=&\frac{1}{2\pi i}\int_\gamma dz{\rm Tr}\>v_uR_\Lambda(z)v_y
R_\Lambda(z)\chi_b({\bf u})P_{{\rm F},\Lambda}\ret
&=&\frac{1}{2\pi i}\int_\gamma dz{\rm Tr}\>v_uR_\Lambda(z)v_y
\chi_\Lambda^\delta
R_\Lambda(z)\chi_b({\bf u})P_{{\rm F},\Lambda}\ret
&+&\frac{1}{2\pi i}\int_\gamma dz{\rm Tr}\>v_uR_\Lambda(z)v_y
(1-\chi_\Lambda^\delta)
R_\Lambda(z)\chi_b({\bf u})P_{{\rm F},\Lambda},
\label{TrvuPychiP}
\end{eqnarray}
where $\chi_\Lambda^\delta$ is the $C^2$, positive cutoff function 
which was also introduced in Section~\ref{Conplateaus}. By the same argument as in 
the proof of Lemma~\ref{lemma:sigmaxyinrep}, the absolute value of the second term 
in the right-hand side has a stretched exponentially decaying bound as  
in Lemma~\ref{lemma:ARBdecay}. 
Since the number of ${\bf u}$ for the summation in the right-hand side of 
(\ref{vsPyP}) is of order of the volume $L_xL_y={\cal O}(L^2)$, 
the corresponding contribution is vanishing 
in the infinite volume limit $L\uparrow\infty$. 
Using the identity, 
$v_y\chi_\Lambda^\delta=(i/\hbar)[y,z-H_{\omega,\Lambda}]\chi_\Lambda^\delta$, 
one has 
\begin{eqnarray}
& &\frac{1}{2\pi i}\int_\gamma dzR_\Lambda(z)v_y
\chi_\Lambda^\delta R_\Lambda(z)\chi_b({\bf u})\ret
&=&\frac{i}{\hbar}[P_{{\rm F},\Lambda},y\chi_\Lambda^\delta]\chi_b({\bf u})
-\frac{1}{2\pi \hbar}\int_\gamma dzR_\Lambda(z)yW(\chi_\Lambda^\delta)
R_\Lambda(z)\chi_b({\bf u})
\label{intRvychiRchi}
\end{eqnarray}
for the first term in the right-hand side of (\ref{TrvuPychiP}). 
This second term in the right-hand side also gives a small correction. 
In consequence, only the first term in the right-hand side of (\ref{intRvychiRchi}) 
may lead to a nonvanishing contribution in the infinite-volume limit. 

Since the second term in the summand in the right-hand side 
of (\ref{vsPyP}) can handled in the same way, we get 
\begin{eqnarray}
& &{\rm Tr}\left[v_uP_{y,\Lambda}\chi_b({\bf u})P_{{\rm F},\Lambda}
+v_uP_{{\rm F},\Lambda}\chi_b({\bf u})P_{y,\Lambda}\right]\ret
&=&
\frac{i}{\hbar}{\rm Tr}\left\{v_u[P_{{\rm F},\Lambda},y\chi_\Lambda^\delta]
\chi_b({\bf u})P_{{\rm F},\Lambda}+v_uP_{{\rm F},\Lambda}\chi_b({\bf u})
[P_{{\rm F},\Lambda},y\chi_\Lambda^\delta]\right\}+{\rm corrections}\ret
&=&\frac{i}{\hbar}{\rm Tr}\left\{-v_uy\chi_\Lambda^\delta
P_{{\rm F},\Lambda}\chi_b({\bf u})P_{{\rm F},\Lambda}+
v_uP_{{\rm F},\Lambda}\chi_b({\bf u})
P_{{\rm F},\Lambda}y\chi_\Lambda^\delta\right\}+{\rm corrections}\ret
&=&-\frac{1}{m_e}\delta_{u,y}{\rm Tr}\>P_{{\rm F},\Lambda}\chi_b({\bf u})
P_{{\rm F},\Lambda}+{\rm corrections},
\end{eqnarray}
where we have used $v_uy=-(i\hbar/m_e)\delta_{u,y}+yv_u$ 
for getting the last equality. Substituting this and (\ref{vsPyP}) 
into the expression (\ref{gammasy}) of $\gamma_{uy}$, we obtain 
\begin{equation}
\lim_{\Lambda^{\rm sys}\uparrow{\bf R}^2}\gamma_{uy}=0\quad\mbox{for }\ 
u=x,y\quad\mbox{with probability one.} 
\end{equation}

Next consider the corrections $\delta\sigma_{uy}(t)$ due to the initial 
adiabatic process in (\ref{sigmatot}). 
We begin with recalling the expression \cite{Koma3}, 
\begin{equation}
\delta\sigma_{uy}(t)=\frac{ie^2}{L_xL_y}
\left\langle\Phi_{0,\Lambda}^{(N)},v_u^{(N)}[1-P_{0,\Lambda}^{(N)}]
{\cal M}_t(E_{0,\Lambda}^{(N)}-H_{\omega,\Lambda}^{(N)})v_y^{(N)}
\Phi_{0,\Lambda}^{(N)}\right\rangle
+{\rm c.c.},
\end{equation}
with 
\begin{equation} 
{\cal M}_t({\cal E})=\left\{\left[\frac{i T}{{\cal E}+i\hbar\eta}
-\frac{\hbar}{\left({\cal E}+i\hbar\eta\right)^2}\right]
e^{-\eta T}e^{i{\cal E}T/\hbar}
+\left[
\frac{\hbar}{{\cal E}^2}
-\frac{\hbar}{\left({\cal E}+i\hbar\eta\right)^2}\right]\right\}e^{i{\cal E}t/\hbar}, 
\end{equation}
where $\Phi_{0,\Lambda}^{(N)}$ is the $N$ electron ground state vector with 
the energy eigenvalue $E_{0,\Lambda}^{(N)}$, $v_u^{(N)}$ the $N$ electron velocity 
operator, and $P_{0,\Lambda}^{(N)}$ the projection onto the $N$ electron 
ground state. 

Since all the contributions can be handled in the same way, 
we consider  
\begin{equation}
{\cal N}_{sy}(t)
:=\frac{1}{L_xL_y}
\left\langle\Phi_{0,\Lambda}^{(N)},v_u^{(N)}P_{\rm ex}\left[
\left(E_{0,\Lambda}^{(N)}-H_{\omega,\Lambda}^{(N)}\right)^{-2}
-\left(E_{0,\Lambda}^{(N)}-H_{\omega,\Lambda}^{(N)}+i\hbar\eta\right)^{-2}\right]
e^{i{\hat \theta}}v_y^{(N)}\Phi_{0,\Lambda}^{(N)}\right\rangle
\end{equation}
with ${\hat \theta}=(E_{0,\Lambda}^{(N)}-H_{\omega,\Lambda}^{(N)})t/\hbar$ 
as an example. Here we have written $P_{\rm ex}=1-P_{0,\Lambda}^{(N)}$ 
for short. In order to eliminate the factor $e^{i{\hat \theta}}$, 
we use Schwarz's inequality. As a result, we obtain  
$\left|{\cal N}_{uy}(t)\right|^2\le{\tilde {\cal N}}_u(t){\tilde {\cal N}}_y(t)$ 
with
\begin{equation}
{\tilde {\cal N}}_u(t):=\frac{1}{L_xL_y}
\left\langle\Phi_{0,\Lambda}^{(N)},
v_u^{(N)}P_{\rm ex}\left|
\left[E_{0,\Lambda}^{(N)}-H_{\omega,\Lambda}^{(N)}\right]^{-1}
+\left[E_0^{(N)}-H_{\omega,\Lambda}^{(N)}+i\hbar\eta\right]^{-1}\right|^2
v_u^{(N)}\Phi_{0,\Lambda}^{(N)}\right\rangle
\end{equation}
and 
\begin{equation}
{\tilde {\cal N}}_y(t):=\frac{1}{L_xL_y}
\left\langle\Phi_{0,\Lambda}^{(N)},
v_y^{(N)}P_{\rm ex}\left|
\left[E_{0,\Lambda}^{(N)}-H_{\omega,\Lambda}^{(N)}\right]^{-1}
-\left[E_{0,\Lambda}^{(N)}-H_{\omega,\Lambda}^{(N)}+i\hbar\eta\right]^{-1}\right|^2
v_y^{(N)}\Phi_{0,\Lambda}^{(N)}\right\rangle.
\end{equation}
Further the application of the inequality $\sqrt{ab}\le(a+b)/2$ for $a,b\ge 0$ yields  
\begin{equation}
\left|{\cal N}_{uy}(t)\right|\le
\left[\eta^{s/4}{\tilde {\cal N}}_u(t)+\eta^{-s/4}{\tilde {\cal N}}_y(t)\right]/2
\quad\mbox{for}\ s\in (0,1/3).
\label{MESCH}
\end{equation}
Since the present system has no electron-electron interaction, 
${\tilde {\cal N}}_y(t)$ is written as 
\begin{eqnarray}
{\tilde {\cal N}}_y(t)&=&\frac{1}{L_xL_y}{\rm Tr}\> P_{{\rm F},\Lambda}
\frac{1}{2\pi i}\int_\gamma dz \frac{1}{z-H_{\omega,\Lambda}}v_y
\left[\frac{1}{z-H_{\omega,\Lambda}}
-\frac{1}{z-H_{\omega,\Lambda}+i\hbar\eta}\right]\ret
& &\qquad\times\frac{1}{2\pi i}\int_\gamma dz' 
\left[\frac{1}{z'-H_{\omega,\Lambda}}
-\frac{1}{z'-H_{\omega,\Lambda}-i\hbar\eta}\right]v_y
\frac{1}{z'-H_{\omega,\Lambda}}\ret
&=&\frac{1}{L_xL_y}\sum_{\bf a}{\rm Tr}\> P_{{\rm F},\Lambda}
\frac{1}{2\pi i}\int_\gamma dz \frac{1}{z-H_{\omega,\Lambda}}v_y
\left[\frac{1}{z-H_{\omega,\Lambda}}-\frac{1}{z
-H_{\omega,\Lambda}+i\hbar\eta}\right]\ret
& &\times\chi_b({\bf a})\frac{1}{2\pi i}\int_\gamma dz' 
\left[\frac{1}{z'-H_{\omega,\Lambda}}
-\frac{1}{z'-H_{\omega,\Lambda}-i\hbar\eta}\right]v_y
\frac{1}{z'-H_{\omega,\Lambda}}.
\label{2ndfctfree}
\end{eqnarray}
Here we have introduced the partition of unity $\{\chi_b({\bf a})\}$. 
In the same way as the above, the summand can be further written as  
\begin{eqnarray}
& &{\rm Tr}\> P_{{\rm F},\Lambda}\frac{1}{2\pi i}\int_{\gamma} dz 
[y,(z-H_{\omega,\Lambda})^{-1}]
\frac{\eta}{z-H_{\omega,\Lambda}+i\hbar\eta}\ret
&\times&\chi_b({\bf a})\frac{1}{2\pi i}\int_{\gamma} dz'
\frac{\eta}{z'-H_{\omega,\Lambda}-i\hbar\eta}
[(z'-H_{\omega,\Lambda})^{-1},y]+\mbox{correction}.
\label{2nsfctfreecor}
\end{eqnarray}
Here the correction vanishes almost surely in the infinite volume 
limit by taking a suitable sequence $\{L_x,L_y\}$ of the system sizes 
as in the proof of Theorem~\ref{theorem:sigmaxyoutzero}. 
If $z$ of the resolvent $R_\Lambda(z)=(z-H_{\omega,\Lambda})^{-1}$ is 
not near to the spectrum $\sigma(H_{\omega,\Lambda})$, then 
the resolvent is bounded and decays exponentially at large distance.  
Therefore we consider only the contributions from the paths 
near the Fermi energy $E_{\rm F}$ in the first term in the right-hand side of 
(\ref{2nsfctfreecor}). 
As a typical example of the corresponding contributions, let us consider  
\begin{eqnarray} 
I_{\bf a}&:=&\sum_{{\bf u},{\bf v},{\bf w},{\bf z}}
\int_{t_-}^{t_+}dt\int_{t_-}^{t_+}dt'\>{\rm Tr}\>
\chi_b({\bf u})P_{{\rm F},\Lambda}\chi_b({\bf v})[y,R_\Lambda(E_{\rm F}+it)]
\chi_b({\bf w})\ret
&\times&\eta R_\Lambda(E_{\rm F}+i(t+\hbar\eta))\chi_b({\bf a})
\eta R_\Lambda(E_{\rm F}+i(t'-\hbar\eta))\chi_b({\bf z})
[R_\Lambda(E_{\rm F}+it'),y]\chi_b({\bf u}).\ret
\label{defIa}
\end{eqnarray}
Note that 
\begin{eqnarray}
\left|{\rm Tr}\>\chi_b({\bf u})P_{{\rm F},\Lambda}\chi_b({\bf v})A\right|
&\le&\sqrt{{\rm Tr}\>\chi_b({\bf u})P_{{\rm F},\Lambda}\chi_b({\bf u})}
\sqrt{{\rm Tr}\>A^\ast\chi_b({\bf v})P_{{\rm F},\Lambda}\chi_b({\bf v})A}\ret
&\le&{\rm Const.}\Vert A\Vert 
\end{eqnarray}
for any bounded operator $A$, where we have used the bound (\ref{localTr}), 
and that 
\begin{equation}
\left\Vert\chi_b({\bf v})[y,R_\Lambda(z)]\chi_b({\bf w})\right\Vert\le
({\rm Const.}+|v_2-w_2|)
\left\Vert\chi_b({\bf v})R_\Lambda(z)\chi_b({\bf w})\right\Vert, 
\end{equation}
where we have used the decomposition $y=y-y^b+y^b$ in the proof of 
Lemma~\ref{lemma:ETrbounded}. 
{From} these observations, we have 
\begin{eqnarray}
|I_{\bf a}|&\le&{\rm Const.}\sum_{{\bf u},{\bf v},{\bf w},{\bf z}}
({\rm Const.}+|v_2-w_2|)({\rm Const.}+|z_2-u_2|)\ret
&\times&\int_{t_-}^{t_+}dt\>
\left\Vert\chi_b({\bf v})R_\Lambda(E_{\rm F}+it)\chi_b({\bf w})\right\Vert
\cdot\eta\left\Vert\chi_b({\bf w})R_\Lambda(E_{\rm F}+i(t+\hbar\eta))
\chi_b({\bf a})\right\Vert\ret
&\times&\int_{t_-}^{t_+}dt'\>\eta 
\left\Vert\chi_b({\bf a})R_\Lambda(E_{\rm F}+i(t'-\hbar\eta))
\chi_b({\bf z})\right\Vert
\left\Vert\chi_b({\bf z})R_\Lambda(E_{\rm F}+it')\chi_b({\bf u})\right\Vert.\ret
\label{Iabound}
\end{eqnarray}
The first integral is decomposed into two parts as 
\begin{equation}
\int_{t_-}^{t_+}dt\cdots=\int_{t_-}^{-\hbar\eta/2}dt\cdots
+\int_{-\hbar\eta/2}^{t_+}dt\cdots.
\end{equation}
The second part of the integral is estimated as 
\begin{eqnarray}
& &\int_{-\hbar\eta/2}^{t_+}dt\>
\left\Vert\chi_b({\bf v})R_\Lambda(E_{\rm F}+it)\chi_b({\bf w})\right\Vert
\cdot\eta\left\Vert\chi_b({\bf w})R_\Lambda(E_{\rm F}+i(t+\hbar\eta))
\chi_b({\bf a})\right\Vert\ret
&\le&{\rm Const.}\times\eta^{s/4}\int_{-\hbar\eta/2}^{t_+}dt\>
\left\Vert\chi_b({\bf v})R_\Lambda(E_{\rm F}+it)\chi_b({\bf w})\right\Vert^{s/4}
|t|^{s/4-1}\ret
& &\qquad\times\left\Vert\chi_b({\bf w})R_\Lambda(E_{\rm F}+i(t+\hbar\eta))
\chi_b({\bf a})\right\Vert^{s/4},
\end{eqnarray}
where we have used the inequality 
$\left\Vert\chi_b({\bf w})R_\Lambda(E_{\rm F}+i(t+\hbar\eta))
\chi_b({\bf a})\right\Vert\le 2/(\hbar\eta)$ for $t\ge-\hbar\eta/2$. 
For the first part of the integral, we can obtain a similar bound by using   
$\left\Vert\chi_b({\bf v})R_\Lambda(E_{\rm F}+it)\chi_b({\bf w})\right\Vert
\le 2/(\hbar\eta)$ for $t\le-\hbar\eta/2$. 
Clearly the second integral about $t'$ in the right-hand side of (\ref{Iabound}) 
can be treated in the same way. Combining these observations with 
H\"oledr inequality, 
\begin{eqnarray}
& &{\bf E}\left[\left\Vert\chi_b({\bf v})R_\Lambda(E_{\rm F}+it)\chi_b({\bf w})
\right\Vert^{s/4}
\left\Vert\chi_b({\bf w})R_\Lambda(E_{\rm F}+i(t+\hbar\eta))
\chi_b({\bf a})\right\Vert^{s/4}\right.\ret
& &\times\left.\left\Vert\chi_b({\bf a})R_\Lambda(E_{\rm F}+i(t'-\hbar\eta))
\chi_b({\bf z})\right\Vert^{s/4}
\left\Vert\chi_b({\bf z})R_\Lambda(E_{\rm F}+it')\chi_b({\bf u})
\right\Vert^{s/4}\right]\ret
&\le&{\bf E}\left[\left\Vert\chi_b({\bf v})R_\Lambda(E_{\rm F}+it)\chi_b({\bf w})
\right\Vert^s\right]^{1/4}
{\bf E}\left[\left\Vert\chi_b({\bf w})R_\Lambda(E_{\rm F}+i(t+\hbar\eta))
\chi_b({\bf a})\right\Vert^s\right]^{1/4}\ret
&\times&{\bf E}\left[\left\Vert\chi_b({\bf a})R_\Lambda(E_{\rm F}+i(t'-\hbar\eta))
\chi_b({\bf z})\right\Vert^s\right]^{1/4}
{\bf E}\left[\left\Vert\chi_b({\bf z})R_\Lambda(E_{\rm F}+it')\chi_b({\bf u})
\right\Vert^s\right]^{1/4},\ret
\end{eqnarray}
we obtain 
\begin{eqnarray}
{\bf E}|I_{\bf a}|&\le&{\rm Const.}\times\eta^{s/2}\sum_{{\bf u},{\bf v},{\bf w},{\bf z}}
({\rm Const.}+|v_2-w_2|)({\rm Const.}+|z_2-u_2|)\ret
& &\quad\times e^{-\mu|{\bf v}-{\bf w}|/4}e^{-\mu|{\bf w}-{\bf a}|/4}
e^{-\mu|{\bf a}-{\bf z}|/4}e^{-\mu|{\bf z}-{\bf u}|/4}\ret
&\le&{\rm Const.}\times\eta^{s/2}, 
\end{eqnarray}
where we have used Fatou's lemma, Fubini-Tonelli theorem and 
the fractional moment bound (\ref{DRI}) as in the proof of 
Lemma~\ref{lemma:PFD}. Combining this bound, 
(\ref{2ndfctfree}), (\ref{2nsfctfreecor}) and (\ref{defIa}), 
the expectation of ${\tilde {\cal N}}_y(t)$ 
of (\ref{MESCH}) is bounded by ${\rm Const.}\times\eta^{s/2}$. 

Since the bound, 
\begin{equation}
\left|{\cal E}^{-1}+({\cal E}+i\hbar\eta)^{-1}\right|^2
\le{4}{\cal E}^{-2},
\end{equation}
holds for ${\cal E}\in{\bf R}$, 
the expectation of ${\tilde {\cal N}}_u(t)$ of (\ref{MESCH}) can be proved to be 
bounded in a easier way. 
As a result, we obtain that there exists a sequence 
$\{L_{x,n},L_{y,n}\}_n$ of 
the system sizes such that the bound, 
\begin{equation}
\lim_{L_x,L_y\rightarrow\infty}
{\cal N}_{uy}(t)\le {\rm Const.}\eta^{s/4}, 
\end{equation}
holds almost surely, where the positive constant may depend on $\omega$. 

Since the rest of the contributions for $\delta\sigma_{uy}(t)$ 
can be handled in the same way, we obtain the desired result that the bound, 
\begin{equation}
|\delta\sigma_{uy}(t)|\le[{\cal C}_1(\omega)+{\cal C}_2(\omega)T]e^{-\eta T}
+{\cal C}_3(\omega)\eta^{s/4},
\end{equation}
holds almost surely for $s\in(0,1/3)$ and for $u=x,y$. Here the positive constants, 
${\cal C}_j(\omega)<\infty$, $j=1,2,3$, may depend on $\omega$. 
Choosing $s=4/13<1/3$, we get the bound (\ref{sigmacorrections}).

%%%%%%%%%%%%%%%%%%%%%%%%%%%%%%%%%%%%%%%%%%%%%%%%%%%%%%%%%%%%%%%%%%%
\appendix
%%%%%%%%%%%%%%%%%%%%%%%%%%%%%%%%%%%%%%%%%%%%%%%%%%%%%%%%%%%%%%%%%%
\Section{Wegner estimate}
\label{appendix:Wegner}

In this appendix,  
we study the Wegner estimate \cite{Wegner} for the density of states 
for a single-electron  
Hamiltonian in a general setting. For this purpose, 
we follow the argument by Barbaroux, Combes and Hislop \cite{BCH}.  
However, the following argument is slightly simplified compared to their 
original one because we are interested in two dimensions only. 

Consider a single spinless electron system with 
a disorder potential $V_\omega$ in $d$ dimensions. 
The Hamiltonian is given by $H_\omega=H_0+V_\omega$ on $L^2({\bf R}^d)$ 
with the unperturbed Hamiltonian, 
\begin{equation}
H_0=\frac{1}{2m_e}({\bf p}+e{\bf A})^2+V_0.
\end{equation}
We assume ${\bf A}\in C^1({\bf R}^d,{\bf R}^d)$ and 
$V_0\in L^\infty({\bf R}^d)$, and assume 
that the Hamiltonian $H_0$ is essentially self-adjoint 
with a boundary condition. As a disorder potential $V_\omega$, 
we consider an Anderson type potential of impurities, 
\begin{equation}
V_\omega({\bf r})=\sum_{{\bf a}\in{\bf L}^d}\lambda_{\bf a}(\omega)
u({\bf r}-{\bf a}), 
\label{Vomega}
\end{equation}
for ${\bf r}=(x_1,x_2,\ldots,x_d)\in{\bf R}^d$. 
The constants $\{\lambda_{\bf a}(\omega)|\ {\bf a}\in{\bf L}^d\}$ form 
a family of independent, identically distributed random variables on 
a $d$-dimensional periodic lattice ${\bf L}^d\subset{\bf R}^d$. 
The common distribution has a density $g\ge 0$ which has compact 
support and satisfies $g\in L^\infty({\bf R})\cap C({\bf R})$. 
We assume that the single-site potential $u$ is non-negative and has 
compact support. We write the sum of the single-site potentials $u$ as 
\begin{equation}
{\cal U}({\bf r}):=
\sum_{{\bf a}\in{\bf L}^d}u({\bf r}-{\bf a})\quad \mbox{for}\ 
{\bf r}\in{\bf R}^d.
\label{defcalU}
\end{equation}
We assume $\Vert {\cal U}\Vert_\infty<+\infty$. Clearly this implies 
$\Vert u\Vert_\infty<+\infty$. 
From these assumptions, 
we have $\Vert V_\omega\Vert_\infty\le v_{\rm R}<+\infty$ with some positive 
constant $v_{\rm R}$ which is independent of the random variables. 

For a bounded region $\Lambda\subset{\bf R}^d$, we denote by 
$H_{\omega,\Lambda}=H_{0,\Lambda}+V_{\omega,\Lambda}$ the Hamiltonian 
$H_\omega$ restricted to $\Lambda$ with a boundary condition. 
Here $V_{\omega,\Lambda}=V_\omega|\Lambda$, i.e., 
$V_{\omega,\Lambda}$ is the restriction of $V_\omega$ to $\Lambda$. 
We assume 
\begin{equation}
{\cal U}_\Lambda:={\cal U}|\Lambda\ge {\cal U}_{\rm min}\chi_\Lambda,
\label{noflatregion}
\end{equation}
where ${\cal U}_{\rm min}$ is a positive constant which is independent 
of the bounded region $\Lambda$, and $\chi_\Lambda$ is 
the characteristic function for $\Lambda$. Namely 
there is no flat potential region satisfying $u=0$. 
Further we assume that  
\begin{equation}
{\rm Tr}\left({H_{0,\Lambda}+{\cal E}_{\rm min}}\right)^{-2} 
\chi_\Omega\le K_0\left|\Omega\right|^{n_0}
\label{TrR02}
\end{equation}
for a finite region $\Omega\subset\Lambda$, with a positive constant 
${\cal E}_{\rm min}>\Vert V_0^-\Vert_\infty$, 
where ${\rm Tr}$ stands for the trace on $L^2(\Lambda)$; 
$K_0$ and $n_0$ are the positive constants which are 
independent of the volumes $\left|\Lambda\right|,\left|\Omega\right|$ 
of the finite regions $\Lambda,\Omega$. 
If the vector potential ${\bf A}$ satisfies the additional 
assumption ${\bf A}\in C^2({\bf R}^d,{\bf R}^d)$, then 
the inequality (\ref{TrR02}) is valid in the dimensions $d\le 3$. 
See ref.~\cite{BCH} for details and also 
for the treatment in the case of higher dimensions $d\ge 4$ in 
which they require a stronger assumption than the above assumption 
(\ref{TrR02}). Consider the present system with 
the unperturbed Hamiltonian (\ref{Ham0}). From the inequality (\ref{Eloweredgebound}) 
for the lower edge of the Landau band, we have 
\begin{equation}
{\rm Tr}\left({H_{0,\Lambda}+{\cal E}_{\rm min}}\right)^{-2} 
\chi_\Omega\le{\rm Const.}\times{|b|}/{B}
\end{equation}
for strong magnetic field strengths $B$.
Here we have assumed that the region $\Omega$ is contained in a rectangular box $b$ 
such that the volume $|b|$ of the box satisfies the flux quantization 
condition $|b|/(2\pi\ell_B^2)\in{\bf N}$. Namely 
\begin{equation}
n_0=1\quad\mbox{and}\quad K_0\sim{\rm Const.}\times B^{-1}\quad\mbox{for a large}\ B. 
\label{K0n0}
\end{equation}

Let $Q_{\omega,\Lambda}(\Delta)$ denote the spectral 
projection for $H_{\omega,\Lambda}$ with an energy interval 
$\Delta\subset{\bf R}$. 
Let $\psi_E$ be an eigenvector of the Hamiltonian $H_{\omega,\Lambda}$, 
i.e., $H_{\omega,\Lambda}\psi_E=E\psi_E$ with an energy eigenvalue 
$E\in{\bf R}$. The Schr\"odinger equation is written as 
\begin{equation}
\left(H_{0,\Lambda}+{\cal E}_{\rm min}\right)\psi_E
=\left(-V_{\omega,\Lambda}+E+{\cal E}_{\rm min}\right)\psi_E.
\end{equation}
Since $H_{0,\Lambda}+{\cal E}_{\rm min}>0$ from the assumption 
${\cal E}_{\rm min}>\Vert V_0^-\Vert_\infty$, one has  
\begin{equation}
\psi_E=\frac{1}{H_{0,\Lambda}+{\cal E}_{\rm min}}
\left(-V_{\omega,\Lambda}+E+{\cal E}_{\rm min}\right)\psi_E
=\frac{1}{H_{0,\Lambda}+{\cal E}_{\rm min}}
\left(-V_{\omega,\Lambda}+H_{\omega,\Lambda}+{\cal E}_{\rm min}\right)\psi_E.
\end{equation}
Using this identity, one obtains 
\begin{eqnarray}
& &{\rm Tr}\ Q_{\omega,\Lambda}(\Delta)\ret
&=&{\rm Tr}\left({H_{0,\Lambda}+{\cal E}_{\rm min}}\right)^{-2}
\left(-V_{\omega,\Lambda}+H_{\omega,\Lambda}+{\cal E}_{\rm min}\right)
Q_{\omega,\Lambda}(\Delta)
\left(-V_{\omega,\Lambda}+H_{\omega,\Lambda}+{\cal E}_{\rm min}\right)\ret
&=&{\rm Tr}\left({H_{0,\Lambda}+{\cal E}_{\rm min}}\right)^{-2}
V_{\omega,\Lambda}Q_{\omega,\Lambda}(\Delta)V_{\omega,\Lambda}\ret
&-&{\rm Tr}\left({H_{0,\Lambda}+{\cal E}_{\rm min}}\right)^{-2}
V_{\omega,\Lambda}\left(H_{\omega,\Lambda}+{\cal E}_{\rm min}\right)
Q_{\omega,\Lambda}(\Delta)\ret
&-&{\rm Tr}\left({H_{0,\Lambda}+{\cal E}_{\rm min}}\right)^{-2}
\left(H_{\omega,\Lambda}+{\cal E}_{\rm min}\right)
Q_{\omega,\Lambda}(\Delta)V_{\omega,\Lambda}\ret
&+&{\rm Tr}\left({H_{0,\Lambda}+{\cal E}_{\rm min}}\right)^{-2}
\left(H_{\omega,\Lambda}+{\cal E}_{\rm min}\right)^2Q_{\omega,\Lambda}(\Delta). 
\label{TrEDelta}
\end{eqnarray}

Let us evaluate the first term in the last line of the right-hand side. 
Substituting the expression (\ref{Vomega}) of $V_\omega$ into the term, 
one has 
\begin{eqnarray}
& &{\rm Tr}\left({H_{0,\Lambda}+{\cal E}_{\rm min}}\right)^{-2}
V_{\omega,\Lambda}Q_{\omega,\Lambda}(\Delta)V_{\omega,\Lambda}\ret
&=&\sum_{{\bf a},{\bf b}}\lambda_{\bf a}\lambda_{\bf b}
{\rm Tr}\left({H_{0,\Lambda}+{\cal E}_{\rm min}}\right)^{-2}
u_{{\bf a},\Lambda}Q_{\omega,\Lambda}(\Delta)u_{{\bf b},\Lambda}\ret
&=&\sum_{{\bf a},{\bf b}}\lambda_{\bf a}\lambda_{\bf b}
{\rm Tr}\ u_{{\bf b},\Lambda}^{1/2}
\left({H_{0,\Lambda}+{\cal E}_{\rm min}}\right)^{-2}u_{{\bf a},\Lambda}^{1/2}
\ u_{{\bf a},\Lambda}^{1/2}Q_{\omega,\Lambda}(\Delta)
u_{{\bf b},\Lambda}^{1/2},
\label{TrR2VEV}
\end{eqnarray}
where we have written $u_{\bf a}({\bf r})=u({\bf r}-{\bf a})$ and 
$u_{{\bf a},\Lambda}=u_{\bf a}|\Lambda$. Since the operator 
\begin{equation}
\Upsilon_{{\bf b},{\bf a}}^{(1)}:=u_{{\bf b},\Lambda}^{1/2}
\left({H_{0,\Lambda}+{\cal E}_{\rm min}}\right)^{-2}u_{{\bf a},\Lambda}^{1/2}
\end{equation}
is compact, there exist a pair of orthonormal bases, 
$\{\varphi_n^{(1)}\}_{n=1}^\infty$ and $\{\psi_n^{(1)}\}_{n=1}^\infty$, 
and nonnegative numbers $\{\mu_n^{(1)}\}_{n=1}^\infty$ 
such that\footnote{See, for 
example, Chapter~VI of the book by Reed and Simon \cite{RSI}.}
\begin{equation}
\Upsilon_{{\bf b},{\bf a}}^{(1)}=\sum_{n=1}^\infty
\mu_n^{(1)}\varphi_n^{(1)}(\psi_n^{(1)},\cdots).
\label{R2badecom}
\end{equation}
The numbers $\mu_n^{(1)}$ are the eigenvalues of 
$|\Upsilon_{{\bf b},{\bf a}}^{(1)}|$. 
Using this representation (\ref{R2badecom}), one has 
\begin{eqnarray}
& &\left|{\rm Tr}\Upsilon_{{\bf b},{\bf a}}^{(1)}u_{{\bf a},\Lambda}^{1/2}
Q_{\omega,\Lambda}(\Delta)u_{{\bf b},\Lambda}^{1/2}\right|\ret
&\le&\sum_{n=1}^\infty \mu_n^{(1)} \left|\left(\psi_n^{(1)},
u_{{\bf a},\Lambda}^{1/2}
Q_{\omega,\Lambda}(\Delta)u_{{\bf b},\Lambda}^{1/2}
\varphi_n^{(1)}\right)\right|\ret
&\le&\frac{1}{2}\sum_{n=1}^\infty \mu_n^{(1)} 
\left[\left(\psi_n^{(1)},u_{{\bf a},\Lambda}^{1/2}
Q_{\omega,\Lambda}(\Delta)u_{{\bf a},\Lambda}^{1/2}\psi_n^{(1)}\right)
+\left(\varphi_n^{(1)}, u_{{\bf b},\Lambda}^{1/2}
Q_{\omega,\Lambda}(\Delta)u_{{\bf b},\Lambda}^{1/2}\varphi_n^{(1)}
\right)\right].
\end{eqnarray}
Therefore the expectation value of the left-hand side of the first inequality 
can be bounded from above as 
\begin{eqnarray}
& &{\bf E}_\Lambda\left[
\left|{\rm Tr}\Upsilon_{{\bf b},{\bf a}}^{(1)}u_{{\bf a},\Lambda}^{1/2}
Q_{\omega,\Lambda}(\Delta)u_{{\bf b},\Lambda}^{1/2}\right|\right]\ret
&\le&\frac{1}{2}\left(\sum_{n=1}^\infty \mu_n^{(1)}\right)
\sup_n{\bf E}_\Lambda\left[\left(\psi_n^{(1)},u_{{\bf a},\Lambda}^{1/2}
Q_{\omega,\Lambda}(\Delta)u_{{\bf a},\Lambda}^{1/2}\psi_n^{(1)}\right)
+\left(\varphi_n^{(1)}, u_{{\bf b},\Lambda}^{1/2}
Q_{\omega,\Lambda}(\Delta)u_{{\bf b},\Lambda}^{1/2}\varphi_n^{(1)}\right)
\right],\ret
\label{expectTrUpsilonE}
\end{eqnarray}
where ${\bf E}_\Lambda[\cdots]$ stands for the expectation with respect to the random 
variables on a region $\Lambda\subset{\bf R}^d$.
The right-hand side can be evaluated by using the following 
Lemma~\ref{lemma:expspectraEbound} which is essentially due to 
Kotani and Simon \cite{KS}. 
In order to make this paper self-contained, we give 
the proof of Lemma~\ref{lemma:expspectraEbound} in 
Appendix~\ref{appendix:expspectraEbound}, following from ref.~\cite{CH1}. 

\begin{lemma}
\label{lemma:expspectraEbound}
Let $v$ be a nonnegative function satisfying $v\le u_{{\bf a},\Lambda}$. 
Then 
\begin{equation}
\left\Vert\int_{\bf R}d\lambda_{\bf a}g(\lambda_{\bf a})
v^{1/2}Q_{\omega,\Lambda}(\Delta)v^{1/2}\right\Vert
\le \Vert g\Vert_\infty|\Delta|.
\label{expspectraEbound}
\end{equation}
\end{lemma}

{From} the bound (\ref{expspectraEbound}) and the inequality (\ref{expectTrUpsilonE}), 
one has 
\begin{equation}
{\bf E}_\Lambda\left[
\left|{\rm Tr}\Upsilon_{{\bf b},{\bf a}}^{(1)}u_{{\bf a},\Lambda}^{1/2}
Q_{\omega,\Lambda}(\Delta)u_{{\bf b},\Lambda}^{1/2}\right|\right]
\le \Vert g\Vert_\infty|\Delta|\left\Vert\Upsilon_{{\bf b},{\bf a}}^{(1)}\right\Vert_1, 
\end{equation}
where $\Vert\cdots\Vert_1:={\rm Tr}\left|\cdots\right|$.
Moreover, combining this bound with (\ref{TrR2VEV}), one gets 
\begin{eqnarray}
{\bf E}_\Lambda
\left[{\rm Tr}\left({H_{0,\Lambda}+{\cal E}_{\rm min}}\right)^{-2}
V_{\omega,\Lambda}Q_{\omega,\Lambda}(\Delta)V_{\omega,\Lambda}\right]
&=&\sum_{{\bf a},{\bf b}}{\bf E}_\Lambda\left[\lambda_{\bf a}\lambda_{\bf b}
{\rm Tr}\ \Upsilon_{{\bf b},{\bf a}}^{(1)}
\ u_{{\bf a},\Lambda}^{1/2}Q_{\omega,\Lambda}(\Delta)
u_{{\bf b},\Lambda}^{1/2}\right]\ret
&\le&{\cal M}^2\sum_{{\bf a},{\bf b}}{\bf E}_\Lambda\left[\left|
{\rm Tr}\ \Upsilon_{{\bf b},{\bf a}}^{(1)}
\ u_{{\bf a},\Lambda}^{1/2}Q_{\omega,\Lambda}(\Delta)
u_{{\bf b},\Lambda}^{1/2}\right|\right]\ret
&\le&{\cal M}^2\Vert g\Vert_\infty|\Delta|\sum_{{\bf a},{\bf b}}
\left\Vert\Upsilon_{{\bf b},{\bf a}}^{(1)}\right\Vert_1,
\label{ETrR2VQV} 
\end{eqnarray}
where ${\cal M}:=\sup_{\lambda\in\ {\rm supp}\ g}|\lambda|$. 

Next consider the second term in the right-hand side of 
the second equality in (\ref{TrEDelta}). From the assumption (\ref{noflatregion}), 
one can easily find a set of nonnegative functions 
$\{{\tilde u}_{{\bf a},\Lambda}\}_{\bf a}$ 
satisfying the following two conditions: 
\begin{equation}
{\tilde u}_{{\bf a},\Lambda}\le u_{{\bf a},\Lambda}\ \ 
\mbox{for any lattice site}\ {\bf a},
\label{conditiontildeu}
\end{equation}
{and} 
\begin{equation}
\sum_{{\bf a}}{\tilde u}_{{\bf a},\Lambda}={\cal U}_{\rm min}\chi_\Lambda.
\label{chitildeu}
\end{equation}
Using the identity (\ref{chitildeu}), one has 
\begin{eqnarray}
& &{\rm Tr}\ \left({H_{0,\Lambda}+{\cal E}_{\rm min}}\right)^{-2}
V_{\omega,\Lambda}(H_{\omega,\Lambda}+{\cal E}_{\rm min})Q_{\omega,\Lambda}(\Delta)\ret
&=&\sum_{\bf a}\lambda_{\bf a}
{\rm Tr}\ \left({H_{0,\Lambda}+{\cal E}_{\rm min}}\right)^{-2}u_{{\bf a},\Lambda}
(H_{\omega,\Lambda}+{\cal E}_{\rm min})Q_{\omega,\Lambda}(\Delta)\chi_\Lambda\ret
&=&\frac{1}{{\cal U}_{\rm min}}\sum_{{\bf a},{\bf b}}\lambda_{\bf a}
{\rm Tr}\ \left({H_{0,\Lambda}+{\cal E}_{\rm min}}\right)^{-2}u_{{\bf a},\Lambda}
(H_{\omega,\Lambda}+{\cal E}_{\rm min})Q_{\omega,\Lambda}(\Delta)
{\tilde u}_{{\bf b},\Lambda}\ret
&=&\frac{1}{{\cal U}_{\rm min}}\sum_{{\bf a},{\bf b}}\lambda_{\bf a}
{\rm Tr}\ \Upsilon_{{\bf b},{\bf a}}^{(2)}u_{{\bf a},\Lambda}^{1/2}
(H_{\omega,\Lambda}+{\cal E}_{\rm min})Q_{\omega,\Lambda}(\Delta)
{\tilde u}_{{\bf b},\Lambda}^{1/2},
\label{TrR2VHE}
\end{eqnarray}
where 
\begin{equation}
\Upsilon_{{\bf b},{\bf a}}^{(2)}:=
{\tilde u}_{{\bf b},\Lambda}^{1/2}
\left({H_{0,\Lambda}+{\cal E}_{\rm min}}\right)^{-2}
u_{{\bf a},\Lambda}^{1/2}. 
\end{equation}
In the same way, 
\begin{eqnarray}
& &\left|{\rm Tr}\ \Upsilon_{{\bf b},{\bf a}}^{(2)}
u_{{\bf a},\Lambda}^{1/2}
(H_{\omega,\Lambda}+{\cal E}_{\rm min})Q_{\omega,\Lambda}(\Delta)
{\tilde u}_{{\bf b},\Lambda}^{1/2}\right|\ret
&\le&\sum_{n=1}^\infty\mu_n^{(2)}\left|
\left(\psi_n^{(2)},u_{{\bf a},\Lambda}^{1/2}
(H_{\omega,\Lambda}+{\cal E}_{\rm min})Q_{\omega,\Lambda}(\Delta)
{\tilde u}_{{\bf b},\Lambda}^{1/2}\varphi_n^{(2)}\right)\right|\ret
&\le&\frac{{\cal E}_{\rm max}(\Delta)}{2}\sum_{n=1}^\infty \mu_n^{(2)}
\left[\left(\psi_n^{(2)},u_{{\bf a},\Lambda}^{1/2}Q_{\omega,\Lambda}(\Delta)
u_{{\bf a},\Lambda}^{1/2}\psi_n^{(2)}\right)+
\left(\varphi_n^{(2)}{\tilde u}_{{\bf b},\Lambda}^{1/2}
Q_{\omega,\Lambda}(\Delta)
{\tilde u}_{{\bf b},\Lambda}^{1/2}\varphi_n^{(2)}\right)\right],\ret
\label{TrUpsilonuHEu}
\end{eqnarray}
where $\{\varphi_n^{(2)}\}_{n=1}^\infty$ and $\{\psi_n^{(2)}\}_{n=1}^\infty$ 
are orthonormal bases such that 
\begin{equation}
\Upsilon_{{\bf b},{\bf a}}^{(2)}=\sum_{n=1}^\infty \mu_n^{(2)}
\varphi_n^{(2)}(\psi_n^{(2)},\cdots)
\end{equation}
with the eigenvalues $\mu_n^{(2)}$ of $|\Upsilon_{{\bf b},{\bf a}}^{(2)}|$, 
and 
\begin{equation}
{\cal E}_{\rm max}(\Delta)=\sup_{E\in\Delta}|E+{\cal E}_{\rm min}|.
\label{def:EmaxDelta}
\end{equation}
Combining (\ref{TrR2VHE}), (\ref{TrUpsilonuHEu}) and 
Lemma~\ref{lemma:expspectraEbound} with the condition (\ref{conditiontildeu}), 
one has 
\begin{equation}
{\bf E}_\Lambda\left[\left|
{\rm Tr}\ \left({H_{0,\Lambda}+{\cal E}_{\rm min}}\right)^{-2}
V_{\omega,\Lambda}(H_{\omega,\Lambda}+{\cal E}_{\rm min})Q_{\omega,\Lambda}(\Delta)
\right|\right]\le {\cal M}\frac{{\cal E}_{\rm max}(\Delta)}{{\cal U}_{\rm min}}
\Vert g\Vert_\infty |\Delta|\sum_{{\bf a},{\bf b}}
\left\Vert\Upsilon_{{\bf b},{\bf a}}^{(2)}\right\Vert_1.
\label{ETrR2VHQ} 
\end{equation}
In the same way, 
\begin{equation}
{\bf E}_\Lambda\left[\left|
{\rm Tr}\ \left({H_{0,\Lambda}+{\cal E}_{\rm min}}\right)^{-2}
(H_{\omega,\Lambda}+{\cal E}_{\rm min})Q_{\omega,\Lambda}(\Delta)V_{\omega,\Lambda}
\right|\right]\le {\cal M}\frac{{\cal E}_{\rm max}(\Delta)}{{\cal U}_{\rm min}}
\Vert g\Vert_\infty |\Delta|\sum_{{\bf a},{\bf b}}
\left\Vert\Upsilon_{{\bf b},{\bf a}}^{(3)}\right\Vert_1 
\label{ETrR2HQV}
\end{equation}
and
\begin{equation}
{\bf E}_\Lambda\left[\left|
{\rm Tr}\ \left({H_{0,\Lambda}+{\cal E}_{\rm min}}\right)^{-2}
(H_{\omega,\Lambda}+{\cal E}_{\rm min})^2Q_{\omega,\Lambda}(\Delta)
\right|\right]\le \left(\frac{{\cal E}_{\rm max}(\Delta)}{{\cal U}_{\rm min}}\right)^2
\Vert g\Vert_\infty |\Delta|\sum_{{\bf a},{\bf b}}
\left\Vert\Upsilon_{{\bf b},{\bf a}}^{(4)}\right\Vert_1, 
\label{ETrR2H2Q}
\end{equation}
where 
\begin{equation}
\Upsilon_{{\bf b},{\bf a}}^{(3)}:=
u_{{\bf b},\Lambda}^{1/2}
\left({H_{0,\Lambda}+{\cal E}_{\rm min}}\right)^{-2}
{\tilde u}_{{\bf a},\Lambda}^{1/2} 
\quad\mbox{and}\quad
\Upsilon_{{\bf b},{\bf a}}^{(4)}:=
{\tilde u}_{{\bf b},\Lambda}^{1/2}
\left({H_{0,\Lambda}+{\cal E}_{\rm min}}\right)^{-2}
{\tilde u}_{{\bf a},\Lambda}^{1/2}. 
\end{equation}

Next let us estimate 
$\sum_{{\bf a},{\bf b}}\left\Vert\Upsilon_{{\bf b},{\bf a}}^{(i)}\right\Vert_1$, 
$i=1,2,3,4$. For simplicity, we consider only the case with $i=1$ because 
all the other cases can be estimated in the same way. We decompose the sum 
into two parts as  
\begin{equation}
\sum_{{\bf a},{\bf b}}\left\Vert\Upsilon_{{\bf b},{\bf a}}^{(1)}\right\Vert_1
=\sum_{{\bf a},{\bf b}\atop{\rm overlap}}
\left\Vert\Upsilon_{{\bf b},{\bf a}}^{(1)}\right\Vert_1+
\sum_{{\bf a},{\bf b}\atop{\rm non-overlap}}
\left\Vert\Upsilon_{{\bf b},{\bf a}}^{(1)}\right\Vert_1,
\label{decomUpsilonsumnorm1}
\end{equation}
where the first sum is over the lattice sites ${\bf a},{\bf b}$ such that 
the corresponding two potentials $u_{{\bf a},\Lambda},u_{{\bf b},\Lambda}$ 
overlap with each other, and the second sum is over those 
for the non-overlapping potentials. Note that 
\begin{equation}
\left\Vert\Upsilon_{{\bf b},{\bf a}}^{(i)}\right\Vert_1
\le\sqrt{\left\Vert\Upsilon_{{\bf b},{\bf b}}^{(i)}\right\Vert_1
\left\Vert\Upsilon_{{\bf a},{\bf a}}^{(i)}\right\Vert_1},
\label{Upsilonbound11}
\end{equation}
where we have used the inequality $\Vert AB\Vert_1\le\Vert A\Vert_2\Vert B\Vert_2$ 
for bounded operators $A,B$. Here the norm $\Vert\cdots\Vert_2$ is defined 
as $\Vert A\Vert_2:=\sqrt{{\rm Tr}A^\ast A}$ for a bounded operator $A$ if 
the right-hand side exists. Using the inequality (\ref{Upsilonbound11}) 
and the assumption (\ref{TrR02}), one has    
\begin{equation}
\sum_{{\bf a},{\bf b}\atop{\rm overlap}}
\left\Vert\Upsilon_{{\bf b},{\bf a}}^{(1)}\right\Vert_1
\le {\rm Const.}\times K_0\Vert u\Vert_\infty
\left|{\rm supp}\ u\right|^{n_0}\left|\Lambda\right|,
\label{sumUpsilonoverlap}
\end{equation}
where the positive constant depends only on the lattice ${\bf L}^d$ 
and on the support of the potential $u$, and so the constant is finite from 
the assumptions on the lattice and the potential. 

Using Proposition~\ref{proUpsilonnorm1} in Appendix~\ref{decayestimateUpsilon}, 
the second sum in the 
right-hand side of (\ref{decomUpsilonsumnorm1}) can be evaluated as 
\begin{eqnarray}
\sum_{{\bf a},{\bf b}\atop{\rm non-overlap}}
\left\Vert\Upsilon_{{\bf b},{\bf a}}^{(1)}\right\Vert_1
&\le&\frac{{\rm Const.}\times\Vert u\Vert_\infty}{(1-\kappa^2)({\cal E}_{\rm min}-
\Vert V_0^-\Vert_\infty)}K_0|{\rm supp}\ u|^{n_0/2}
\sum_{{\bf a},{\bf b}\atop{\rm non-overlap}}r^ne^{-\alpha r}\ret
&\le&\frac{{\rm Const.}\times\Vert u\Vert_\infty}{(1-\kappa^2)({\cal E}_{\rm min}-
\Vert V_0^-\Vert_\infty)}K_0K_2(\alpha)|{\rm supp}\ u|^{n_0/2}|\Lambda|, 
\end{eqnarray}
where $\alpha$ is given by (\ref{def:alpha}), 
$r={\rm dist}({\rm supp}\ u_{\bf a},{\rm supp}\ u_{\bf b})$, 
and the positive constant $K_2(\alpha)$ satisfies the bound:  
\begin{equation}
K_2(\alpha)\le {\rm Const.}\times \sum_{\bf b}r^ne^{-\alpha r}
\end{equation}
with the positive constant and with a fixed lattice site ${\bf a}$. 
Combining this result with the bound (\ref{sumUpsilonoverlap}), one has 
\begin{equation}
\sum_{{\bf a},{\bf b}}
\left\Vert\Upsilon_{{\bf b},{\bf a}}^{(1)}\right\Vert_1
\le{\rm Const.}\times K_0K_1\Vert u\Vert_\infty|{\rm supp}\ u|^{n_0/2}
|\Lambda|
\end{equation}
with the positive constant
\begin{equation}
K_1=|{\rm supp}\ u|^{n_0/2}
+\frac{K_2(\alpha)}{(1-\kappa^2)({\cal E}_{\rm min}-\Vert V_0^-\Vert_\infty)}.
\end{equation}
Clearly this constant $K_1$ is independent of the strength $B$ of the 
magnetic field and of the random potential $V_\omega$. 
{From} the condition (\ref{conditiontildeu}) for ${\tilde u}_{{\bf a},\Lambda}$, 
the following bounds also hold: 
\begin{equation}
\sum_{{\bf a},{\bf b}}
\left\Vert\Upsilon_{{\bf b},{\bf a}}^{(i)}\right\Vert_1
\le{\rm Const.}\times K_0K_1\Vert u\Vert_\infty|{\rm supp}\ u|^{n_0}
|\Lambda|
\end{equation}
for all $i=1,2,3,4$. Combining this, (\ref{TrEDelta}), (\ref{ETrR2VQV}),  
(\ref{ETrR2VHQ}), (\ref{ETrR2HQV}) and (\ref{ETrR2H2Q}), one obtains 
the following theorem:

\begin{theorem}
\label{theorem:Wegner}
Assume the conditions (\ref{noflatregion}) and (\ref{TrR02}). 
Let $\Delta=[E-\delta E,E+\delta E]$ be an interval of the energy with $\delta E>0$. 
Then the following inequality is valid:
\begin{equation}
{\rm Prob}\left[{\rm dist}(\sigma(H_{\omega,\Lambda}),E)<\delta E\right]
\le {\bf E}_\Lambda\left[{\rm Tr}\ Q_{\omega,\Lambda}(\Delta)\right]
\le C_{\rm W}K_3\Vert g\Vert_\infty\delta E
|\Lambda|
\label{Wegnerestimate}
\end{equation}
with some positive constant $C_{\rm W}$ and with 
\begin{equation}
K_3=\left[{\cal M}+{\cal E}_{\rm max}(\Delta)/{\cal U}_{\rm min}\right]^2
K_0K_1\Vert u\Vert_\infty|{\rm supp}\ u|^{n_0}.
\label{defK3}
\end{equation}
Here ${\bf E}_\Lambda[\cdots]$ is the expectation with respect to the random 
variables on a region $\Lambda\subset{\bf R}^d$.
\end{theorem}

\noindent
{\bf Remark:} Consider the present system with 
the unperturbed Hamiltonian (\ref{Ham0}). The positive number $K_0$ behaves as 
$K_0\sim{\rm Const.}\times B^{-1}$ for a large $B$ as in (\ref{K0n0}). 
{From} the definition (\ref{def:EmaxDelta}), 
${\cal E}_{\rm max}(\Delta)\sim{\rm Const.}\times B$ for a large $B$. 
Substituting these into the above expression (\ref{defK3}) of $K_3$, 
one has $K_3\sim{\rm Const.}\times B$ for a large $B$. 
{From} this observation and the above theorem, one notices that 
the upper bound for the number of the states in the energy interval 
with a fixed width $\delta E$ 
is proportional to the strength $B$ of the magnetic field for 
a large $B$.

%%%%%%%%%%%%%%%%%%%%%%%%%%%%%%%%%%%%%%%%%%%%%%%%%%%%%%%%%%%%%%%%%%%%%%
\Section{Proof of Lemma~\ref{lemma:expspectraEbound}}
\label{appendix:expspectraEbound}

Following Combes and Hislop \cite{CH1}, we give the proof of 
the Kotani-Simon lemma \cite{KS}.  

We begin with preparing the following lemma: 

\begin{lemma}
\label{Klambdaintlemma}
Write $H_{\omega,\Lambda}=H_{\omega,\Lambda}'+\lambda_{\bf a}u_{{\bf a},
\Lambda}$. Let $v$ be a nonnegative function satisfying 
$u_{{\bf a},\Lambda}\ge v$, and define 
\begin{equation}
K(\lambda,E-i\delta)=v^{1/2}\left(H_{\omega,\Lambda}'+\lambda 
u_{{\bf a},\Lambda}-E+i\delta\right)^{-1}v^{1/2}
\label{defK}
\end{equation}
for $E\in{\bf R}$ and $\delta>0$. Then 
\begin{equation}
\left\Vert\int_{\bf R}d\lambda \frac{\lambda_0^2}{\lambda^2+\lambda_0^2}
K(\lambda,E-i\delta)\right\Vert\le \pi
\label{Klambdaintbound}
\end{equation}
for $\lambda_0>0$. 
\end{lemma}

\begin{proof}{Proof} Since $K(\lambda,E-i\delta)$ is holomorphic in $\lambda$ 
in the upper half-plane, one has 
\begin{equation}
\int_{\bf R}d\lambda \frac{\lambda_0^2}{\lambda^2+\lambda_0^2}
K(\lambda,E-i\delta)
=\pi\lambda_0K(i\lambda_0,E-i\delta). 
\label{Klambdaint}
\end{equation}
Note that 
\begin{eqnarray}
& &-{\rm Im}K(i\lambda_0,E-i\delta)\ret
&=&v^{1/2}\left(H_{\omega,\Lambda}'-i\lambda_0 u_{{\bf a},\Lambda}
-E-i\delta\right)^{-1}(\lambda_0 u_{{\bf a},\Lambda}+\delta)
\left(H_{\omega,\Lambda}'+i\lambda_0 u_{{\bf a},\Lambda}-E+i\delta\right)^{-1}
v^{1/2}\ret
&\ge&\lambda_0K(i\lambda_0,E-i\delta)^\ast K(i\lambda_0,E-i\delta), 
\end{eqnarray}
where we have used the assumptions $u_{{\bf a},\Lambda}\ge v$ and $\delta>0$. 
This implies $\left\Vert K(i\lambda_0,E-i\delta)\right\Vert\le \lambda_0^{-1}$. 
Combining this with the above (\ref{Klambdaint}), one has 
the desired result (\ref{Klambdaintbound}). 
\end{proof}

\begin{proof}{Proof of Lemma~\ref{lemma:expspectraEbound}}
Let ${\tilde \Delta}\supset\Delta$ and ${\tilde \Delta}\ne\Delta$. 
Using Stone's formula, one has 
\begin{equation}
\left(\varphi,v^{1/2}Q_{\omega,\Lambda}(\Delta)v^{1/2}\varphi\right)
\le\frac{1}{\pi}\lim_{\delta\downarrow 0}{\rm Im}
\int_{\tilde \Delta}dE\left(\varphi,v^{1/2}
\left(H_{\omega,\Lambda}-E+i\delta\right)^{-1}v^{1/2}\varphi\right)
\end{equation}
for any vector $\varphi$. Further, 
\begin{eqnarray}
& &\int_{\bf R}d\lambda_{\bf a}\frac{\lambda_0^2}{\lambda_{\bf a}^2+
\lambda_0^2}\left(\varphi,v^{1/2}Q_{\omega,\Lambda}(\Delta)v^{1/2}
\varphi\right)\ret
&\le&\frac{1}{\pi}\lim_{\delta\downarrow 0}
{\rm Im}\int_{\tilde \Delta}dE
\int_{\bf R}d\lambda_{\bf a}\frac{\lambda_0^2}{\lambda_{\bf a}^2+
\lambda_0^2}\left(\varphi,K(\lambda_{\bf a},E-i\delta)\varphi\right)
\le|{\tilde \Delta}|\Vert\varphi\Vert^2
\end{eqnarray}
by using Fubini's theorem and Lemma~\ref{Klambdaintlemma}. 
Here $\lambda_0>0$, and $K(\lambda,E-i\delta)$ is given by (\ref{defK}). 
Since $g\in L_\infty({\bf R})$ with compact support from the assumption, 
one has 
\begin{equation}
\left\Vert\int_{\bf R}d\lambda_{\bf a}g(\lambda_{\bf a})v^{1/2}
Q_{\omega,\Lambda}(\Delta)v^{1/2}\right\Vert
\le \sup_\lambda g(\lambda)\frac{\lambda^2+\lambda_0^2}{\lambda_0^2}
|\Delta|
\end{equation}
for any $\lambda_0>0$. This proves the bound (\ref{expspectraEbound}). 
\end{proof}

%%%%%%%%%%%%%%%%%%%%%%%%%%%%%%%%%%%%%%%%%%%%%%%%%%%%%%%%%%%%%%%%%
\Section{A decay estimate of $\Upsilon_{{\bf b},{\bf a}}^{(i)}$}
\label{decayestimateUpsilon}

In this appendix, we follow Barbaroux, Combes and Hislop \cite{BCH},  
in order to estimate $\Upsilon_{{\bf b},{\bf a}}^{(i)}$ 
that appear in Appendix~\ref{appendix:Wegner}. 
The result is summarized as the following proposition: 

\begin{pro}
\label{proUpsilonnorm1}
Let $v,w$ be bounded functions with a compact support, and 
suppose ${\rm dist}({\rm supp}\ v,{\rm supp}\ w)=r$ 
with  a positive distance $r$. Then 
\begin{equation}
\left\Vert v \left(R_{0,\Lambda}\right)^2 w\right\Vert_1
\le\frac{{\rm Const.}\times\Vert v\Vert_\infty\Vert w\Vert_\infty}
{(1-\kappa^2)({\cal E}_{\rm min}-\Vert V_0^-\Vert_\infty)}
K_0\left(|{\rm supp}\ v|^{n_0/2}+|{\rm supp}\ w|^{n_0/2}\right)r^ne^{-\alpha r} 
\label{vR2wdecay}
\end{equation}
with 
\begin{equation}
\alpha=\frac{\sqrt{2m_e({\cal E}_{\rm min}-\Vert V_0^-\Vert_\infty)}}{3\hbar}\kappa
\quad \mbox{with}\ \kappa\in(0,1)
\label{def:alpha} 
\end{equation}
and with some positive number $n$,   
where $R_{0,\Lambda}=(H_{0,\Lambda}+{\cal E}_{\rm min})^{-1}$, and 
$\Vert\cdots\Vert_1:={\rm Tr}|\cdots|$. The constants $K_0$ and $n_0$ are given 
in (\ref{TrR02}).   
\end{pro}

Let $\Omega$ be a region in ${\bf R}^d$. Then we denote by $\partial\Omega$ 
the boundary of $\Omega$, and define the subset 
$\Omega_{\rm in}$ of $\Omega$ as 
$\Omega_{\rm in}=\left\{{\bf r}\in\Omega|\ {\rm dist}({\bf r},\partial\Omega)>\delta
\right\}$ with a positive $\delta$. With a small $\delta$, one can take 
three regions $\Omega^{(i)}$, $i=1,2,3$ satisfying the following conditions: 
\begin{equation}
{\rm supp}\ v\subset \Omega_{\rm in}^{(1)}\subset \Omega^{(1)}
\subset \Omega_{\rm in}^{(2)}\subset \Omega^{(2)}\subset \Omega_{\rm in}^{(3)}
\subset \Omega^{(3)}\subset\Lambda,
\end{equation}
\begin{equation}
\left|\Omega^{(2)}\right|\le{\rm Const.}\times r^d,
\label{assumptionOmega2}
\end{equation}
\begin{equation}
{\rm dist}({\rm supp}\ w,\Omega^{(3)})>r/3,
\label{distwO3}
\end{equation}
and 
\begin{equation}
{\rm dist}({\rm supp}\ v,\Gamma^{(1)})>r/3,
\label{distvG1}
\end{equation}
where $\Gamma^{(1)}=\Omega^{(1)}\backslash\Omega_{\rm in}^{(1)}$, 
and we also write $\Gamma^{(i)}=\Omega^{(i)}\backslash\Omega_{\rm in}^{(i)}$ 
for $i=2,3$. 

Let ${\tilde \chi}_i\in C^2(\Lambda)$, $i=1,2,3$, be three nonnegative 
functions satisfying 
\begin{equation}
\left.{\tilde \chi}_i\right|_{\Omega_{\rm in}^{(i)}}=1\quad
\mbox{and}\quad
\left.{\tilde \chi}_i\right|_{\Lambda\backslash\Omega^{(i)}}=0.
\label{conditionchitilde}
\end{equation}
In the following, we denote by $\chi_i'$ the characteristic function 
$\chi_{\Gamma^{(i)}}$ of the region $\Gamma^{(i)}$ for $i=1,2,3$,  
and write $R_{0,i}=R_{0,\Omega^{(i)}}=(H_{0,\Omega^{(i)}}+{\cal E}_{\rm min})^{-1}$ 
for $i=1,2,3$. Next introduce the geometric resolvent equation, 
\begin{equation}
{\tilde \chi}_i R_{0,\Lambda}=R_{0,i}{\tilde \chi}_i
+R_{0,i}W({\tilde \chi}_i)R_{0,\Lambda}
\label{GREOmegai}
\end{equation}
for $i=1,2,3$, where $W(\cdots)$ is given by (\ref{defW0}). 

Using (\ref{GREOmegai}) and $v{\tilde \chi}_2=v$, one has 
\begin{equation}
v\left(R_{0,\Lambda}\right)^2w=v{\tilde \chi}_2\left(R_{0,\Lambda}\right)^2w
=vR_{0,2}{\tilde \chi}_2R_{0,\Lambda}w+vR_{0,2}W({\tilde \chi}_2)
\left(R_{0,\Lambda}\right)^2w.
\label{vR2w}
\end{equation}
The first term in the right-hand side can be rewritten as 
\begin{eqnarray}
vR_{0,2}{\tilde \chi}_2R_{0,\Lambda}w&=&v\left(R_{0,2}\right)^2
W({\tilde \chi}_2)R_{0,\Lambda}w\ret
&=&v\left(R_{0,2}\right)^2W({\tilde \chi}_2){\tilde \chi}_3R_{0,\Lambda}w\ret
&=&v\left(R_{0,2}\right)^2W({\tilde \chi}_2)R_{0,3}W({\tilde \chi}_3)
R_{0,\Lambda}w
\end{eqnarray}
by using ${\tilde \chi}_2w={\tilde \chi}_3w=0$, 
$W({\tilde \chi}_2){\tilde \chi}_3=W({\tilde \chi}_2)$ and 
the geometric resolvent equation (\ref{GREOmegai}). 
Therefore 
\begin{eqnarray}
\left\Vert vR_{0,2}{\tilde \chi}_2R_{0,\Lambda}w\right\Vert_1
&=&
\left\Vert v\left(R_{0,2}\right)^2W({\tilde \chi}_2)R_{0,3}W({\tilde \chi}_3)
R_{0,\Lambda}w\right\Vert_1\ret
&\le&\left\Vert vR_{0,2}\right\Vert_2
\left\Vert R_{0,2}W({\tilde \chi}_2)R_{0,3}W({\tilde \chi}_3)
R_{0,\Lambda}w\right\Vert_2\ret
&\le&\left\Vert vR_{0,2}\right\Vert_2
\left\Vert wR_{0,\Lambda}W^\ast({\tilde \chi}_3)R_{0,3}W^\ast({\tilde \chi}_2)
R_{0,2}\right\Vert_2\ret
&=&\left\Vert vR_{0,2}\right\Vert_2
\left\Vert wR_{0,\Lambda}\chi_3'W^\ast({\tilde \chi}_3)R_{0,3}
W^\ast({\tilde \chi}_2)
R_{0,2}\right\Vert_2\ret
&\le&\left\Vert vR_{0,2}\right\Vert_2
\left\Vert wR_{0,\Lambda}\chi_3'\right\Vert\ 
\left\Vert W^\ast({\tilde \chi}_3)R_{0,3}W^\ast({\tilde \chi}_2)
R_{0,2}\right\Vert_2,
\label{est1vR2w}
\end{eqnarray}
where we have used $\chi_3'W^\ast({\tilde \chi}_3)=W^\ast({\tilde \chi}_3)$, 
the equality $\Vert A^\ast\Vert_2=\Vert A\Vert_2$ and 
the inequality $\Vert AB\Vert_1\le\Vert A\Vert_2\Vert B\Vert_2$ 
for bounded operators $A,B$. The norm $\Vert\cdots\Vert_2$ is defined as 
$\Vert A\Vert_2:=\sqrt{{\rm Tr}A^\ast A}$ for a bounded operator $A$.

The second term in the right-hand side of (\ref{vR2w}) can be written as 
\begin{eqnarray}
vR_{0,2}W({\tilde \chi}_2)\left(R_{0,\Lambda}\right)^2w
&=&v{\tilde \chi}_1R_{0,2}W({\tilde \chi}_2)\left(R_{0,\Lambda}\right)^2w\ret
&=&vR_{0,1}W({\tilde \chi}_1)R_{0,2}W({\tilde \chi}_2)
\left(R_{0,\Lambda}\right)^2w
\end{eqnarray}
by using $v{\tilde \chi}_1=v$, ${\tilde \chi}_1W({\tilde \chi}_2)=0$, and 
the geometric resolvent equation 
\begin{equation}
{\tilde \chi}_1R_{0,2}=R_{0,1}{\tilde \chi}_1+R_{0,1}W({\tilde \chi}_1)
R_{0,2}. 
\end{equation}
Therefore the norm can be evaluated as 
\begin{eqnarray}
\left\Vert vR_{0,2}W({\tilde \chi}_2)\left(R_{0,\Lambda}\right)^2w\right\Vert_1
&=&\left\Vert
vR_{0,1}W({\tilde \chi}_1)R_{0,2}W({\tilde \chi}_2)
\left(R_{0,\Lambda}\right)^2w\right\Vert_1\ret
&=&\left\Vert
vR_{0,1}\chi_1'W({\tilde \chi}_1)R_{0,2}W({\tilde \chi}_2)
\left(R_{0,\Lambda}\right)^2w\right\Vert_1\ret
&\le&\left\Vert vR_{0,1}\chi_1'\right\Vert
\left\Vert W({\tilde \chi}_1)R_{0,2}W({\tilde \chi}_2)
\left(R_{0,\Lambda}\right)^2w\right\Vert_1\ret
&\le&\left\Vert vR_{0,1}\chi_1'\right\Vert
\left\Vert W({\tilde \chi}_1)R_{0,2}W({\tilde \chi}_2)
R_{0,\Lambda}\right\Vert_2
\left\Vert R_{0,\Lambda}w\right\Vert_2,
\label{est2vR2w}
\end{eqnarray}
where we have used the identity 
$\chi_1'W({\tilde \chi}_1)=W({\tilde \chi}_1)$. 

Combining (\ref{vR2w}), (\ref{est1vR2w}) and (\ref{est2vR2w}), one has 
\begin{eqnarray}
\left\Vert v\left(R_{0,\Lambda}\right)^2w\right\Vert_1&\le&
\left\Vert vR_{0,2}{\tilde \chi}_2R_{0,\Lambda}w\right\Vert_1
+\left\Vert vR_{0,2}W({\tilde \chi}_2)\left(R_{0,\Lambda}\right)^2w
\right\Vert_1\ret
&\le&\Vert vR_{0,2}\Vert_2\left\Vert wR_{0,\Lambda}\chi_3'\right\Vert
\left\Vert W^\ast({\tilde \chi}_3)R_{0,3}W^\ast({\tilde \chi}_2)
R_{0,2}\right\Vert_2\ret
&+&\Vert R_{0,\Lambda}w\Vert_2\left\Vert vR_{0,1}\chi_1'\right\Vert
\left\Vert W({\tilde \chi}_1)R_{0,2}W({\tilde \chi}_2)
R_{0,\Lambda}\right\Vert_2.  
\label{vR2west}
\end{eqnarray}
In order to estimate the right-hand side, we use the following lemma: 

\begin{lemma}
\label{lemmaWRWR2}
Let $\varphi$ be a vector in the domain of the Hamiltonian 
$H_{0,\Omega^{(2)}}$. Then 
\begin{equation}
\left\Vert W^\ast({\tilde \chi}_3)R_{0,3}W^\ast({\tilde \chi}_2)\varphi\right\Vert
\le{\rm Const.}\times\Vert\varphi\Vert, 
\label{WastR03Wastvarphibound}
\end{equation}
where the positive constant in the right-hand side depends only on 
the cut-off functions, ${\tilde \chi}_2$ and ${\tilde \chi}_3$.
\end{lemma}

\begin{proof}{Proof}
Note that
\begin{equation}
\left\Vert W^\ast({\tilde \chi}_3)R_{0,3}W^\ast({\tilde \chi}_2)\varphi\right\Vert
\le
\left\Vert W^\ast({\tilde \chi}_3)R_{0,3}^{1/2}\right\Vert
\left\Vert R_{0,3}^{1/2}W^\ast({\tilde \chi}_2)\varphi\right\Vert.
\label{WastR03Wastvarphi2decom}
\end{equation}
Using (\ref{defW}), one has 
\begin{equation}
\left\Vert W^\ast({\tilde \chi}_3)R_{0,3}^{1/2}\right\Vert
\le\frac{\hbar}{m_e}\left\Vert(\nabla{\tilde \chi}_3)\cdot({\bf p}+e{\bf A})
R_{0,3}^{1/2}\right\Vert+\frac{\hbar^2}{2m_e}
\left\Vert(\Delta{\tilde \chi}_3)R_{0,3}^{1/2}\right\Vert.
\label{WastR1/2bound}
\end{equation}
The first term in the right-hand side can be estimated as follows: 
Using the Schwarz inequality, one has 
\begin{eqnarray}
& &\left(\psi,R_{0,3}^{1/2}({\bf p}+e{\bf A})\cdot(\nabla{\tilde \chi}_3)
(\nabla{\tilde \chi}_3)\cdot({\bf p}+e{\bf A})R_{0,3}^{1/2}\psi\right)\ret
&\le&\sqrt{\left(\psi,R_{0,3}^{1/2}({\bf p}+e{\bf A})^2R_{0,3}^{1/2}\psi\right)}\ret
&\times&\sqrt{\left(\psi,R_{0,3}^{1/2}({\bf p}+e{\bf A})\cdot(\nabla{\tilde \chi}_3)
|\nabla{\tilde \chi}_3|^2(\nabla{\tilde \chi}_3)\cdot({\bf p}+e{\bf A})
R_{0,3}^{1/2}\psi\right)}\ret
&\le&\sqrt{2m_e}\left\Vert|\nabla{\tilde \chi}_3|\right\Vert_\infty
\left\Vert(\nabla{\tilde \chi}_3)\cdot({\bf p}+e{\bf A})R_{0,3}^{1/2}\psi\right\Vert
\Vert\psi\Vert
\end{eqnarray}
for any vector $\psi$, where we have used 
\begin{equation}
R_{0,3}^{1/2}\frac{1}{2m_e}({\bf p}+e{\bf A})^2R_{0,3}^{1/2}\le 1
\label{R1/2p2R1/2inequality} 
\end{equation}
which is derived from the assumption ${\cal E}_{\rm min}>\Vert V_0^-\Vert_\infty$.
As a result, one obtain 
\begin{equation}
\left\Vert(\nabla{\tilde \chi}_3)\cdot({\bf p}+e{\bf A})R_{0,3}^{1/2}\right\Vert
\le\sqrt{2m_e}\left\Vert|\nabla{\tilde \chi}_3|\right\Vert_\infty.
\end{equation}
Substituting this into the right-hand side of (\ref{WastR1/2bound}), one gets 
\begin{equation}
\left\Vert W^\ast({\tilde \chi}_3)R_{0,3}^{1/2}\right\Vert
\le\hbar\sqrt{\frac{m_e}{2}}\left\Vert|\nabla{\tilde \chi}_3|\right\Vert_\infty
+\frac{\hbar^2}{2m_e}
\frac{\Vert\Delta{\tilde \chi}_3\Vert_\infty}{\sqrt{{\cal E}_{\rm min}
-\Vert V_0^-\Vert_\infty}}.
\label{WastR031/2bound} 
\end{equation}
Using (\ref{defW}) again, one has 
\begin{equation}
\left\Vert R_{0,3}^{1/2}W^\ast({\tilde \chi}_2)\varphi\right\Vert
\le\frac{\hbar}{m_e}\left\Vert R_{0,3}^{1/2}({\bf p}+e{\bf A})\cdot
(\nabla{\tilde \chi}_2)\varphi\right\Vert
+\frac{\hbar^2}{2m_e}\left\Vert R_{0,3}^{1/2}(\Delta{\tilde\chi}_2)\varphi\right\Vert.
\label{R1/2Wastvarphinorm}
\end{equation}
The norm of the first term in the right-hand side can be evaluated as 
\begin{eqnarray}
& &\left(\varphi,(\nabla{\tilde \chi}_2)\cdot({\bf p}+e{\bf A})R_{0,3}
({\bf p}+e{\bf A})\cdot(\nabla{\tilde \chi}_2)\varphi\right)\ret
&\le&
\left\Vert|\nabla{\tilde\chi}_2|\right\Vert_\infty\Vert\varphi\Vert
\sqrt{\left(\varphi,(\nabla{\tilde \chi}_2)\cdot({\bf p}+e{\bf A})R_{0,3}
({\bf p}+e{\bf A})^2R_{0,3}({\bf p}+e{\bf A})\cdot
(\nabla{\tilde \chi}_2)\varphi\right)}\ret
&\le&\sqrt{2m_e}\left\Vert|\nabla{\tilde\chi}_2|\right\Vert_\infty\Vert\varphi\Vert
\left\Vert R_{0,3}^{1/2}({\bf p}+e{\bf A})\cdot
(\nabla{\tilde \chi}_2)\varphi\right\Vert,
\end{eqnarray}
where we have used the Schwarz inequality and the inequality 
(\ref{R1/2p2R1/2inequality}). Therefore,   
\begin{equation}
\left\Vert R_{0,3}^{1/2}({\bf p}+e{\bf A})\cdot
(\nabla{\tilde \chi}_2)\varphi\right\Vert
\le\sqrt{2m_e}\left\Vert|\nabla{\tilde\chi}_2|\right\Vert_\infty\Vert\varphi\Vert. 
\end{equation}
Substituting this into the right-hand side of (\ref{R1/2Wastvarphinorm}), 
one gets 
\begin{equation}
\left\Vert R_{0,3}^{1/2}W^\ast({\tilde \chi}_2)\varphi\right\Vert
\le\left\{\hbar\sqrt{\frac{m_e}{2}}\left\Vert|\nabla{\tilde\chi}_2|\right\Vert_\infty
+\frac{\hbar^2}{2m_e}
\frac{\Vert\Delta{\tilde \chi}_2\Vert_\infty}{\sqrt{{\cal E}_{\rm min}
-\Vert V_0^-\Vert_\infty}}\right\}\Vert\varphi\Vert.
\label{R03Wastvarphibound}
\end{equation}
The bound (\ref{WastR03Wastvarphibound}) follows from (\ref{WastR03Wastvarphi2decom}), 
(\ref{WastR031/2bound}) and (\ref{R03Wastvarphibound}).  
\end{proof}

{From} this lemma, immediately one gets 
\begin{equation}
\left\Vert W^\ast({\tilde \chi}_3)R_{0,3}W^\ast({\tilde \chi}_2)R_{0,2}
\right\Vert_2\le {\rm Const.}\times\left\Vert \chi_2'R_{0,2}\right\Vert_2.
\end{equation}
In the same way, 
\begin{equation}
\left\Vert W({\tilde \chi}_1)R_{0,2}W({\tilde \chi}_2)R_{0,\Lambda}
\right\Vert_2\le{\rm Const.}\times\left\Vert \chi_2'R_{0,\Lambda}\right\Vert_2.
\end{equation}
Substituting these two bounds into (\ref{vR2west}),  one gets 
\begin{equation}
\left\Vert v\left(R_{0,\Lambda}\right)^2w\right\Vert_1 
\le {\rm Const.}\left[\Vert vR_{0,2}\Vert_2
\left\Vert\chi_2'R_{0,2}\right\Vert_2
\left\Vert wR_{0,\Lambda}\chi_3'\right\Vert
+\Vert R_{0,\Lambda}w\Vert_2\left\Vert \chi_2'R_{0,\Lambda}\right\Vert_2
\left\Vert vR_{0,1}\chi_1'\right\Vert\right].
\label{vRw1bound} 
\end{equation}

Note that the ground state energy $E_0$ of the Hamiltonian $H_{0,\Lambda}$ 
satisfies $E_0\ge-\Vert V_0^-\Vert_\infty$. 
Taking 
\begin{equation}
\beta=\frac{\sqrt{2m_e({\cal E}_{\rm min}-\Vert V_0^-\Vert_\infty)}}{\hbar}\kappa\quad
\mbox{with}\ \kappa\in(0,1)
\label{beta0}
\end{equation}
in the bound (\ref{Rdecayboundbotom}) in the next Appendix~\ref{section:decayR}, 
one has 
\begin{equation}
\Vert vR_{0,\Omega}w\Vert\le 
\frac{\Vert v\Vert_\infty\Vert w\Vert_\infty}{(1-\kappa^2)({\cal E}_{\rm min}-
\Vert V_0^-\Vert_\infty)}e^{-\beta r}\quad\mbox{for a region}\ \Omega,
\end{equation}
where $r$ is the distance between the supports of the two functions $v$ and $w$.
Combining this inequality, (\ref{distwO3}), (\ref{distvG1}) and (\ref{vRw1bound}), 
one has  
\begin{eqnarray}
\left\Vert v\left(R_{0,\Lambda}\right)^2w\right\Vert_1 
&\le&\frac{{\rm Const.}\times\Vert v\Vert_\infty\Vert w\Vert_\infty}
{(1-\kappa^2)({\cal E}_{\rm min}-\Vert V_0^-\Vert_\infty)}\ret
&\times&\left[\frac{\Vert vR_{0,2}\Vert_2}{\Vert v\Vert_\infty}
\left\Vert\chi_2'R_{0,2}\right\Vert_2 
+\frac{\Vert R_{0,\Lambda}w\Vert_2}{\Vert w\Vert_\infty}
\left\Vert \chi_2'R_{0,\Lambda}\right\Vert_2
\right] e^{-\alpha r},
\end{eqnarray}
where $\alpha=\beta/3$ with the above $\beta$ of (\ref{beta0}). 
Further, 
\begin{eqnarray}
\left\Vert v\left(R_{0,\Lambda}\right)^2w\right\Vert_1 
&\le&\frac{{\rm Const.}\times\Vert v\Vert_\infty\Vert w\Vert_\infty}
{(1-\kappa^2)({\cal E}_{\rm min}-\Vert V_0^-\Vert_\infty)}\ret
&\times&K_0\left(|{\rm supp}\ v|^{n_0/2}+|{\rm supp}\ w|^{n_0/2}\right)
\left|\Omega^{(2)}\backslash\Omega_{\rm in}^{(2)}\right|^{n_0/2}e^{-\alpha r} 
\end{eqnarray}
by using the assumption (\ref{TrR02}). Thus one gets the desired bound 
(\ref{vR2wdecay}) from the bound (\ref{assumptionOmega2}) 
on the region $\Omega^{(2)}$.  

%%%%%%%%%%%%%%%%%%%%%%%%%%%%%%%%%%%%%%%%%%%%%%%%
\Section{Decay estimates of resolvents} 
\label{section:decayR}

In this appendix, the exponential decay bound for the resolvent 
$(H_\omega-z)^{-1}$ is obtained by using the Combes-Thomas method \cite{CT}. 

%%%%%%%%%%%%%%%%%%%%%%%%%%%%%%%%
\subsection{The general case}
\label{Decaygeneral}

For the general case with ${\bf A}_{\rm P}\ne 0$, 
we use the improved version \cite{BCH} of the Combes-Thomas method \cite{CT}. 
The results are given by 
Theorem~\ref{theorem:Rdecaybound} and the bound (\ref{Rdecayboundbotom}) below. 

Consider the $d$-dimensional Hamiltonian, 
\begin{equation}
H_\Lambda=\frac{1}{2m_e}({\bf p}+e{\bf A})^2+V_\Lambda, 
\end{equation}
with a general vector potential ${\bf A}\in C^1(\Lambda,{\bf R}^d)$ and 
a general electrostatic potential $V_\Lambda\in L^\infty(\Lambda)$. 
Let $j_\delta\in C_0^\infty({\bf R}^d)$  
satisfying $j_\delta\ge 0$, 
${\rm supp}\ j_\delta\subset\{{\bf r}|\> |{\bf r}|\le \delta\}$ 
with a small $\delta$ and $\int_{{\bf R}^d}j_\delta({\bf r})dx_1\cdots dx_d=1$.
Let $\Omega$ be a bounded region with smooth boundary, 
and define ${\tilde \rho}({\bf r})={\rm dist}({\bf r},\Omega)$. 
Following \cite{CH}, we introduce the smooth distance function 
$\rho({\bf r})=j_\delta\ast {\tilde \rho}({\bf r})$.  
Note that, for $\beta>0$,  
\begin{eqnarray}
e^{-\beta\rho}H_\Lambda e^{\beta\rho}&=&
\frac{1}{2m_e}\left({\bf p}+e{\bf A}-i\hbar\beta\nabla\rho\right)^2+V_\Lambda\ret
&=&H_\Lambda-\frac{\hbar^2\beta^2}{2m_e}(\nabla\rho)^2-\frac{i\hbar\beta}{2m_e}
\left[\nabla\rho\cdot({\bf p}+e{\bf A})+({\bf p}+e{\bf A})\cdot\nabla\rho\right].
\end{eqnarray}
We write
\begin{equation}
e^{-\beta\rho}H_\Lambda e^{\beta\rho}={\tilde H}_\Lambda+i\beta J
\label{expHrho}
\end{equation}
with  
\begin{equation}
{\tilde H}_\Lambda=H_\Lambda-\frac{\hbar^2\beta^2}{2m_e}(\nabla\rho)^2
\quad\mbox{and}\quad
J=-\frac{\hbar}{2m_e}
\left[\nabla\rho\cdot({\bf p}+e{\bf A})+({\bf p}+e{\bf A})\cdot\nabla\rho\right]. 
\label{expJ} 
\end{equation}
We take ${\tilde C}_0>0$ satisfying  
\begin{equation}
-\Vert V_\Lambda^-\Vert_\infty-\frac{\hbar^2\beta^2}{2m_e}
\Vert|\nabla\rho|\Vert_\infty^2
+{\tilde C}_0>0, 
\label{defC0}
\end{equation}
where $V_\Lambda^\pm=\max\{\pm V_\Lambda,0\}$.  
Then one has 
\begin{equation}
{\tilde H}_\Lambda+{\tilde C}_0\ge C_0>0
\label{def:C0} 
\end{equation}
with some constant 
$C_0$. We define 
\begin{equation}
X_{E+i\varepsilon}=\frac{{\tilde H}_\Lambda-E-i\varepsilon}{{\tilde H}_\Lambda+{\tilde C}_0}
\quad\mbox{for}\ E,\varepsilon\in{\bf R}, 
\end{equation}
and define 
\begin{equation}
Y=({\tilde H}_\Lambda+{\tilde C}_0)^{-1/2}J({\tilde H}_\Lambda+{\tilde C}_0)^{-1/2}. 
\end{equation}

Let us estimate the norm $\Vert Y\Vert$ of this operator. 
{From} the expression (\ref{expJ}) of the operator $J$, one has 
\begin{equation}
\Vert Y\Vert \le \frac{\hbar}{2m_e}\left[
\Vert {\tilde R}^{1/2}\nabla\rho\cdot({\bf p}+e{\bf A})
{\tilde R}^{1/2}\Vert
+\Vert {\tilde R}^{1/2}({\bf p}+e{\bf A})\cdot\nabla\rho
{\tilde R}^{1/2}\Vert\right],
\label{normboundY1} 
\end{equation}
where we have written 
${\tilde R}=({\tilde H}_\Lambda+{\tilde C}_0)^{-1}$. 
Take $\psi={\tilde R}^{1/2}\varphi$ with $\varphi\in L^2(\Lambda)$. 
Using the Schwarz inequality and (\ref{def:C0}), one has
\begin{eqnarray}
& &\left(\psi,({\bf p}+e{\bf A})\cdot\nabla\rho{\tilde R}
\nabla\rho\cdot({\bf p}+e{\bf A})\psi\right)\ret
&\le&\sqrt{(\psi,({\bf p}+e{\bf A})^2\psi)}\times
\sqrt{(\psi,({\bf p}+e{\bf A})\cdot\nabla\rho{\tilde R}|\nabla\rho|^2
{\tilde R}\nabla\rho\cdot({\bf p}+e{\bf A})\psi)}\ret
&\le&\sqrt{2m_e}\Vert|\nabla\rho|\Vert_\infty\Vert\varphi
\Vert\sqrt{(\psi,({\bf p}+e{\bf A})\cdot\nabla\rho{\tilde R}^2
\nabla\rho\cdot({\bf p}+e{\bf A})\psi)}\ret
&\le&\sqrt{\frac{2m_e}{C_0}}\Vert|\nabla\rho|\Vert_\infty\Vert\varphi
\Vert\sqrt{(\psi,({\bf p}+e{\bf A})\cdot\nabla\rho{\tilde R}
\nabla\rho\cdot({\bf p}+e{\bf A})\psi)}.
\end{eqnarray}
Therefore 
\begin{equation}
\left(\psi,({\bf p}+e{\bf A})\cdot\nabla\rho{\tilde R}
\nabla\rho\cdot({\bf p}+e{\bf A})\psi\right)\le\frac{2m_e}{C_0}
\Vert|\nabla\rho|\Vert_\infty^2
\Vert \varphi\Vert^2. 
\end{equation}
Similarly 
\begin{eqnarray}
& &\left(\psi,\nabla\rho\cdot({\bf p}+e{\bf A}){\tilde R}
({\bf p}+e{\bf A})\cdot\nabla\rho\psi\right)\ret
&\le&\Vert|\nabla\rho|\Vert_\infty
\Vert\psi\Vert\sqrt{\left(\psi,\nabla\rho\cdot({\bf p}+e{\bf A})
{\tilde R}({\bf p}+e{\bf A})^2{\tilde R}({\bf p}+e{\bf A})\cdot 
\nabla\rho\psi\right)}\ret
&\le& \sqrt{2m_e}\Vert|\nabla\rho|\Vert_\infty
\Vert\psi\Vert\sqrt{\left(\psi,\nabla\rho\cdot({\bf p}+e{\bf A})
{\tilde R}({\bf p}+e{\bf A})\cdot\nabla\rho\psi\right)}. 
\end{eqnarray}
This implies 
\begin{equation}
\left(\psi,\nabla\rho\cdot({\bf p}+e{\bf A})
{\tilde R}({\bf p}+e{\bf A})\cdot\nabla\rho\psi\right)\le 
2m_e\Vert|\nabla\rho|\Vert_\infty^2
\Vert\psi\Vert^2\le \frac{2m_e}{C_0}
\Vert|\nabla\rho|\Vert_\infty^2
\Vert\varphi\Vert^2. 
\end{equation}
Substituting these bounds into (\ref{normboundY1}), one has 
\begin{equation}
\Vert Y\Vert \le \frac{\sqrt{2}\hbar}{\sqrt{m_eC_0}}\Vert|\nabla\rho|\Vert_\infty.
\label{normYbound}
\end{equation}

Consider the situation that the Hamiltonian $H_\Lambda$ has a spectral gap 
$(E_-,E_+)$, and we take $E\in(E_-,E_+)$. 
We define $d_\pm:={\rm dist}(\sigma(X_E)\cap{\bf R}^\pm,0)$, and 
$u_\pm=P_\pm u$, where $P_\pm$ is the spectral projections for $X_E$ 
onto the subspaces corresponding to the sets $\sigma(X_E)\cap{\bf R}^\pm$, 
respectively. We take $\beta$ satisfying $E_+-E>\hbar^2\beta^2
\Vert|\nabla\rho|\Vert_\infty^2/(2m_e)$. Then one has
\begin{equation}
d_+>\frac{E_+-E-\hbar^2\beta^2\Vert|\nabla\rho|\Vert_\infty^2
/(2m_e)}{E_++{\tilde C}_0}=:\delta_+
\label{d+bound}
\end{equation}
and 
\begin{equation}
d_->\frac{E-E_-}{E_-+{\tilde C}_0}=:\delta_-. 
\label{d-bound}
\end{equation}
 
\begin{lemma}
Suppose that 
the parameter $\beta$ satisfies the condition,  
\begin{equation}
0<\beta<\frac{\sqrt{2m_eC_0}}{\hbar\Vert|\nabla\rho|\Vert_\infty}
\min\left\{\frac{1}{4}
\sqrt{\delta_+\delta_-}, \sqrt{\frac{E_+-E}{C_0}}\right\}.
\label{betacondition}
\end{equation}
Then 
\begin{equation}
\Vert(X_{E+i\varepsilon}+i\beta Y)u\Vert\ge\frac{1}{2}\min\{d_+,d_-\}\Vert u\Vert
\label{XYbound}
\end{equation}
for $\varepsilon\in{\bf R}$. 
\end{lemma}

\begin{proof}{Proof}
{From} the bound (\ref{normYbound}) for $\Vert Y\Vert$ and the assumption on 
$\beta$, one has 
\begin{equation}
\beta\Vert Y\Vert\le \frac{1}{2}\sqrt{\delta_+\delta_-}\le
\frac{1}{2}\sqrt{d_+d_-}.
\end{equation}
Note that $X_{E+i\varepsilon}=X_E-i\varepsilon{\tilde R}$ 
with ${\tilde R}=({\tilde H}_\Lambda+{\tilde C}_0)^{-1}$. 
Using this and the Schwarz inequality, one gets 
\begin{eqnarray}
\Vert u\Vert\Vert(X_{E+i\varepsilon}+i\beta Y)u\Vert&\ge&{\rm Re}
\left((u_+-u_-),(X_E-i\varepsilon{\tilde R}+i\beta Y)(u_++u_-)\right)\ret 
&\ge&d_+\Vert u_+\Vert^2+d_-\Vert u_-\Vert^2-2\beta{\rm Im}(u_+,Yu_-)\ret
&\ge&\frac{1}{2}\left(d_+\Vert u_+\Vert^2+d_-\Vert u_-\Vert^2\right).
\end{eqnarray}
This implies the desired bound. 
\end{proof}

We write 
\begin{equation}
\beta=\frac{\sqrt{2m_e}}{\hbar\Vert|\nabla\rho|\Vert_\infty}
\sqrt{E_+-E}\times\kappa
\end{equation}
in terms of the parameter $\kappa\in(0,1)$. Substituting this into (\ref{d+bound}), 
one has 
\begin{equation}
\delta_+=\frac{E_+-E}{E_++{\tilde C}_0}(1-\kappa^2).
\end{equation}
Further, by substituting these into the bound (\ref{betacondition}), 
the maximum value of $\kappa$ satisfying the bound is 
obtained as
\begin{equation}
\kappa=\sqrt{\frac{C_0(E-E_-)}{C_0(E-E_-)+16(E_++{\tilde C}_0)
(E_-+{\tilde C}_0)}}<1. 
\label{maxkappa}
\end{equation}
As a result, we can take   
\begin{equation}
\beta=\frac{\sqrt{2m_e}}{\hbar\Vert|\nabla\rho|\Vert_\infty}
\sqrt{\frac{C_0(E_+-E)(E-E_-)}{C_0(E-E_-)+16(E_++{\tilde C}_0)
(E_-+{\tilde C}_0)}}.
\label{generalbeta} 
\end{equation}

\begin{theorem}
\label{theorem:Rdecaybound}
Let $E$ be in the spectral gap $(E_-,E_+)\subset{\bf R}$
of the Hamiltonian $H_\Lambda$.   
Let $v,w$ be bounded functions with a compact support. 
Suppose that the boundary of ${\rm supp}\> v$ is smooth. 
Then 
\begin{equation}
\left\Vert v\left(H_\Lambda-E-i\varepsilon\right)^{-1}w\right\Vert\le C_1
\Vert v\Vert_\infty\Vert w\Vert_\infty e^{-\beta r}\quad 
\mbox{for}\ \varepsilon\in{\bf R},
\label{Rdecaybound}
\end{equation} 
where $r={\rm dist}({\rm supp}\ v,{\rm supp}\ w)$, 
$\beta$ is given by (\ref{generalbeta}), and 
\begin{equation}
C_1={\rm Const.}\times C_0^{-1}
\max\left\{\frac{E_++{\tilde C}_0}{(1-\kappa^2)(E_+-E)},
\frac{E_-+{\tilde C}_0}{E-E_-}\right\}
\end{equation}
with $\kappa$ given by (\ref{maxkappa}). 
The two real constants $C_0$ and ${\tilde C}_0$ satisfy  
the conditions ${\tilde H}+{\tilde C}_0>C_0>0$ and (\ref{defC0}). 
\end{theorem}

\begin{proof}{Proof}
Let $\varphi$ be in the domain of the operator ${\tilde H}_\Lambda$. Then 
\begin{eqnarray}
\left\Vert \left({\tilde H}_\Lambda+i\beta J-E-i\varepsilon\right)\varphi\right\Vert
&=&\left\Vert({\tilde H}_\Lambda+{\tilde C}_0)^{1/2}(X_{E+i\varepsilon}+i\beta Y)
({\tilde H}_\Lambda+{\tilde C}_0)^{1/2}\varphi\right\Vert\ret
&\ge&C_0^{1/2}\left\Vert (X_{E+i\varepsilon}+i\beta Y)
({\tilde H}_\Lambda+{\tilde C}_0)^{1/2}\varphi\right\Vert\ret
&\ge&\frac{1}{2}C_0^{1/2}\min\{d_+,d_-\}
\Vert ({\tilde H}_\Lambda+{\tilde C}_0)^{1/2}\varphi\Vert\ret
&\ge&\frac{1}{2}C_0\min\{d_+,d_-\}\Vert \varphi\Vert,
\end{eqnarray}
where we have used the inequality (\ref{XYbound}) and 
${\tilde H}_\Lambda+{\tilde C}_0>C_0$. 
Taking 
\begin{equation}
\varphi=e^{-\beta\rho}\frac{1}{H_\Lambda-E-i\varepsilon}e^{\beta\rho}u, 
\end{equation}
one has  
\begin{equation}
\left\Vert e^{-\beta\rho}\left(H_\Lambda-E-i\varepsilon\right)^{-1}
e^{\beta\rho}u\right\Vert
\le C_1\Vert u\Vert,
\label{Hrhobound}
\end{equation}
where we have used (\ref{expHrho}), (\ref{d+bound}) and (\ref{d-bound}). 
Choosing $\Omega={\rm supp}\ v$ in the definition of the distance function 
$\rho({\bf r})$ and using this bound (\ref{Hrhobound}), 
the desired bound (\ref{Rdecaybound}) is obtained as  
\begin{eqnarray}
{\rm Const.}\left\Vert v\left({H_\Lambda-E-i\varepsilon}\right)^{-1}w\right\Vert 
&\le& \left\Vert ve^{-\beta\rho}\left(H_\Lambda-E-i\varepsilon\right)^{-1}
w\right\Vert\ret
&\le&\Vert v\Vert_\infty \left\Vert e^{-\beta\rho}
\left(H_\Lambda-E-i\varepsilon\right)^{-1}
e^{\beta\rho}e^{-\beta\rho}w\right\Vert\ret
&\le&C_1 \Vert v\Vert_\infty\left\Vert e^{-\beta\rho}w\right\Vert\ret
&\le&{\rm Const.}\times 
C_1\Vert v\Vert_\infty\Vert w\Vert_\infty e^{-\beta r}. 
\end{eqnarray}
\end{proof}

Next consider the case with $H_\Lambda>E$. Then the Schwarz inequality yields 
\begin{eqnarray}
\Vert\varphi\Vert \left\Vert({\tilde H}_\Lambda-E+i\beta J)\varphi\right\Vert
&\ge&{\rm Re}(\varphi,({\tilde H}_\Lambda-E+i\beta J)\varphi)\ret
&=&(\varphi,({\tilde H}_\Lambda-E)\varphi)\ret
&\ge&
\left[E_0-E-\hbar^2\beta^2\Vert|\nabla\rho|\Vert_\infty^2/(2m_e)\right]
\Vert\varphi\Vert
\end{eqnarray}  
for $\varphi$ in the domain of $H_\Lambda$. Here $E_0$ is the ground state energy of 
$H_\Lambda$. Therefore, in the same way as in the proof of 
Theorem~\ref{theorem:Rdecaybound}, one has 
\begin{equation}
\left\Vert v\frac{1}{H_\Lambda-E}w\right\Vert\le 
\frac{{\rm Const.}\times
\Vert v\Vert_\infty\Vert w\Vert_\infty}{E_0-E-\hbar^2\beta^2
\Vert|\nabla\rho|\Vert_\infty^2/(2m_e)}
e^{-\beta r}
\quad \mbox{for}\ 
0<\beta<\frac{\sqrt{2m_e(E_0-E)}}{\hbar\Vert|\nabla\rho|\Vert_\infty},
\label{Rdecayboundbotom} 
\end{equation} 
where $v,w$ and $r$ are the same as in Theorem~\ref{theorem:Rdecaybound}.  

Finally, we consider the decay of $(H_\Lambda-E-iy)^{-1}$ with $y\ne 0$. 
To begin with, we note that 
\begin{eqnarray}
& &\Vert\varphi\Vert\left\Vert\left[
H_\Lambda-(\hbar^2\beta^2/(2m_e))(\nabla\rho)^2-E-iy+i\beta J
\right]\varphi\right\Vert\ret
&\ge&\left|\left(\varphi,\left[
H_\Lambda-(\hbar^2\beta^2)/(2m_e))(\nabla\rho)^2-E-iy+i\beta J
\right]\varphi\right)\right|\ret
&=&\sqrt{\left|\left(\varphi,\left[
H_\Lambda-(\hbar^2\beta^2/(2m_e))(\nabla\rho)^2-E\right]\varphi\right)\right|^2
+\left|\left(\varphi,(y-\beta J)\varphi\right)\right|^2}.
\label{modHybound}
\end{eqnarray}
Let $E'$ be a real number satisfying 
\begin{equation}
E'>E+\frac{\hbar^2\beta^2}{2m_e}\Vert|\nabla\rho|\Vert_\infty^2+C_2
\end{equation}
with a positive constant $C_2$. Then if the vector $\varphi$ satisfies 
$(\varphi,H_\Lambda\varphi)\ge E'\Vert\varphi\Vert^2$, 
the right-hand side of (\ref{modHybound}) is bounded from below by 
$C_2\Vert\varphi\Vert^2$. 

Thus it is sufficient to consider the case that 
$(\varphi,H_\Lambda\varphi)<E'\Vert\varphi\Vert^2$. 
Note that 
\begin{equation}
|(\varphi,(y-\beta J)\varphi)|\ge
\left||y|\Vert\varphi\Vert^2-\beta|(\varphi,J\varphi)|\right|.
\end{equation}
The expectation value in the right-hand side is evaluated as 
\begin{eqnarray}
|(\varphi,J\varphi)|&\le&\frac{\hbar}{m_e}
|(\varphi,\nabla\rho\cdot
({\bf p}+e{\bf A})\varphi)|\ret
&\le&\frac{\hbar}{m_e}\Vert|\nabla\rho|\Vert_\infty
\Vert\varphi\Vert\sqrt{(\varphi,({\bf p}+e{\bf A})^2\varphi)}\ret
&\le&\hbar\sqrt{\frac{2}{m_e}}\Vert|\nabla\rho|\Vert_\infty\Vert\varphi\Vert
\sqrt{(\varphi,(H_\Lambda+\Vert V_\Lambda\Vert_\infty)\varphi)}\ret
&\le&\hbar\Vert|\nabla\rho|\Vert_\infty
\sqrt{2(E'++\Vert V_\Lambda\Vert_\infty)/m_e}\Vert\varphi\Vert^2.
\end{eqnarray}
For a given $y$, we choose a small $\beta$ and a small $C_2$ to satisfy  
\begin{equation}
|y|>\beta\hbar\Vert|\nabla\rho|\Vert_\infty
\sqrt{2(E'++\Vert V_\Lambda\Vert_\infty)/m_e}+C_2. 
\end{equation}
Then these bounds yield 
\begin{equation}
|(\varphi,(y-\beta J)\varphi)|\ge C_2\Vert\varphi\Vert^2.
\end{equation}
This implies that 
the right-hand side of (\ref{modHybound}) is bounded from below by 
the same $C_2\Vert\varphi\Vert^2$. Thus, in both of the cases, one obtains  
\begin{equation}
\left\Vert\left[
H_\Lambda-(\hbar^2\beta^2/(2m_e))(\nabla\rho)^2-E-iy+i\beta J
\right]\varphi\right\Vert\ge C_2\Vert\varphi\Vert. 
\end{equation}
In the same way as the above, this leads to the decay bound,  
\begin{equation}
\left\Vert v(H_\Lambda-E-iy)^{-1}w\right\Vert\le{\rm Const.}\times 
C_2^{-1}\Vert v\Vert_\infty\Vert w\Vert_\infty
e^{-\beta r}.
\end{equation}
Clearly, for a small $|y|$, both of the parameters $\beta$ and $C_2$ must be small. 
But the resolvent always decays exponentially at large distance for any $y\ne 0$. 

%%%%%%%%%%%%%%%%%%%%%%%%%%%%%%%%%%%%%%%%%%%%%%%%%%%%%%%%%%%%%%%%%%%%%%%%%%%
\subsection{The Landau Hamiltonian with ${\bf A}_{\rm P}=0$}
\label{appendix:decayresolventLandau}

Following \cite{CH,Wang}, we obtain the exponential decay bound 
(\ref{iniresolventbound0}) for the resolvent 
in Theorem~\ref{theorem:decayLandauresolvent} below. 

Consider the two-dimensional single electron in the uniform magnetic field 
and with a electrostatic potential $V_\Lambda$.  
The Hamiltonian has the form, $H_\Lambda=H_{\rm L}+V_\Lambda$, 
on the rectangular box $\Lambda$ with the periodic boundary conditions 
as in Section~\ref{model}, where $H_{\rm L}$ is the Landau Hamiltonian (\ref{hamHL}). 
We also impose the condition of the flux quantization, 
$|\Lambda|/(2\pi\ell_B^2)\in{\bf N}$. We assume $V_\Lambda\in C^2(\Lambda)$. 
We denote by $Q_{0,\Lambda}^{(n)}$ the spectral projection onto 
the $n+1$-th Landau level whose energy eigenvalue of $H_{\rm L}$ 
is given by ${\cal E}_n=(n+1/2)\hbar\omega_c$ with $\omega_c=eB/m_e$.   
We introduce one-parameter families of operators as  
\begin{equation}
H_{\rm L}(\beta)=e^{-\beta\rho}H_{\rm L}e^{\beta\rho},\quad 
H_\Lambda(\beta)=H_{\rm L}(\beta)+V_\Lambda
\quad
\mbox{and} 
\quad
Q_{0,\Lambda}^{(n)}(\beta)=e^{-\beta\rho}Q_{0,\Lambda}^{(n)}e^{\beta\rho}
\end{equation}
for $\beta\in{\bf R}$. 
Here the distance function $\rho({\bf r})$ is given in the preceding 
subsection. 
We write $H_{\rm L}(\beta)={\tilde H}_{\rm L}+i\beta J$, 
where the operator $J$ is given by (\ref{expJ}) with ${\bf A}={\bf A}_0$, and 
\begin{equation}
{\tilde H}_{\rm L}=H_{\rm L}-\frac{\hbar^2\beta^2}{2m_e}(\nabla\rho)^2. 
\end{equation}  

\begin{lemma}
\label{lemma:tilderesolventbound}
Let $z$ be a complex number satisfying 
${\rm dist}(\sigma({\tilde H}_{\rm L}),z)\ge\hbar\omega_c/4$, where  
$\sigma({\tilde H}_{\rm L})$ is the spectrum of 
the Hamiltonian ${\tilde H}_{\rm L}$.  
Then there exists $\kappa(z)\in(0,1)$ which depends only on $z$ such that 
the following bound is valid:   
\begin{equation}
\left\Vert[z-H_{\rm L}(\beta)]^{-1}\right\Vert
\le\frac{8}{\hbar\omega_c}
\qquad\mbox{for any real $\beta$ satisfying}\ |\beta|\le\kappa(z)\ell_B^{-1}.
\label{RHbetabound}
\end{equation}
\end{lemma}

\begin{proof}{Proof}
Note that, for a vector $\varphi$ in the domain of the Hamiltonian and for 
a small $\beta$,  
\begin{eqnarray}
\left\Vert({\tilde H}_{\rm L}-z+i\beta J)\varphi\right\Vert&=&
\left\Vert\left[1+i\beta J({\tilde H}_{\rm L}-z)^{-1}\right]
({\tilde H}_{\rm L}-z)\varphi\right\Vert\ret
&\ge&\left[1-|\beta|\left\Vert J({\tilde H}_{\rm L}-z)^{-1}\right\Vert\right]
\left\Vert({\tilde H}_{\rm L}-z)\varphi\right\Vert\ret
&\ge&\left[1-|\beta|\left\Vert J({\tilde H}_{\rm L}-z)^{-1}\right\Vert\right]
{\rm dist}(\sigma({\tilde H}_{\rm L}),z)\Vert\varphi\Vert\ret
&\ge&\frac{1}{4}\hbar\omega_c
\left[1-|\beta|\left\Vert J({\tilde H}_{\rm L}-z)^{-1}\right\Vert\right]
\Vert\varphi\Vert. 
\end{eqnarray}
Therefore it is sufficient to show 
$|\beta|\left\Vert J({\tilde H}_{\rm L}-z)^{-1}\right\Vert<1/2$ for a small $\beta$.
Since 
\begin{equation}
J=\frac{i\hbar^2}{2m_e}\Delta\rho-\frac{\hbar}{m_e}\nabla\rho\cdot
({\bf p}+e{\bf A}_0), 
\label{expressJ}
\end{equation}
one has
\begin{equation}
\left\Vert J({\tilde H}_{\rm L}-z)^{-1}\right\Vert\le
\frac{\hbar^2}{2m_e}\Vert\Delta\rho\Vert_\infty
\frac{1}{{\rm dist}(\sigma({\tilde H}_{\rm L}),z)}
+\frac{2\hbar}{m_e}\Vert|\nabla\rho|\Vert_\infty
\max_s\Vert(p_s+eA_{0,s})({\tilde H}_{\rm L}-z)^{-1}\Vert.
\label{JRnormbound}
\end{equation}
The norm of the operator in the second term is evaluated as 
\begin{eqnarray}
& &\left(\varphi,({\tilde H}_{\rm L}-z^\ast)^{-1}(p_s+eA_{0,s})^2
({\tilde H}_{\rm L}-z)^{-1}\varphi\right)\ret
&\le&2m_e\left(\varphi,({\tilde H}_{\rm L}-z^\ast)^{-1}
\left({\tilde H}_{\rm L}+\frac{\hbar^2\beta^2}{2m_e}\Vert|\nabla\rho|\Vert_\infty^2\right)
({\tilde H}_{\rm L}-z)^{-1}\varphi\right)\ret
&\le&2m_e\left[\frac{1}{{\rm dist}(\sigma({\tilde H}_{\rm L}),z)}
+\frac{|z|+\hbar^2\beta^2\Vert|\nabla\rho|\Vert_\infty^2/(2m_e)}
{{\rm dist}(\sigma({\tilde H}_{\rm L}),z)^2}\right]
\Vert\varphi\Vert^2.
\end{eqnarray}
Combining this, (\ref{JRnormbound}) and the assumption on $z$, 
the desired condition for $\beta$ in (\ref{RHbetabound}) can be obtained. 
\end{proof}

We introduce the integral representation of the projection $Q_{0,\Lambda}^{(n)}$ as 
\begin{equation}
Q_{0,\Lambda}^{(n)}=\frac{1}{2\pi i}\int_\gamma\ dz'\frac{1}{z'-H_{\rm L}},
\end{equation}
where the closed path $\gamma$ encircles the spectrum of the $n+1$-th Landau 
level. Further we choose the path $\gamma$ such that the length of the path 
is bounded as $|\gamma|\le 3\hbar\omega_c$, and 
that the distance from the spectrum of the two 
Hamiltonians $H_{\rm L},{\tilde H}_{\rm L}$ satisfies 
\begin{equation}
{\rm dist}(\gamma,\sigma(H_{\rm L}))\ge\hbar\omega_c/4\quad 
\mbox{and}\quad{\rm dist}(\gamma,\sigma({\tilde H}_{\rm L}))\ge\hbar\omega_c/4 
\end{equation}
for any real $\beta$ satisfying $|\beta|\Vert|\nabla\rho|\Vert_\infty\le\ell_B^{-1}$. 
Then, from the above lemma, there exists $\kappa_n\in(0,1)$ which depends only on 
the index $n$ such that the representation, 
\begin{equation}
Q_{0,\Lambda}^{(n)}(\beta)=
\frac{1}{2\pi i}\int_\gamma\ dz'\frac{1}{z'-H_{\rm L}(\beta)},
\end{equation}
is well defined for any $\beta$ satisfying 
$|\beta|\Vert|\nabla\rho|\Vert_\infty\le\kappa_n\ell_B^{-1}$. 

\begin{lemma}
Assume the above condition $|\beta|\Vert|\nabla\rho|\Vert_\infty\le\kappa_n\ell_B^{-1}$. 
Then the following bound is valid: 
\begin{equation}
\left\Vert\left[Q_{0,\Lambda}^{(n)}(\beta),V_\Lambda\right]\right\Vert
\le C_{0,0}^{(n)}\ell_B
\label{QVcommubound}
\end{equation}
with
\begin{equation}
C_{0,0}^{(n)}=\frac{48}{\pi}\left[\ell_B\Vert\Delta V_\Lambda\Vert_\infty 
+2(4+\sqrt{8n+29})\max_{j=x,y}\Vert \partial_jV_\Lambda\Vert_\infty\right], 
\end{equation}
where we have written $\nabla=(\partial_x,\partial_y)$.  
\end{lemma}

\begin{proof}{Proof}
Note that 
\begin{eqnarray}
\left[Q_{0,\Lambda}^{(n)}(\beta),V_\Lambda\right]&=&\frac{1}{2\pi i}\int_\gamma\ dz'
\left[\frac{1}{z'-H_{\rm L}(\beta)}V_\Lambda-V_\Lambda
\frac{1}{z'-H_{\rm L}(\beta)}\right]\ret
&=&\frac{1}{2\pi i}\int_\gamma\ dz'
\frac{1}{z'-H_{\rm L}(\beta)}[H_{\rm L}(\beta),V_\Lambda]\frac{1}{z'-H_{\rm L}(\beta)}.
\end{eqnarray}
The commutator in the right-hand side is computed as 
\begin{equation}
[H_{\rm L}(\beta),V_\Lambda]=-\frac{\hbar^2}{2m_e}\Delta V_\Lambda
-\frac{i\hbar}{m_e}\nabla V_\Lambda\cdot{\bf \Pi},
\end{equation}
where we have written 
${\bf \Pi}=(\Pi_x,\Pi_y)={\bf p}+e{\bf A}_0-i\hbar\beta\nabla\rho$. 
{From} these observations, one has 
\begin{eqnarray}
\left\Vert\left[Q_{0,\Lambda}^{(n)}(\beta),V_\Lambda\right]\right\Vert
&\le&\frac{3\hbar^2\omega_c}{2\pi m_e}\left[\frac{\hbar}{2}\Vert R(\beta)\Vert^2
\Vert\Delta V_\Lambda\Vert_\infty+\Vert R(\beta)\Vert
\max_{\ell=x,y}\Vert \partial_\ell V_\Lambda\Vert_\infty\sum_{j=x,y}
\Vert\Pi_jR(\beta)\Vert\right]\ret
&\le&\frac{48}{\pi}\left[\ell_B^2\Vert\Delta V_\Lambda\Vert_\infty
+\frac{\hbar}{2m_e}\max_{\ell=x,y}\Vert \partial_\ell V_\Lambda\Vert_\infty
\max_{j=x,y}\Vert\Pi_jR(\beta)\Vert\right],
\label{commuQVnormbound} 
\end{eqnarray}
where we have written $R(\beta)=(z'-H_{\rm L}(\beta))^{-1}$ for short, 
and used the bound (\ref{RHbetabound}).
Therefore it is sufficient to estimate $\Vert\Pi_jR(\beta)\Vert$. 
Note that 
\begin{eqnarray}
\sum_{j=x,y}\left\Vert\Pi_jR(\beta)\varphi\right\Vert^2&=&
\sum_{j=x,y}\left(R(\beta)\varphi,\Pi_j^\ast\Pi_jR(\beta)\varphi\right)\ret
&=&2m_e\left(R(\beta)\varphi,H_{\rm L}(\beta)R(\beta)\varphi\right)
+2i\hbar\beta\sum_{j=x,y}\left(R(\beta)\varphi,(\partial_j\rho)
\Pi_jR(\beta)\varphi\right).\ret
\end{eqnarray}
Further  
\begin{eqnarray}
\max_{j=x,y}\left\Vert\Pi_jR(\beta)\varphi\right\Vert^2
&\le&2m_e\left(\Vert\varphi\Vert\Vert R(\beta)\varphi\Vert+|z'|\Vert R(\beta)\varphi\Vert^2
\right)\ret&+&4\hbar\beta\Vert|\nabla\rho|\Vert_\infty
\Vert R(\beta)\varphi\Vert\max_{j=x,y}\left\Vert\Pi_jR(\beta)\varphi\right\Vert\ret 
&\le&\frac{16m_e^2}{\hbar eB}(8n+13)\Vert\varphi\Vert^2+
\frac{32m_e}{\sqrt{\hbar eB}}\Vert\varphi\Vert
\max_{j=x,y}\left\Vert\Pi_jR(\beta)\varphi\right\Vert,
\end{eqnarray}
where we have used (\ref{RHbetabound}), $|z'|\le{\rm dist}({\cal E}_n,z')+{\cal E}_n$ 
and $\beta\Vert|\nabla\rho|\Vert_\infty<\ell_B^{-1}$. Solving this, one has 
\begin{equation}
\left\Vert\Pi_jR(\beta)\right\Vert\le\frac{4m_e}{\sqrt{\hbar eB}} 
(4+\sqrt{8n+29})\quad\mbox{for}\ j=x,y.
\end{equation}
Substituting this into (\ref{commuQVnormbound}), the bound (\ref{QVcommubound}) 
is obtained. 
\end{proof}

\begin{lemma}
For any given $\epsilon\in(0,1)$, there exists 
$\kappa_{n,\epsilon}\in(0,1)$ such that $\kappa_{n,\epsilon}$ depends only on 
$\epsilon$ and the index $n$ of the Landau level, and that, 
for any real $\beta$ 
satisfying $|\beta|\Vert|\nabla\rho|\Vert_\infty\le\kappa_{n,\epsilon}\ell_B^{-1}$, 
the following bounds are valid: 
\begin{equation}
\left\Vert Q_{0,\Lambda}^{(n)}(\beta)-Q_{0,\Lambda}^{(n)}\right\Vert
\le\epsilon
\label{Qdifboundepsilon}
\end{equation}
and 
\begin{equation}
\left\Vert(H_{\rm L}(\beta)-z)\left[Q_{0,\Lambda}^{(n)}(\beta)-Q_{0,\Lambda}^{(n)}
\right]\right\Vert
\le\epsilon\hbar\omega_c\quad\mbox{for}\ z\ \mbox{satisfying}\ 
{\rm dist}(z,{\cal E}_n)\le\Delta{\cal E}_{\rm max}
\label{HbetaQdiffbound}  
\end{equation}
with a positive constant $\Delta{\cal E}_{\rm max}$. 
\end{lemma}

\begin{proof}{Proof}
Note that  
\begin{eqnarray}
Q_{0,\Lambda}^{(n)}(\beta)-Q_{0,\Lambda}^{(n)}&=&
\frac{1}{2\pi i}\int_\gamma\ dz'\left[\frac{1}{z'-H_{\rm L}(\beta)} 
-\frac{1}{z'-H_{\rm L}}\right]\ret
&=&\frac{1}{2\pi i}\int_\gamma\ dz'\frac{1}{z'-H_{\rm L}(\beta)}
(H_{\rm L}(\beta)-H_{\rm L})\frac{1}{z'-H_{\rm L}}\ret
&=&\frac{1}{2\pi i}\int_\gamma\ dz'\frac{1}{z'-H_{\rm L}(\beta)}
\left[-\frac{\hbar^2\beta^2}{2m_e}(\nabla\rho)^2+i\beta J\right]
\frac{1}{z'-H_{\rm L}} 
\end{eqnarray}
for any $\beta$ satisfying $|\beta|\le\kappa_n\ell_B^{-1}$. 
Therefore the norm of the left-hand side is evaluated as  
\begin{eqnarray}
& &\left\Vert Q_{0,\Lambda}^{(n)}(\beta)-Q_{0,\Lambda}^{(n)}\right\Vert\ret
&\le&
\frac{3\hbar\omega_c}{2\pi}\sup_{z'\in\Gamma}
\left\Vert\frac{1}{z'-H_{\rm L}(\beta)}\right\Vert
\left(\frac{\hbar^2\beta^2}{2m_e}\Vert|\nabla\rho|\Vert_\infty^2
\left\Vert\frac{1}{z'-H_{\rm L}}\right\Vert
+\beta\left\Vert J\frac{1}{z'-H_{\rm L}}\right\Vert\right)\ret
&\le&\frac{12}{\pi}\left(2\ell_B^2\beta^2\Vert|\nabla\rho|\Vert_\infty^2+\epsilon'\right), 
\label{Qdifbound}
\end{eqnarray}
where we have used the bound (\ref{RHbetabound}) and  
$\Vert(z'-H_{\rm L})^{-1}\Vert\le 4/(\hbar\omega_c)$, and we have chosen 
$\beta$ so that, for a given small $\epsilon'$,  
\begin{equation}
|\beta|\left\Vert J\frac{1}{z'-H_{\rm L}}\right\Vert\le \epsilon' 
\end{equation}
which can be proved in the same way as in Lemma~\ref{lemma:tilderesolventbound}. 
The resulting bound (\ref{Qdifbound}) with a small $\beta$ implies the 
desired bound (\ref{Qdifboundepsilon}).
 
In order to obtain the second bound, we note that 
\begin{equation}
(H_{\rm L}(\beta)-z)\left[Q_{0,\Lambda}^{(n)}(\beta)-Q_{0,\Lambda}^{(n)}\right]
=\frac{1}{2\pi i}\int_\gamma\ dz'\left[-1+\frac{z'-z}{z'-H_{\rm L}(\beta)}\right]
(H_{\rm L}(\beta)-H_{\rm L})\frac{1}{z'-H_{\rm L}}.
\end{equation}
In the same way as in the above, the norm is estimated as 
\begin{equation}
\left\Vert(H_{\rm L}(\beta)-z)
\left[Q_{0,\Lambda}^{(n)}(\beta)-Q_{0,\Lambda}^{(n)}\right]\right\Vert
\le\frac{3\hbar\omega_c}{2\pi}
\left(9+8\frac{\Delta{\cal E}_{\rm max}}{\hbar\omega_c}\right)
\left\Vert(H_{\rm L}(\beta)-H_{\rm L})\frac{1}{z'-H_{\rm L}}\right\Vert.
\end{equation}
Here we have used $|z'-z|\le{\rm dist}({\cal E}_n,z')
+{\rm dist}({\cal E}_n,z)\le\hbar\omega_c+\Delta{\cal E}_{\rm max}$. 
The norm of the operator in the right-hand side is already estimated 
in the above.  
\end{proof}

We write $z=E+i\varepsilon$ with $E,\varepsilon\in{\bf R}$. 

\begin{lemma}
\label{lemma:(1-Q)bound} 
Suppose ${\cal E}_{n-1}+\Vert V_\Lambda^+\Vert_\infty+{\hat \delta}_-\hbar\omega_c
\le E\le{\cal E}_{n+1}-\Vert V_\Lambda^-\Vert_\infty-2{\hat\delta}_+\hbar\omega_c$ 
for $n=0,1,2,\ldots$ with some positive constants ${\hat\delta}_\pm$ and with  
${\cal E}_{-1}=-\infty$. 
Let $\varphi$ be a vector in the domain of the Hamiltonian. 
Then the following bound is valid: 
\begin{equation}
\left\Vert(H_\Lambda(\beta)-z)[1-Q_{0,\Lambda}^{(n)}]\varphi\right\Vert
\ge C_{0,1}^{(n)}\hbar\omega_c\left\Vert[1-Q_{0,\Lambda}^{(n)}]\varphi\right\Vert
\label{Hbeta1Qbound}
\end{equation}
for any $\beta$ satisfying 
$|\beta|\Vert|\nabla\rho|\Vert_\infty\le\kappa_n'\ell_B^{-1}$, where 
\begin{equation}
C_{0,1}^{(n)}=\cases{\qquad\displaystyle{\hat\delta}_+, & for $n=0$;\cr
                                               &   \cr    
\displaystyle\min\{{\hat\delta}_+,{\hat\delta}_-\}/2, & for $n=1,2,\ldots$,\cr}
\label{defC01}
\end{equation}
and
\begin{equation}
\kappa_n'=
\cases{\sqrt{2{\hat\delta}_+}, & for $n=0$;\cr 
                                             & \cr 
\displaystyle\min\left\{\frac{\sqrt{{\hat\delta}_+{\hat\delta}_-}}{
\ell_B\Vert\Delta\rho\Vert_\infty/\Vert|\nabla\rho|\Vert_\infty+2\sqrt{2n-1}},
\sqrt{2{\hat\delta}_+}\right\}, 
& for $n=1,2,\ldots$.\cr}
\end{equation}
\end{lemma}

\begin{proof}{Proof}
We write $\psi=[1-Q_{0,\Lambda}^{(n)}]\varphi$, and decompose the vector $\psi$ 
into two parts as 
\begin{equation}
\varphi_+=\sum_{j>n}Q_{0,\Lambda}^{(j)}\varphi\quad\mbox{and}\quad
\varphi_-=\sum_{j<n}Q_{0,\Lambda}^{(j)}\varphi. 
\end{equation}
Then one has 
\begin{eqnarray}
\Vert\psi\Vert\left\Vert(H_\Lambda(\beta)-z)\psi\right\Vert&\ge&
{\rm Re}\left(\varphi_+-\varphi_-,\left(H_\Lambda-E-\frac{\hbar^2\beta^2}{2m_e}
(\nabla\rho)^2-i\varepsilon+i\beta J\right)\varphi_++\varphi_-\right)\ret
&\ge&\hbar\omega_c{\hat\delta}_+\Vert\varphi_+\Vert^2
+\hbar\omega_c{\hat\delta}_-\Vert\varphi_-\Vert^2
-2\beta{\rm Im}(\varphi_+,J\varphi_-),
\label{HLbetapsibound} 
\end{eqnarray}
where we have used 
\begin{equation}
{{\cal E}_{n+1}-\Vert V_\Lambda^-\Vert_\infty-E
-\hbar^2\beta^2\Vert|\nabla\rho|\Vert_\infty^2/(2m_e)}\ge{\hat\delta}_+{\hbar\omega_c}
\end{equation}
and
\begin{equation}
{E-{\cal E}_{n-1}-\Vert V_\Lambda^+\Vert_\infty}\ge{\hat\delta}_-{\hbar\omega_c}
\end{equation}
which are easily derived from the assumptions. 
Clearly, in the case with $n=0$, one has $\varphi_-=0$, and so the statement holds. 
For the rest of the cases, we use the following bound: 
\begin{eqnarray}
|(\varphi_+,J\varphi_-)|&\le&\frac{\hbar^2}{2m_e}\Vert\Delta\rho\Vert_\infty
\Vert\varphi_+\Vert\Vert\varphi_-\Vert+\frac{\hbar}{m_e}
|(\varphi_+,\nabla\rho\cdot({\bf p}+e{\bf A}_0)\varphi_-)|\ret
&\le&\frac{\hbar^2}{2m_e}\Vert\Delta\rho\Vert_\infty\Vert\varphi_+\Vert\Vert\varphi_-\Vert
+\frac{\hbar}{m_e}\sqrt{(\varphi_+,|\nabla\rho|^2\varphi_+)
(\varphi_-,({\bf p}+e{\bf A}_0)^2\varphi_-)}\ret
&\le&\left(\frac{\hbar^2}{2m_e}\Vert\Delta\rho\Vert_\infty
+\hbar\sqrt{\frac{2{\cal E}_{n-1}}{m_e}}\Vert|\nabla\rho|\Vert_\infty\right)
\Vert\varphi_+\Vert\Vert\varphi_-\Vert,
\end{eqnarray}
where we have used (\ref{expressJ}) and the Schwarz inequality. 
Combining this, (\ref{HLbetapsibound}) and the assumption on $\beta$, the desired 
bound is obtained.
\end{proof}

Let us estimate $\Vert(H_\Lambda(\beta)-z)\varphi\Vert$ for a vector $\varphi$ 
in the domain of the Hamiltonian and for $z\in{\bf C}$. We take $\beta$ satisfying  
$|\beta|\Vert|\nabla\rho|\Vert_\infty
\le\ell_B^{-1}\min\{\kappa_{n,\epsilon},\kappa_n'\}$ for a given 
$\epsilon$. Note that 
\begin{eqnarray}
\left\Vert Q_{0,\Lambda}^{(n)}(H_\Lambda(\beta)-z)\varphi\right\Vert
&\ge&\left\Vert e^{-\beta\rho}(H_\Lambda-z)
Q_{0,\Lambda}^{(n)}e^{\beta\rho}\varphi\right\Vert\ret
&-&\left\Vert Q_{0,\Lambda}^{(n)}(H_\Lambda(\beta)-z)\varphi-
e^{-\beta\rho}(H_\Lambda-z)Q_{0,\Lambda}^{(n)}e^{\beta\rho}\varphi\right\Vert. 
\label{QHbetavarphibound}
\end{eqnarray}
The first term in the right-hand side can be evaluated as 
\begin{eqnarray}
\left\Vert e^{-\beta\rho}(H_\Lambda-z)
Q_{0,\Lambda}^{(n)}e^{\beta\rho}\varphi\right\Vert
&=& \left\Vert({\cal E}_n+V_\Lambda-z)e^{-\beta\rho}
Q_{0,\Lambda}^{(n)}e^{\beta\rho}\varphi\right\Vert\ret
&\ge&\Delta E\left\Vert Q_{0,\Lambda}^{(n)}(\beta)
\varphi\right\Vert\ret
&\ge&\Delta E\left\Vert Q_{0,\Lambda}^{(n)}\varphi\right\Vert
-\Delta E\left\Vert\left[Q_{0,\Lambda}^{(n)}(\beta)-Q_{0,\Lambda}^{(n)}\right]
\varphi\right\Vert\ret
&\ge&\Delta E\left\Vert Q_{0,\Lambda}^{(n)}\varphi\right\Vert-\epsilon
\Delta E\Vert\varphi\Vert, 
\end{eqnarray}
where 
\begin{equation}
\Delta E=\inf |{\cal E}_n+V_\Lambda-{\rm Re}\, z|
\label{defDE}
\end{equation}
and we have used the inequality (\ref{Qdifboundepsilon}). 
The second term in the right-hand side of (\ref{QHbetavarphibound}) 
can be evaluated as 
\begin{eqnarray}
& &\left\Vert Q_{0,\Lambda}^{(n)}(H_\Lambda(\beta)-z)\varphi-
e^{-\beta\rho}(H_\Lambda-z)Q_{0,\Lambda}^{(n)}e^{\beta\rho}\varphi\right\Vert\ret
&=&\left\Vert Q_{0,\Lambda}^{(n)}(H_\Lambda(\beta)-z)\varphi-
(H_\Lambda(\beta)-z)Q_{0,\Lambda}^{(n)}(\beta)\varphi\right\Vert\ret
&\le&\left\Vert\left[Q_{0,\Lambda}^{(n)}-Q_{0,\Lambda}^{(n)}(\beta)\right]
(H_\Lambda(\beta)-z)\varphi\right\Vert
+\left\Vert\left[[Q_{0,\Lambda}^{(n)}(\beta),H_\Lambda(\beta)\right]
\varphi\right\Vert\ret
&\le&\left\Vert\left[Q_{0,\Lambda}^{(n)}-Q_{0,\Lambda}^{(n)}(\beta)\right]
(H_\Lambda(\beta)-z)\varphi\right\Vert
+\left\Vert\left[[Q_{0,\Lambda}^{(n)}(\beta),V_\Lambda\right]
\varphi\right\Vert\ret
&\le&\epsilon\Vert(H_\Lambda(\beta)-z)\varphi\Vert+C_{0,0}^{(n)}\ell_B\Vert\varphi\Vert, 
\end{eqnarray}
we have used the inequalities (\ref{QVcommubound}) and (\ref{Qdifboundepsilon}).
Substituting these bounds into (\ref{QHbetavarphibound}), one has 
\begin{equation}
(1+\epsilon)\Vert(H_\Lambda(\beta)-z)\varphi\Vert\ge
\Delta E\left\Vert Q_{0,\Lambda}^{(n)}\varphi\right\Vert
-\left(\epsilon\Delta E+C_{0,0}^{(n)}\ell_B\right)\Vert\varphi\Vert. 
\label{1QHbetabound}
\end{equation}

Note that 
\begin{eqnarray}
\left\Vert(1-Q_{0,\Lambda}^{(n)})(H_\Lambda(\beta)-z)\varphi\right\Vert
&\ge&\left\Vert[1-Q_{0,\Lambda}^{(n)}(\beta)](H_\Lambda(\beta)-z)\varphi\right\Vert\ret
&-&\left\Vert[Q_{0,\Lambda}^{(n)}(\beta)-Q_{0,\Lambda}^{(n)})]
(H_\Lambda(\beta)-z)\varphi\right\Vert\ret
&\ge&\left\Vert(H_\Lambda(\beta)-z)[1-Q_{0,\Lambda}^{(n)}(\beta)]\varphi\right\Vert
-\left\Vert[Q_{0,\Lambda}^{(n)}(\beta),H_\Lambda(\beta)]\varphi\right\Vert\ret
&-&\left\Vert[Q_{0,\Lambda}^{(n)}(\beta)-Q_{0,\Lambda}^{(n)})]
(H_\Lambda(\beta)-z)\varphi\right\Vert\ret
&\ge&\left\Vert(H_\Lambda(\beta)-z)[1-Q_{0,\Lambda}^{(n)}]\varphi\right\Vert\ret
&-&\left\Vert(H_\Lambda(\beta)-z)[Q_{0,\Lambda}^{(n)}(\beta)-Q_{0,\Lambda}^{(n)}]
\varphi\right\Vert
-\left\Vert[Q_{0,\Lambda}^{(n)}(\beta),V_\Lambda]\varphi\right\Vert\ret
&-&\left\Vert[Q_{0,\Lambda}^{(n)}(\beta)-Q_{0,\Lambda}^{(n)})]
(H_\Lambda(\beta)-z)\varphi\right\Vert.
\end{eqnarray}
Using the inequalities (\ref{QVcommubound}), (\ref{Qdifboundepsilon}), 
(\ref{HbetaQdiffbound}) and (\ref{Hbeta1Qbound}) for this right-hand side, one has 
\begin{equation}
(1+\epsilon)\Vert(H_\Lambda(\beta)-z)\varphi\Vert\ge C_{0,1}^{(n)}\hbar\omega_c
\left\Vert[1-Q_{0,\Lambda}^{(n)}]\varphi\right\Vert
-\left(\epsilon\hbar\omega_c+\epsilon\Vert V_\Lambda\Vert_\infty+C_{0,0}^{(n)}\ell_B\right)
\Vert\varphi\Vert
\end{equation}
Combining this with the above inequality (\ref{1QHbetabound}), one has 
\begin{eqnarray}
& &(1+\epsilon)\left(\frac{1}{\Delta E}+\frac{1}{C_{0,1}^{(n)}\hbar\omega_c}\right)
\Vert(H_\Lambda(\beta)-z)\varphi\Vert\ret
&\ge&\left\{1-\epsilon\left[1+\frac{1}{C_{0,1}^{(n)}}
\left(1+\frac{\Vert V_\Lambda\Vert_\infty}{\hbar\omega_c}\right)\right]
-C_{0,0}^{(n)}\ell_B\left(\frac{1}{\Delta E}+\frac{1}{C_{0,1}^{(n)}\hbar\omega_c}
\right)\right\}\Vert\varphi\Vert.
\label{(Hbeta-z)bound}
\end{eqnarray}

Assume that the energy $E$ satisfies the condition in Lemma~\ref{lemma:(1-Q)bound} 
with the positive constants ${\hat \delta}_\pm$ which are independent of 
the strength $B$ of the uniform magnetic field. 
Under this assumption, $C_{0,1}^{(n)}$ and $\kappa'_n$ in Lemma~\ref{lemma:(1-Q)bound} 
can be chosen to be independent of $B$ except for a small $B$. 
Further assume that  
$\Delta E$ satisfies $\Delta E\ge\Delta{\cal E}>0$ with some constant 
$\Delta{\cal E}$ which is independent of $B$. 
Then there exists $B_{0,1}^{(n)}$ such that, for any $B>B_{0,1}^{(n)}$, 
\begin{equation}
C_{0,0}^{(n)}\ell_B\left[\left(\Delta E\right)^{-1}
+\left(C_{0,1}^{(n)}\hbar\omega_c\right)^{-1}
\right]\le{1}/{3}. 
\end{equation}
Moreover we can choose $\epsilon$ satisfying  
\begin{equation}
\epsilon\left\{1+\left(C_{0,1}^{(n)}\right)^{-1}
\left[1+{\Vert V_\Lambda\Vert_\infty}/(\hbar\omega_c)\right]\right\}\le{1}/{3}. 
\end{equation}
Substituting these into the above bound (\ref{(Hbeta-z)bound}), one has  
\begin{equation}
{C_{0,2}^{(n)}}(\Delta {\cal E})^{-1}\Vert(H_\Lambda(\beta)-z)\varphi\Vert
\ge\Vert\varphi\Vert\quad\mbox{for any}\ B\ge B_{0,1}^{(n)},
\end{equation}
where the positive constant $C_{0,2}^{(n)}$ depends only on the index $n$. 
Therefore a similar decay estimate for the resolvent is obtained 
in the same way as in Theorem~\ref{theorem:Rdecaybound}. We summarize the result as 

\begin{theorem}
\label{theorem:decayLandauresolvent}
Let $v,w$ be bounded functions with a compact support, 
and suppose that the boundary of the region ${\rm supp}\ v$ is smooth. 
Write $z=E+i\varepsilon$ with $E,\varepsilon\in{\bf R}$. 
Assume that the energy $E$ satisfies the condition in Lemma~\ref{lemma:(1-Q)bound} 
with the positive constants ${\hat \delta}_\pm$ which are independent of 
the strength $B$ of the uniform magnetic field. 
Further assume that $\Delta E$ of (\ref{defDE}) satisfies 
$\Delta E\ge\Delta{\cal E}>0$ with 
some constant $\Delta{\cal E}$ which is independent of $B$.
Then there exist $B_{0,1}^{(n)}$ and ${\tilde \kappa}_n$ which depend only on 
the index $n$ of the Landau level such that 
\begin{equation}
\left\Vert v\left(H_\Lambda-E-i\varepsilon\right)^{-1}w\right\Vert
\le\frac{C_{0,3}^{(n)}}{\Delta {\cal E}}
\Vert v\Vert_\infty\Vert w\Vert_\infty
\exp[-{\tilde\kappa}_n\ell_B^{-1}r]\quad\mbox{for any}\ B\ge B_{0,1}^{(n)}.
\label{iniresolventbound0} 
\end{equation}
Here $r={\rm dist}({\rm supp}\ v,{\rm supp}\ w)$ and 
the positive constant $C_{0,3}^{(n)}$ depends only on the index $n$.
\end{theorem}

%%%%%%%%%%%%%%%%%%%%%%%%%%%%%%%%%%%%%%%%%%%%%%%%%%%%%%%%%%%%%
\Section{Proof of Lemma~\ref{lemma:pRbounds}}
\label{Proof5.1}

The first inequality (\ref{chipRbound}) can be obtained as 
\begin{eqnarray}
\left\Vert\alpha_i(p_i+eA_i)R\psi\right\Vert^2
&=&\left(\psi,R^\ast(p_i+eA_i)|\alpha_i|^2(p_i+eA_i)R\psi\right)\ret
&\le&\Vert\alpha_i\Vert_\infty^2\left(\psi,R^\ast(p_i+eA_i)^2R\psi\right)\ret
&\le&2m_e
\Vert\alpha_i\Vert_\infty^2
\left\{\Vert R\Vert +[|E|+\Vert(V_0^-+V_\omega^-)\Vert_\infty]\Vert R\Vert^2\right\}
\Vert\psi\Vert^2\ret
\label{alphaipiRpsibound}
\end{eqnarray}
for $i=x,y$ and for any vector $\psi$. Here we have used 
\begin{eqnarray}
\sum_{i=x,y}R^\ast(p_i+eA_i)^2R&\le&2m_eR^\ast
\left[H_\omega+\Vert(V_0^-+V_\omega^-)\Vert_\infty\right]R\ret
&\le&m_e
\left\{(R+R^\ast)+2[|E|+\Vert(V_0^-+V_\omega^-)\Vert_\infty]R^\ast R\right\}.
\label{RpRbound} 
\end{eqnarray}

In order to get the second inequality, we first note that 
\begin{eqnarray}
& &\left\Vert(p_i+eA_i)R({\bf p}+e{\bf A})\cdot\mbox{\boldmath $\alpha$}
\varphi\right\Vert^2\ret
&=&\left(\varphi,\mbox{\boldmath $\alpha$}\cdot({\bf p}+e{\bf A})R^\ast(p_i+eA_i)^2
R({\bf p}+e{\bf A})\cdot\mbox{\boldmath $\alpha$}\varphi\right)\ret
&\le&m_e\left(\varphi,\mbox{\boldmath $\alpha$}\cdot({\bf p}+e{\bf A})(R+R^\ast)
({\bf p}+e{\bf A})\cdot\mbox{\boldmath $\alpha$}\varphi\right)\ret
&+&2m_e[|E|+\Vert(V_0^-+V_\omega^-)\Vert_\infty]
\left(\varphi,\mbox{\boldmath $\alpha$}\cdot({\bf p}+e{\bf A})R^\ast R
({\bf p}+e{\bf A})\cdot\mbox{\boldmath $\alpha$}\varphi\right).
\label{pRpchibound0}
\end{eqnarray}
Using the Schwarz inequality, one has
\begin{eqnarray}
& &\left(\varphi,\mbox{\boldmath $\alpha$}\cdot({\bf p}+e{\bf A})R^\ast R
({\bf p}+e{\bf A})\cdot\mbox{\boldmath $\alpha$}\varphi\right)\ret
&\le&\Vert|\mbox{\boldmath $\alpha$}|\Vert_\infty\Vert\varphi\Vert
\sqrt{\left(\varphi,\mbox{\boldmath $\alpha$}\cdot({\bf p}+e{\bf A})
R^\ast R({\bf p}+e{\bf A})^2
R^\ast R({\bf p}+e{\bf A})\cdot\mbox{\boldmath $\alpha$}\varphi\right)}.   
\end{eqnarray}
Combining this with the inequality (\ref{RpRbound}), one obtains 
the bound (\ref{pRRpbound}). Similarly,  
\begin{eqnarray}
& &\left|\left(\varphi,\mbox{\boldmath $\alpha$}\cdot({\bf p}+e{\bf A})R
({\bf p}+e{\bf A})\cdot\mbox{\boldmath $\alpha$}\varphi\right)\right|\ret
&\le&\Vert|\mbox{\boldmath $\alpha$}|\Vert_\infty\Vert\varphi\Vert
\sqrt{\left(\varphi,\mbox{\boldmath $\alpha$}\cdot({\bf p}+e{\bf A})
R^\ast({\bf p}+e{\bf A})^2
R({\bf p}+e{\bf A})\cdot\mbox{\boldmath $\alpha$}\varphi\right)}.   
\end{eqnarray}
Combining this with the inequality (\ref{RpRbound}), one obtains 
\begin{eqnarray}
& &\left|\left(\varphi,\mbox{\boldmath $\alpha$}\cdot({\bf p}+e{\bf A})R
({\bf p}+e{\bf A})\cdot\mbox{\boldmath $\alpha$}\varphi\right)\right|^2\ret
&\le&2m_e\Vert|\mbox{\boldmath $\alpha$}|\Vert_\infty^2\Vert\varphi\Vert^2
\left\{\left|\left(\varphi,\mbox{\boldmath $\alpha$}\cdot({\bf p}+e{\bf A})R
({\bf p}+e{\bf A})\cdot\mbox{\boldmath $\alpha$}\varphi\right)\right|\right.\ret
&+&\left.[|E|+\Vert(V_0^-+V_\omega^-)\Vert_\infty]
\left(\varphi,\mbox{\boldmath $\alpha$}\cdot({\bf p}+e{\bf A})R^\ast R
({\bf p}+e{\bf A})\cdot\mbox{\boldmath $\alpha$}\varphi\right)\right\}\ret
&\le&2m_e\Vert|\mbox{\boldmath $\alpha$}|\Vert_\infty^2\Vert\varphi\Vert^2
\left|\left(\varphi,\mbox{\boldmath $\alpha$}\cdot({\bf p}+e{\bf A})R
({\bf p}+e{\bf A})\cdot\mbox{\boldmath $\alpha$}\varphi\right)\right|\ret
&+&4m_e^2\Vert|\mbox{\boldmath $\alpha$}|\Vert_\infty^4
f_{E,R}(1+f_{E,R})\Vert\varphi\Vert^4, 
\end{eqnarray} 
where we have used the bound (\ref{pRRpbound}). 
Solving this quadratic inequality, one has 
\begin{equation}
\left|\left(\varphi,\mbox{\boldmath $\alpha$}\cdot({\bf p}+e{\bf A})R
({\bf p}+e{\bf A})\cdot\mbox{\boldmath $\alpha$}\varphi\right)\right|
\le 2m_e\Vert|\mbox{\boldmath $\alpha$}|\Vert_\infty^2(1+f_{E,R})\Vert\varphi\Vert^2.  
\end{equation}
Substituting this and (\ref{pRRpbound}) into the right-hand side of (\ref{pRpchibound0}),  
the desired bound (\ref{pRpchibound}) is obtained. 

%%%%%%%%%%%%%%%%%%%%%%%%%%%%%%%%%%%%%%%%%%%%%%%%%%%%%%%%%%%%%%%%
\Section{Proofs of Lemmas~\ref{lemma:relationell'ell}\ and \ref{lemma:lksequence}}
\label{MSAProofs}

For the purpose of this appendix, we prepare the following three lemmas: 

\begin{lemma}
\label{expdecayR}
(i) Let $\ell,\ell'$ be odd integers larger than $1$ such that 
$\ell'$ is a multiple of $\ell$. 
Let $A^{\rm good}$ be the event that no two disjoint $\gamma$-bad boxes 
of size $3\ell$ with center in $\Gamma_\ell\cap\Lambda_{5\ell'}({\bf z})$ 
exist. Assume ${\rm Prob}[\Lambda_{3\ell}(\cdots)\mbox{ is $\gamma$-good}]\ge 
1-\eta$ with a small $\eta>0$. 
Then ${\rm Prob}(A^{\rm good})\ge 1-(5\ell'/\ell)^4\eta^2$. 

\noindent
(ii) Assume that the event $A^{\rm good}$ occurs.   
Let ${\bf u}, {\bf v}\in\Gamma_\ell$ such that  
$\Lambda_\ell({\bf u})\subset\Lambda_{\ell'}({\bf z})$ and that 
$\Lambda_\ell({\bf v})\cap(\Lambda_{3\ell'}({\bf z})\backslash
\Lambda_{3\ell'}^\delta({\bf z}))\ne\emptyset$. 
Then 
\begin{equation}
\left\Vert\chi_\ell({\bf u})R_{5\ell',{\bf z}}(E+i\varepsilon)
\chi_\ell({\bf v})\right\Vert\le
\left(8e^{-\gamma\ell}\right)^{\ell'/\ell-4}
\left\Vert R_{5\ell',{\bf z}}(E+i\varepsilon)\right\Vert
\label{uR5vbound}
\end{equation}
\end{lemma}

\begin{proof}{Proof}
The statement (i) follows from elementary combinatorics. 

\noindent (ii) Using the geometric resolvent equation, 
\begin{equation}
\chi_{3\ell}^\delta({\bf u})R_{5\ell',{\bf z}}
=R_{3\ell,{\bf u}}\chi_{3\ell}^\delta({\bf u})
+R_{3\ell,{\bf u}}{\tilde W}_{3\ell}^\delta({\bf u})
R_{5\ell',{\bf z}},
\end{equation}
one has 
\begin{eqnarray}
\chi_\ell({\bf u})R_{5\ell',{\bf z}}
\chi_\ell({\bf v})&=&\chi_\ell({\bf u})\chi_{3\ell}^\delta({\bf u})
R_{5\ell',{\bf z}}
\chi_\ell({\bf v})\ret
&=&\chi_\ell({\bf u})R_{3\ell,{\bf u}}{\tilde W}_{3\ell}^\delta({\bf u})
R_{5\ell',{\bf z}}\chi_\ell({\bf v})\ret
&=&\chi_\ell({\bf u})R_{3\ell,{\bf u}}{\tilde W}_{3\ell}^\delta({\bf u})
\sum_{{\tilde{\bf u}}\in\Gamma_\ell\cap
(\Lambda_{3\ell}({\bf u})\backslash\Lambda_{\ell}({\bf u}))}
\chi_\ell({\tilde {\bf u}})
R_{5\ell',{\bf z}}\chi_\ell({\bf v}),
\end{eqnarray}
where we have written $R_{\ell,{\bf z}}$ for $R_{\ell,{\bf z}}(E+i\varepsilon)$ 
for short. We can choose ${\bf u}_1$ from the set of ${\tilde{\bf u}}$ so that 
$\left\Vert\chi_\ell({\tilde {\bf u}})R_{5\ell',{\bf z}}\chi_\ell({\bf v})
\right\Vert$ becomes maximal. Thus one has 
\begin{equation}
\left\Vert\chi_\ell({\bf u})R_{5\ell',{\bf z}}
\chi_\ell({\bf v})\right\Vert\le 8e^{-\gamma\ell}
\left\Vert\chi_\ell({\bf u}_1)R_{5\ell',{\bf z}}\chi_\ell({\bf v})\right\Vert
\end{equation}
when $\Lambda_{3\ell}({\bf u})$ is $\gamma$-good. Since 
the norm of the operator in the right-hand side can be estimated 
in the same way, one can repeat this procedure and construct 
the points, ${\bf u}_1,{\bf u}_2,\ldots,{\bf u}_k\in\Gamma_\ell$, 
as long as $\Lambda_{3\ell}({\bf u}_{k-1})$ is $\gamma$-good and 
does not hit $\Lambda_\ell({\bf v})$ or $\partial\Lambda_{5\ell'}({\bf z})$.  

The same type of estimate can be applied to ${\bf v}$ as a starting point as 
follows: Using the adjoint of the geometric resolvent equation, 
\begin{equation}
R_{5\ell',{\bf z}}\chi_{3\ell}^\delta({\bf v})=\chi_{3\ell}^\delta({\bf v})
R_{3\ell,{\bf v}}+R_{5\ell',{\bf z}}\left({\tilde W}_{3\ell}^\delta({\bf v})\right)^\ast
R_{3\ell,{\bf v}},
\label{AGRE}
\end{equation}
one has 
\begin{eqnarray}
\chi_\ell({\bf u}_k)R_{5\ell',{\bf z}}\chi_\ell({\bf v})
&=&\chi_\ell({\bf u}_k)R_{5\ell',{\bf z}}\chi_{3\ell}^\delta({\bf v})
\chi_\ell({\bf v})\ret
&=&\chi_\ell({\bf u}_k)R_{5\ell',{\bf z}}\left({\tilde W}_{3\ell}^\delta({\bf v})\right)^\ast
R_{3\ell,{\bf v}}
\chi_\ell({\bf v})\ret
&=&\chi_\ell({\bf u}_k)R_{5\ell',{\bf z}}
\sum_{{\tilde{\bf v}}\in\Gamma_\ell\cap
(\Lambda_{3\ell}({\bf v})\backslash\Lambda_{\ell}({\bf v}))}
\chi_\ell({\tilde {\bf v}})
\left({\tilde W}_{3\ell}^\delta({\bf v})\right)^\ast
R_{3\ell,{\bf v}}
\chi_\ell({\bf v}).\ret
\end{eqnarray}
Thus  
\begin{equation}
\left\Vert\chi_\ell({\bf u}_k)R_{5\ell',{\bf z}}
\chi_\ell({\bf v})\right\Vert\le 8e^{-\gamma\ell}
\left\Vert\chi_\ell({\bf u}_k)R_{5\ell',{\bf z}}\chi_\ell({\bf v}_1)\right\Vert
\end{equation}
when $\Lambda_{3\ell}({\bf v})$ is $\gamma$-good. In the same way, 
the procedure yields the points, ${\bf v}_1,{\bf v}_2,\ldots,{\bf v}_j$, and 
one obtains the bound,
\begin{equation}
\left\Vert\chi_\ell({\bf u})R_{5\ell',{\bf z}}\chi_\ell({\bf v})\right\Vert
\le \left(8e^{-\gamma\ell}\right)^{k+j}
\left\Vert\chi_\ell({\bf u}_k)R_{5\ell',{\bf z}}\chi_\ell({\bf v}_j)\right\Vert
\le \left(8e^{-\gamma\ell}\right)^{k+j}
\left\Vert R_{5\ell',{\bf z}}\right\Vert.
\end{equation}
This process moves in steps of $\ell$. 
The assumption that $A^{\rm good}$ occurs implies that there may be only one 
cluster of overlapping $\gamma$-bad boxes. The diameter of 
such a bad cluster is at most $5\ell$. From these observations, 
one has that $k+j\ge|{\bf u}-{\bf v}|/\ell-4$ iterations can be 
performed before the process stops on the both sides.   
Since $|{\bf u}-{\bf v}|\ge\ell'$ from the assumption, 
the desired bound (\ref{uR5vbound}) is obtained.    
When the process hits the boundary of $\Lambda_{5\ell'}({\bf z})$ 
without hitting a $\gamma$-bad box, we have $\ell k\ge 2\ell'$ or $\ell j\ge\ell'$. 
Therefore the bound (\ref{uR5vbound}) remains valid.  
\end{proof}

\begin{lemma}
Let $\ell,\ell'$ be odd integers larger than $1$ such that $\ell'$ is a multiple 
of $\ell$ and satisfies $\ell'>4\ell$.
Assume that the event $A^{\rm good}$ given in the preceding lemma occurs.
Then
\begin{eqnarray}
& &\left\Vert\chi_{\ell'}({\bf z})R_{3\ell',{\bf z}}
\left({\tilde W}_{3\ell'}^\delta({\bf z})\right)^\ast\right\Vert\ret&\le&
6\left(\frac{\ell'}{\ell}\right)^3\left(8e^{-\gamma\ell}\right)^{\ell'/\ell-4}
\left[f_4(|E|,\left\Vert R_{5\ell',{\bf z}}\right\Vert)
+f_5(|E|,\left\Vert R_{3\ell',{\bf z}}\right\Vert)\right]
\left\Vert R_{5\ell',{\bf z}}\right\Vert,
\label{chiR3Wvarphibound}
\end{eqnarray} 
where the functions, $f_4$ and $f_5$, are given by 
\begin{eqnarray}
f_4(|E|,\Vert R\Vert)&=&2C_{\delta,\omega}
+\frac{16\hbar^2}{m_e}
\max_{m,n=x,y}\{\left\Vert\partial_m\phi_{n,3\ell'}({\bf z})\right\Vert_\infty\}
(1+f_{E,R})\ret
&+&2\sqrt{2}\left(\frac{\hbar}{\sqrt{m_e}}\right)^3
\max_{n=x,y}\{\left\Vert\Delta\phi_{n,3\ell'}({\bf z})\right\Vert_\infty\}
(1+f_{E,R})^{1/2}\Vert R\Vert^{1/2}
\end{eqnarray}
and
\begin{eqnarray}
& &\hspace{-1cm}f_5(|E|,\Vert R\Vert)\ret
&=&C_{\delta,\omega}^2
\left\Vert R\right\Vert
+\frac{2\sqrt{2}\hbar}{\sqrt{m_e}}C_{\delta,\omega}
\left[\left\Vert|\nabla\chi_{3\ell'}^\delta({\bf z})|\right\Vert_\infty
+2\max_{m=x,y}\{\Vert\partial_m\chi_{3\ell'}^\delta({\bf z})
\Vert_\infty\}\right](1+f_{E,R})^{1/2}\Vert R\Vert^{1/2}\ret
&+&\frac{4\hbar^2}{m_e}
\max_{m=x,y}\{\Vert\partial_m\chi_{3\ell'}^\delta({\bf z})
\Vert_\infty\}\left\Vert|\nabla\chi_{3\ell'}^\delta({\bf z})|\right\Vert_\infty
(1+f_{E,R}).  
\end{eqnarray}
Here 
\begin{equation}
C_{\delta,\omega}=\frac{\hbar^2}{2m_e}\left\Vert\Delta\chi_{3\ell'}^\delta({\bf z})
\right\Vert_\infty+\Vert V_\omega\Vert_\infty,\qquad
\phi_{i,3\ell'}({\bf z})=\chi_{3\ell'}^\delta({\bf z})
\partial_i\chi_{3\ell'}^\delta({\bf z}),
\end{equation}
and the function $f_{E,R}$ is given by (\ref{deffER}). 
\end{lemma}

\begin{proof}{Proof}
Using the adjoint of the geometric resolvent equation (\ref{AGRE}), one has  
\begin{eqnarray}
& &\chi_{\ell'}({\bf z})R_{5\ell',{\bf z}}\chi_{3\ell'}^\delta({\bf z})
\left({\tilde W}_{3\ell'}^\delta({\bf z})\right)^\ast\ret
&=&\chi_{\ell'}({\bf z})R_{3\ell',{\bf z}}
\left({\tilde W}_{3\ell'}^\delta({\bf z})\right)^\ast
+\chi_{\ell'}({\bf z})R_{5\ell',{\bf z}}
\left({\tilde W}_{3\ell'}^\delta({\bf z})\right)^\ast
R_{3\ell',{\bf z}}\left({\tilde W}_{3\ell'}^\delta({\bf z})\right)^\ast.
\end{eqnarray}
Therefore
\begin{eqnarray}
& &\left\Vert\chi_{\ell'}({\bf z})R_{3\ell',{\bf z}}
\left({\tilde W}_{3\ell'}^\delta({\bf z})\right)^\ast\varphi\right\Vert\ret
&\le&\left\Vert\chi_{\ell'}({\bf z})R_{5\ell',{\bf z}}\chi_{3\ell'}^\delta({\bf z})
\left({\tilde W}_{3\ell'}^\delta({\bf z})\right)^\ast\varphi\right\Vert
+\left\Vert\chi_{\ell'}({\bf z})R_{5\ell',{\bf z}}
\left({\tilde W}_{3\ell'}^\delta({\bf z})\right)^\ast
R_{3\ell',{\bf z}}\left({\tilde W}_{3\ell'}^\delta({\bf z})\right)^\ast
\varphi\right\Vert\ret
\label{chiR5Wvarphibound}
\end{eqnarray}
for any vector $\varphi$ in the domain of the operator ${\bf p}+e{\bf A}$. 

Let us estimate the first term in the right-hand side. Using the expression 
(\ref{defW}) of $W(\cdots)$, one has 
\begin{eqnarray}
& &\left\Vert\chi_{\ell'}({\bf z})R_{5\ell',{\bf z}}\chi_{3\ell'}^\delta({\bf z})
\left({\tilde W}_{3\ell'}^\delta({\bf z})\right)^\ast\varphi\right\Vert\ret
&\le&\left\Vert
\chi_{\ell'}({\bf z})R_{5\ell',{\bf z}}\chi_{3\ell'}^\delta({\bf z})
\upsilon_{3\ell'}({\bf z})\varphi\right\Vert
+\frac{\hbar}{m_e}\sum_{i=x,y}\left\Vert\chi_{\ell'}({\bf z})R_{5\ell',{\bf z}}
\phi_{i,3\ell'}({\bf z})(p_i+eA_i)\varphi\right\Vert,\ret
\label{chiR5chiphiWbound}
\end{eqnarray}
where we have written 
\begin{equation}
\upsilon_{3\ell'}({\bf z})=\frac{\hbar^2}{2m_e}\Delta\chi_{3\ell'}^\delta({\bf z})
+\delta V_{\omega,3\ell',3\ell'}\chi_{3\ell'}^\delta({\bf z}).
\end{equation}
Using the bound (\ref{uR5vbound}), the first term in the right-hand side 
can be estimated as 
\begin{eqnarray}
\left\Vert\chi_{\ell'}({\bf z})R_{5\ell',{\bf z}}\chi_{3\ell'}^\delta({\bf z})
\upsilon_{3\ell'}({\bf z})\varphi\right\Vert
&\le&\sum_{{\bf u},{\bf v}}
\left\Vert\chi_\ell({\bf u})R_{5\ell',{\bf z}}\chi_\ell({\bf v})\right\Vert
\left\Vert\upsilon_{3\ell'}({\bf z})\right\Vert_\infty
\Vert\varphi\Vert\ret
&\le&12\left(\frac{\ell'}{\ell}\right)^3
\left(8e^{-\gamma\ell}\right)^{\ell'/\ell-4}
C_{\delta,\omega}
\left\Vert R_{5\ell',{\bf z}}\right\Vert\Vert\varphi\Vert.\ret
\label{chiR5chiphiWbound1}
\end{eqnarray}
The summand in the right-hand side in (\ref{chiR5chiphiWbound}) 
can be written as 
\begin{eqnarray}
& &\left\Vert\chi_{\ell'}({\bf z})R_{5\ell',{\bf z}}
\phi_{i,3\ell'}({\bf z})(p_i+eA_i)\varphi\right\Vert\ret
&=&\left\Vert\chi_{\ell'}({\bf z})R_{5\ell',{\bf z}}
W(\phi_{i,3\ell'}({\bf z}))R_{5\ell',{\bf z}}(p_i+eA_i)\varphi\right\Vert\ret
&\le&\frac{\hbar^2}{2m_e}\left\Vert\chi_{\ell'}({\bf z})R_{5\ell',{\bf z}}
(\Delta\phi_{i,3\ell'}({\bf z}))R_{5\ell',{\bf z}}(p_i+eA_i)\varphi\right\Vert\ret
&+&\frac{\hbar}{m_e}\sum_{j=x,y}\left\Vert\chi_{\ell'}({\bf z})R_{5\ell',{\bf z}}
(\partial_j\phi_{i,3\ell'}({\bf z}))(p_j+eA_j)R_{5\ell',{\bf z}}(p_i+eA_i)
\varphi\right\Vert\ret
&\le&6\left(\frac{\ell'}{\ell}\right)^3\left(8e^{-\gamma\ell}\right)^{\ell'/\ell-4}
\left\Vert R_{5\ell',{\bf z}}\right\Vert\left[
\frac{\hbar^2}{m_e}
\max_{n=x,y}\{\left\Vert\Delta\phi_{n,3\ell'}({\bf z})\right\Vert_\infty\}
\left\Vert R_{5\ell',{\bf z}}(p_i+eA_i)\varphi\right\Vert\right.\qquad\quad\ret
& &\qquad\qquad\quad+\left.\frac{4\hbar}{m_e}
\max_{m,n=x,y}\{\left\Vert\partial_m\phi_{n,3\ell'}({\bf z})\right\Vert_\infty\}
\left\Vert(p_j+eA_j)R_{5\ell',{\bf z}}(p_i+eA_i)\varphi\right\Vert\right],
\label{chiR5chiphiWbound2}
\end{eqnarray}
where we have used (\ref{defW}) and (\ref{uR5vbound}) again. 
{From} (\ref{pRpchibound}) and (\ref{pRRpbound})  
with $\mbox{\boldmath $\alpha$}=(1,0)$ or $(0,1)$, one has
\begin{equation}
\Vert R(p_i+eA_i)\varphi\Vert\le\sqrt{2m_e}(1+f_{E,R})^{1/2}\Vert R\Vert^{1/2}
\Vert\varphi\Vert
\end{equation}
and  
\begin{equation}
\Vert(p_j+eA_j)R(p_i+eA_i)\varphi\Vert\le 2m_e(1+f_{E,R})\Vert\varphi\Vert.
\end{equation}
Combining these, (\ref{chiR5chiphiWbound}), (\ref{chiR5chiphiWbound1}) and 
(\ref{chiR5chiphiWbound2}), we have 
\begin{equation}
\left\Vert\chi_{\ell'}({\bf z})R_{5\ell',{\bf z}}\chi_{3\ell'}^\delta({\bf z})
\left({\tilde W}_{3\ell'}^\delta({\bf z})\right)^\ast\varphi\right\Vert
\le6\left(\frac{\ell'}{\ell}\right)^3\left(8e^{-\gamma\ell}\right)^{\ell'/\ell-4}
f_4(|E|,\left\Vert R_{5\ell',{\bf z}}\right\Vert)
\left\Vert R_{5\ell',{\bf z}}\right\Vert\Vert\varphi\Vert.
\label{chiRchiWvarphibound}
\end{equation}

Next let us estimate the second term in the right-hand side of 
(\ref{chiR5Wvarphibound}). Note that 
\begin{eqnarray}
& &\left\Vert\left({\tilde W}_{3\ell'}^\delta({\bf z})\right)^\ast
R_{3\ell',{\bf z}}\left({\tilde W}_{3\ell'}^\delta({\bf z})\right)^\ast
\varphi\right\Vert\ret
&\le&C_{\delta,\omega}^2
\left\Vert R_{3\ell',{\bf z}}\right\Vert\Vert\varphi\Vert
+\frac{\hbar}{m_e}C_{\delta,\omega}
\left\Vert R_{3\ell',{\bf z}}({\bf p}+e{\bf A})\cdot
(\nabla\chi_{3\ell'}^\delta({\bf z}))\varphi\right\Vert\ret
&+&\frac{\hbar}{m_e}C_{\delta,\omega}
\max_{m=x,y}\{\Vert\partial_m\chi_{3\ell'}^\delta({\bf z})
\Vert_\infty\}\sum_{i=x,y}\left\Vert(p_i+eA_i)R_{3\ell',{\bf z}}\right\Vert
\Vert\varphi\Vert\ret
&+&\frac{\hbar^2}{m_e^2}\max_{m=x,y}\{\Vert\partial_m\chi_{3\ell'}^\delta({\bf z})
\Vert_\infty\}\sum_{i=x,y}\left\Vert(p_i+eA_i)R_{3\ell',{\bf z}}({\bf p}+e{\bf A})\cdot
(\nabla\chi_{3\ell'}^\delta({\bf z}))\varphi\right\Vert,
\end{eqnarray}
where we have used (\ref{defW}). Thus we have 
\begin{eqnarray}
& &\left\Vert\chi_{\ell'}({\bf z})R_{5\ell',{\bf z}}
\left({\tilde W}_{3\ell'}^\delta({\bf z})\right)^\ast
R_{3\ell',{\bf z}}\left({\tilde W}_{3\ell'}^\delta({\bf z})\right)^\ast
\varphi\right\Vert\ret
&\le&6\left(\frac{\ell'}{\ell}\right)^3\left(8e^{-\gamma\ell}\right)^{\ell'/\ell-4}
f_5(|E|,\left\Vert R_{3\ell',{\bf z}}\right\Vert)
\left\Vert R_{5\ell',{\bf z}}\right\Vert\Vert\varphi\Vert
\end{eqnarray}
in the same way. Substituting this and (\ref{chiRchiWvarphibound}) 
into (\ref{chiR5Wvarphibound}), the desired bound (\ref{chiR3Wvarphibound}) 
is obtained. 
\end{proof}

Similarly, one has the following lemma:

\begin{lemma}
\label{lemma:WRchibound}
Let $\ell,\ell'$ be odd integers larger than $1$ such that $\ell'$ is a multiple 
of $\ell$ and satisfies $\ell'>4\ell$.
Assume that the event $A^{\rm good}$ given in Lemma~\ref{expdecayR} occurs.
Then
\begin{eqnarray}
& &\left\Vert
{\tilde W}_{3\ell'}^\delta({\bf z})R_{3\ell',{\bf z}}\chi_{\ell'}({\bf z})
\right\Vert\ret
&\le&6\left(\frac{\ell'}{\ell}\right)^3\left(8e^{-\gamma\ell}\right)^{\ell'/\ell-4}
\left[f_6(|E|,\left\Vert R_{5\ell',{\bf z}}\right\Vert)
+f_5(|E|,\left\Vert R_{3\ell',{\bf z}}\right\Vert)\right]
\left\Vert R_{5\ell',{\bf z}}\right\Vert,
\label{WRchi3ellpbound}
\end{eqnarray}
where the function $f_6$ is given by 
\begin{eqnarray}
f_6(|E|,\Vert R\Vert)&=&2C_{\delta,\omega}
+\frac{8\hbar^2}{m_e}
\max_{m=x,y}\{\left\Vert|\nabla\phi_{m,3\ell'}({\bf z})|\right\Vert_\infty\}
(1+f_{E,R})\ret
&+&2\sqrt{2}\left(\frac{\hbar}{\sqrt{m_e}}\right)^3
\max_{n=x,y}\{\left\Vert\Delta\phi_{n,3\ell'}({\bf z})\right\Vert_\infty\}
(1+f_{E,R})^{1/2}\Vert R\Vert^{1/2}.
\end{eqnarray}
\end{lemma}

\begin{proof}{Proof}
Using the geometric resolvent equation, one has  
\begin{eqnarray}
& &{\tilde W}_{3\ell'}^\delta({\bf z})\chi_{3\ell'}^\delta({\bf z})
R_{5\ell',{\bf z}}\chi_{\ell'}({\bf z})\ret
&=&{\tilde W}_{3\ell'}^\delta({\bf z})R_{3\ell',{\bf z}}\chi_{\ell'}({\bf z})
+{\tilde W}_{3\ell'}^\delta({\bf z})R_{3\ell',{\bf z}}{\tilde W}_{3\ell'}^\delta({\bf z})
R_{5\ell',{\bf z}}\chi_{\ell'}({\bf z}).
\end{eqnarray}
Therefore
\begin{eqnarray}
\left\Vert
{\tilde W}_{3\ell'}^\delta({\bf z})R_{3\ell',{\bf z}}\chi_{\ell'}({\bf z})
\right\Vert
&\le&\left\Vert{\tilde W}_{3\ell'}^\delta({\bf z})\chi_{3\ell'}^\delta({\bf z})
R_{5\ell',{\bf z}}\chi_{\ell'}({\bf z})\right\Vert\ret
&+&\left\Vert {\tilde W}_{3\ell'}^\delta({\bf z})R_{3\ell',{\bf z}}
{\tilde W}_{3\ell'}^\delta({\bf z})
R_{5\ell',{\bf z}}\chi_{\ell'}({\bf z})\right\Vert. 
\end{eqnarray}
Since the second term in the right-hand side can be estimated 
in the same way as in the proof of 
the preceding lemma, it is enough to estimate the first term. Using (\ref{defW}), one has 
\begin{eqnarray}
\left\Vert{\tilde W}_{3\ell'}^\delta({\bf z})\chi_{3\ell'}^\delta({\bf z})
R_{5\ell',{\bf z}}\chi_{\ell'}({\bf z})\right\Vert
&\le&\left\Vert\upsilon_{3\ell'}({\bf z})
\chi_{3\ell'}^\delta({\bf z})R_{5\ell',{\bf z}}\chi_{\ell'}({\bf z})\right\Vert\ret
&+&\frac{\hbar}{m_e}\sum_{i=x,y}\left\Vert
(p_i+eA_i)\left(\partial_i\chi_{3\ell'}^\delta({\bf z})\right)
\chi_{3\ell'}^\delta({\bf z})R_{5\ell',{\bf z}}\chi_{\ell'}({\bf z})\right\Vert.\ret
\label{WchiR5decompose}
\end{eqnarray}
The first term in the right-hand side can be estimated by using 
the bound (\ref{uR5vbound}). 
The operators in the sum in the right-hand side can be written as 
\begin{eqnarray}
(p_i+eA_i)\phi_{i,3\ell'}({\bf z})R_{5\ell',{\bf z}}\chi_{\ell'}({\bf z})
&=&(p_i+eA_i)R_{5\ell',{\bf z}}W(\phi_{i,3\ell'}({\bf z}))
R_{5\ell',{\bf z}}\chi_{\ell'}({\bf z})\ret
&=&\frac{\hbar^2}{2m_e}(p_i+eA_i)R_{5\ell',{\bf z}}
\left(\Delta\phi_{i,3\ell'}({\bf z})\right)R_{5\ell',{\bf z}}\chi_{\ell'}({\bf z})\ret
&-&\frac{i\hbar}{m_e}(p_i+eA_i)R_{5\ell',{\bf z}}({\bf p}+e{\bf A})\cdot
(\nabla\phi_{i,3\ell'}({\bf z}))R_{5\ell',{\bf z}}\chi_{\ell'}({\bf z}),\ret
\end{eqnarray}
where we have used (\ref{defW}) again. 
Similarly the norm of the operators in this right-hand side can be estimated 
by using (\ref{pRpchibound}), (\ref{alphaipiRpsibound}) and (\ref{uR5vbound}). 
\end{proof}

\begin{proof}{Proof of Lemma~\ref{lemma:relationell'ell}}
Assume that the event $A^{\rm good}$ given in Lemma~\ref{expdecayR} occurs.
Then, from the preceding three lemmas, one has 
\begin{eqnarray}
\left\Vert\chi_{\ell'}R_{3\ell,{\bf z}}
\left({\tilde W}_{3\ell'}^\delta({\bf z})\right)^\ast
\right\Vert&\le&{\rm Const.}\times(\ell')^3
\exp\left[-\ell'\{\gamma(1-4\ell/\ell')-3\log 2/\ell\}\right]\ret
&\times&|E|(\Vert R_{3\ell,{\bf z}}\Vert+\Vert R_{5\ell,{\bf z}}\Vert)
\Vert R_{5\ell,{\bf z}}\Vert
\end{eqnarray}
and 
\begin{eqnarray}
\left\Vert{\tilde W}_{3\ell'}^\delta({\bf z})R_{3\ell,{\bf z}}\chi_{\ell'}
\right\Vert&\le&{\rm Const.}\times(\ell')^3
\exp\left[-\ell'\{\gamma(1-4\ell/\ell')-3\log 2/\ell\}\right]\ret
&\times&|E|(\Vert R_{3\ell,{\bf z}}\Vert+\Vert R_{5\ell,{\bf z}}\Vert)
\Vert R_{5\ell,{\bf z}}\Vert
\label{Lambda3ell'gamma'good}
\end{eqnarray}
with the probability larger than $1-(5\ell'/\ell)^4\eta^2$ for 
a large $|E|$ and for large $\Vert R_{3\ell,{\bf z}}\Vert$, 
$\Vert R_{5\ell,{\bf z}}\Vert$. 

In the Wegner estimate (\ref{Wegnerestimate}), we choose 
\begin{equation}
(\delta E)^{-1}=C_{\rm W}K_3\Vert g\Vert_\infty|\Lambda_{5\ell'}|\times 4(\ell')^\xi. 
\end{equation}
Then, for $q=3,5$, one has 
\begin{equation}
\Vert R_{q\ell,{\bf z}}\Vert\le(\delta E)^{-1}
\end{equation}
with the probability larger than $1-(\ell')^{-\xi}/4$. 
Clearly, one has 
\begin{equation}
\Vert R_{q\ell,{\bf z}}\Vert\le{\rm Const.}\times K_3(\ell')^{\xi+2}
\end{equation}
for a large $\ell'$.  
Since the probability that each event occurs is larger than $1-(\ell')^{-\xi}/4$, 
the probability that the two events simultaneously occur is larger than 
$1-(\ell')^{-\xi}/2$. 

{From} these observations, the right-hand side of (\ref{Lambda3ell'gamma'good}) 
can be bounded from above by 
\begin{equation}
\exp\left[-\ell'\{\gamma(1-4\ell/\ell')-3\log 2/\ell
-\log({\rm Const.}\times K_3^2|E|)/\ell'-(2\xi+7)\log \ell'/\ell'
\}\right]
\end{equation}
with the probability larger than $1-(5\ell'/\ell)^4\eta^2-(\ell')^{-\xi}/2$ for 
a large $|E|$ and for a large $\ell'$. 
Since one has  
\begin{equation}
\frac{3\log 2}{\ell}+\frac{\log({\rm Const.}\times K_3^2|E|)}{\ell'}
\le\frac{\log(c_0 K_3^2|E|)}{\ell} 
\end{equation}
with a positive constant $c_0$, the proof of the lemma is completed. 
\end{proof}

\begin{proof}{Proof of Lemma~\ref{lemma:lksequence}}
Take $\ell'=\ell_{k+1}$ and $\ell=\ell_k$ in Lemma~\ref{lemma:relationell'ell}, 
and assume that $\Lambda_{3\ell_k}(\cdots)$ is a $\gamma_k$-good box 
with the probability larger than $1-\eta$ with $\eta=(\ell_k)^{-\xi}$. 
{From} the definition (\ref{defellk+1}) of $\ell_{k+1}$, we have
\begin{equation}
\frac{\ell_{k+1}}{\ell_k}=\left[\ell_k^{1/2}\right]_{\rm odd}^\ge
=\ell_k^{1/2}+\delta\ell_k\quad\mbox{with }\ 0\le\delta\ell_k<2.
\label{fracellk} 
\end{equation} 
Using this identity, $\eta'$ of (\ref{eta'}) can be written as 
\begin{eqnarray}
\eta'=\eta_{k+1}&=&5^4\left(\frac{\ell_{k+1}}{\ell_k}\right)^4(\ell_k)^{-2\xi}
+\frac{1}{2}(\ell_{k+1})^{-\xi}\ret
&=&(\ell_{k+1})^{-\xi}
\left[5^4\left(\frac{\ell_{k+1}}{\ell_k}\right)^{4+\xi}(\ell_k)^{-\xi}
+\frac{1}{2}\right]\ret
&=&(\ell_{k+1})^{-\xi}\left[5^4(\ell_k)^{2-\xi/2}
\left(1+\frac{\delta\ell_k}{\ell_k^{1/2}}\right)^{4+\xi}+\frac{1}{2}\right].
\end{eqnarray}
Therefore, if the initial scale $\ell_0$ satisfies 
\begin{equation}
5^4(\ell_0)^{2-\xi/2}
\left(1+{2}{\ell_0^{-1/2}}\right)^{4+\xi}\le{1}/{2},
\end{equation}
then we have $\eta_{k+1}\le(\ell_{k+1})^{-\xi}$. Actually this inequality holds for 
a large $\ell_0$ because of the assumption, $\xi>4$. 

Next we define $\gamma_k$ inductively according to (\ref{gamma'}) as 
\begin{equation}
\gamma_{k+1}=\gamma_k\left(1-{4\ell_k}/{\ell_{k+1}}\right)-d_k
\end{equation}
with 
\begin{equation}
d_k=\frac{\log(c_0K_3^2|E|)}{\ell_k}+\frac{(2s+7)\log\ell_{k+1}}{\ell_{k+1}}.
\end{equation}
One can easily find 
\begin{eqnarray}
\gamma_{k+1}&=&\left(1-\frac{4\ell_k}{\ell_{k+1}}\right)
\left(1-\frac{4\ell_{k-1}}{\ell_k}\right)\cdots
\left(1-\frac{4\ell_1}{\ell_2}\right)
\left(1-\frac{4\ell_0}{\ell_1}\right)\gamma_0\ret
& &-d_k-\left(1-\frac{4\ell_k}{\ell_{k+1}}\right)d_{k-1}
-\left(1-\frac{4\ell_k}{\ell_{k+1}}\right)
\left(1-\frac{4\ell_{k-1}}{\ell_k}\right)d_{k-2}\cdots\ret
& &-\left(1-\frac{4\ell_k}{\ell_{k+1}}\right)
\left(1-\frac{4\ell_{k-1}}{\ell_k}\right)\cdots
\left(1-\frac{4\ell_1}{\ell_2}\right)d_0\ret
&\ge&\gamma_0\prod_{j=0}^k\left(1-\frac{4\ell_j}{\ell_{j+1}}\right)
-\sum_{j=0}^k d_j.
\label{estimategammak}
\end{eqnarray}
Note that 
\begin{equation}
\ell_k\ge(\ell_{k-1})^{3/2}\ge(\ell_{k-2})^{(3/2)^2}\cdots\ge
(\ell_0)^{(3/2)^k}, 
\end{equation}
and one has 
\begin{equation}
\frac{\ell_k}{\ell_{k+1}}\le\frac{1}{\ell_k^{1/2}}
\le\exp\left[-\frac{1}{2}\left({3}/{2}\right)^k\log\ell_0\right].
\end{equation}
Using this inequality, 
the product in the right-hand side of (\ref{estimategammak})
can be evaluated as 
\begin{equation}
\prod_{j=0}^k\left(1-\frac{4\ell_j}{\ell_{j+1}}\right)
\ge\prod_{j=0}^k\left\{1-4\exp\left[-\frac{1}{2}\left({3}/{2}\right)^j
\log\ell_0\right]\right\}.
\end{equation}
This right-hand side is uniformly bounded from below by some positive 
constant if $\ell_0>16$. 
The sum of $d_k$ in the right-hand side of (\ref{estimategammak})
can also be evaluated as 
\begin{eqnarray}
\sum_{j=0}^kd_j&\le&\log(c_0K_3^2|E|)\sum_{j=0}^k
\exp\left[-\left({3}/{2}\right)^j\log\ell_0\right]\ret
&+&(2s+7)\log\ell_0\sum_{j=0}^k\left({3}/{2}\right)^{j+1}
\exp\left[-\left({3}/{2}\right)^{j+1}\log\ell_0\right].
\end{eqnarray}
This right-hand side becomes small for $B$ large enough because of 
$K_3={\cal O}(B)$ or ${\cal O}(1)$, $|E|={\cal O}(B)$ and 
$\ell_0={\cal O}(B^{1/2})$. As a result, $\gamma_k$ is uniformly bounded 
from below by some positive constant $\gamma_\infty$.  
\end{proof}

%%%%%%%%%%%%%%%%%%%%%%%%%%%%%%%%%%%%%%%%%%%%%%%%%%%%%%%%%%%%%
\Section{Proof of Lemma~\ref{lemma:Index2}}
\label{IndexProof}

The difference between (\ref{Index1}) and (\ref{Index2}) is estimated by 
\begin{equation}
\Delta I:=\frac{1}{{\cal V}_\ell}
\left(\sum_{{\bf a}\in\Lambda_\ell}\sum_{{\bf u}\in({\bf Z}\varepsilon^2)^\ast
\backslash\Lambda_\ell^\ast}
\sum_{{\bf v},{\bf w}}
+\sum_{{\bf a}\in{\bf Z}_\varepsilon^2\backslash\Lambda_\ell}
\sum_{{\bf u}\in\Lambda_\ell^\ast}\sum_{{\bf v},{\bf w}}
\left|t_{{\bf u},{\bf v}}t_{{\bf v},{\bf w}}t_{{\bf w},{\bf u}}\right|
\left|S_{\bf u,v,w,u}\right|\right). 
\end{equation}
In the same way as in the proof of Theorem~\ref{theorem:sigmaxyoutzero}, 
it is sufficient to show ${\bf E}[\Delta I]\rightarrow 0$ as $\ell\rightarrow\infty$. 

To begin with, we note that 
\begin{eqnarray}
& &\left|{\rm Tr}\>\chi_\varepsilon({\bf u})P_{\rm F}\chi_\varepsilon({\bf v})
P_{\rm F}\chi_\varepsilon({\bf w})P_{\rm F}\chi_\varepsilon({\bf u})\right|\ret
&\le&\sqrt{{\rm Tr}\>\chi_\varepsilon({\bf u})P_{\rm F}\chi_\varepsilon({\bf v})
P_{\rm F}\chi_\varepsilon({\bf v})P_{\rm F}\chi_\varepsilon({\bf u})}
\sqrt{{\rm Tr}\>\chi_\varepsilon({\bf u})P_{\rm F}\chi_\varepsilon({\bf w})
P_{\rm F}\chi_\varepsilon({\bf w})P_{\rm F}\chi_\varepsilon({\bf u})}\ret
&\le&\left\Vert\chi_\varepsilon({\bf u})P_{\rm F}\chi_\varepsilon({\bf v})\right\Vert
\left\Vert\chi_\varepsilon({\bf u})P_{\rm F}\chi_\varepsilon({\bf w})\right\Vert
\sqrt{{\rm Tr}\>\chi_\varepsilon({\bf v})P_{\rm F}\chi_\varepsilon({\bf v})}
\sqrt{{\rm Tr}\>\chi_\varepsilon({\bf w})P_{\rm F}\chi_\varepsilon({\bf w})}\ret
&\le&{\rm Const.}
\left\Vert\chi_\varepsilon({\bf u})P_{\rm F}\chi_\varepsilon({\bf v})\right\Vert
\left\Vert\chi_\varepsilon({\bf u})P_{\rm F}\chi_\varepsilon({\bf w})\right\Vert, 
\label{SD} 
\end{eqnarray}
where we have used (\ref{localTr}). Further, Schwarz's inequality yields 
\begin{eqnarray}
& &{\bf E}\left[\left\Vert\chi_\varepsilon({\bf u})
P_{\rm F}\chi_\varepsilon({\bf v})\right\Vert
\left\Vert\chi_\varepsilon({\bf u})P_{\rm F}\chi_\varepsilon({\bf w})\right\Vert
\right]\ret
&\le&{\bf E}\left[\left\Vert\chi_\varepsilon({\bf u})P_{\rm F}
\chi_\varepsilon({\bf v})\right\Vert^2\right]^{1/2}
{\bf E}\left[\left\Vert\chi_\varepsilon({\bf u})P_{\rm F}
\chi_\varepsilon({\bf w})\right\Vert^2\right]^{1/2}\ret
&\le&{\bf E}\left[\left\Vert\chi_\varepsilon({\bf u})P_{\rm F}
\chi_\varepsilon({\bf v})\right\Vert\right]^{1/2}
{\bf E}\left[
\left\Vert\chi_\varepsilon({\bf u})P_{\rm F}\chi_\varepsilon({\bf w})
\right\Vert\right]^{1/2},
\end{eqnarray}
where we have used 
$\left\Vert\chi_\varepsilon({\bf u})P_{\rm F}\chi_\varepsilon({\bf v})\right\Vert
\le 1$ for any ${\bf u,v}$. From these observations and the decay bound (\ref{PFD}) 
for the Fermi sea projection, it is sufficient to estimate  
\begin{equation}
\frac{1}{{\cal V}_\ell}
\left(\sum_{{\bf a}\in\Lambda_\ell}\sum_{{\bf u}\in({\bf Z}_\varepsilon^2)^\ast
\backslash\Lambda_\ell^\ast}
\sum_{{\bf v},{\bf w}}
+\sum_{{\bf a}\in{\bf Z}_\varepsilon^2\backslash\Lambda_\ell}
\sum_{{\bf u}\in\Lambda_\ell^\ast}\sum_{{\bf v},{\bf w}}
\left|t_{{\bf u},{\bf v}}t_{{\bf v},{\bf w}}t_{{\bf w},{\bf u}}\right|
e^{-\mu|{\bf u}-{\bf v}|/2}e^{-\mu|{\bf u}-{\bf w}|/2}\right).
\end{equation}

Consider first the case with $|{\bf u-a}|\le\varepsilon_1\ell^\delta$ 
with $\delta\in(0,1)$. 
Using the bound,  
\begin{equation}
\left|t_{{\bf u},{\bf v}}t_{{\bf v},{\bf w}}t_{{\bf w},{\bf u}}\right|
\le 2^3\frac{|{\bf u}-{\bf v}||{\bf u}-{\bf w}|}{|{\bf u-a}|^2},
\end{equation}
which is derived from (\ref{2sinthetadiffbound}), we have 
\begin{equation}
\sum_{{\bf v},{\bf w}}
\left|t_{{\bf u},{\bf v}}t_{{\bf v},{\bf w}}t_{{\bf w},{\bf u}}\right|
e^{-\mu|{\bf u}-{\bf v}|/2}e^{-\mu|{\bf v}-{\bf w}|/2}
<{\rm Const.}\frac{1}{|{\bf u-a}|^2}. 
\end{equation}
Therefore the corresponding error is estimated by  
\begin{equation}
\frac{1}{\ell^2}
\left(
\mathop{\sum_{{\bf a}\in\Lambda_\ell,{\bf u}\in({\bf Z}_\varepsilon^2)^\ast
\backslash\Lambda_\ell^\ast :}}_{|{\bf u-a}|\le \varepsilon_1\ell^\delta}
+\mathop{\sum_{{\bf a}\in{\bf Z}_\varepsilon^2\backslash\Lambda_\ell,
{\bf u}\in\Lambda_\ell^\ast :}}_{|{\bf u-a}|\le\varepsilon_1\ell^\delta}
\frac{1}{|{\bf u-a}|^2}\right)
\le\frac{{\rm Const.}\ell\cdot\ell^\delta({\rm Const.}
+{\rm Const.}\log \ell)}{\ell^2}. 
\end{equation}
This vanishes as $\ell\rightarrow\infty$. 
 
When $|{\bf u-a}|>\varepsilon_1\ell^\delta$, we further decompose it into 
two cases: (i) both ${\bf v}$ and ${\bf w}$ fall into inside  
the ball with radius $|{\bf u-a}|$ around ${\bf u}$, 
i.e., $|{\bf v-u}|<|{\bf u-a}|$ and $|{\bf w-u}|<|{\bf u-a}|$,
(ii) one of ${\bf v}$ or ${\bf w}$ falls into outside the ball, i.e.,  
$|{\bf v-u}|\ge|{\bf u-a}|$ or $|{\bf w-u}|\ge|{\bf u-a}|$. 
The latter contribution is exponentially small in $\ell^\delta$. Actually, 
one has   
\begin{eqnarray}
& &\sum_{{\bf v},{\bf w}\ {\rm satisfy\ (ii)}}
\left|t_{{\bf u},{\bf v}}t_{{\bf v},{\bf w}}t_{{\bf w},{\bf u}}\right|
e^{-\mu|{\bf u}-{\bf v}|/2}e^{-\mu|{\bf u}-{\bf w}|/2}\ret
&\le&
2^3\left[\mathop{\sum_{{\bf v}:|{\bf v-u}|\ge|{\bf u-a}|,}}_{\bf w}
+\mathop{\sum_{{\bf w}:|{\bf w-u}|\ge|{\bf u-a}|,}}_{\bf v}
e^{-\mu|{\bf u}-{\bf v}|/2}e^{-\mu|{\bf u}-{\bf w}|/2}\right]
\le{\rm Const.}e^{-\mu'|{\bf u-a}|}
\end{eqnarray}
with a positive constant $\mu'$. 

Finally, consider the former case (i). To begin with, we note that 
\begin{equation}
t_{{\bf u},{\bf v}}t_{{\bf v},{\bf w}}t_{{\bf w},{\bf u}}
=2i\left\{\sin\angle({\bf u},{\bf a},{\bf v})
+\sin\angle({\bf v},{\bf a},{\bf w})
+\sin\angle({\bf w},{\bf a},{\bf u})\right\}.
\end{equation}
We write $\alpha=\angle({\bf u},{\bf a},{\bf v})$, 
$\beta=\angle({\bf v},{\bf a},{\bf w})$ 
and $\gamma=\angle({\bf w},{\bf a},{\bf u})$ for short. 
In this case, one notices that $\alpha,\beta\in(-\pi/2,\pi/2)$ and $\alpha+\beta+\gamma=0$. 
{From} these, one has
\begin{eqnarray}
|\sin\alpha+\sin\beta+\sin\gamma|&\le&
2\left(|\sin\alpha|\sin^2\frac{\beta}{2}+|\sin\beta|\sin^2\frac{\alpha}{2}\right)\ret
&\le&2(|\sin\alpha|\sin^2\beta+|\sin\beta|\sin^2\alpha)\ret
&\le&\frac{2}{|{\bf u}-{\bf a}|^3}\left(|{\bf u-v}||{\bf w-u}|^2
+|{\bf v-u}|^2|{\bf w-u}|\right),
\end{eqnarray}
where we have used 
\begin{equation}
|\sin\alpha|\le\frac{|{\bf v-u}|}{|{\bf u-a}|}\quad\mbox{and}\quad
|\sin\beta|\le\frac{|{\bf w-u}|}{|{\bf u-a}|}
\end{equation}
for getting the third inequality. From these observations, we obtain 
\begin{equation}
\sum_{{\bf v},{\bf w}}
\left|t_{{\bf u},{\bf v}}t_{{\bf v},{\bf w}}t_{{\bf w},{\bf u}}\right|
e^{-\mu|{\bf u}-{\bf v}|/2}e^{-\mu|{\bf u}-{\bf w}|/2}
\le{\rm Const.}\frac{1}{|{\bf u}-{\bf a}|^3}. 
\end{equation}
The corresponding contribution is estimated by  
\begin{equation}
\frac{1}{\ell^2}
\left(
\mathop{\sum_{{\bf a}\in\Lambda_\ell,{\bf u}\in({\bf Z}_\varepsilon^2)^\ast
\backslash\Lambda_\ell^\ast:}}_{|{\bf u-a}|>\varepsilon_1\ell^\delta}
+\mathop{\sum_{{\bf a}\in{\bf Z}_\varepsilon^2\backslash\Lambda_\ell,
{\bf u}\in\Lambda_\ell^\ast:}}_{|{\bf u-a}|>\varepsilon_1\ell^\delta}
\frac{1}{|{\bf u-a}|^3}\right)\ret
\le\frac{{\rm Const.}}{\ell^\delta}. 
\end{equation}
This vanishes as $\ell\rightarrow\infty$.
%%%%%%%%%%%%%%%%%%%%%%%%%%%%%%%%%%%%%%%%%%%%%%%%%%
\Section{Proof of Lemma~\ref{lemma:indexRIF}}
\label{Proof9.7}

In order to prove Lemma~\ref{lemma:indexRIF}, we introduce a partition 
$\{\chi_b^\delta({\bf u})\}_{\bf u}$ of unity 
and prepare Lemma~\ref{lemma:pchiRchibound} below. 
Here $\chi_b^\delta({\bf u})$ are $C^2$ positive functions 
with a compact support such that $\sum_{\bf u}\chi_b^\delta({\bf u})=1$. 
Let ${\tilde\chi}_b({\bf u})$ denote the characteristic function 
of the support of $\chi_b^\delta({\bf u})$. 

\begin{lemma}
\label{lemma:pchiRchibound}
The following bound is valid:
\begin{eqnarray}
\left\Vert(p_s+eA_s)\chi_b^\delta({\bf u})R(z)
\chi_\varepsilon({\bf v})\right\Vert&\le&
[{\rm Const.}+(2m_e|z|)^{1/2}]
\left\Vert{\tilde\chi}_b({\bf u})R(z)\chi_\varepsilon({\bf v})\right\Vert\ret
&+&\sqrt{2m_e}\left\Vert\chi_b^\delta({\bf u})R(z)
\chi_\varepsilon({\bf v})\right\Vert^{1/2},
\end{eqnarray}
where the positive constant depends only on the strengths of the potentials, 
$V_0,V_\omega$ and on the cutoff functions $\chi_b^\delta({\bf u})$. 
\end{lemma}

\begin{proof}{Proof} 
Let $\varphi$ be a vector on ${\bf R}^2$. Then one has 
\begin{eqnarray}
& &\left(\varphi,\chi_\varepsilon({\bf v})R^\ast(z)
\chi_b^\delta({\bf u})(p_s+eA_s)^2\chi_b^\delta({\bf u})R(z)
\chi_\varepsilon({\bf v})\varphi\right)\ret
&\le&2m_e(\Vert V_0\Vert_\infty+\Vert V_\omega\Vert_\infty)
\left\Vert\chi_b^\delta({\bf u})R(z)\chi_\varepsilon({\bf v})\varphi\right\Vert^2\ret
&+&2m_e\left(\varphi,\chi_\varepsilon({\bf v})R^\ast(z)
\chi_b^\delta({\bf u})H_\omega\chi_b^\delta({\bf u})R(z)
\chi_\varepsilon({\bf v})\varphi\right)
\end{eqnarray}
by using the inequality $(p_s+eA_s)^2/(2m_e)\le H_\omega+\Vert V_0\Vert_\infty
+\Vert V_\omega\Vert_\infty$. Further, the second term in the right-hand side 
is evaluated as 
\begin{eqnarray}
& &2m_e\left(\varphi,\chi_\varepsilon({\bf v})R^\ast(z)
\chi_b^\delta({\bf u})H_\omega\chi_b^\delta({\bf u})R(z)
\chi_\varepsilon({\bf v})\varphi\right)\ret
&\le&2m_e|z|
\left\Vert\chi_b^\delta({\bf u})R(z)\chi_\varepsilon({\bf v})\varphi\right\Vert^2\ret
&+&2\hbar\sum_{s=x,y}\left\Vert[\partial_s\chi_b^\delta({\bf u})]R(z)
\chi_\varepsilon({\bf v})\varphi\right\Vert
\left\Vert\chi_b^\delta({\bf u})(p_s+eA_s)R(z)\chi_\varepsilon({\bf v})\varphi
\right\Vert\ret
&+&\hbar^2\left\Vert\chi_b^\delta({\bf u})R(z)\chi_\varepsilon({\bf v})\varphi\right\Vert
\left\Vert[\Delta\chi_b^\delta({\bf u})]R(z)\chi_\varepsilon({\bf v})\varphi
\right\Vert\ret
&+&2m_e\left\Vert\chi_b^\delta({\bf u})R(z)\chi_\varepsilon({\bf v})\varphi\right\Vert
\left\Vert\chi_b^\delta({\bf u})\chi_\varepsilon({\bf v})\varphi\right\Vert
\end{eqnarray}
by using 
\begin{equation}
H_\omega\chi_b^\delta({\bf u})
=-\frac{i\hbar}{m_e}\nabla\chi_b^\delta({\bf u})\cdot({\bf p}+e{\bf A})
-\frac{\hbar^2}{2m_e}\Delta\chi_b^\delta({\bf u})+\chi_b^\delta({\bf u})H_\omega.
\end{equation}
Combing these two bounds, we obtain 
\begin{eqnarray}
& &\left\Vert(p_s+eA_s)\chi_b^\delta({\bf u})R(z)\chi_\varepsilon({\bf v})
\right\Vert^2\ret
&\le&\left[2m_e(\Vert V_0\Vert_\infty+\Vert V_\omega\Vert_\infty+|z|)
+2\hbar^2\sum_{s=x,y}\Vert\partial_s\chi_b^\delta({\bf u})\Vert_\infty^2
+\hbar^2\Vert\Delta\chi_b^\delta({\bf u})\Vert_\infty\right]\ret
& &\qquad\qquad\qquad\times
\left\Vert{\tilde\chi}_b({\bf u})R(z)\chi_\varepsilon({\bf v})\right\Vert^2\ret
&+&2\hbar\sum_{s=x,y}\Vert\partial_s\chi_b^\delta({\bf u})\Vert_\infty\left\Vert
{\tilde\chi}_b({\bf u})R(z)\chi_\varepsilon({\bf v})\right\Vert
\left\Vert(p_s+eA_s)\chi_b^\delta({\bf u})R(z)\chi_\varepsilon({\bf v})
\right\Vert\ret
&+&2m_e\left\Vert\chi_b^\delta({\bf u})R(z)\chi_\varepsilon({\bf v})\varphi\right\Vert
\left\Vert\chi_b^\delta({\bf u})\chi_\varepsilon({\bf v})\right\Vert.
\end{eqnarray}
Solving this quadratic inequality and using the inequality 
$\sqrt{a+b}\le\sqrt{a}+\sqrt{b}$ for $a,b\ge 0$, the desired bound is obtained. 
\end{proof}

\begin{proof}{Proof of Lemma~\ref{lemma:indexRIF}}
Note that  
\begin{eqnarray}
{\bf E}\left[\left|{\cal I}(P_{\rm F};\Omega,\ell_{\rm P})
-{\cal I}(P_{{\rm F},\Lambda};\Omega)\right|\right]
&=&{\bf E}\left[\left|{\cal I}(P_{\rm F};\Omega,\ell_{\rm P})
-{\cal I}(P_{{\rm F},\Lambda};\Omega,\ell_{\rm P})\right|{\bf I}(M_\Lambda)\right]\ret
&+&{\bf E}\left[\left|{\cal I}(P_{\rm F};\Omega,\ell_{\rm P})
-{\cal I}(P_{{\rm F},\Lambda};\Omega,\ell_{\rm P})\right|{\bf I}(M_\Lambda^c)\right],
\label{diffindexIM}
\end{eqnarray}
where $M_\Lambda$ is the event which was introduced in the proof 
of Lemma~\ref{lemma:ARBdecay}, 
and ${\bf I}(A)$ is the indicator function of an event $A$. The second term 
in the right-hand side is vanishing in the limit $L\uparrow\infty$ as 
\begin{eqnarray}
& &{\bf E}\left[\left|{\cal I}(P_{\rm F};\Omega,\ell_{\rm P})
-{\cal I}(P_{{\rm F},\Lambda};\Omega)\right|{\bf I}(M_\Lambda^c)\right]\ret
&\le&
{\bf E}\left[\left|{\cal I}(P_{\rm F};\Omega,\ell_{\rm P})
-{\cal I}(P_{{\rm F},\Lambda};\Omega)\right|^2\right]^{1/2}
 {\rm E}\left[{\bf I}(M_\Lambda^c)\right]^{1/2}
\le{\rm Const.}L^{-2[\kappa(\xi+2)-3]/6}, 
\end{eqnarray}
where we have used Schwarz's inequality for getting the first inequality,   
and we have used\footnote{The bound (\ref{TrDCPF2}) 
of Lemma~\ref{lemma:ETrbound2} holds also 
for $\Lambda={\bf R}^2$.}
Lemma~\ref{lemma:ETrbound2} and 
${\rm Prob}(M_\Lambda^c)\le{\rm Const.}L^{-2[\kappa(\xi+2)-3]/3}$ 
for the second inequality. 

In order to estimate the first term in the right-hand side of (\ref{diffindexIM}), 
we first note that 
\begin{equation}
{\rm Tr}\>\chi_\Omega P_{\rm F}[[P_{\rm F},x],[P_{\rm F},y]]\chi_\Omega
={\rm Tr}\>\chi_\Omega(P_{\rm F}xP_{\rm F}yP_{\rm F}-
P_{\rm F}yP_{\rm F}xP_{\rm F})\chi_\Omega.
\end{equation}
We want to rewrite this right-hand side. We have 
\begin{eqnarray}
{\rm Tr}\>\chi_\Omega P_{\rm F}xP_{\rm F}yP_{\rm F}\chi_\Omega 
&=&{\rm Tr}\>\chi_\Omega P_{\rm F}x(\chi_\Lambda^\delta+1-\chi_\Lambda^\delta)
P_{\rm F}y(\chi_\Lambda^\delta+1-\chi_\Lambda^\delta)P_{\rm F}\chi_\Omega \ret
&=&{\rm Tr}\>\chi_\Omega P_{\rm F}x\chi_\Lambda^\delta
P_{\rm F}y\chi_\Lambda^\delta P_{\rm F}\chi_\Omega +{\rm corrections}.
\end{eqnarray}
The contributions from the corrections decay exponentially in the distance 
between $\Omega$ and the support of $1-\chi_\Lambda^\delta$ by the bound 
(\ref{PFD}) for the Fermi sea projection $P_{\rm F}$. 
The rest is written 
\begin{equation}
{\rm Tr}\>\chi_\Omega P_{\rm F}x\chi_\Lambda^\delta
P_{\rm F}y\chi_\Lambda^\delta P_{\rm F}\chi_\Omega \ret
=\frac{1}{2\pi i}\int dz
{\rm Tr}\>\chi_\Omega R(z)x\chi_\Lambda^\delta
P_{\rm F}y\chi_\Lambda^\delta P_{\rm F}\chi_\Omega.
\end{equation}
Note that one has 
\begin{eqnarray}
\chi_\Omega R(z)&=&\chi_\Omega \chi_\Lambda^\delta R(z)
=\chi_\Omega R_\Lambda(z)\chi_\Lambda^\delta+
\chi_\Omega R_\Lambda(z){\tilde W}(\chi_\Lambda^\delta)R(z)
\label{chiRtoGRE}
\end{eqnarray}
{from} the geometric resolvent equation, 
$\chi_\Lambda^\delta R(z)=R_\Lambda(z)\chi_\Lambda^\delta
+R_\Lambda(z){\tilde W}(\chi_\Lambda^\delta)R(z)$, 
where ${\tilde W}(\chi_\Lambda^\delta)
=W(\chi_\Lambda^\delta)+({\tilde V}_{\omega,\Lambda}-V_\omega)\chi_\Lambda^\delta$, 
and ${\tilde V}_{\omega,\Lambda}$ is the slightly modified potential 
near the boundary of $\Lambda$. (The precise definition of 
${\tilde V}_{\omega,\Lambda}$ is given in Section~\ref{model}.)  
The contribution from the second term in the right-hand side of (\ref{chiRtoGRE}) 
is 
\begin{equation}
\frac{1}{2\pi i}\int dz{\rm Tr}\>
\chi_\Omega R_\Lambda(z){\tilde W}(\chi_\Lambda^\delta)R(z)
x\chi_\Lambda^\delta
P_{\rm F}y\chi_\Lambda^\delta P_{\rm F}\chi_\Omega .
\end{equation}
The integrand is estimated by  
\begin{equation}
\sum_{{\bf u}:s_b({\bf u})\cap\Omega\ne\emptyset}
\sum_{{\bf v},{\bf w}}\left|{\rm Tr}\>
\chi_b({\bf u})R_\Lambda(z){\tilde W}(\chi_\Lambda^\delta)R(z)
x\chi_\Lambda^\delta\chi_b({\bf v})
P_{\rm F}\chi_b({\bf w})y\chi_\Lambda^\delta P_{\rm F}\chi_b({\bf u})\right|.
\end{equation}
For any bounded operators $A,B$, 
\begin{eqnarray}
\left|{\rm Tr}\>A\chi_b({\bf v})P_{\rm F}\chi_b({\bf w})B\right|
&\le&\sqrt{{\rm Tr}\>A\chi_b({\bf v})P_{\rm F}\chi_b({\bf v})A^\ast}
\cdot\sqrt{{\rm Tr}\>B^\ast\chi_b({\bf w})P_{\rm F}\chi_b({\bf w})B}\ret
&\le&{\rm Const.}\Vert A\Vert\Vert B\Vert,
\end{eqnarray}
where we have used the bound (\ref{localTr}). Using this inequality, one has 
\begin{eqnarray}
& &\left|{\rm Tr}\>\chi_b({\bf u})R_\Lambda(z){\tilde W}(\chi_\Lambda^\delta)
R(z)x\chi_\Lambda^\delta\chi_b({\bf v})P_{\rm F}\chi_b({\bf w})
y\chi_\Lambda^\delta P_{\rm F}\chi_b({\bf u})\right|\ret
&\le&{\rm Const.}\left\Vert\chi_b({\bf u})R_\Lambda(z){\tilde W}(\chi_\Lambda^\delta)
R(z)x\chi_\Lambda^\delta\chi_b({\bf v})\right\Vert
\left\Vert\chi_b({\bf w})
y\chi_\Lambda^\delta P_{\rm F}\chi_b({\bf u})\right\Vert\ret
&\le&{\rm Const.}\sum_{{\bf u}'}
L^2\left\Vert\chi_b({\bf u})R_\Lambda(z){\tilde\chi}_b({\bf u}')\right\Vert
\left\Vert{\tilde W}(\chi_\Lambda^\delta)\chi_b^\delta({\bf u}')
R(z)\chi_b({\bf v})\right\Vert
\left\Vert\chi_b({\bf w})P_{\rm F}\chi_b({\bf u})\right\Vert,\ret
\end{eqnarray}
where the sum is over ${\bf u}'$ such that 
${\rm supp}\>\chi_b^\delta({\bf u}')\cap{\rm supp}\>|\nabla\chi_\Lambda^\delta|
\ne\emptyset$. 
Because of the existence of the indicator function ${\bf I}(M_\Lambda)$ in 
(\ref{diffindexIM}), the first factor in the summand is estimated as  
\begin{equation}
\left\Vert\chi_b({\bf u})R_\Lambda(z){\tilde\chi}_b({\bf u}')\right\Vert
{\bf I}(M_\Lambda)\le{\rm Const.}L^{\kappa(\xi-2)+4}\exp[-\mu_\infty L^{2\kappa/3}]
\end{equation}
from Lemma~\ref{lemma:ARBdecay}. The second factor can be estimated 
by using Lemma~\ref{lemma:pchiRchibound}. Therefore it is sufficient to 
estimate the following quantity near the Fermi energy: 
\begin{equation}
{\bf E}\int_{y_-}^{y_+}dy\>\left\Vert{\tilde\chi}_b({\bf u}')
R(E_{\rm F}+iy)\chi_b({\bf v})\right\Vert
\left\Vert\chi_b({\bf w})P_{\rm F}\chi_b({\bf u})\right\Vert. 
\end{equation}
Note that 
\begin{equation}
\int_{y_-}^{y_+}dy
\left\Vert{\tilde\chi}_b({\bf u}')R(E_{\rm F}+iy)\chi_b({\bf v})\right\Vert
\le\int_{y_-}^{y_+}dy
\left\Vert{\tilde\chi}_b({\bf u}')R(E_{\rm F}+iy)\chi_b({\bf v})\right\Vert^{s/2} 
|y|^{s/2-1}.
\end{equation}
Substituting this into the above, we obtain
\begin{eqnarray}
& &{\bf E}\int_{y_-}^{y_+}dy\left\Vert{\tilde\chi}_b({\bf u}')
R(E_{\rm F}+iy)\chi_b({\bf v})\right\Vert
\left\Vert\chi_b({\bf w})P_{\rm F}\chi_b({\bf u})\right\Vert\ret
&\le&\mathop{\lim\inf}_{\varepsilon_n\rightarrow 0}
\int_{I_n}dy|y|^{s/2-1}
{\bf E}\left[\left\Vert{\tilde\chi}_b({\bf u}')R(E_{\rm F}+iy)\chi_b({\bf v})\right\Vert^{s/2}
\left\Vert\chi_b({\bf w})P_{\rm F}\chi_b({\bf u})\right\Vert\right]\ret
&\le&\mathop{\lim\inf}_{\varepsilon_n\rightarrow 0}
\int_{I_n}dy|y|^{s/2-1}
{\bf E}\left[\left\Vert{\tilde\chi}_b({\bf u}')R(E_{\rm F}+iy)\chi_b({\bf v})
\right\Vert^s\right]^{1/2}{\bf E}\left[
\left\Vert\chi_b({\bf w})P_{\rm F}\chi_b({\bf u})\right\Vert^2\right]^{1/2}\ret
&\le&{\rm Const.}\exp[-\mu|{\bf u}'-{\bf v}|/2]\exp[-\mu|{\bf w}-{\bf u}|/2],
\end{eqnarray}
where we have written 
$I_n=[y_-,y_+]\backslash(-\varepsilon_n,\varepsilon_n)$, and 
we have used Fatou's lemma, Fubini-Tonelli theorem, Schwarz's inequality, 
the bounds (\ref{DRI}) and (\ref{PFD}). Thus, the corresponding contribution is 
vanishing as $L\uparrow\infty$. 

Consequently, it is enough to consider 
${\rm Tr}\>\chi_\Omega P_{{\rm F},\Lambda}x(\chi_\Lambda^\delta)^2P_{\rm F}
y\chi_\Lambda^\delta P_{\rm F}\chi_\Omega$ 
which comes from the first term in the right-hand side of (\ref{chiRtoGRE}). 
Using the adjoint of the geometric resolvent equation, 
$R(z)\chi_\Lambda^\delta=\chi_\Lambda^\delta R_\Lambda(z)
-R(z){\tilde W}(\chi_\Lambda^\delta)R_\Lambda(z)$, we have 
\begin{equation}
{\rm Tr}\>\chi_\Omega P_{{\rm F},\Lambda}x(\chi_\Lambda^\delta)^2P_{\rm F}
y\chi_\Lambda^\delta P_{\rm F}\chi_\Omega
={\rm Tr}\>\chi_\Omega P_{{\rm F},\Lambda}x(\chi_\Lambda^\delta)^2P_{\rm F}
y(\chi_\Lambda^\delta)^2P_{{\rm F},\Lambda}\chi_\Omega
+{\rm correction}
\end{equation}
in the same way. The correction is vanishing as $L\uparrow\infty$. 
Using the geometric resolvent equation again, the first term in the right-hand side 
is written 
\begin{eqnarray}
& &{\rm Tr}\>\chi_\Omega P_{{\rm F},\Lambda}x(\chi_\Lambda^\delta)^2P_{\rm F}
y(\chi_\Lambda^\delta)^2P_{{\rm F},\Lambda}\chi_\Omega\ret&=&
{\rm Tr}\>\chi_\Omega P_{{\rm F},\Lambda}x\chi_\Lambda^\delta
P_{{\rm F},\Lambda}
y(\chi_\Lambda^\delta)^3P_{{\rm F},\Lambda}\chi_\Omega\ret
&+&\frac{1}{2\pi i}\int_\gamma dz\>
{\rm Tr}\>\chi_\Omega P_{{\rm F},\Lambda}x\chi_\Lambda^\delta
R_\Lambda(z){\tilde W}(\chi_\Lambda^\delta)R(z)
y(\chi_\Lambda^\delta)^2P_{{\rm F},\Lambda}\chi_\Omega.
\end{eqnarray}
The integrand in the second term in the right-hand side is written 
\begin{eqnarray}
& &{\rm Tr}\>\chi_\Omega P_{{\rm F},\Lambda}x\chi_\Lambda^\delta
R_\Lambda(z){\tilde W}(\chi_\Lambda^\delta)R(z)
y(\chi_\Lambda^\delta)^2P_{{\rm F},\Lambda}\chi_\Omega\ret&=&
{\rm Tr}\>\chi_\Omega P_{{\rm F},\Lambda}x\chi_\Lambda^\delta\chi_{\Lambda'}
R_\Lambda(z){\tilde W}(\chi_\Lambda^\delta)R(z)
y(\chi_\Lambda^\delta)^2P_{{\rm F},\Lambda}\chi_\Omega\ret
&+&{\rm Tr}\>\chi_\Omega P_{{\rm F},\Lambda}x\chi_\Lambda^\delta
(1-\chi_{\Lambda'})
R_\Lambda(z){\tilde W}(\chi_\Lambda^\delta)R(z)
y(\chi_\Lambda^\delta)^2P_{{\rm F},\Lambda}\chi_\Omega,
\end{eqnarray}
where $\chi_{\Lambda'}$ is the characteristic function of the region 
$\Lambda'$ which satisfies the conditions of (\ref{distOmegaLambdaLambdap}). 
This right-hand side can be shown to be vanishing as $L\uparrow\infty$ 
in the same way. Consequently, we obtain  
\begin{equation}
{\rm Tr}\>\chi_\Omega P_{{\rm F},\Lambda}x\chi_\Lambda^\delta
P_{{\rm F},\Lambda}
y(\chi_\Lambda^\delta)^3P_{{\rm F},\Lambda}\chi_\Omega
={\rm Tr}\>\chi_\Omega P_{{\rm F},\Lambda}x
P_{{\rm F},\Lambda}yP_{{\rm F},\Lambda}\chi_\Omega+{\rm correction}.
\end{equation}
The first term in the right-hand side is nothing but the desired form.  
\end{proof}

%%%%%%%%%%%%%%%%%%%%%%%%%%%%%%%%%%%%%%%%%%%%%%%%%%%%%%%%%%%%%%%%%%%%%%%%%
\Section{The index formula for the switch functions}
\label{IndexSwitch}

The aim of this appendix is to give a proof of the following theorem: 

\begin{theorem}
\label{theorem:Indexs}
For a fixed period $\ell_P$ of the potentials ${\bf A}^{\rm LP}$ and $V_0^{\rm LP}$ 
in the Hamiltonian $H_\omega^{\rm LP}$ of (\ref{HamLP}) on the whole plane ${\bf R}^2$,  
the following relation is valid almost surely:
\begin{equation}
{\rm Index}(P_{\rm F}U_{\bf a}P_{\rm F})=2\pi i{\rm Tr}\>P_{\rm F}
[[P_{\rm F},\lambda_{1,{\bf a}}],[P_{\rm F},\lambda_{2,{\bf a}}]],
\label{Indexswitch}
\end{equation}
where $\lambda_{j,{\bf a}}$, $j=1,2$, are two switch functions given by 
\begin{equation}
\lambda_{1,{\bf a}}({\bf r}):=
\cases{1, & for $x-a_1\ge 0$;\cr
       0, & for $x-a_1<0$\cr}
\quad
\mbox{and} 
\quad
\lambda_{2,{\bf a}}({\bf r}):=
\cases{1, & for $y-a_2\ge 0$;\cr
       0, & for $y-a_2<0$\cr}
\end{equation}
with the locations ${\bf a}=(a_1,a_2)\in{\bf R}^2$ of the steps. 
\end{theorem}

\noindent
{\bf Remark:} 1. The right-hand side of (\ref{Indexswitch}) is equal to 
the form of another Hall conductance which was discussed in \cite{ASS}.  
Elgart and Schlein \cite{ES} justified this Hall conductance formula 
within the linear response approximation under the assumption that 
the Fermi energy lies in a spectral gap. They also proved that 
the value of (\ref{Indexswitch}) takes the desired integer 
under the same gap assumption. As mentioned above, 
Germinet, Klein and Schenker proved the constancy of (\ref{Indexswitch}) 
in the localization regime, for a random Landau Hamiltonian 
with translation ergodicity, by using a consequence of the multiscale 
analysis . 
\smallskip

\noindent
2. From Theorem~\ref{theorem:standindcond}, we obtain that 
the Hall conductance using the position operator is equal to that using 
the switch functions. 
\medskip

We write the index as   
${\cal I}_s(P_{\rm F};\ell_P)=2\pi i{\rm Tr}\>P_{\rm F}
[[P_{\rm F},\lambda_{1,{\bf a}}],[P_{\rm F},\lambda_{2,{\bf a}}]]$. 
First, we shall show that the index ${\cal I}_s(P_{\rm F};\ell_P)$ is 
well defined for almost every $\omega$. 
Note that 
\begin{eqnarray}
& &{\rm Tr}\>
\left|[P_{\rm F},\lambda_{1,{\bf a}}][P_{\rm F},\lambda_{2,{\bf a}}]\right|\ret
&\le&\sum_{{\bf u},{\bf v},{\bf w}\in({\bf Z}_\varepsilon^2)^\ast}
{\rm Tr}\>\left|\chi_\varepsilon({\bf u})
[P_{\rm F},\lambda_{1,{\bf a}}]\chi_\varepsilon({\bf v})
[P_{\rm F},\lambda_{2,{\bf a}}]\chi_\varepsilon({\bf w})\right|\ret
&\le&\sum_{{\bf u},{\bf v},{\bf w}\in({\bf Z}_\varepsilon^2)^\ast}
|\lambda_{1,{\bf a}}({\bf u})
-\lambda_{1,{\bf a}}({\bf v})|
|\lambda_{2,{\bf a}}({\bf v})
-\lambda_{2,{\bf a}}({\bf w})|
{\rm Tr}\>|\chi_\varepsilon({\bf u})P_{\rm F}\chi_\varepsilon({\bf v})
P_{\rm F}\chi_\varepsilon({\bf w})|,\ret
\label{Trabso2commu}
\end{eqnarray}
where $\chi_\varepsilon({\bf u})$ is the characteristic function of 
the $\varepsilon_1\times\varepsilon_2$ rectangular box $s_\varepsilon({\bf u})$ 
centered at ${\bf u}$, and we have chosen the set $({\bf Z}_\varepsilon^2)^\ast$ 
of the centers ${\bf u}$ of 
the boxes $s_\varepsilon({\bf u})$ so that ${\bf a}$ becomes a vertex 
of a rectangular box, i.e., ${\bf a}\in{\bf Z}_\varepsilon^2$. 
Using Schwarz's inequality, we have  
\begin{eqnarray}
& &{\bf E}\left[{\rm Tr}\>\left|\chi_\varepsilon({\bf u})P_{\rm F}\chi_\varepsilon({\bf v})
P_{\rm F}\chi_\varepsilon({\bf w})\right|\right]\ret
&\le&\sqrt{{\bf E}\left[{\rm Tr}\>\chi_\varepsilon({\bf u})
P_{\rm F}\chi_\varepsilon({\bf v})
P_{\rm F}\chi_\varepsilon({\bf u})\right]}
\sqrt{{\bf E}\left[{\rm Tr}\>\chi_\varepsilon({\bf w})
P_{\rm F}\chi_\varepsilon({\bf v})
P_{\rm F}\chi_\varepsilon({\bf w})\right]}
\end{eqnarray}
and
\begin{eqnarray}
{\rm Tr}\>\chi_\varepsilon({\bf u})
P_{\rm F}\chi_\varepsilon({\bf v})
P_{\rm F}\chi_\varepsilon({\bf u})&\le&
\sqrt{{\rm Tr}\>\chi_\varepsilon({\bf u})P_{\rm F}\chi_\varepsilon({\bf u})
\cdot{\rm Tr}\>\chi_\varepsilon({\bf u})P_{\rm F}\chi_\varepsilon({\bf v})
P_{\rm F}\chi_\varepsilon({\bf v})P_{\rm F}\chi_\varepsilon({\bf u})}\ret
&\le&{\rm Const.}\Vert \chi_\varepsilon({\bf u})P_{\rm F}
\chi_\varepsilon({\bf v})\Vert,
\end{eqnarray}
where we have used the bound (\ref{localTr}). From these bounds, 
we obtain 
\begin{eqnarray}
{\bf E}\left[{\rm Tr}\>|\chi_\varepsilon({\bf u})P_{\rm F}\chi_\varepsilon({\bf v})
P_{\rm F}\chi_\varepsilon({\bf w})|\right] 
&\le&{\rm Const.}\sqrt{{\bf E}\left[\Vert \chi_\varepsilon({\bf u})P_{\rm F}
\chi_\varepsilon({\bf v})\Vert\right]}
\sqrt{{\bf E}\left[\Vert \chi_\varepsilon({\bf w})P_{\rm F}
\chi_\varepsilon({\bf v})\Vert\right]}\ret
&\le&{\rm Const.}e^{-\mu|{\bf u}-{\bf v}|/2}e^{-\mu|{\bf w}-{\bf v}|/2},
\label{Trchi3P2bound}
\end{eqnarray}
where we have used the decay bound (\ref{PFD}) for the Fermi sea projection.  
Note that 
\begin{equation}
|\lambda_{j,{\bf a}}({\bf u})-\lambda_{j,{\bf a}}({\bf v})|
=\cases{0, & for $(u_j-a_j)(v_j-a_j)>0$;\cr
        1, & for $(u_j-a_j)(v_j-a_j)<0$,\cr}
\label{difflambda}
\end{equation}
and 
\begin{equation}
\sqrt{|x_1-y_1|^2+|x_2-y_2|^2}\ge|x_1-y_1|/2+|x_2-y_2|/2.
\label{Edistlowbound}
\end{equation}
Combining these, (\ref{Trabso2commu}) and (\ref{Trchi3P2bound}), we obtain 
\begin{eqnarray}
{\bf E}\left[{\rm Tr}\>
\left|[P_{\rm F},\lambda_{1,{\bf a}}][P_{\rm F},\lambda_{2,{\bf a}}]\right|\right]
&\le&{\rm Const.}\sum_{{\bf u},{\bf v},{\bf w}}
e^{-\mu|u_1-a_1|/4}e^{-\mu|v_1-a_1|/4}e^{-\mu|u_2-v_2|/4}\ret
& &\quad
\times e^{-\mu|w_2-a_2|/4}e^{-\mu|v_2-a_2|/4}e^{-\mu|w_1-v_1|/4}<\infty. 
\end{eqnarray}
Thus the operator $[P_{\rm F},\lambda_{1,{\bf a}}][P_{\rm F},
\lambda_{2,{\bf a}}]$ is trace class for almost every $\omega$. 

Next we show that the index ${\cal I}_s(P_{\rm F};\ell_P)$ is independent of 
the locations $a_1,a_2$ of the steps of the switch functions $\lambda_{j,{\bf a}}$. 
Let ${\bf a}',{\bf a}\in{\bf R}^2$. Then we have  
\begin{eqnarray}
& &{\rm Tr}\>P_{\rm F}[[P_{\rm F},\lambda_{1,{\bf a}'}],
[P_{\rm F},\lambda_{2,{\bf a}'}]]
-{\rm Tr}\>P_{\rm F}[[P_{\rm F},\lambda_{1,{\bf a}}],
[P_{\rm F},\lambda_{2,{\bf a}}]]\ret
&=&{\rm Tr}\>P_{\rm F}[[P_{\rm F},(\lambda_{1,{\bf a}'}-\lambda_{1,{\bf a}})],
[P_{\rm F},\lambda_{2,{\bf a}'}]]
+{\rm Tr}\>P_{\rm F}[[P_{\rm F},\lambda_{1,{\bf a}}],
[P_{\rm F},(\lambda_{2,{\bf a}'}-\lambda_{2,{\bf a}})]].\ret
\label{decomdiffindex}
\end{eqnarray}
We will prove that the first term in the right-hand side is 
vanishing because the second term can be handled in the same way. 
We choose $\varepsilon=(\varepsilon_1,\varepsilon_2)$ so that 
both of ${\bf a}'$ and ${\bf a}$ satisfy 
${\bf a}',{\bf a}\in{\bf Z}_\varepsilon^2({\bf b}):=
{\bf Z}_\varepsilon^2-{\bf b}$ with some ${\bf b}\in{\bf R}^2$. 
We denote by $({\bf Z}_\varepsilon^2({\bf b}))^\ast$ the dual lattice of 
${\bf Z}_\varepsilon^2({\bf b})$.

\begin{lemma}
\label{lemma:PF1normdecaybound}
For ${\bf u},{\bf v}\in({\bf Z}_\varepsilon^2({\bf b}))^\ast$, 
the following bound is valid: 
\begin{equation}
{\bf E}\left[{\rm Tr}\>\left|\chi_\varepsilon({\bf u})P_{\rm F}
\chi_\varepsilon({\bf v})\right|
\right]\le{\rm Const.}e^{-\mu'|{\bf u}-{\bf v}|}
\end{equation}
with some positive constant $\mu'$. 
\end{lemma}

\begin{proof}{Proof}
Note that  
\begin{eqnarray}
{\bf E}\left[{\rm Tr}\>\left|\chi_\varepsilon({\bf u})P_{\rm F}
\chi_\varepsilon({\bf v})\right|
\right]&\le&\sum_{{\bf w}\in({\bf Z}_\varepsilon^2({\bf b}))^\ast}
{\bf E}\left[{\rm Tr}\>\left|\chi_\varepsilon({\bf u})P_{\rm F}
\chi_\varepsilon({\bf w})\chi_\varepsilon({\bf w})P_{\rm F}
\chi_\varepsilon({\bf v})\right|\right]\ret
&\le&\sum_{{\bf w}\in({\bf Z}_\varepsilon^2({\bf b}))^\ast}
\sqrt{{\bf E}\left[{\rm Tr}\>
\chi_\varepsilon({\bf u})P_{\rm F}\chi_\varepsilon({\bf w})P_{\rm F}
\chi_\varepsilon({\bf u})\right]}\ret
& &\quad\times\sqrt{{\bf E}\left[{\rm Tr}\>
\chi_\varepsilon({\bf v})P_{\rm F}\chi_\varepsilon({\bf w})P_{\rm F}
\chi_\varepsilon({\bf v})\right]}.
\label{ETrabsoPbound}
\end{eqnarray} 
Further, we have 
\begin{eqnarray}
{\rm Tr}\>
\chi_\varepsilon({\bf u})P_{\rm F}\chi_\varepsilon({\bf w})P_{\rm F}
\chi_\varepsilon({\bf u})
&\le&\sqrt{{\rm Tr}\>\chi_\varepsilon({\bf u})
P_{\rm F}\chi_\varepsilon({\bf u})}
\sqrt{{\rm Tr}\>\chi_\varepsilon({\bf u})P_{\rm F}\chi_\varepsilon({\bf w})
P_{\rm F}\chi_\varepsilon({\bf w})
P_{\rm F}\chi_\varepsilon({\bf u})}\ret
&\le&{\rm Const.}\Vert\chi_\varepsilon({\bf u})P_{\rm F}\chi_\varepsilon({\bf w})
\Vert,
\end{eqnarray}
where we have used the bound (\ref{localTr}). Combining this, 
the decay bound (\ref{PFD}) for the Fermi sea projection, 
(\ref{ETrabsoPbound}), we obtain 
\begin{equation}
{\bf E}\left[{\rm Tr}\>\left|\chi_\varepsilon({\bf u})P_{\rm F}
\chi_\varepsilon({\bf v})\right|
\right]\le{\rm Const.}\sum_{{\bf w}\in({\bf Z}_\varepsilon^2({\bf b}))^\ast}
e^{-\mu|{\bf u}-{\bf w}|/2}
e^{-\mu|{\bf w}-{\bf v}|/2}
\le{\rm Const.}e^{-\mu'|{\bf u}-{\bf v}|}.
\end{equation}
\end{proof}

Now let us consider the first term in the right-hand side of 
(\ref{decomdiffindex}). We write $\Delta\lambda$ for 
$\lambda_{1,{\bf a}'}-\lambda_{1,{\bf a}}$ for short. 

\begin{lemma}
\label{lemma:ETrabDlamPf}
We have 
\begin{equation}
{\bf E}\left[{\rm Tr}\>\left|\Delta\lambda[P_{\rm F},\lambda_{2,{\bf a}'}]
\right|\right]<\infty
\quad\mbox{and}\quad
{\bf E}\left[{\rm Tr}\>\left|\Delta\lambda P_{\rm F}
[P_{\rm F},\lambda_{2,{\bf a}'}]
\right|\right]<\infty.
\end{equation}
\end{lemma}

\begin{proof}{Proof}
Without loss of generality, we can assume $a_1'>a_1$. Then we obtain 
\begin{eqnarray}
{\bf E}\left[{\rm Tr}\>\left|\Delta\lambda[P_{\rm F},\lambda_{2,{\bf a}'}]
\right|\right]&\le&\sum_{{\bf u},{\bf v}\in({\bf Z}_\varepsilon^2({\bf b}))^\ast}
{\bf E}\left[{\rm Tr}\>\left|\Delta\lambda\chi_\varepsilon({\bf u})
[P_{\rm F},\lambda_{2,{\bf a}'}]\chi_\varepsilon({\bf v})
\right|\right]\ret
&\le&\sum_{{\bf u},{\bf v}\in({\bf Z}_\varepsilon^2({\bf b}))^\ast}
|\lambda_{1,{\bf a}'}({\bf u})
-\lambda_{1,{\bf a}}({\bf u})||\lambda_{2,{\bf a}'}({\bf v})
-\lambda_{2,{\bf a}'}({\bf u})|\ret
& &\quad\times{\bf E}\left[{\rm Tr}\>|\chi_\varepsilon({\bf u})P_{\rm F}
\chi_\varepsilon({\bf v})|\right]\ret
&\le&\sum_{a_1<u_1<a_1',u_2}\sum_{v_1,v_2}e^{-\mu'|u_1-v_1|/2}
e^{-\mu'|u_2-a_2'|/2}e^{-\mu'|v_2-a_2'|/2}<\infty,\ret
\end{eqnarray}
where we have used (\ref{difflambda}), (\ref{Edistlowbound}) 
and Lemma~\ref{lemma:PF1normdecaybound}. 

Similarly, we have 
\begin{eqnarray}
{\bf E}\left[{\rm Tr}\>\left|\Delta\lambda P_{\rm F}[P_{\rm F},\lambda_{2,{\bf a}'}]
\right|\right]&\le&\sum_{{\bf u},{\bf v},{\bf w}}
{\bf E}\left[{\rm Tr}\>\left|\Delta\lambda\chi_\varepsilon({\bf u})
P_{\rm F}\chi_\varepsilon({\bf v})
[P_{\rm F},\lambda_{2,{\bf a}'}]\chi_\varepsilon({\bf w})
\right|\right]\ret
&\le&\sum_{{\bf u},{\bf v},{\bf w}}|\lambda_{1,{\bf a}'}({\bf u})
-\lambda_{1,{\bf a}}({\bf u})||\lambda_{2,{\bf a}'}({\bf w})
-\lambda_{2,{\bf a}'}({\bf v})|\ret
& &\quad\times{\bf E}\left[{\rm Tr}\>|\chi_\varepsilon({\bf u})P_{\rm F}
\chi_\varepsilon({\bf v})P_{\rm F}\chi_\varepsilon({\bf w})|\right]\ret
&\le&{\rm Const.}\sum_{{\bf u},{\bf v},{\bf w}}|\lambda_{1,{\bf a}'}({\bf u})
-\lambda_{1,{\bf a}}({\bf u})||\lambda_{2,{\bf a}'}({\bf w})
-\lambda_{2,{\bf a}'}({\bf v})|\ret
& &\quad\times e^{-\mu|{\bf u}-{\bf v}|/2}e^{-\mu|{\bf w}-{\bf v}|/2}\ret
&\le&{\rm Const.}\sum_{a_1<u_1<a_1',v_1,w_1}e^{-\mu|u_1-v_1|/4}e^{-\mu|v_1-w_1|/4}\ret
& &\times\sum_{u_2,v_2,w_2}e^{-\mu|u_2-v_2|/4}e^{-\mu|v_2-a_2'|/4}
e^{-\mu|w_2-a_2'|/4}<\infty,
\end{eqnarray}
where we have used the bound (\ref{Trchi3P2bound}).   
\end{proof}

Relying on this Lemma~\ref{lemma:ETrabDlamPf}, we have 
\begin{equation}
{\rm Tr}\>P_{\rm F}[[P_{\rm F},\Delta\lambda],[P_{\rm F},\lambda_{2,{\bf a}'}]]
={\rm Tr}\>P_{\rm F}\Delta\lambda(1-P_{\rm F})[P_{\rm F},\lambda_{2,{\bf a}'}]
+{\rm Tr}\>[P_{\rm F},\lambda_{2,{\bf a}'}](1-P_{\rm F})\Delta\lambda P_{\rm F}.
\end{equation}
Further, the first term in the right-hand side is written 
\begin{equation}
{\rm Tr}\>P_{\rm F}\Delta\lambda(1-P_{\rm F})[P_{\rm F},\lambda_{2,{\bf a}'}]
={\rm Tr}\>\Delta\lambda(1-P_{\rm F})[P_{\rm F},\lambda_{2,{\bf a}'}]P_{\rm F}
={\rm Tr}\>\Delta\lambda(1-P_{\rm F})[P_{\rm F},\lambda_{2,{\bf a}'}],
\end{equation}
where we have used 
$(1-P_{\rm F})[P_{\rm F},\lambda_{2,{\bf a}'}](1-P_{\rm F})=0$. 
The second term becomes 
\begin{eqnarray}
{\rm Tr}\>[P_{\rm F},\lambda_{2,{\bf a}'}](1-P_{\rm F})\Delta\lambda P_{\rm F}
&=&{\rm Tr}\>[P_{\rm F},\lambda_{2,{\bf a}'}](1-P_{\rm F})
\chi_{{\rm supp}\Delta\lambda}\Delta\lambda P_{\rm F}\ret
&=&{\rm Tr}\>\Delta\lambda P_{\rm F}[P_{\rm F},\lambda_{2,{\bf a}'}](1-P_{\rm F})
\chi_{{\rm supp}\Delta\lambda}\ret
&=&{\rm Tr}\>\Delta\lambda P_{\rm F}[P_{\rm F},\lambda_{2,{\bf a}'}](1-P_{\rm F})\ret
&=&{\rm Tr}\>\Delta\lambda P_{\rm F}[P_{\rm F},\lambda_{2,{\bf a}'}],
\end{eqnarray}
where $\chi_{{\rm supp}\Delta\lambda}$ is the characteristic function of 
the support of $\Delta\lambda$, and we have used 
$P_{\rm F}[P_{\rm F},\lambda_{2,{\bf a}'}]P_{\rm F}=0$.
As a result, we obtain 
\begin{equation}
{\rm Tr}\>P_{\rm F}[[P_{\rm F},\Delta\lambda],[P_{\rm F},\lambda_{2,{\bf a}'}]]
={\rm Tr}\>\Delta\lambda [P_{\rm F},\lambda_{2,{\bf a}'}].
\end{equation}
This right-hand side is decomposed into two parts as 
\begin{equation}
{\rm Tr}\>\Delta\lambda [P_{\rm F},\lambda_{2,{\bf a}'}]
={\rm Tr}\>\Delta\lambda \chi_\ell[P_{\rm F},\lambda_{2,{\bf a}'}]+
{\rm Tr}\>\Delta\lambda(1-\chi_\ell)[P_{\rm F},\lambda_{2,{\bf a}'}]
\end{equation}
with the characteristic function $\chi_\ell$ of the square box centered 
at ${\bf r}=0$ with a sufficiently large sidelength $\ell$. 
Since we have 
\begin{equation}
{\bf E}\left[{\rm Tr}\>|\chi_\ell P_{\rm F}|\right]
\le\sum_{\bf v}{\bf E}\left[{\rm Tr}\>|\chi_\ell P_{\rm F}
\chi_\varepsilon({\bf v})|\right]<\infty
\end{equation}
from Lemma~\ref{lemma:PF1normdecaybound}, the first term in the right-hand 
side is vanishing by cyclicity of the trace. The second term can be 
evaluated in the same way as in the proof of Lemma~\ref{lemma:ETrabDlamPf}.
In consequence, it vanishes as $\ell\uparrow\infty$. Thus 
we obtain 
${\rm Tr}\>P_{\rm F}[[P_{\rm F},\Delta\lambda],[P_{\rm F},\lambda_{2,{\bf a}'}]]=0$. 
Since the second term in the right-hand side of (\ref{decomdiffindex}) can 
be handled in the same way, the index ${\cal I}_s(P_{\rm F};\ell_P)$ is independent 
of the locations $a_1,a_2$ of the steps of the switch functions. 

Using this property, the index is written 
\begin{equation}
{\cal I}_s(P_{\rm F};\ell_P)=\frac{2\pi i}{{\cal V}_\ell}\sum_{{\bf a}\in\Lambda_\ell}
\sum_{{\bf u},{\bf v},{\bf w}\in({\bf Z}_\varepsilon^2)^\ast}
\left[D_{12}({\bf v},{\bf w},{\bf u};{\bf a})
-D_{21}({\bf v},{\bf w},{\bf u};{\bf a})\right]
S({\bf u},{\bf v},{\bf w},{\bf u})
\end{equation}
with
\begin{equation}
D_{12}({\bf v},{\bf w},{\bf u};{\bf a}):=
[\lambda_{1,{\bf a}}({\bf v})-\lambda_{1,{\bf a}}({\bf w})]
[\lambda_{2,{\bf a}}({\bf w})-\lambda_{2,{\bf a}}({\bf u})]
\end{equation}
and
\begin{equation}
D_{21}({\bf v},{\bf w},{\bf u};{\bf a}):=
[\lambda_{2,{\bf a}}({\bf v})-\lambda_{2,{\bf a}}({\bf w})]
[\lambda_{1,{\bf a}}({\bf w})-\lambda_{1,{\bf a}}({\bf u})],
\end{equation}
where both $\Lambda_\ell$ and ${\cal V}_\ell$ are the same as in (\ref{Index1}), 
and $S({\bf u},{\bf v},{\bf w},{\bf u})$ is given by (\ref{defS}).
We also write 
\begin{equation}
{\cal I}_s^\varepsilon(P_{\rm F};\Omega,\ell_P)
=\frac{2\pi i}{{\cal V}_\ell}\sum_{{\bf u}\in\Lambda_\ell^\ast}
\sum_{{\bf v},{\bf w}\in({\bf Z}_\varepsilon^2)^\ast}
\sum_{{\bf a}\in{\bf Z}_\varepsilon^2}
\left[D_{12}({\bf v},{\bf w},{\bf u};{\bf a})
-D_{21}({\bf v},{\bf w},{\bf u};{\bf a})\right]
S({\bf u},{\bf v},{\bf w},{\bf u}),
\end{equation}
where $\Lambda_\ell^\ast$ is given by (\ref{Lambdaellast}). 

\begin{lemma}
\label{lemma:diffIsIseps}
The following holds: ${\bf E}\left[\left|{\cal I}_s(P_{\rm F};\ell_P)
-{\cal I}_s^\varepsilon(P_{\rm F};\Omega,\ell_P)\right|\right]
\rightarrow 0$ as $|\Omega|\uparrow\infty$. 
\end{lemma}

\begin{proof}{Proof}
To begin with, we note that 
\begin{eqnarray}
\left|{\bf E}\left[{\rm Tr}\>
\chi_\varepsilon({\bf u})P_{\rm F}\chi_\varepsilon({\bf v})
P_{\rm F}\chi_\varepsilon({\bf w})P_{\rm F}\chi_\varepsilon({\bf u})
\right]\right|&\le&
{\bf E}\left[{\rm Tr}\>\left|\chi_\varepsilon({\bf v})
P_{\rm F}\chi_\varepsilon({\bf w})P_{\rm F}\chi_\varepsilon({\bf u})\right|
\right]\ret
&\le&{\rm Const.}e^{-\mu|{\bf v}-{\bf w}|/2}
e^{-\mu|{\bf w}-{\bf u}|/2}
\end{eqnarray}
which is derived from (\ref{Trchi3P2bound}). Using this, (\ref{difflambda}) 
and (\ref{Edistlowbound}), we have 
\begin{eqnarray}
& &{\bf E}\left[\left|{\cal I}_s(P_{\rm F};\ell_P)
-{\cal I}_s^\varepsilon(P_{\rm F};\Omega,\ell_P)\right|\right]\ret
&\le&
\frac{{\rm Const.}}{\ell^2}\sum_{{\bf a}\in{\bf Z}_\varepsilon^2\backslash
\Lambda_\ell}\sum_{{\bf u}\in\Lambda_\ell^\ast}
+\sum_{{\bf a}\in\Lambda_\ell}\sum_{{\bf u}\in({\bf Z}_\varepsilon^2)^\ast
\backslash\Lambda_\ell^\ast}
\sum_{v_1,w_1}e^{-\mu|v_1-a_1|/4}e^{-\mu|w_1-a_1|/4}e^{-\mu|w_1-u_1|/4}\ret
& &\qquad\qquad\qquad\times
\sum_{v_2,w_2}e^{-\mu|v_2-w_2|/4}e^{-\mu|w_2-a_2|/4}e^{-\mu|u_2-a_2|/4}\ret
&\le&\frac{{\rm Const.}}{\ell^2}\sum_{{\bf a}\in{\bf Z}_\varepsilon^2\backslash
\Lambda_\ell}\sum_{{\bf u}\in\Lambda_\ell^\ast}
+\sum_{{\bf a}\in\Lambda_\ell}\sum_{{\bf u}\in({\bf Z}_\varepsilon^2)^\ast
\backslash\Lambda_\ell^\ast}
e^{-\mu'|u_1-a_1|}e^{-\mu|u_2-a_2|/4}
\end{eqnarray}
with a positive constant $\mu'$. 
This right-hand side is easily shown to vanish as $\ell\uparrow\infty$.  
\end{proof}

Using the identity,  
$\sum_{{\bf a}\in{\bf Z}_\varepsilon^2}
\left[D_{12}({\bf v},{\bf w},{\bf u};{\bf a})
-D_{21}({\bf v},{\bf w},{\bf u};{\bf a})\right]
=-({\bf v}-{\bf w})\times({\bf w}-{\bf u})$, one has 
\begin{eqnarray}
{\cal I}_s^\varepsilon(P_{\rm F};\Omega,\ell_P)
&=&-\frac{2\pi i}{{\cal V}_\ell}\sum_{{\bf u}\in\Lambda_\ell^\ast}
\sum_{{\bf v},{\bf w}}({\bf v}-{\bf w})\times({\bf w}-{\bf u}){\rm Tr}\>
\chi_\varepsilon({\bf u})P_{\rm F}\chi_\varepsilon({\bf v})P_{\rm F}
\chi_\varepsilon({\bf w})P_{\rm F}\chi_\varepsilon({\bf u})\ret
&=&{\cal I}^\varepsilon(P_{\rm F};\Omega,\ell_P),
\end{eqnarray}
where we have used the expression (\ref{Index3}) of 
${\cal I}^\varepsilon(P_{\rm F};\Omega,\ell_P)$ and (\ref{idexpro}).  
Combining this, (\ref{EIndexIvarepsilonapprox}) and Lemma~\ref{lemma:diffIsIseps}, 
we obtain Theorem~\ref{theorem:Indexs}.

%%%%%%%%%%%%%%%%%%%%%%%%%%%%%%%%%%%%%%%%%%%%%%%%%%%%%%%%%%%%%%%%%%%%%%%%%%%
\bigskip

\noindent
{\bf Acknowledgements:} I would like to thank Shu Nakamura, 
Hermann Schulz-Baldes and Hal Tasaki 
for helpful discussions. 
%%%%%%%%%%%%%%%%%%%%%%%%%%%%%%%%%%%%%%%%%%%%%%%%%%%%%%%%%%%%%
%\newpage

\end{document}